\documentclass[12pt,a4paper]{report}

\usepackage{geometry}
\usepackage[colorlinks=true,urlcolor=blue,citecolor=blue,linkcolor=red]{hyperref}
\usepackage{quoting}
\quotingsetup{font=small}
\usepackage[utf8]{inputenc} 
\usepackage[english]{babel}
\usepackage{graphicx}
\usepackage{epstopdf}
\usepackage{hyphenat}
\usepackage{amsmath}
\usepackage{wrapfig}
\usepackage{setspace}
\usepackage{hyperref}
\usepackage{textcomp}
\usepackage{subfigure}
\usepackage{rotating}
\usepackage{wrapfig}
\usepackage{braket}
\usepackage{amssymb}
\usepackage{amsmath}
\usepackage{xcolor}
\usepackage{bbold}
\usepackage{lmodern}

\allowdisplaybreaks

\begin{document}
	
	\begin{titlepage}
\label{frontespizio}

\vspace*{2.0cm}

\begin{flushleft}
\Large{Tommaso Favalli\\}
\vspace*{0.5cm}
\vspace*{2.0cm}
\Huge{\textsc{On the Emergence of Time and Space~in~Closed~Quantum~Systems}\\}
\vspace*{1.0cm}
\Large{Doctoral Thesis accepted by\\University of Naples \lq\lq Federico II\rq\rq, Naples, Italy}
\end{flushleft}

\end{titlepage}
	\relax 
\thispagestyle{empty}
\begin{quote}
\label{bianca}

\end{quote}
	
	\pagenumbering{roman}
	
	\relax 
\thispagestyle{empty}
\begin{quote}
	\label{dedica}
	
	\begin{flushright}
		\null\vspace{\stretch{1}}
		
		{\textit{To my wife, Chiara}}\\[0.5em]
		{\textit{To my sons, Gabriele and Matteo,}}\\
		{\textit{and to the one we could not meet}}
		
		\vspace{\stretch{2}}\null
	\end{flushright}
\end{quote}

	\relax 
\thispagestyle{empty}
\begin{quote}
\label{bianca}

\end{quote}

	\chapter*{Abstract}
\label{abstract}

In this work we revisit, generalize and extend the Page and Wootters (PaW) theory that was originally introduced in order to describe the emergence of time through entanglement between subsystems in a globally static Universe. In the theory, the total Hilbert space is divided in two entangled subsystems and one of them is equipped with an appropriate clock observable: through a constraint on total energy, the temporal dynamics (namely the Schrödinger equation) is recovered for the rest of the Universe with respect to the eigenvalues of the clock observable. 

We start providing a complete review of the salient aspects of the theory, summarizing the original works of Page and Wootters, looking at the criticisms that were initially directed toward the theory and showing how they have been overcome. Then we focus on the choice of the clock observable. For clock Hamiltonians with equally-spaced energy spectrum we introduce a time observable described by an Hermitian operator, complement of the Hamiltonian. Furthermore, we generalize the mechanism showing that it is possible to consider bounded, discrete Hamiltonians with generic spectrum as clock Hamiltonians. In this case time is described by a POVM (complement of the clock Hamiltonian) and we demonstrate that the POVM states provide a consistent dynamical evolution of the rest of the Universe even if they are not orthogonal. A continuous flow of time for the rest of the Universe is recovered in this framework still maintaining a discrete energy spectrum for the clock Hamiltonian and the case where an interaction is present between the clock and the rest of the Universe is treated and discussed.

Being able to use any generic Hamiltonian as a clock Hamiltonian allow us to merge the principle of Canonical Typicality with PaW quantum time. Considering indeed a quantum Universe composed by a small subsystem and a large environment, it has been demonstrated that, for the vast majority of randomly chosen wave-functions of the Universe satisfying a total energy constraint, the reduced density matrix of the small subsystem is given by the canonical statistical distribution. We show that time and non-equilibrium dynamics can emerge as a consequence of the entanglement between the small subsystem and the environment present in the (randomly chosen) global wave-function of the Universe. The clock is provided by the environment, which ticks the temporal evolution of the small subsystem. The~paradox~of the peaceful coexistence of statistical equilibrium and non-equilibrium dynamics is solved by identifying the trace over the environment degrees of freedom with the temporal trace over the entire history of the small subsystem.

The PaW mechanism can be interpreted as an internalization of the temporal reference frame. A natural evolution is therefore to extend the approach to the spatial degree of freedom. Considering a global quantum Universe satisfying a double constraint, both on total energy and total momentum, we generalize the PaW quantum clock formalism providing a model of 3+1 dimensional, quantum spacetime emerging from entanglement among different subsystems in a globally ``timeless" and  ``positionless" Universe. With our fully relational approach, space and time are treated on equal footing within quantum theory and the dynamics of quantum systems is derived with respect to a 3+1 dimensional quantum reference frame. The case of the dynamics of relativistic particles is also considered.

As a last point we study the evolution of quantum clocks interacting with a Newtonian gravitational field produced by a spherical mass. The time states of our quantum clocks are chosen again to belongs to the complement of an Hamiltonian. In the framework of PaW approach, we derive a time dilation effect for the time states of the clocks. The delay is in agreement up to first order with the gravitational time dilation obtained from the Schwarzschild metric. The result can be extended by considering the relativistic gravitational potential: in this case we obtain the agreement with the exact Schwarzschild solution.

	\onehalfspacing
	
	\tableofcontents
	\newpage

\thispagestyle{empty}
\begin{quote}
\label{esergo}

\leftskip=1.1in
My temper was sometimes violent, and my passions vehement; but by some law in my temperature they were turned not towards childish pursuits but to an eager desire to learn, and not to learn all things indiscriminately. I confess that neither the structure of languages, nor the code of governments, nor the politics of various states possessed attractions for me. It was the secrets of heaven and earth that I desired to learn; and whether it was the outward substance of things or the inner spirit of nature and the mysterious soul of man that occupied me, still my inquiries were directed to the metaphysical, or in it highest sense, the physical secrets of the world.\

\begin{flushright}
from Mary Shelley, \textit{Frankenstein}
\end{flushright}
\vspace*{1.5cm}


Remember to look up at the stars and not down at your feet. Try to make sense of what you see and wonder~about what makes the universe exist. Be curious. And however difficult life may seem, there is always something you can do and succeed at. It matters that you don’t just give~up.\

\begin{flushright}
Stephen W. Hawking
\end{flushright}

\end{quote}

	\relax 
\thispagestyle{empty}
\begin{quote}
\label{bianca}

\end{quote}
	\setcounter{page}{1}
	
	\pagenumbering{arabic}
	
	\chapter{Introduction}
\label{intro}

Time, space and entanglement are the main characters in this work. Their nature is still a great mystery in physics and here we study the possibility that these three phenomena are strikly connected. We will show indeed that entanglement can be at the basis of the emergence of time and space within closed quantum systems. 

Entanglement is a phenomenon that has a relatively recent history within physics, because it was formulated soon after the advent of quantum mechanics and then it was increasingly studied. Despite this, the nature of quantum correlations remains a deep mystery and their implications in physics are still matter of research. Basically, in 1935, Einstein, Podolsky and Rosen designed a thought experiment (called then the \textit{EPR Paradox}) to demonstrate the incompleteness of quantum mechanics \cite{EPR}. This paradox was solved about thirty years later, when J. S. Bell showed that at least one of the quantum mechanics assumptions of locality\footnote{We consider \lq\lq local\rq\rq a theory where the outcomes of an experiment on a system are independent of the actions performed on a different system which has no causal connection with the first.} and counterfactual definiteness\footnote{Let us define \lq\lq counterfactual-definite\rq\rq a theory whose experiments uncover properties that are pre-existing. In other words, in a counterfactual-definite theory it is meaningful to assign a property to a system independently of whether the measurement of such property is carried out.} (or reality) must be false \cite{bell}. Bell introduced inequalities (that now carry his name) that satisfy the assumptions of locality and counterfactual definiteness and then showed that for certain quantum states they are violated, indicating that quantum mechanics cannot be both local and counterfactual-definite (or equivalently that quantum mechanics is either non-local or non counterfactual-definite). After these events, experimental violation of Bell's inequalities was confirmed many times by many quantum systems. The quantum states that violate Bell's inequalities are some (not all) among the entangled states.
 
Let's quickly review what is meant by entangled states. Considering a quantum system in a Hilbert space that is product of more than one subspaces, the structure of quantum mechanics allows for states of $\mathcal{H}$ that cannot be written in the separable form. That is, consider $\mathcal{H}_{12} = \mathcal{H}_1 \otimes \mathcal{H}_2$, then there are quantum states $\ket{\psi}_{12}$ that can not be written as a product of the states $\ket{\phi}_1$ and $\ket{\chi}_2$, i. e.
\begin{equation}
\ket{\psi}_{12} \ne \ket{\phi}_1 \otimes \ket{\chi}_2
\end{equation}
where $\ket{\phi}_1$ and $\ket{\chi}_2$ are the states spanned on the spaces $\mathcal{H}_1$ and $\mathcal{H}_2$ respectively, and $\ket{\psi}_{12}$ is the global state in the space $\mathcal{H}_{12}$. This kind of states are called entangled\footnote{We have here described the case of a pure state, but the same argument applies to mixed states. In this latter case we should simply use the formalism of density matrix instead of state vector in describing the state of the system.}. Between two entangled particles there is a deep quantum connection that goes beyond time and space and we will see, throughout this work, how entanglement may even be responsible for the emergence of a quantum spacetime.

Now we start focusing on the concept of time: it is a not recent mystery, but since the beginning of the human history, because it is one of the most obvious things we experience every day, but it is still difficult physically to establish what it is. Moreover, as we will see, it is time that was initially connected to entanglement.

\section{What \lq\lq time" is?}
\begin{quoting}
	Quid est ergo tempus? Si nemo ex me quaerat, scio; si quaerenti explicare velim, nescio\footnote{From Aurelius Augustinus Hipponensis, \textit{Confessiones}, XI, 14.}. 
\end{quoting}As Augustinus Hipponensis many centuries ago, physicists are still in the situation in which the notion of time is intuitively clear but several problems arise when trying to give an exact definition. The concept of time takes indeed different meanings in the various branches of physics and many conceptions have been attributed to time during science history, while new and diversified theories arose.

In classical physics time is considered, together with space, as a fixed background, a container of events. That is, time and space constitute a sort of theatre stage where the events of the Universe could perform as actors. With the advent of the Special Relativity and then with the General Relativity, the concept of time has changed dramatically, becoming inextricably linked to space. After Einstein, in relativistic physical systems, we can no longer speak separately of space and time but of events; time is no longer absolute but each observer measures his own.
In the years in which the theories of relativity were born another theory arose and developed: this is Quantum Mechanics. From the beginning in this new theory the concept of time has given some problems. Every observable in the quantum theory is represented by an Hermitian operator acting on a specific Hilbert space, but this was not easily applicable to the concept of time. The fact that the generator of temporal translations is the Hamiltonian of the system implies that a possible time operator has to be conjugated to Hamiltonian. From this fact it follows that the time operator and the Hamiltonian should have the same spectral structure, because unitarily equivalent. This is clearly not always true, given that within quantum theory Hamiltonians are often bounded and have a discrete spectrum, while time seems to flow continuously. This fact led W. Pauli to formulate his objection in which he stated that time could not be described by an Hermitian operator \cite{pauli} (see also \cite{peres,per,jan1}). Therefore time, inside quantum theory, remained a classical parameter, an external factor that inexorably flows while each quantum state evolves with respect to it following the well-known Schrödinger equation \cite{diracbook,shankar}:
\begin{equation}
i\hslash \frac{\partial}{\partial t} \ket{\psi (t)} = \hat{H} \ket{\psi (t)} ,
\end{equation}
where $\hat{H}$ and $\ket{\psi (t)}$ are the Hamiltonian and the state of the system respectively, and $t$ is the classical parameter which indeed marks the the passage of time. 

In this sense, there is a problem in interpreting the well-known uncertainty relation between time and energy $\Delta t \Delta E \ge \frac{\hslash}{2}$ because, although it is accepted and confirmed when considering certain phenomena, remains a controversial issue since the advent of quantum theory with respect to its appropriate formalization and possible meanings \cite{jan2}. P. Bush, in \cite{bush}, distinguishes three aspects of time in quantum theory, which consequently lead to three possible interpretations of the uncertainty relation. He asserts that the problem about the uncertainty relation between time and energy is linked to the ambiguity about the role of time in quantum theory. The first aspect he emphasizes is about the issue we have already stated, namely time as a classical parameter that enters in the Schrödinger equation and measured by an external clock. He calls this aspect of time as \textit{external time}. By contrast, he calls \textit{dynamical time} the time when defined and measured in terms of physical systems undergoing changes. Indeed it can be seen that each dynamical variable in physics marks the passage of time as well as giving a (at least approximate) quantitative measure of the length of time interval between two events. As a last aspect, Bush considers whether and when time can be considered as an observable within quantum theory. In this context, the problem returns to what we have already observed, i. e. whether it is possible to define a time operator that is conjugated to Hamiltonian. On this point, we will come back several times during our work.

For our purposes it is useful now to introduce a problem known as \lq\lq the problem of time\rq\rq. In the context of canonical quantization of gravity J. A. Wheeler and B. DeWitt encountered what was later called the Wheeler-DeWitt equation (see for example \cite{dewitt,isham}), that is
\begin{equation}
\hat{H} \ket{\Psi} = 0
\end{equation}
where $\hat{H}$ is the global Hamiltonian of the Universe\footnote{We call \lq\lq Universe\rq\rq any closed quantum system. All our considerations apply indeed to any closed quantum system and consequently also to our Universe, which is clearly the largest closed system we know about. We will return to this point later.} and $\ket{\Psi}$ is its state. According to this equation the Universe should be in an eigenfunction of the total Hamiltonian, so it would not show any dynamical behavior. It is interesting to note that a time shift of the state of the Universe would be unobservable from physical considerations. If indeed we shift the whole state of the Universe there is nothing external that can keep track of this shift, and in this way we could not observe such shift. The same considerations can be made for the spatial degrees of freedom. A global shift in the position of the Universe would be in the same way unobservable, and it could  be in an eigenstate of its momentum operator. These observations lead us to say that the Wheeler-DeWitt equation is in some way compatible with our observations, but then the question arises: why do we experience an evolving Universe?

\section{Inside the Story of a Timeless Universe}
\begin{quoting}
	$\left[...\right]$ if we seek to examine Time, we find ourselves examining Reality\footnote{From Friedel Weinert, \textit{The March of Time: Evolving Conceptions of Time in the Light of Scientific Discoveries}, edited by Springer (2013).}. 
\end{quoting}In 1983 D. N. Page and W. K. Wootters proposed a theory to solve the \lq\lq problem of time" \cite{pagewootters,wootters} (see also \cite{pagearrow}), modeling a protocol in which time emerges from quantum correlations, and which will be the starting point of our work. The Page and Wootter (PaW) theory essentially consists in dividing the total Hilbert space of the Universe into two subsystems and assigning one of it to time. The flow of time then consists simply in the correlation between the quantum degree of freedom of time and the rest, a correlation present in a time-independent global state. In this framework we do not consider time as an abstract, external coordinate, but as \lq\lq what is shown on a quantum clock”, where the clock is some physical system that is taken as time reference. Time and dynamics then emerge in a subsystem of the Universe that is entangled with the clock subspace, equipped with an appropriate observable that we will call \textit{clock observable}. We will see that, considering a constrained Universe consisting of two non-interacting subsystems and choosing an appropriate clock observable, the global state of the Universe will be in a superposition of different clock eigenstates. As a result, the Schrödinger equation can be written for the relative state (in Everett sense \cite{everett,libroMWI,libroMWI2}) of the rest of the Universe with respect to the clock values. In this sense both time and dynamics emerge through the entanglement among the different subsystems in a globally \lq\lq timeless\rq\rq Universe.

This elegant model had promising features considering its operational nature (which encourages experiments with quantum clocks), but had never been developed beyond the toy model stage. In 2014, for the first time, an experimental illustration of the PaW mechanism was made \cite{esp1} (see also \cite{esp2}): after this experiment, many theoretical works were published and a new research has started. The fundamental criticisms addressed to the mechanism have been overcome and the PaW interpretation of time has come to life again. Among the first works released, one of the most important contribution is certainly the work of V. Giovannetti, S. Lloyd and L. Maccone (GLM) \cite{lloydmaccone} in which the authors proposed a revision and an extension of the PaW mechanism, choosing a clock space isomorphic to the Hilber space of a particle on a line. With this choice they are able to obtain a clock observable that is conjugated to the clock Hamiltonian (in the approximation of considering the clock energy equal to its momentum), and they can give a description of the PaW mechanism that overcomes some criticisms. In particular they show how the model allows to reproduce the correct statistics of sequential measurements performed on a system at different times, thus overcoming the most crucial objection that had been made by Kuchar in \cite{kuchar} (a previous solution to the problem was addressed by R. Gambini, R. A. Porto, J. Pullin, and S. Torterolo in \cite{gppt}: we will discuss this issue in detail in Chapter 2). At present, the PaW mechanism attracts a large amount of interest and several generalizations has been proposed (see for example \cite{nostro,nostro2,vedral,vedraltemperature,leonmaccone,macconeoptempoarrivo,interacting,timedilation,simile,simile2,review,review2,scalarparticles,dirac,foti,brukner,wigner,indefinite,asimmetry,pinto1,pinto2,peggtimeqm}), including a first review \cite{altaie}. In Chapter 2 we will present the basic aspects of the PaW theory, we will show how the main objections were overcome, and we will make some considerations concerning the interpretation of the mechanism.

A very interesting question is to understand which systems can be chosen as clocks for PaW theory: which are the ideal systems and also which are the most physical. In the original work of Page and Wootters the clock is provided by a quantum spin rotating under the action of an applied magnetic field. Considering recent works, the most widely used clock is the system proposed by GLM that, as already mentioned, consists of an unbounded clock Hamiltonian with continuous spectrum with the clock energy equal to its momentum. Therefore, an important part of our work will focus on which systems can be chosen as clock observables (namely which Hamiltonians can be used as clock Hamiltonians), and our intention is to generalize as much as possible the PaW mechanism in order to use a large number of physical systems as clock. We will see that, through our study, it will be possible to use any generic bounded Hamiltonian with discrete spectrum as clock Hamiltonian, also recovering a continuous flow of time for the rest of the Universe while maintaining a finite and discrete the clock observable.

For this purpose in Chapter 3 we present a work of D. T. Pegg in which he explores the possibility of the existence of a quantity that can be regarded as the complement of the Hamiltonian for a quantum system with discrete energy levels \cite{pegg} (see also \cite{opfase,peggbar}). He finds that such a quantity exists and can be represented by a probability-operator measure (POVM). The quantity Pegg finds has dimensions of time, but it is not time. Rather this quantity is a property of the system and its behavior profoundly depends on the state of the system. So Pegg suggests to describe such a quantity with the name \lq\lq Age\rq\rq. We will show that the Pegg's formalism can be accomodating and find a physical justification within the PaW framework. We will introduce in the PaW theory generic bounded clock Hamiltonians and we consider time as described by the POVM. We demonstrate that Pegg's POVM states provide a consistent dynamical evolution for the rest of the Universe even if they are not orthogonal, and therefore partially undistinguishables. Furthermore we notice that, considering an Hamiltonian with equally spaced energy eigenvalues, Pegg's Age can be described by an Hermitian operator that is conjugate to the Hamiltonian of the system\footnote{Age is conjugate to the Hamiltonian of the system in the sense that Age is the generator of the energy shift and the Hamiltonian is the generator of translation in the eigenvalues of Age.}. In this particular case, we will see that the Pauli objection is overcome and the Hermitian operator Age proposed by Pegg can be considered as a real time operator \cite{nostro}. We will show this framework in Chaper 4. 

In Chapter 5 we connect the PaW mechanism with a new formulation of statistical mechanics through entanglement that has recently been proposed \cite{canonical,popescu1} (see also \cite{popescu2,zurek,cantyp}). It was realized that ensemble averages and subjective ignorance are not needed, because individual quantum states of systems can exhibit statistical properties. This seems to be a purely quantum phenomenon, and the key is entanglement, which leads to objective lack of knowledge. Considering indeed our quantum Universe composed by a small subsystem and a large environment, it has been demonstrated that, for the vast majority of randomly chosen wave-functions of the Universe satisfying a total energy constraint, the reduced density matrix of the small subsystem is given by the canonical statistical distribution. We will show that time and non-equilibrium dynamics can emerge through the PaW mechanism between the small subsystem and the environment present in the (randomly chosen) global wave-function of the Universe. The clock is provided by the environment, which ticks the temporal evolution of the rest of the Universe (we recall that we can use any generic Hamiltonian as a clock Hamiltonian \cite{nostro}). In this way we demonstrate the peaceful coexistence of statistical equilibrium and non-equilibrium dynamics by identifying the trace over the environment degrees of freedom with the temporal trace over the entire history of the small subsystem \cite{nostro2}.

In Chapter 6 we introduce the spatial degree of freedom into the discussion. It is possible to read the PaW approach as a sort of internalization of the temporal reference frame with the clock being an appropriately chosen physical system and time is considered as \lq\lq what is shown on a quantum clock\rq\rq. Here we study the possibility to extend this protocol in order to internalizing, together, the temporal and the spatial reference frames. In this approach also space will be an emerging property of entangled subsystems and the concept of position is recovered relative to \lq\lq what is shown on a quantum rod\rq\rq. Quantum reference frames for the spatial degree of freedom have been extensively studied in quantum information and quantum foundations (see for example \cite{burnett,QRF1,QRF2,QRF3,QRF4,QRF5,QRF6,QRF7,QRF8,QRF9,QRF10,QRF11,QRF12,QRF13,QRF14,QRF15,QRF16,QRF17,QRF18,QRF19,QRF20,QRF21,ultimamail}). In the quantum gravity literature, it has been suggested that quantum reference frames are needed to formulate a quantum theory of gravity \cite{dewitt,afundamental,QG1,QG2}. In \cite{change1} (see also \cite{change2,change3,change4,change5,change6,change7,change8,change9}) has also been introduced a formalism for describing transformations to change the description between different quantum reference frames in various contexts. Just as the PaW mechanism has been extensively studied in order to describe the temporal degree of freedom, all these works have dealt only with internalization of the space reference frame leaving time as an external parameter in the theory. Also a generalization of the PaW mechanism for the spatial degree of freedom alone has already been addressed in \cite{hoehn1,hoehn2,hoehn3} (see also \cite{change2,change3}). Only recently, in \cite{giacomini}, has been introduced a fully relational formulation of a $1+1$ dimensional spacetime for the case of a system of $N$ relativistic quantum particles in a weak gravitational field. In Chapter 6 we start by considering the Universe satisfying a constraint on total momentum and divided into two subsystems where the first acts as a spatial reference frame for the second. In this framework we show that a well-defined relative position emerges between the two entangled subsystems. Then we introduce into the Universe a further subsystem that acts as a clock and we consider the Universe satisfying a double constraint: both on total energy and total momentum. In this way we give a model of 3+1 dimensional, non-relativistic, quantum spacetime emerging from entanglement in a globally \lq\lq timeless\rq\rq and \lq\lq positionless\rq\rq Universe \cite{nostro3}. 

Finally, in Chapter 7, we will investigate the behavior of quantum clocks when interacting with a Newtonian gravitational field produced by a non-rotating, spherical mass. As quantum clocks we will consider quantum systems governed by an Hamiltonian with bounded, discrete spectrum and equally-spaced energy levels. Through the PaW approach, we will examine the evolution of the clocks and we will find a time dilation effect in agreement with the first order expansion of the gravitational time dilation as obtained from the theory of General Relativity. The result can be extended by considering the \textit{relativistic gravitational potential} \cite{gravpot1,gravpot2}: in this case we will obtain the agreement with the exact Schwarzschild solution.

 
 As a last consideration we note that an objection can certainly arise concerning the difficulty in considering the whole Universe as described by a pure quantum state. As already mentioned, our study applies in general to all closed quantum systems that can be divided into subsystems and can certainly be used in this sense. Nevertheless, leaving apart the difficulties in writing the quantum wave function of the Universe (that is of course a closed macroscopic object), it is interesting to notice that recent observations on the cosmic microwave background (CMB) \cite{planck} toghether with the inflationary paradigm indicate that at the beginning of cosmic inflation the Universe was in a pure state with highly-correlated quantum fluctuations \cite{bianchi}. Furthermore, it has been also suggested that the assumption that the observable and the non-observable Universe might be entangled provides an argument in support of inflation \cite{vedraltemperature}, \cite{vedralinflation}. Therefore, it seems somewhat natural for us to speculate whether our discussion may also apply to the entire Universe. We will return to this point in Chapter 4 and Chapter 5.

\begin{quote}
\label{bianca2}

\end{quote}

	\chapter{Page and Wootters Theory}
\label{PaW}

As we mentioned in the Introduction, we start our discussion recalling the so-called \lq\lq problem of time\rq\rq, that has led to the \lq\lq timeless approch to time\rq\rq proposed by Page and Wootters (PaW) in 1983 \cite{pagewootters} (see also \cite{wootters,pagearrow}). The \lq\lq problem of time\rq\rq emerged in the context of canonical quantization of gravity. The usual quantization procedure yields the quantum mechanical result that the system is in a zero energy eigenstate of the Hamiltonian operator and the application of the Schrödinger time evolution equation involving the corresponding Hamiltonian operator then leads to a \lq\lq frozen\rq\rq dynamics for which nothing seems to happen. This aspect of the problem of time was introduced by Wheeler and DeWitt, who formalized the problem into an equation that has become to be know as the Wheeler-DeWitt equation: 

\begin{equation}\label{2WDW}
\hat{H} \ket{\psi} = 0
\end{equation}
where $\hat{H}$ is the Hamiltonian constraint of general relativity but elavated to the status of quantum operator and $\ket{\Psi}$ represents the state of our quantum Universe. As already mentioned, expression (\ref{2WDW}) along with the Schrödinger equation make it clear that physical states experience no time evolution. This may seem to be a rather strange requirement to impose on the Universe but it is the starting point for the interpretation of time that was proposed by Page and Wootters. 

In this Chapter we will proceed as follows: in the Section 2.1 we will introduce the PaW theory following the approach as it was originally proposed by the authors; in the Section 2.2 we will provide a formal description of the mechanism; in the Section 2.3 we will present a revision of the PaW mechanism as proposed by V. Giovannetti, S. Lloyd and L. Maccone (GLM) in \cite{lloydmaccone}; in the Section 2.4 we will show how the main criticisms of the theory have been overcome; finally in Section 2.5 we will provide some important considerations.

\section{Introduction to the Theory}
The idea that PaW proposed to solve the \lq\lq problem of time" is both simple and powerful, and invokes one of the most fascinating mysteries of physics, namely quantum entanglement. Their theory consists in dividing the total Hilbert space into two subsystems and assigning one of them to time. The \lq\lq flow of time" then emerges simply through the correlation between the temporal degree of freedom and the rest of the system; a correlation present in a globally static state. In this framework we do not consider time as an abstract, external coordinate, but as \lq\lq what is shown on a clock", where the clock is some physical system that is taken as time reference. The PaW mechanism aims to replace the statement \lq\lq the state of the rest of the system at time t" with the statement \lq\lq the state of the rest of the system given that the clock shows t". Using the words of Page and Wootters:\begin{quoting}
	We shall argue that the temporal behavior we observe is actually a dependence on an internal clock time, not on an external coordinate time. It is perfectly consistent with our observation to assume that any closed system is in an eigenstate of energy and thus stationary with respect to coordinate time, since coordinate time translation are unobservable. Such a state can be decomposed into states of definite clock time. The dependence of these component states upon the clock time labeling them can then represent the observed temporal behavior of the system\footnote{From D. N. Page and W. K. Wootters, \textit{Evolution whitout evolution: Dynamics described by stationary observables}, Physical Review D \textbf{27}, 2885 (1983).}.
\end{quoting}
In the PaW framework we therefore have a single stationary state in which the entire history of the system is \lq\lq condensed". Wootters, in \cite{wootters}, explains that, as well as solving the \lq\lq problem of time", one motivation for considering such a condensation of history <<is the desire for economy as regards the number of basic elements of the theory>>. Indeed he points out that <<quantum correlations are an integral part of quantum theory already, and yet an old element, time, is being eliminated becoming a secondary and even approximate concept>>. The second reason he mentions is the fact that, as time, also space can be perhaps replaced by quantum correlations. Thus a new formulation of quantum spacetime <<would obviously provide a new starting point for attemps to construct a complete and workable quantum theory of gravity>>. 

As mentioned in the Introduction, we underline that the PaW approach, not surprisingly, has not been without criticism. Most noteworthy is the one of K. V. Kuchar \cite{kuchar} who questioned the possibility of constructing two-time propagator within PaW model. Kuchar's criticism have been firstly investigated by Dolby in \cite{dolby}. Further investigations has been carried out by R. Gambini, R. A. Porto, J. Pullin, and S. Torterolo (GPPT) in \cite{gppt} and GLM in \cite{lloydmaccone} (see also \cite{esp2}). Following GPPT and GLM, we will show how to reproduce the correct statistics of sequential measurements made on a system at different times in the Section 2.4. Very recently an additional solution to the problem has also been proposed in \cite{review}. Another criticism was made by A. Albrecht and A. Iglesias in \cite{iglesias}, where they stressed how the possibility for different choices of the clock inexorably leads to an ambiguity in the dynamics of the rest of the Universe. This problem was addressed and solved by C. Marletto and V. Vedral in \cite{vedral}.

\subsection{Two-Particle Universe}
We start considering a simple model of the Universe composed by two non-interacting spin-1/2 particles in an uniform magnetic field. In doing this, we follow the example proposed originally by Wootters in \cite{wootters}; we think in this way to simplify the comprehension of the mechanism, following the path that one of its authors wanted to outline. It is important to notice that we will assume for now that there is an observer outside of the Universe who can make measurements on it. We take the magnetic field uniform and parallel to the positive $z$ axis, $\vec B = \left(0,0,B\right)$, and the two particles starting at $t=0$ with both their spins pointing in the positive $x$ direction. In this way we have, at $t=0$, the particles in the state ($\hslash = 1$):

\begin{equation}
\ket{\psi_1}=\ket{\psi_2}=\frac{1}{\sqrt{2}}\left(\ket{\uparrow} + \ket{\downarrow}\right) 
\end{equation}
where $\ket{\uparrow}$ and $\ket{\downarrow}$ are the eigenstates $\hat{S}_{z}$. The equation of motion then requires that the spins of the particles precess togheter around the vertical axes, that is formally 

\begin{equation}
\ket{\psi_1(t)}=\ket{\psi_2(t)}=\frac{1}{\sqrt{2}}\left(e^{i \alpha t}\ket{\uparrow} +e^{-i \alpha t } \ket{\downarrow}\right)
\end{equation}
where $\alpha = \mu_{B} B$ and $\mu_{B}$ is the magnetic moment of the particles. If we make some measurements on the spins of the two particles in any direction along the plane perpendicular to $z$, we will find with high probability that the outcomes of the two measurements are the same. This does not surprise us as we know that the particles precede together around the $z$ axis. Let us now consider a different initial state of the same two-particle system, namely the global stationary state 

\begin{equation}
\begin{split}
\ket{\Psi}  & \propto 	\int_{0}^{\frac{2 \pi}{\alpha}} \ket{\psi_1(t)}\otimes\ket{\psi_2(t)} dt =\\ 
&= \frac{1}{2} \int_{0}^{\frac{2 \pi}{\alpha}} \left( e^{i\alpha t} \ket{\uparrow^{(1)}} + e^{-i\alpha t} \ket{\downarrow^{(1)}} \right) \otimes \left(  e^{i\alpha t} \ket{\uparrow^{(2)}} + e^{-i\alpha t} \ket{\downarrow^{(2)}}  \right) dt =  \\ 
&=  \frac{1}{2} \int_{0}^{\frac{2 \pi}{\alpha}} \left( e^{i 2\alpha t} \ket{\uparrow^{(1)}\uparrow^{(2)}} + \ket{\uparrow^{(1)}\downarrow^{(2)}} + \ket{\downarrow^{(1)}\uparrow^{(2)}} + e^{-i2\alpha t} \ket{\downarrow^{(1)}\downarrow^{(2)}} \right) dt =  \\ 
&= \frac{1}{2} \left[ \frac{e^{i 2\alpha t}}{i 2\alpha t}\biggr\rvert^{\frac{2\pi}{\alpha}}_{0} \ket{\uparrow^{(1)}\uparrow^{(2)}} +t\biggr\rvert^{\frac{2\pi}{\alpha}}_{0}\left(\ket{\uparrow^{(1)}\downarrow^{(2)}}+\ket{\downarrow^{(1)}\uparrow^{(2)}}\right) + \frac{e^{-i 2\alpha t}}{-i 2\alpha t}\biggr\rvert^{\frac{2\pi}{\alpha}}_{0} \ket{\downarrow^{(1)}\downarrow^{(2)}}  \right]    = \\ 
&= \frac{1}{2} \left[ \frac{e^{i 4\pi} - 1}{i 2\alpha t} \ket{\uparrow^{(1)}\uparrow^{(2)}} +\frac{2\pi}{\alpha} \left(\ket{\uparrow^{(1)}\downarrow^{(2)}}+\ket{\downarrow^{(1)}\uparrow^{(2)}}\right) + \frac{1 - e^{-i 4\pi}}{i 2\alpha t} \ket{\downarrow^{(1)}\downarrow^{(2)}}  \right] \\
&= \frac{\pi}{\alpha} \left(\ket{\uparrow^{(1)}\downarrow^{(2)}}+\ket{\downarrow^{(1)}\uparrow^{(2)}}\right).
\end{split}
\end{equation} 
Thus the normalized state of the two-particle system is 

\begin{equation}
\ket{\Psi} = \frac{1}{\sqrt{2}}\left(\ket{\uparrow^{(1)}\downarrow^{(2)}}+\ket{\downarrow^{(1)}\uparrow^{(2)}}\right),
\end{equation}
that is an eigenstate of the global Hamiltonian $\hat{H} \propto \hat{S}^{(1)}_z + \hat{S}^{(2)}_z$, with energy eigenvalue equal to zero (indeed it is easy to verify that $\hat{H} \ket{\Psi} = 0$). Now we reguard one of the particle as a clock, considering its direction of spin analogous to the direction of the pointer of an ordinary clock, and we consider the other particle as the object particle. Once again if we simoultaneusly make measurement on the spins of the two particles in any direction along the plane perpendicular to $z$, we will find with high probability that outcomes of the two measurements are the same. Therefore, even though the state $\ket{\Psi}$ is stationary, it has the same property that we saw in the time-dependent state. This is the reason for which Wootters suggests to consider the two particles as \lq\lq precessing together" and, in particular, he suggests to consider the object particle as \lq\lq precessing with respect to clock time", since the state of the object particle is correlated with the clock's pointer position. For example, if the clock has been found to be pointing on the positive $x$ direction, then the other particle, when subjected to a spin measurament along the $x$ axis, will certainly also be found pointing on the positive direction. This can be seen by noting that 

	\begin{multline}
	\braket{S^{(1)}_x = \frac{1}{2},S^{(2)}_x = - \frac{1}{2} | \Psi}  =\\= \left[\frac{1}{\sqrt{2}} \left( \bra{\uparrow^{(1)}} + \bra{\downarrow^{(1)}}    \right) \frac{1}{\sqrt{2}} \left(   \bra{\uparrow^{(2)}} - \bra{\downarrow^{(2)}}    \right)\right] \times \left[\frac{1}{\sqrt{2}}\left(\ket{\uparrow^{(1)}\downarrow^{(2)}}+\ket{\downarrow^{(1)}\uparrow^{(2)}}\right) \right] =\\ 
	= \frac{1}{2\sqrt{2}} \left( \bra{\uparrow^{(1)}} + \bra{\downarrow^{(1)}} \right) \left(   \ket{\downarrow^{(1)}}-\ket{\uparrow^{(1)}}\right) = 0   
	\end{multline}
	where $\ket{S_x = \frac{1}{2}}$ and $\ket{S_x = -\frac{1}{2}}$ are the eigenstates of the operator $\hat{S}_x$ with eigenvalue $1/2$ and $-1/2$ rispectively. Such perfect agreement between the two particles holds for any direction perpendicular to the $z$ axis. Wootters says:
\begin{quoting}
	In this sense one can speak of a kind of evolution, namely, evolution with respect to clock time, even within a single stationary state\footnote{ From W. K. Wootters, \textit{\lq\lq Time" replaced by quantum correlations}, International Journal of Theoretical Physics \textbf{23}, 701–711 (1984).}.  
\end{quoting}
We now therefore formalize these concepts mathematically. We call \lq\lq clock" (C) the particle $1$, and \lq\lq system" (S) the particle $2$, and we ask the question: if the clock particle is found with its spin pointing on the positive $x$ direction, by which we mean that is found in the state $\ket{S_x = \frac{1}{2}}$, then what is the probability $P(S^{(s)}_x=\frac{1}{2} | S^{(c)}_x =  \frac{1}{2})$ of finding the system particle also pointing in the positive $x$ direction? $P(a|b)$ is a conditional probability\footnote{As we will see conditional probabilities constitute an important part of the PaW theory and they will return several times in the following. This is the reason why the PaW mechanism is sometimes called \lq\lq conditional probability interpretation of time\rq\rq. }, namely the probability of finding $a$ conditioned to the fact of having previously found $b$. Then the calculation is:

\begin{equation}
\begin{split}
P(S^{(s)}_x = \frac{1}{2} | S^{(c)}_x =  \frac{1}{2}) & = \frac{P\left(\textit{Both pointing in the positive $x$ direction}\right)}{P\left(\textit{Clock pointing in the positive $x$ direction}\right)} = \\&= \frac{\left|   \braket{S^{(c)}_x = \frac{1}{2},S^{(s)}_x =  \frac{1}{2} | \Psi}    \right|^2}{ \sum_{m=-1/2}^{1/2} \left|    \braket{S^{(c)}_x = \frac{1}{2},S^{(s)}_x = m | \Psi}   \right|^2}.
\end{split}
\end{equation}
We have already seen that in the case of $\frac{1}{2}$-spin particles we found $P(S^{(s)}_x = \frac{1}{2} | S^{(c)}_x = \frac{1}{2}) = 1$. Wootters repeats the calculations by considering spins grater than $1/2$, until considering the limit $s \longrightarrow \infty$. In this limiting case we have

\begin{equation}
P(S^{(s)}_x = S_{max} | S^{(c)}_x = S_{max}) = \frac{\sqrt3}{2} \simeq 0,866
\end{equation} 
and it is the lower bound. This bound is quite high when one considers that there are $2S_{max} + 1$ possible orthogonal states in which one could find the system particle. As before, with each spin value $(s=1/2,1,3/2,....,\infty)$, this high degree of agreement holds for any other direction in the plane $(x,y)$ perpendicular to the $z$ axis, and we therefore say that the system particle is precessing with respect to clock time.

\subsection{Evolution Within a Single State}
We are now ready to formally see the \textit{evolution without evolution} and, in doing this, we start considering one single stationary state $\ket{\Psi}$ describing the whole history of the Universe. We recall that so far we had assumed that there was an observer outside the universe who could make measurements on it. We now change our point of view and we discuss the interpretation of $\ket{\Psi}$ in terms of measurement made by observers within the universe. Using Wootters words:\begin{quoting}
	An observer within the universe is somewhat analogous to one of the particles in the above [...] example, in the sense that his own state is highly correlated with the state of the rest of the world. When he makes a measurement on the world around him, he also makes a measurement, without even trying, on some of his own internal variables. This combined measurement, being made entirely within the universe, does not collapse the state of the unverse as a whole but only gives the observer the experience of being in one of the many different states which are possible for him within the state $\ket{\Psi}$. Given that he is in this state, the correlations inherent in $\ket{\Psi}$ place strong restrictions on the result of his measurement of the world around him. The correlations he thus observes between his own state and the state of the world around him, as well as the correlations among the various parts of the outside world, are interpreted by him as the passage of time\footnote{We will discuss the meaning and implications of these Wootters statements in the last section of this Chapter, where we will address our considerations.}.
\end{quoting}
It is clear that, in order to see evolution, we need to consider at least two particles and let one of them serve as a clock, and this is what we do in real life: when we speak of a system changing in time we are always really comparing two systems, an object system and a clock. Therefore in this picture $\ket{\Psi}$ is fixed once for all, but within the global state the observed evolution is determined by correlation between subsystems, and one can construct a $\ket{\Psi}$ with whatever correlation pleases. Then, formally, we consider the Universe as formed by two non-interacting parts, the clock $C$ and the system $S$. The Hamiltonian takes the form:

\begin{equation}
\hat{H}=\hat{H}_C + \hat{H}_S.
\end{equation}
The state of the Universe is an eignestate of $\hat{H}$, and we take for semplicity the eigenvalue to be zero. Now let $\ket{\psi_c (0)}$ be a state of the clock chosen to corrispond to the zero of clock time, and for each $\tau$ we define

\begin{equation}
\ket{\psi_c (\tau)} = e^{-i\hat{H}_c \tau} \ket{\psi_c (0)}
\end{equation}
as the state possessing the value $\tau$ of clock time. Finally let $\ket{\psi_s (\tau)}$ be the conditioned state of the system $S$ given that the clock is in the state $\ket{\psi_c (\tau)}$, that is: $\ket{\psi_s (\tau)}$ is the coefficient of $\ket{\psi_c (\tau)}$ in an expansion of $\ket{\Psi}$ as a linear combination of clock states. So we can write 

\begin{equation}\label{2asmannave}
\ket{\psi_s (\tau)} = \braket{\psi_c (\tau) | \Psi}	
\end{equation} 
where we have to remember that $\ket{\Psi}$ is in a larger Hilbert space than $\ket{\psi_c (\tau)}$ and $\ket{\psi_s (\tau)}$. We notice that the operation in the definition (\ref{2asmannave}) is not a projective measurement but the definition of the Everett relative state \cite{everett}. The meaning and the implications of this operation will be discussed later in the next Section. Into this framework the evolution within a single state follows from:

\begin{equation}\label{2schro1}
\begin{split}
i \frac{d}{d\tau} \ket{\psi_s (\tau)} & = i \frac{d}{d\tau} \braket{\psi_c (\tau) | \Psi} = - \bra{\psi_c (\tau)} \hat{H}_C \ket{\Psi} =\\&= -\bra{\psi_c (\tau)} \hat{H} - \hat{H}_S \ket{\Psi} = \bra{\psi_c (\tau)} \hat{H}_S\ket{\Psi}= \\&= \hat{H}_S \braket{\psi_c (\tau) | \Psi} =  \hat{H}_S \ket{\psi_s (\tau)}. 
\end{split}
\end{equation}
In the (\ref{2schro1}) we recognise the Schrödinger equation for the state $\ket{\psi_s (\tau)}$ that evolve with its Hamiltonian $\hat{H}_S$ with respect to $\tau$. Thus, as long as $\ket{\Psi}$ is an eigenstate of $\hat{H}$ and as long as there exists a clock, then the rest of the Universe automatically follows the usual law of evolution with respect to clock time. In conclusion of his work Wootters says that <<it is not necessary to include time as an a priori element of physics because \textit{clock time}, which makes sense even within a stationary state, is the only kind of time that can be observed>>.

\section{Evolution Without Evolution}
We shall now rewiev the PaW approach summarizing its most important aspects and looking at the conditions under which it is applicable. 

\subsection{Conditions and Definitions}
We start considering a \lq\lq timeless" Universe and we therefore consider a global state $\ket{\Psi}$ that is eigenstate of the Hamiltonian $\hat{H}$, namely:

\begin{equation}\label{2constraint}
\hat{H}\ket{\Psi} = 0
\end{equation}
that is a constraint compatible with the Wheeler-DeWitt equation. We want to emphasize here that, as pointed out in \cite{lloydmaccone}, in the PaW approach the zero eigenvalue of $\hat{H}$ does not play a special role in identifying the global state. Indeed, up to an irrelevant global phase in the dynamics of the rest of the Universe, the global state $\ket{\Psi}$ can be obtained also by imposing the constraint $\hat{H}\ket{\Psi} = E \ket{\Psi}$ with real $E$.

Then, by supposing that the Hamiltonian is sufficiently local, it is always possible to regard the Universe as consisting of two subsystems, which we will call \lq\lq the clock" $C$ and \lq\lq the system" $S$. In dividing the Universe into the two parts $C$ and $S$ we have to be careful about a further condition, namely the clock has to be a \lq\lq good" clock. In the PaW theory a good clock must have a much larger dimension than the system $S$ and also has to interact weakly with $S$ (in the ideal case it should not interact at all). So, we need a tensor-product structure $\mathcal{H} = \mathcal{H}_C \otimes \mathcal{H}_S$, where $\mathcal{H}_C$ and $\mathcal{H}_S$ are the Hilbert spaces of the clock and the system respectively. In terms of Hamiltonians and identity operators for the respective subsystems we can write: 

\begin{equation}\label{2h+hcaso1}
\hat{H}=\hat{H}_C \otimes\mathbb{1}_S +\mathbb{1}_C \otimes \hat{H}_S
\end{equation}  
where $\hat{H}_C$ and $\hat{H}_S$ act on $C$ and $S$ respectively. We underline here that one is free to consider the time quantum degree of freedom as an abstract space without any physical significance instead of a dynamical degree of freedom connected to a system that represents the clock, that is a part of the Universe. In the first case, the PaW mechanism provides only a mathematical model to justify the emergence of time, while the second point of view could describe an operational definition of time, i. e. time is \lq\lq what is read on the clock", where the clock is a specific physical system. We are more interested in the second option and for this reason we are interested in which Hamiltonians can be chosen as clock Hamiltonians. Therefore we now focus on which operator can be used to be the clock operator. 

For an ideal clock, the observable to choose would be an operator $\hat{T}$ conjugated to the Hamiltonian $\hat{H}_C$, in the sense that $\hat{T}$ is the generator of energy shifts and that $\hat{H}_c$ is the generator of translations in the $\hat{T}$ eigenvalues. Eigenstates and eigenvalues of the operator $\hat{T}$, defined by the relation $\hat{T}\ket{t}_C = t\ket{t}_C$, will thus be respectively the clock states and the values that can be read as time values. An important consideration is that until this moment $\hat{T}$ must not be absolutely considered a time operator, but simply an observable choosen to be the clock observable\footnote{In the original work of Wootters we have seen previously the operator choosen to be the clock observable was the spin along the $x$ direction while the clock Hamiltonian was proportional to the spin along the $z$ direction. With this choice the clock observable is not conjugated to the Hamiltonian, but however the clock Hamiltonian generates the shift on the clock states.}.

The remaining condition on PaW's approach is the requirement of entanglement between $C$ and $S$. We note that this happens necessarily by considering the Universe in a pure global state satisfying the energy constraint.
 As already mentioned, the request that $C$ and $S$ are entangled is the property that allows the appearance of a dynamical evolution for the system $S$ with respect to the clock values, while no evolution appears in the global state of the Universe with respect to the external time. Formally, the state $\ket{\Psi}$ must satisfy equation (\ref{2constraint}) and must have the form:
\begin{equation}
\ket{\Psi} = \sum_{t} c_t \ket{t}_C \otimes \ket{\phi(t)}_S,
\end{equation}
where $c_t$ are the coefficients of the expansion and $\ket{\phi(t)}_S$ is a generic state defined on $\mathcal{H}_S$. We will see that, in order to have a good description of the dynamics, it is necessary that the $c_t$ values are different from zero for a sufficiently large number of $t$. In our work we make a radical simplification by taking all the coefficients constant, $c_t= c = 1/\sqrt{d_C}$ where $d_C$ in the dimension of the Hilbert space $\mathcal{H}_C$. This assumption basically means to consider each time value as equally probable. We do this in order to simplify the treatment considering that some problems arise in considering non-constant coefficients and considering also that this choice does not affect our contributions regarding the PaW mechanism. Therefore, for our quantum Universe, we consider the state:

\begin{equation}
\ket{\Psi} =\frac{1}{\sqrt{d_C}} \sum_{t} \ket{t}_C \otimes \ket{\phi(t)}_S.
\end{equation}
In order to approach the central point of the PaW mechanism showing again the Schrödinger evolution of the subsystem $S$, the first step is to define the state of $S$ at time $t$, namely the state of the system $S$ given that the clock shows $t$. This is made by conditioning the state $\ket{\Psi}$ on having the time $t$ via projection with the eigenvectors of the operator $\hat{T}$:
\begin{equation}\label{23vedral}
\ket{\phi(t)}_S = \frac{\braket{t|\Psi}}{1/\sqrt{d_C}} .
\end{equation}
As we have mentioned in the previous Section, the definition (\ref{23vedral}) is the Everett \textit{relative state} definition for the subsystem $S$ with respect to the clock $C$ \cite{everett}. Therefore, as clearly pointed out by Marletto and Vedral in \cite{vedral}, <<this kind of projection has nothing to do with the measurement and does not require one to be performed on the clock: rather, the relative states are a 1-parameter family of $\ket{\phi(t)}_S$ states, labelled by $t$>>, each of them describing the state of the subsystem $S$ with respect to the clock given that the clock is in the state $\ket{t}_C$.

\subsection{Deriving Schrödinger Equation}
We can now derive the Schrödinger evolution for the state $\ket{\phi(t)}_S$ with respect to $t$. By using the constraint (\ref{2constraint}), the form of $\hat{H}$ (\ref{2h+hcaso1}) and the important fact that $\ket{t}_C = e^{-i\hat{H}_C (t-t_0)} \ket{t_0}_C$ (where $\ket{t_0}_C$ is one of the clock eigenstates choosen as the initial time), we can easily obtain:

\begin{equation}\label{2minchia}
\begin{split}
\ket{\phi(t)}_S & = \frac{\braket{t|\Psi} }{1/ \sqrt{d_C}}  = \sqrt{d_C} \bra{t_0} e^{i\hat{H}_C (t-t_0)}  \ket{\Psi} =\\&= \sqrt{d_C} \bra{t_0} e^{i (\hat{H} - \hat{H}_S) (t-t_0)}  \ket{\Psi} = \sqrt{d_C} \bra{t_0} e^{-i\hat{H}_S (t-t_0)} e^{i \hat{H} (t-t_0)} \ket{\Psi} =\\&= \sqrt{d_C} \bra{t_0} e^{-i\hat{H}_S (t-t_0)}\ket{\Psi} = \sqrt{d_C} ~ e^{-i\hat{H}_S (t-t_0)}\braket{t_0|\Psi} =\\&= \sqrt{d_C} ~ e^{-i\hat{H}_S (t-t_0)}\frac{\ket{\phi(t_0)}_S}{\sqrt{d_C}} = e^{-i\hat{H}_S (t-t_0)}\ket{\phi(t_0)}_S
\end{split}
\end{equation}
where $\ket{\phi(t_0)}_S$ is the state of the system $S$ choosen to be the state at $t=t_0$. We can clearly recognize in equation (\ref{2minchia}) the Schrödinger evolution for the state $\ket{\phi(t)}_S$ with its Hamiltonian $\hat{H}_S$. Thus the parameter $t$ can be interpreted as time, and the evolution of the subsystem $S$ has been recovered from no evolution at all. 

In conclusion of this Section we emphatize that, assuming the eigenstates of the clock satisfying $\ket{t}_C = e^{-i \hat{H}_C (t-t_0)} \ket{t_0}_C$ together with the fact that the clock $C$ and $S$ are entangled, directly imply that the evolution on the system $S$ has the exponential form solution of the Schrödinger equation. It is clear that the first assumption may seem too strong. Part of our work will indeed focus later on which Hamltonians and what kind of states can be used to be time states (we will give our contribution on this issue in Chapter 4). A choice of time states is made by GLM in \cite{lloydmaccone} where they consider the clock Hilbert space isomorphic to the Hilbert space of a particle on a line. Their model will be the subject of the next Section.

\section{Clock as Particle on a Line}
In this Section we examine the model of clock observable proposed by GLM in \cite{lloydmaccone}. GLM have formalized the PaW mechanism in a very detailed and complete way, proposing a review and an extension of the mechanism. As clock observable GLM choose an Hilbert space isomorphic to the Hilbert space of a particle on a line. Through this choice we will see that they are able to get the Schrödinger equation for the system $S$ in the differential form. So, they consider the total Hilbert space $\mathcal{H} = \mathcal{H}_C \otimes \mathcal{H}_S$. In the clock subspace $\mathcal{H}_C$ they introduce the position operator $\hat{T}$ and its conjugate momentum $\hat{\Omega}$, with $\left[\hat{T},\hat{\Omega}\right] = i$.  Associating the momentum $\hat{\Omega}$ to the energy of the clock,  they introduce the constraint operator of the model as:

\begin{equation}
\hat{\mathcal{J}} =  \hat{\Omega}  +  \hat{H}_S 
\end{equation}
with $\hat{H}_S$ the Hamiltonian of $S$. In this framework the set of vectors $\ket{\Psi}$ which provides the condensed representation of the full history of the system $S$ are identified by the eigenvector equation associated with the null eigenvalue of $\hat{\mathcal{J}}$, i. e.

\begin{equation}\label{2attraversare1}
\hat{\mathcal{J}} \ket{\Psi} = \left( \hat{\Omega}  +   \hat{H}_S \right)\ket{\Psi} = 0
\end{equation}
where $\ket{\Psi}$ is clearly defined on $\mathcal{H}_C \otimes \mathcal{H}_S$. One may interpret equation (\ref{2attraversare1}) as a constraint that forces the vector $\ket{\Psi}$ to be eigenstate of the total \lq\lq Hamiltonian” $\hat{\mathcal{J}}$ with null eigenvalue, consistently with the Wheeler-DeWitt equation. The states $\ket{\Psi}$ are static, in the sense that they do not evolve with respect to the external time. As we have already seen, the system $S$ evolves with respect to the clock time thanks to entanglement between the system and the clock. 
Here, considering the fact that $\left \{ \ket{t}_C\right \}$ provide a decomposition for the identity operator on $\hat{H}_C$, all the solutions to equation (\ref{2attraversare1}) can be written as:

\begin{equation}\label{2grass1}
\ket{\Psi} = \int_{- \infty}^{+ \infty} dt \ket{t}_C \otimes \ket{\psi(t)}_S   
\end{equation}
where $\ket{\psi (t)}_S$ is the system state at time $t$ in $\mathcal{H}_S$ with normalization $\braket{\psi (t)| \psi (t)}$ for all $t$ and $\ket{t}$ is the eigenstate of $\hat{T}$ in $\mathcal{H}_C$. Equation (\ref{2grass1}) shows that the vectors $\ket{\Psi}$ provide a complete description of the temporal evolution of the system $S$, condensing the history of $S$ through the correlations with the clock $C$.

Now we are ready to reiterate the central point of PaW theory showing the \lq\lq evolution within a single state\rq\rq. We define once again the state $\ket{\psi (t)}_S$ of the system $S$ at time $t$ (i. e. the state of the system $S$ when the clock is in the state $\ket{t}_C$) conditioning the solution $\ket{\Psi}$ of equation (\ref{2attraversare1}) on having the time $t$ via projection with the eigenvectors of the time operator $\hat{T}$, that is

\begin{equation}\label{2condizionamento}
\ket{\psi (t)}_S = \braket{t|\Psi} .
\end{equation}
Starting from equation (\ref{2attraversare1}) and conditioning the clock system to the state $\ket{t}$ we can recover the Schrödinger equation for the state of the subsystem $S$. Indeed, writing the momentum in the \lq\lq position" rappresentation ($\bra{t} \hat{\Omega} = -i\frac{d}{dt} \bra{t}$) and using the definition (\ref{2condizionamento}), we can write

\begin{equation}\label{2schroe}
\begin{split}
& \bra{t} \hat{\mathcal{J}} \ket{\Psi} = 0  \\& \Rightarrow \bra{t}   \hat{\Omega}  +   \hat{H}_S  \ket{\Psi} = 0  \\& \Rightarrow - i \frac{\partial}{\partial t} \braket{t|\Psi} +   \hat{H}_S \braket{t|\Psi} = 0  \\& \Rightarrow i \frac{\partial}{\partial t} \ket{\psi (t)}_S = \hat{H}_S  \ket{\psi (t)}_S.
\end{split}
\end{equation}
This is clearly Schrödinger's equation for the system $S$ that evolves with respect clock time $t$. Since the spectrum of the clock observable $\hat{T}$ is continuous, the equation can be written in the usual differential form $i \frac{\partial}{\partial t} \ket{\psi (t)}_S = \hat{H}_S  \ket{\psi (t)}_S$.

\section{Overcoming Criticisms}
We discuss here two of the solutions that have been proposed to the main criticism of the PaW theory, namely Kuchar's objection \cite{kuchar}. An important point of the PaW proposal concerns conditional probabilities (that we have already discussed in Section 2.1). Indeed the probability to obtain the outcome $a$ when measuring the observable $\hat{A}$ (with $\hat{A}\ket{a}=a\ket{a}$) on the subspace $S$ at a certain clock time $t$ is: 
\begin{equation}
\begin{split}
P(a \: on \: S \: | \: t \: on \: C) = \frac{P(a \: on \: S, \: t \: on \: C)}{P( t \: on \: C)}
\end{split}
\end{equation}
that is the conditional probability of obtaining $a$ on $S$ given that the clock $C$ shows $t$. Kuchar criticised in \cite{kuchar} emphasizing that the PaW framework is not able to provide the correct propagators when considering multiple measurements. Indeed measurements of the system at two times will give the wrong statistics because the first measurement \lq\lq collapses\rq\rq the time state and freezes the system. 

Among the solutions to this problem (the most recent given in \cite{review}), we discuss here the GPPT proposal \cite{gppt} (see also \cite{esp1}) and the GLM theory \cite{lloydmaccone} (see also \cite{esp2}). We discuss this issue assuming the framework of previous Section, namely we assume the clock observable as particle on a line with continuous spectrum.

\subsection{The GPPT Proposal}
As pointed out in \cite{esp1} one of the main ingredients in the GPPT theory 
is the averaging over the abstract coordinate time (the \lq\lq external time\rq\rq) in order to eliminate any external time dependence in the observables. So, in the GPPT proposal, the probability to obtain the outcome $a$ when measuring the observable $\hat{A}$ on the subsystem $S$ conditioned to having the outcome $t$ on $C$ is given by \cite{gppt}:

\begin{equation}\label{2cpGPPT}
P(a \: on \: S \: | \: t \: on \: C) = \frac{\int d\theta \: Tr \left[ \hat{\Pi}_{a,t}(\theta) \hat{\rho} \right]}{\int d\theta \: Tr \left[ \hat{\Pi}_{t}(\theta) \hat{\rho} \right]}
\end{equation}
where $\theta$ is the external time, $\hat{\rho}=\ket{\Psi}\bra{\Psi}$ is the global state of the Universe, $\hat{\Pi}_{t}(\theta)=\hat{U}^{\dagger} (\theta) \hat{\Pi}_{t} \hat{U}(\theta)$ (with $\hat{U}(\theta) = e^{-i\hat{H}\theta}$) is the projector relative to the result $t$ for a clock measurement at external time $\theta$ and $\hat{\Pi}_{a,t}(\theta)=\hat{U}^{\dagger} (\theta) \hat{\Pi}_{a,t} \hat{U}(\theta)$ is the projector relative to the result $a$ for a measurement on $S$ and $t$ for a measurement on $C$ at external time $\theta$ (we are working here in the Heisenberg picture with respect to the external time $\theta$). The generalization of equation (\ref{2cpGPPT}) to multiple time measurements is given by \cite{gppt}:

\begin{equation}\label{2cpGPPT2}
\begin{split}
 P(a_f \: on \: S \: | \:t_f \: on & \: C,a_i,t_i) 
=\\&= \frac{\int d\theta \int d\theta' \: Tr \left[\hat{\Pi}_{a_f,t_f}(\theta)  \hat{\Pi}_{a_i,t_i}(\theta') \hat{\rho} \hat{\Pi}_{a_i,t_i}(\theta') \right]}{\int d\theta \int d\theta' \: Tr \left[\hat{\Pi}_{t_f}(\theta) \hat{\Pi}_{a_i,t_i}(\theta') \hat{\rho} \hat{\Pi}_{a_i,t_i}(\theta') \right]}
\end{split}
\end{equation}
which provides the conditional probability of obtaining $a_f$ on the system $S$ at clock time $t_f$, given that a \lq\lq previous\rq\rq joint measurement of $S$ and $C$ returns $a_i$, $t_i$. Equation (\ref{2cpGPPT2}) provides the the correct propagator, indeed proceeding with the calculations we would find $\left| \bra{a_f} e^{-i\hat{H}_S(t_{f} - t_i)}\ket{a_i}\right|^2$. We will return to GPPT theory in the following, when we will show that it also applies to the case of non-orthogonal time states. We will indeed encounter such states in the next Chapters, when we will use Pegg's POVM in describing the clock observable.

\subsection{GLM's Multiple Measurements}
We focus now on the GLM proposal \cite{esp2,lloydmaccone}. We start considering the global state of the Universe written as

\begin{equation}\label{2GLM}
\ket{\Psi} = \int_{- \infty}^{+ \infty} dt \ket{t}_C \otimes \ket{\psi(t)}_S =  \int_{- \infty}^{+ \infty} dt \ket{t}_C \otimes \hat{U}_{S}(t - t_0)\ket{\psi(t_0)}_{S}
\end{equation}
where $\ket{\psi(t_0)}_{S}$ is the state of $S$ conditioned on $t_0$ that is taken as initial time and where, thanks to the (\ref{2schroe}), we have defined $\hat{U}_{S}(t - t_0)=e^{-i\hat{H}_S(t - t_0)}$.

Let's initially consider to perform a measurement within $S$ at time $t_1$. The authors suggest to divide the system $S$ into two subsystems, including both the system to be measured ($Q$, the \textit{observed}) and an ancillary memory system ($M$, the \textit{observer}). In this framework they use von Neumann’s prescription for measurements \cite{vonneumann}, where a measurement apparatus essentially consists in an (ideally instantaneous) interaction between the system $Q$ and the memory degree of freedom $M$. The interaction correlates the system $Q$ and the memory $M$ along the eigenbasis $ \left \{ \ket{a}  \right \}$ of on observable $A$ to be measured, that is

\begin{equation}\label{2mapping}
\ket{\psi(t)}_Q \otimes \ket{r}_M \longrightarrow \sum_{a} (\braket{a|\psi(t)}) \ket{a}_Q \otimes \ket{a}_M
\end{equation}
where $\ket{r}_M$ is the state of the memory before the interaction and $\braket{a|\psi(t)}$ is the probability amplitude of obtaining $a$ measuring the observable $A$. The Hamiltonian $\hat{H}_S$ of the system $S$ can be written as

\begin{equation}\label{2Hdip}
\hat{H}_S (t) = \hat{H}_Q + \delta_{t,t_{1}} \hat{h}_{QM}
\end{equation}
where $\hat{h}_{QM}$ is responsible for the mapping equation (\ref{2mapping}). So we can write the global state $\ket{\Psi}$ including the measurement, thus obtaining

\begin{multline}\label{2meas}
\ket{\Psi} = \int_{-\infty}^{t_1} dt \ket{t}_c \otimes \hat{U}_Q (t-t_0) \ket{\psi(t_0)}_Q \otimes \ket{r}_M +\\+ \int_{t_1}^{+\infty}dt \ket{t}_c  \otimes \sum_{a} (\braket{a|\psi(t_1)}) \hat{U}_{Q} (t-t_1) \ket{a}_Q \otimes \ket{a}_M .
\end{multline}
The first integral describes the evolution of $S$ prior to the measurement (when the memory is in the state $\ket{r}_M$) whereas the second integral describes the evolution after the measurement when the memory is correlated with the subsystem $Q$.

Now the probability that, at a given time $t$, the values $a$ will be registered by the memory $M$ can be expressed as \cite{lloydmaccone}:

\begin{equation}\label{2probGLM}
P(a \: on \: S \: | \: t \: on \: C) = || \left( _{C}\bra{t}\otimes _{M}\bra{a}\right)\ket{\Psi}||^2 
\end{equation}
where we use the norm of a vector as $||\ket{v}||^2 = \braket{v|v}$. GLM notice here that for smaller values of $t<t_1$, the probability of getting a certain outcome $a$ on $M$ does not depend upon $Q$ yielding $P(a \: on \: S \: | \: t \: on \: C)= |\braket{a|r}|^2$, the resulting statistics being only associated with the ready state of the memory. For $t>t_1$ instead, $P(a \: on \: S \: | \: t \: on \: C) =  \left| \braket{a|\psi(t)} \right|^2 $: it only depends upon the statistical uncertainty of the state of the system $Q$ at time $t_1$ and it remains constant in time due to the fact that we have explicitly suppressed any dynamical evolution on $M$.

It is now possible to extend equations (\ref{2Hdip}), (\ref{2meas}) and (\ref{2probGLM}) in order to perform multiple measurements. What we have to do is simply add additional memories ($M_1 , M_2,...,M_N$), each initialized into a ready state $\ket{r}_{M_i}$ and which couple with $Q$ through the time-dependent Hamiltonian 

\begin{equation}\label{2mmm}
\hat{H}_S (t) = \hat{H}_Q + \sum_{i=1}^{N}\delta_{t,t_{i}} \hat{h}_{QM_{i}} .
\end{equation}
We consider (as an example) the case of a sequence of two measurements at times $t_1$ and $t_2$. Equation (\ref{2mmm}) becomes

\begin{equation}
\hat{H}_S (t) = \hat{H}_Q + \delta_{t,t_{1}} \hat{h}_{QM_1} + \delta_{t,t_{2}} \hat{h}_{QM_2}
\end{equation}
and the global state (\ref{2GLM}) including the double measurement can then be written

\begin{multline}\label{2doppiamisuraGLM}
\ket{\Psi} = \int_{-\infty}^{t_1} dt \ket{t}_c \otimes \hat{U}_{Q} (t-t_0) \ket{\psi(t_0)}_Q \otimes \ket{r}_{M_1} \otimes \ket{r}_{M_2} + \\  
+ \int_{t_1}^{t_2} dt \ket{t}_c \otimes \sum_{a} (\braket{a|\psi(t_1)}) \hat{U}_{Q} (t-t_1) \ket{a}_Q \otimes\ket{a}_{M_1}\otimes \ket{r}_{M_2} + \\  
+ \int_{t_2}^{+\infty} dt \ket{t}_c \otimes \sum_{a}\sum_{b} (\bra{b} \hat{U}_{Q} (t_2 - t_1) \ket{a}) (\braket{a|\psi(t_1)}) \hat{U}_{Q} (t-t_2) \ket{b}_Q  \otimes \ket{a}_{M_1} \otimes \ket{b}_{M_2} 
\end{multline} 
where $\ket{a}$ and $\ket{b}$ are the eigenstates of the observables $A$ and $B$ that are measured at times $t_1$ and $t_2$ (referred to the clock time).Through the state (\ref{2doppiamisuraGLM}) we search now the probability of obtaining $b$ on $S$ at time $t_{2}$, given that a \lq\lq previous\rq\rq measurement at time $t_1$ returns $a$. This can be formally expressed as follows:

\begin{multline}\label{2probdoppioGLM}
P\left(  \left( b \: on \: S \: | \: t_2 \: on \: C \right)\:|\:\left( a \: on \: S \: | \: t_1 \: on \: C \right)  \right) =\\= \frac{P(b,a\:|\:t_2)}{P(a \:  | \: t_1)}
= \frac{||(_{C}\bra{t_2}\otimes _{M_1}\bra{a}\otimes _{M_2}\bra{b})\ket{\Psi} ||^2}{|| (_{C}\bra{t_1}\otimes _{M_1}\bra{a})\ket{\Psi} ||^2} 
\end{multline}
which returns the correct result for a two-times measurement:
\begin{equation}\label{2pf} 
P\left(  \left( b \: on \: S \: | \: t_2 \: on \: C \right)\:|\:\left( a \: on \: S \: | \: t_1 \: on \: C \right)  \right)
= \left| \bra{b}\hat{U}_{Q}(t_2 -t_1)\ket{a}\right|^2  .
\end{equation}
This result follows from the fact that $P(b,a\:|\:t_2) = \left| \left( \bra{b}\hat{U}_{Q}(t_2 -t_1)\ket{a}\right) \left( \braket{a|\psi(t_1)}\right) \right|^2$ obtained from the third summation and $P(a \:  | \: t_1) = \left| \braket{a|\psi(t_1)} \right|^2$ obtained from the second summation in equation (\ref{2doppiamisuraGLM}). We emphasize here that the GLM proposal for multiple measurements needs orthogonal time states and will not be applicable when we will represent the clock observable with Pegg's POVMs \cite{nostro}.

So, following GLM, we showed how an observer within the PaW Universe (represented using the memories $M_1$ and $M_2$) can make measurements on a subsystem and record its results. This argument is used by the authors to answer the question of how the flow of time is perceived by an observer within the Universe.  In the next section we will better analyze the question of the flow of time in the PaW framework.

\section{Discussion}
In this section we focus on the physical meaning of PaW quantum time. An operational definition of proper time as \lq\lq what is shown on a clock" requires some physical system that acts as a clock. The PaW formalism naturally accommodates it: $\mathcal{H}_C$ is the Hilbert space of such system. Clearly, for this reason, it becomes important to understand what kind of Hamiltonians and what kind of systems are good clocks. From what we have seen throughout this Section, a good clock must have a greater number of states compared to those of the system (namely we need $d_C\gg d_S$) and must interact weakly with the system $S$ (in the ideal case should not interact at all). We will focus on this point in the next Chapter. The aspect we want to enphatize here is that in the PaW approach there are evidently two points of view, namely: a point of view within the Universe where the observer is part of the system entangled with the clock system and a point of view outside the Universe, in which an hypothetical external entity is able to see the stationary global state. In the following paragraph, we will look more closely at these two possibilities. 

\subsection{The Flow of Time: Two Points of View}
Now we briefly comment the two point of view emerging from the PaW mechanism. 

\begin{itemize}
\item An hypotetical \textit{external observer} is an observer outside the Universe. He has access to abstract coordinate time and he sees Hamiltonian constraint. So he can consider the whole Universe as a static system whose state is in an superposition of all time states. An important aspect reguarding the external observer is that he can make measurements on the Universe. He can make measurements simultaneously on the two subspaces, the clock $C$ and the system $S$, and he can therefore test the validity of the rules of conditional probabilities.

\item On the other hand the \textit{internal observer} does not have access to abstract coordinate time. According to our point of view, it is evident that the confusion about the perception of the flow of time for an observer within the Universe derives from the absence of clarity with which the relative state of the subsystem $S$ (\ref{23vedral}) is often defined. This topic is clearly related to the problem of considering the wave function of the whole Universe that must include internal observers, which is a issue related to the Everett's interpretation of quantum mechanics \cite{everett} (see also \cite{libroMWI} and \cite{libroMWI2}). For this reason we emphasize that the conditioned state (\ref{23vedral}) is a Everett relative state. The operation involved in this definition has nothing to do with the measurement and does not require one to be performed on the clock. In our opinion this was also the original thought of Page and Wootters. Indeed, as already mentioned in the first Section of this Chapter, Wootter himself in \cite{wootters} says that for an internal observer the own state is correlated with the state of the rest of the Universe and, when he makes a measurement on any object around him, he also makes a measurement on his own internal variables. <<This combined measurement, being made entirely within the Universe, does not collapse the state of the Universe as a whole but only gives the observer the experience of being in one of the many different states which are possible for him within the state $\ket{\Psi}$>>. A little different is the interpretation of GLM \cite{lloydmaccone}. Again the flow of time emerges as a result of the fact that the $S$ system is correlated to $C$, but here there are subsystems within the Hilbert space $\mathcal{H}_S$ that can perform measurement and store their results, thus constructing an history (as we have seen in the previous Section). Also in \cite{vedral} the space $\mathcal{H}_S$ is partioned into three subspaces, that are the \lq\lq observer" (assumed to be made by a memory only), the \lq\lq observed", and a sequence of ancillas. In this framework, by treating the ancillas semiclassically, the authors show how it is possible to describe the observer and the observed as undergoing an effective evolution generated by a time-dependent Hamiltonian, in turn generated by interactions with the ancillas\footnote{For further analysis on this argument, we directly refer to the original work \cite{vedral}.}.
\end{itemize}

In conclusion, we note that the problem of the interpretation of the mechanism is clearly still open and we think that there are also questions left unanswered. In any case, the issue is beyond the scope of this work, and we do not look at it further.

\subsection{Experimental Illustration}
In 2014 an experimental illustration of the PaW mechanism has been realized for the first time. In \cite{esp1} the authors implement the PaW theory using an entangled state of two photons, one of which is used as a clock to gauge the evolution of the second. They make the experiment in two different modes: the \lq\lq observer" mode in which they test the point of view of an internal observer that becomes correlated with the clock photon and sees the other subsystem evolve, and the \lq\lq super-observer" mode in which they test the point of view of an external observer that only observes global properties of the two photons and can prove that the global system is static. They choose a global state $\ket{\Psi}$ as an entangled state of the vertical $V$ and horizontal $H$ polarization of the two photons, that is

\begin{equation}
\ket{\Psi} = \frac{1}{\sqrt{2}} \left(\ket{H}_C \ket{V}_S - \ket{V}_C \ket{H}_S \right)
\end{equation} 
and they enforce the Wheeler-DeWitt constraint by taking $\hat{H}_C =  i \omega \left( \ket{H}\bra{V} - \ket{V}\bra{H}\right)$ and $\hat{H}_S =  i \omega \left( \ket{H}\bra{V} - \ket{V}\bra{H}\right)$ as Hamiltonians of the subsystems (where $\omega$ is a parameter which defines the time scale of the model). At this point they induce the rotations of the polarization of the two photons by forcing them to travel through identical birefringent plates. As they say: \begin{quoting}
	This allows us to consider a setting where everything can be decoupled from the \lq\lq flow of time”, i.e. when the photons are traveling outside the plates. Nonetheless, the clock photon is a true (albeit extremely simple) clock: its polarization rotation is proportional to the time it spends crossing the plates. Although extremely simple, our model captures the two, seemingly contradictory, properties of the PaW mechanism: the evolution of the subsystems relative to each other, and the staticity of the global system\footnote{E. Moreva, G. Brida, M. Gramegna, V. Giovannetti, L. Maccone and M. Genovese, \textit{Time from quantum entanglement: an experimental illustration}, Physical Review A \textbf{89}, 052122 (2014).}. 
\end{quoting}
Indeed when they run the experiment in the \lq\lq observer” mode, the experimenter uses the readings of the clock photon to gauge the evolution of the other because, measuring the clock photon polarization, he correlates with the subsystems and can study their evolution. But when they run the experiment in the \lq\lq super-observer” mode, they avoids measuring the properties of the subsystems in the entangled state and they only measure global properties. In this way they can determine that the global system is static. The authors underline that <<this mode describes what an observer external to the Universe would see by measuring global properties of the state $\ket{\Psi}$: such an observer has access to abstract coordinate time>> (which in the experiment means being able to measure the thickness of the plates) <<and he can prove that the global state is static, as it will not evolve even when the thickness of the plates is varied>>. As expected from the PaW mechanism, in this experiment they prove that the evolution arise within their toy Universe in terms of conditional probability and it is independent of the abstract coordinate time (represented by different thicknesses of birefringent plates). Furthermore they prove that the fidelity $F = \bra{\Psi} \rho_{out} \ket{\Psi}$ (which measures the overlap between the theoretical initial state $\ket{\Psi}$ and the final state $\rho_{out}$ after its evolution through the plates) is constant and close to one and this proves that the state of the global system is static as required by the Wheeler-DeWitt constraint. 

We notice here that in \cite{esp1} also the GPPT theory for multiple time measurements has been tested. Furthermore, in \cite{esp2} the authors provide an new experimental illustration of PaW mechanism that is able to describe multiple two-time quantum correlation functions using the GLM prescription. We do not discuss further this issue and for additional details we refer to the original works.

\section{Conclusions}
In this Chapter we gave a brief summary of the PaW mechanism starting from the original work of Page and Wootters. We outlined the salient aspects of the theory and we addressed the work of GLM who first proposed a clock observable conjugate to the clock Hamiltonian, with the approximation of taking the clock Hamiltonian represented by the momentum. We then saw how the main criticism of the theory has been overcome and we provided some considerations. 

To conclude the Chapter we notice that the PaW mechanism currently attracts a large amount of interest and several generalizations has been proposed, including a first review \cite{altaie}. As examples we indicate some works that we regard as significant in addition to those we have already discussed. In \cite{nostro,nostro2} is given a generalization of the PaW framework in order to use any generic bounded, discrete Hamiltonian as clock Hamiltonian. Furthermore a connection is established with a new formulation of statistical mechanics through entanglement that has recently been proposed \cite{canonical,popescu1} (see also \cite{popescu2,zurek,cantyp}) where ensemble averages and subjective ignorance are not needed, because individual quantum states of systems exhibit statistical properties. We will present these works in Chapter 4 and Chapter 5. In \cite{nostro} (see also \cite{leonmaccone}) is illustrated the possibility of obtaining a time operator within the PaW framework, overcoming Pauli's objection, as we will see in Chapter 4. In \cite{scalarparticles,dirac} the PaW theory is generalized in order to consider relativistic particles. In \cite{foti} is taken the classical limit of the mechanism either of the clock only, or of the clock and the evolving system $S$ together. In this latter case the Hamilton equations of motion are recovered for the system $S$ with repect to the clock. In \cite{interacting} is described the case of interacting clock and system while in \cite{timedilation} a relativistic clock is considered and a classical time dilation (corrected by quantum terms) is observed. In \cite{brukner} the authors study the behavior of quantum clocks in presence of gravitating quantum systems and in \cite{indefinite} noncausal Page and Wootters circuits are treated. In \cite{review,review2} is shown the equivalence between three approaches to relational quantum dynamics: relational observables in the clock-neutral picture of Dirac quantization, the PaW formalism and the relational Heisenberg picture obtained via symmetry reduction. As already mentioned these authors also show how their framework resolve Kuchar’s criticism without invoking approximations, ideal clocks or ancilla systems. A generalization of the PaW mechanism for the spatial degree of freedom has been addressed in \cite{hoehn1,hoehn2,hoehn3} (see also \cite{change2,change3}) and in \cite{giacomini}, through the PaW formalism, has been introduced a fully relational formulation of a $1+1$ dimensional spacetime for the case of a system of $N$ relativistic quantum particles in a weak gravitational field. Finally, in \cite{nostro3} PaW theory was extended providing a model of 3+1 quantum spacetime and this will be the main argument of Chapter 6.

	\chapter{Complement of  the Hamiltonian}
\label{Pegg}

The first goal of our work is to maximize as much as possible the set of physical systems that can be used as clock. As we have seen in the previous Chapter, GLM reviewed the PaW mechanism using a clock observable that is conjugated to the clock Hamiltonian assuming the clock space isomorphic to the Hilber space of a particle on a line and also considering the approximation of associating the momentum of the particle to its energy \cite{lloydmaccone}. We want to extend this result to the case of bounded clock systems with discrete energy levels. In order to do this, it is interesting to look at the work of D. T. Pegg \cite{pegg} (see also \cite{opfase,peggbar}) in which he investigates about the existence of a quantity that could be complement of Hamiltonians with bounded, discrete spectrum. In this Chapter we will summarize and revisit Pegg's work. 

We have seen that the question of an energy-time uncertainty principle has been one of the earliest issues in quantum mechanics. The key aspect of this problem is that in quantum mechanics energy is a dynamical variable of the system, represented by an Hermitian operator but time is a parameter, as it is in classical physics. Indeed Pegg notices that, <<whereas the momentum-position uncertainty relation can be derived from the commutation relation 
($\hslash=1$)

\begin{equation}\label{3xp}
[\hat{X}, \hat{P}] = i  ,
\end{equation}
there is apparently no time operator that is canonically conjugate to the Hamiltonian operator>> of a quantum system in the sense of equation (\ref{3xp}). The question therefore arises as to whether or not there is an operator conjugate or complement to the Hamiltonian in a more general sense. So Pegg explores the possibility of the existence of a quantity that can be regarded as the complement of the Hamiltonian for a quantum system with discrete energy levels, noting that such a quantity would have dimensions of time but would be a property of the system itself. As an operator, it would not represent time in the coordinate sense (as read on an external clock), but it would represent an observable of the quantum system. Its eigenstates would describe the state of the quantum system and measurament performed on the system should tell us something about the quantity involved. 

We will proceed as follows: in Section 3.2 (following \cite{pegg}) we will consider a system with discrete energy eigenstates for wich the ratios of the energy differences are rational\footnote{We will see in the following that this limitation may be overcome and we will be able to consider any generic point-like (i.e. non-degenerate) spectrum. Indeed, since any real number can be approximated with arbitrary precision by a ratio between two rational numbers, all the small corrections we will find may be arbitrarily reduced.} and we will find that such a quantity (the \lq\lq $\alpha$ quantity\rq\rq) does exist and can be represented by a POVM\footnote{In quantum measurement theory, a Probability-Operator Measure or Positive-Operator Valued Measure (POVM) is defined as a collection ${G_k}$ of positive operators on a Hilbert space that sum to the identity: $ \sum_{k} G_k = \mathbb{1} $. $G_k$ positive, $G_k \ge 0$, means that $G_k$ is Hermitian with no negative eigenvalues, or, equivalently, that $\bra{\psi} G_k\ket{\psi} \ge 0$ for all $\ket{\psi}$.}. The uncertainty relation between the Hamiltonian and the $\alpha$ quantity is also suggested. In Section 3.3 we investigate systems with equally-spaced energy levels and we will see that in this particular case it will be possible to define an Hermitian operator complement of the Hamiltonian.

\section{The $\alpha$ Quantity}
We consider a quantum system with $p+1$, non-degenerate energy states $\ket{E_i}$, with $i=0,1,2,...,p$. For simplicity and without loss of generality we choose the zero of energy $\ket{E_0}$ so that the eigenvalue $E_0=0$ and we choose the other energies eigenvalues $E_i$ increasing with $i$. We assume that these include all the accessible states of the system. Let $\mathcal{H}_p$ be the Hilbert space of dimension $p+1$ spanned by the energy states $\ket{E_i}$, so the Hamiltonian of the system is

\begin{equation}
\hat{H}_p=\sum_{i=0}^{p}E_i\ket{E_i}\bra{E_i} .
\end{equation}       
We can write the general state of the system as:

\begin{equation}\label{3genericsystem}
\ket{\psi}=\sum_{i=0}^{p}f_i\ket{E_i}
\end{equation}
and it will evolve (with respect to the external time) according to the relation

\begin{equation}
e^{-i \hat{H}_p t}\ket{\psi}=\sum_{i=0}^{p}f_ie^{-i E_it}\ket{E_i}.
\end{equation}
We are looking for a quantity $\alpha$ of the system which has dimensions of time and which be complement to the Hamiltonian in the sense that $\hat{H}_p$ is the generator of translations in the $\alpha$ values. Thus we need a state $\ket{\alpha}$ for which

\begin{equation}\label{3tras}
e^{-i \hat{H}_p \delta\alpha}\ket{\alpha}=\ket{\alpha + \delta\alpha} .
\end{equation}
Considering a generic expansion of the state $\ket{\alpha}$ on the basis of energy $\ket{\alpha}=\sum_{i=0}^{p}c_i(\alpha)\ket{E_i}$, we can write

\begin{equation}\label{3cc}
\begin{split}
\ket{\alpha + \delta\alpha} & = e^{-i \hat{H}_p \delta\alpha}\ket{\alpha}= \\& = e^{-i \hat{H}_p \delta \alpha}\sum_{i=0}^{p}c_i(\alpha)\ket{E_i}= \\& =\sum_{i=0}^{p}c_i(\alpha)e^{-i E_i \delta\alpha}\ket{E_i}.    
\end{split}
\end{equation}
The equation (\ref{3cc}) must be equal to $\sum_{i=0}^{p}c_i(\alpha + \delta\alpha)\ket{E_i}$ and in this way we obtain

\begin{equation}
c_i(\alpha) \propto e^{-i E_i\alpha}.
\end{equation}
We choose all $c_i(\alpha)$ values equal to each other apart from the phase factor and we can therefore write

\begin{equation}\label{3aa}
\ket{\alpha}=\frac{1}{\sqrt{p+1}}\sum_{i=0}^{p}e^{-i E_i\alpha}\ket{E_i}.
\end{equation}
By substituting $\delta\alpha$ on the left side of the equation (\ref{3tras}) for $\delta t$, we obtain a time translation expression. This shows us that the state $\ket{\alpha + \delta\alpha}$ is the state in which $\ket{\alpha}$ would evolve in $\delta t = \delta\alpha$. Thus, although the quantity represented by $\alpha$ is not time (because the $\alpha$ quantity is a property of the system), it presents some relation with time. The states $\ket{\alpha}$ are not orthogonal and their number is greater than the dimensions of the space $\mathcal{H}_p$, 
so these $\ket{\alpha}$ can not be eigenstates of an Hermitian operator on $\mathcal{H}_p$. Therefore we impose some restrictions in order to be able to proceed and we split the problem in two cases: first we treat the case of unequally-spaced energy levels and then we study the case of equally-spaced energy levels.

\section{Unequally-Spaced Energy Levels}
\subsection{General Framework}
For unequally-spaced energy levels we can not even find a subset of $p+1$ states $\ket{\alpha}$ that are orthogonal but we can make progress by requiring that the ratios $E_i / E_1$ are rational numbers. So, for $E_i / E_1$ rational, we can write 

\begin{equation}\label{3numraz1} 
	\frac{E_i}{E_1} = \frac{C_i}{B_i} 
\end{equation}
where $C_i$ and $B_i$ are integers with no common factors. We define $r_i = r_1 C_i/B_i$ for $i>1$, where $r_1$ is choosen to be the lowest common multiple of the values of $B_i$ with $i>1$, and we take $r_0=0$. In this framework the values $r_i$ are integers for all $i \ge 0$. Now we can define


\begin{equation}
T=\frac{2\pi r_1}{E_1}
\end{equation} 
and then, using equation (\ref{3numraz1}), we can write

\begin{equation}
E_i = r_i \frac{2\pi }{T}.
\end{equation} 
We can now select $s+1$ states $\ket{\alpha}$ for which the values of $\alpha$ are uniformly spread over the range $T$. This means considering values of $\alpha$ denoted by

\begin{equation}\label{3alpha1}
\alpha_m = \alpha_0 + m \frac{T}{s+1}
\end{equation} 
with $m=0,1,2,...,s$. With this choice the states (\ref{3aa}) become the states $\ket{\alpha_m}$ which can be written as

\begin{equation}
\ket{\alpha_{m}} = \frac{1}{\sqrt{p+1}} \sum_{i=0}^{p} e^{-i E_i \alpha_{m}} \ket{E_i}
\end{equation}
and they have the following key property:

\begin{equation}\label{311111}
\begin{split}
\sum_{m=0}^{s}\ket{\alpha_m}\bra{\alpha_m} & = \frac{1}{p+1} \left \{ \sum_{m=0}^{s}\sum_{i=0}^{p}\sum_{k=0}^{p} e^{-i  (E_i - E_k) \alpha_m} \ket{E_i}\bra{E_k} \right \}  = \\& = \frac{1}{p+1} \left \{ \sum_{m=0}^{s}\sum_{i=0}^{p}\sum_{k=0}^{p} e^{-i (r_i - r_k) \alpha_m 2\pi  / T} \ket{E_i}\bra{E_k} \right \} = \\& = \frac{1}{p+1} \left \{ \sum_{m=0}^{s}\sum_{i=k}\ket{E_i}\bra{E_i}  + \sum_{i \ne k} \sum_{m=0}^{s} e^{-i  (r_i - r_k) \alpha_m 2\pi  / T} \ket{E_i}\bra{E_k} \right \}.
\end{split}
\end{equation} 
Replacing the expression of $\alpha_m$ (\ref{3alpha1}) in the second term on the rigth-hand side  of the equation (\ref{311111}), we obtain

\begin{equation}\label{32222}
\begin{split}
\sum_{m=0}^{s}\ket{\alpha_m}\bra{\alpha_m} & = \frac{1}{p+1} \left \{ \sum_{m=0}^{s}\sum_{i=k}\ket{E_i}\bra{E_i}  + \sum_{i \ne k} \sum_{m=0}^{s} e^{-i (r_i - r_k) (\alpha_0 + m \frac{T}{s+1}) 2\pi  / T} \ket{E_i}\bra{E_k} \right \} = \\& =  \frac{1}{p+1} \left \{ \sum_{m=0}^{s}\sum_{i=k}\ket{E_i}\bra{E_i}  + \sum_{i \ne k} e^{i  (r_i - r_k)\alpha_0} \sum_{m=0}^{s} e^{-i (r_i - r_k) \frac{2 \pi m}{s+1}} \ket{E_i}\bra{E_k} \right \}.
\end{split}
\end{equation} 
For $E_i/E_1$ rational, and thus $r_i - r_k$ an integer, we have 

\begin{equation}
\sum_{i \ne k} e^{i  (r_i - r_k)\alpha_0} \sum_{m=0}^{s} e^{-i (r_i - r_k) \frac{2 \pi m}{s+1}} \ket{E_i}\bra{E_k} = 0 ,
\end{equation}
because  

\begin{equation}
\sum_{m=0}^{s} e^{-i (r_i - r_k) \frac{2 \pi m}{s+1}} = (s+1)\delta_{i,k} ,
\end{equation}
and therefore equation (\ref{32222}) becomes:

\begin{equation}\label{33333}
\frac{p+1}{s+1}\sum_{m=0}^{s}\ket{\alpha_m}\bra{\alpha_m} = \mathbb{1}_p
\end{equation} 
where $\mathbb{1}_p$ is the unit operator on the space $\mathcal{H}_p$. We can ensure $r_k - r_i$ is not a multiple of $s+1$ by taking $s+1 > r_p$, that is the largest value for $r_i$. Pegg underlines here that the quantity $T$ has a simple physical interpretation:<<$T$ is equal to the smallest time taken for the system to return to its initial state. Thus the state $\ket{\alpha}$ will be the same as the state $\ket{\alpha + T}$. It follows that restricting our selection of states ${\ket{\alpha}}$ to those for which the values of $\alpha$ are uniformly distributed over the range $T$ then prevent us from including the same state twice>>. We can easily see that the values of the $\alpha$ quantity hold a cyclical repetition, indeed we have

\begin{equation}\label{3stepriportaallinizio}
\begin{split}
e^{-i  \hat{H}_p \frac{T}{s+1}} \ket{\alpha_{m=s}} & =  \frac{1}{\sqrt{p+1}} \sum_{i=0}^{p} e^{-i  E_i \frac{T}{s+1}} e^{-i  E_i \alpha_s} \ket{E_i}   =  \frac{1}{\sqrt{p+1}} \sum_{i=0}^{p} e^{-i  E_i (\alpha_0 + \frac{T(s+1)}{s+1}) } \ket{E_i} =\\& =  \frac{1}{\sqrt{p+1}} \sum_{i=0}^{p} e^{-i E_i (\alpha_0 + T)} \ket{E_i}   = \ket{\alpha_0 + T} = \ket{\alpha_0} .
\end{split}	
\end{equation}
Equation (\ref{3stepriportaallinizio}) demonstrates that the translation of a minimal step applied to the state $\ket{\alpha_m}$ with $m=s$ bring the state to the initial state $\ket{\alpha_0}$.

Now we focus on the relation (\ref{33333}): this expression is a resolution of the identity. Thus, although the $\alpha$ quantity in not an observable represented by a Hermitian operator on the space $\mathcal{H}_p$ of dimension $p+1$, it can be represented by a POVM. The $s+1$ nonorthogonal elements of this POVM are

\begin{equation}
\frac{p+1}{s+1} \ket{\alpha_m}\bra{\alpha_m} 
\end{equation}
and using equation (\ref{33333}) we can expand a general state of the system (given from equation (\ref{3genericsystem})) as

\begin{equation}
\ket{\psi} = \frac{p+1}{s+1} \sum_{m=0}^{s}\ket{\alpha_m}\braket{\alpha_m | \psi}.
\end{equation}
Considering the normalization of the $\ket{\psi}$ ($\braket{\psi|\psi} = 1$), we obtain

\begin{equation}\label{3povm}
\frac{p+1}{s+1} \sum_{m=0}^{s} \left| \braket{\alpha_m|\psi} \right|^2 = 1.
\end{equation} 	
As Pegg emphasizes, each term in equation (\ref{3povm}) is positive and <<represent a probability with the total probability correctly normalized>>. According to quantum detection theory, this is the probability that the application of the POVM (through an appropriate measuring instrument) will give the result $\alpha_m$.
\begin{figure} 
	\centering
	\includegraphics [height=11.5cm, angle=-90]{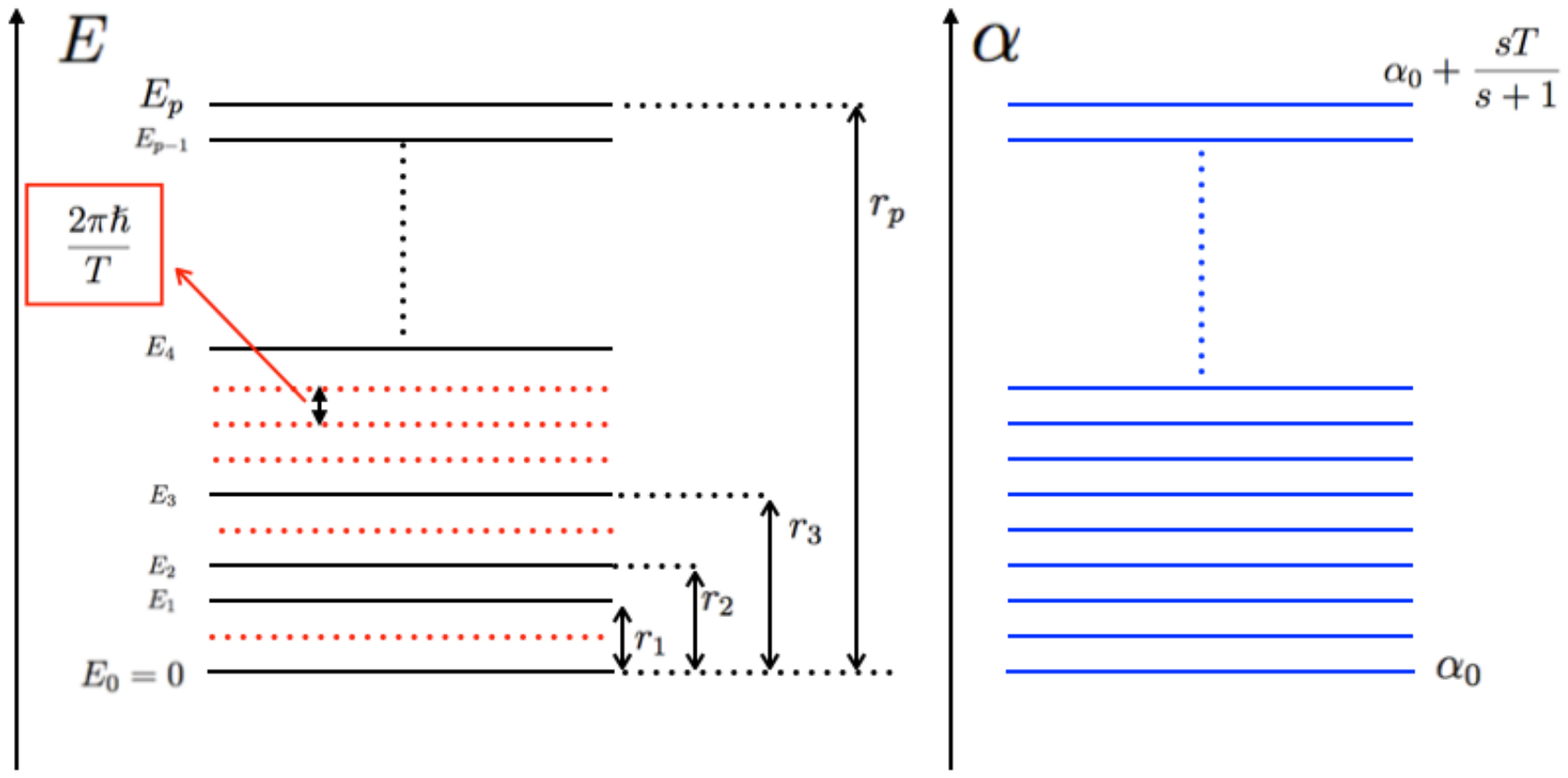} 
	\caption{Schematic representation of the $p+1$ values of the energy spectrum (on the left) and the $s+1 > p+1$ values of $\alpha_{m}$ (on the right) in a system system with unequally-spaced energy levels. The red dashed lines in the energy spectrum indicate the minimum distance between two successive values in the eigenvalues of $\hat{H}_p$ so that all other energy values can be considered as multiples of this minimum step, they are not energy levels.} 
	\label{3peggunequally} 
\end{figure}

\subsection{The $s \longrightarrow \infty$ Limit}
In order for the $\alpha$ quantity to be indipendent of an arbitrary choice of $s$, we can define the $\alpha$ quantity as represented by the above POVM in the limit $s \longrightarrow \infty$. In this limit, the difference between successive $\alpha_m$ values tends to zero and the probability for $\alpha$ being in the interval $\left[\alpha, \alpha + \delta\alpha \right]$ is $P(\alpha)\delta\alpha$, where the probability density $P(\alpha)$ is 
\begin{figure} 
	\centering
	\includegraphics [height=11cm, angle=-90]{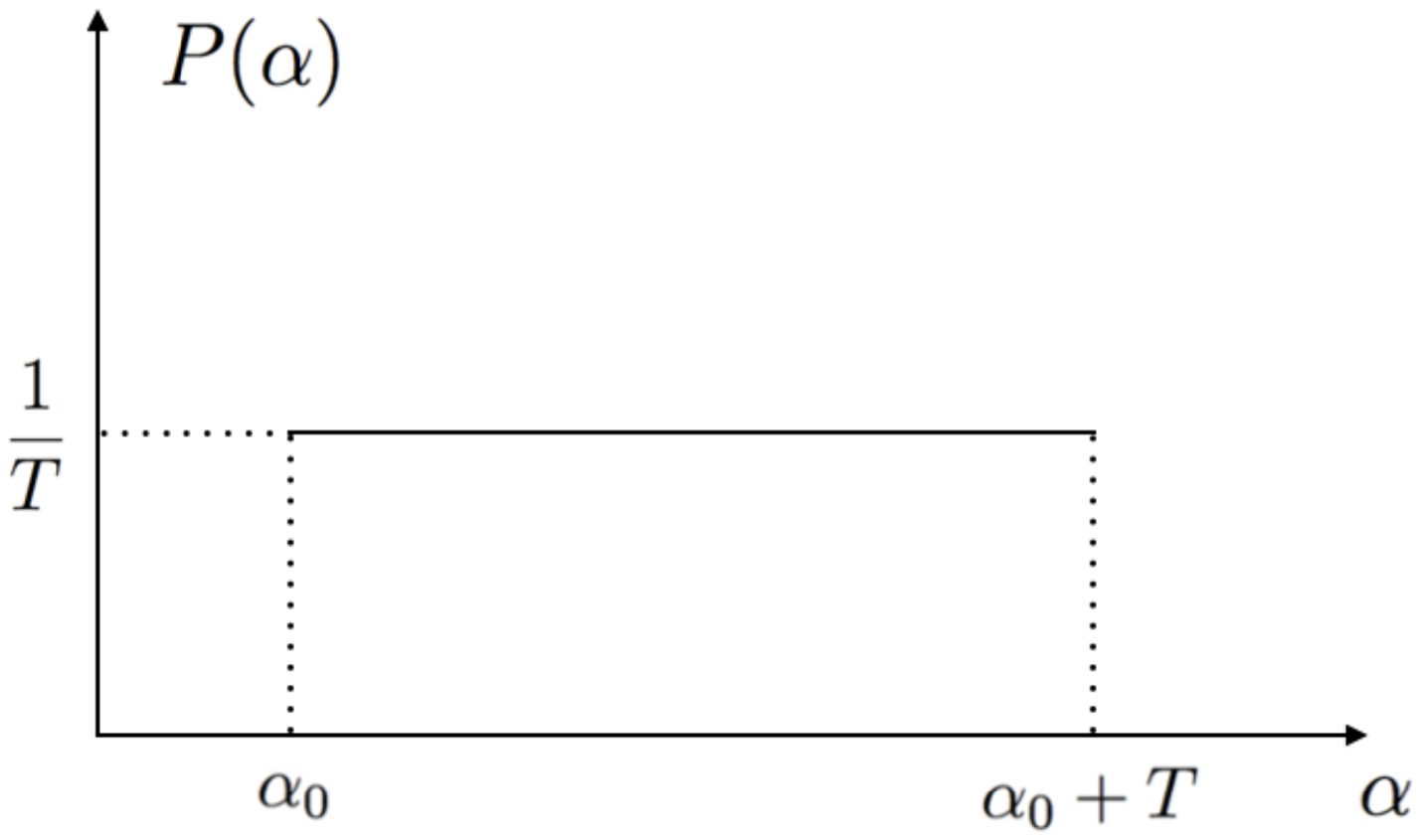} 
	\caption{Rappresentation of the probability $P(\alpha)$ when the system is in an energy eigenstate. The $\alpha$ quantity distribution is constant across the whole period $T$ according to the uncertainty relation.} 
	\label{3graficopiattezzaprob} 
\end{figure}
\begin{equation}\label{318pegg}
P(\alpha) = \frac{1}{T} \left| \braket{\tilde{\alpha}|\psi} \right|^2 
\end{equation}   
with
\begin{equation}\label{3alfatilde}
\ket{\tilde{\alpha}} = \sqrt{p+1}\ket{\alpha} = \sum_{i=0}^{p} e^{-i E_i \alpha} \ket{E_i}.
\end{equation}   
The resolution of the identity (\ref{33333}) now becomes 
\begin{equation}\label{3identitacontinuo}
\frac{1}{T}\int_{\alpha_0}^{\alpha_0 + T} \ket{\tilde{\alpha}}\bra{\tilde{\alpha}}d\alpha = \mathbb{1}_p
\end{equation} 
and the $\alpha$ quantity is now represented by the POVM generated by the operators $\frac{1}{T}\ket{\tilde{\alpha}}\bra{\tilde{\alpha}}d\alpha $. For the system in the state $\ket{\psi}$, the probability density of obtaining the value $\alpha$ for the $\alpha$ quantity is 
\begin{equation}
P(\alpha) = \frac{1}{T} \left| \sum_{i=0}^{p}f_ie^{i E_i \alpha}   \right|^2 .
\end{equation}
This expression clearly shows the complementarity between energy and the $\alpha$ quantity: if the system is in an energy eigenstate, 
we have $P(\alpha) = \frac{1}{T}$, namely the distribution of the $\alpha$ quantity is constant for the entire period $T$. Thus, if the energy is determined exactly, the $\alpha$ quantity is totally random.

\subsection{The Uncertainty Relation} 
In order to find the uncertainty relation between $\Delta E$ and $\Delta \alpha$, Pegg introduces a new Hermitian operator $\hat{A}$, defined by

\begin{equation}\label{3A}
\hat{A} = \frac{1}{T} \int_{\alpha_0}^{\alpha_0 + T} \alpha \ket{\tilde{\alpha}}\bra{\tilde{\alpha}} d\alpha,
\end{equation} 
that can be expressed, using (\ref{3alfatilde}), as

\begin{equation}\label{3A2}
\hat{A} =  \alpha_0 + \frac{T}{2} + i  \sum_{i \ne j} \frac{e^{-i (E_i - E_j) \alpha_0 }}{E_i - E_j} \ket{E_i}\bra{E_j}.
\end{equation} 
The operator $\hat{A}$ has the following property:

\begin{equation}\label{3jkugv}
\begin{split}
\bra{\psi} \hat{A} \ket{\psi} & = \frac{1}{T} \bra{\psi} \int_{\alpha_0}^{\alpha_0 + T} \alpha \ket{\tilde{\alpha}} \bra{\tilde{\alpha}}d\alpha \ket{\psi} = \\& =  \frac{1}{T} \int_{\alpha_0}^{\alpha_0 + T} \alpha \left| \braket{\tilde{\alpha}|\psi}  \right|^2 d\alpha  = \\& = \int_{\alpha_0}^{\alpha_0 + T} \alpha P(\alpha) d\alpha = \:\: <\alpha> .
\end{split}
\end{equation} 
Pegg notices that the operator $\hat{A}$ can not be taken as the operator complement to the Hamiltonian. <<The essential reason is that the eigenstates of $\hat{A}$ are not the states $\ket{\alpha}$ and the Hamiltonian does not generate shifts from one eigenstate of $\hat{A}$ to another>> how instead of occurs in equation (\ref{3tras}). $\hat{A}$ is interesting because <<its expectation value is equal to the expectation value of the $\alpha$ quantity, so it might provide a way of measuring this value, but it can be considered only as the operator acting on $\mathcal{H}_p$ that is nearest to an operator conjugate to $\hat{H}_p$>>. 
Thus, as the $\alpha$ quantity is represented by a POVM, we can not write down immediately its uncertainty relation, as in the case for observables represented by Hermitian operators. Nevertheless, using the resolution of the identity (\ref{3identitacontinuo}), Pegg writes the variance of the energy as

\begin{equation}\label{3pollo}
\begin{split}
<\Delta E^2> & = \bra{\psi}(\hat{H}_p - <E>)(\hat{H}_p - <E>)\ket{\psi} = \\& = \frac{1}{T} \int_{\alpha_0}^{\alpha_0 + T} \bra{\psi}(\hat{H}_p - <E>)  \ket{\tilde{\alpha}} \bra{\tilde{\alpha}} (\hat{H}_p - <E>) \ket{\psi} d\alpha
\end{split}
\end{equation} 
and the variance of the $\alpha$ quantity as

\begin{equation}\label{3pollo2}
\begin{split}
<\Delta\alpha^2> & = \bra{\psi}(\alpha - <\alpha>)(\alpha - <\alpha>)\ket{\psi} = \\& = \frac{1}{T} \int_{\alpha_0}^{\alpha_0 + T} \bra{\psi}(\alpha - <\alpha>)  \ket{\tilde{\alpha}} \bra{\tilde{\alpha}} (\alpha - <\alpha>) \ket{\psi} d\alpha.
\end{split}
\end{equation} 
Combining these two results, we obtain

\begin{multline}\label{3pollo3}
<\Delta E^2><\Delta\alpha^2>  = \frac{1}{T^2} \int_{\alpha_0}^{\alpha_0 + T}\left| \bra{\psi}  (\hat{H}_p - <E>) \ket{\tilde{\alpha}}\right|^2d\alpha \\ \times \int_{\alpha_0}^{\alpha_0 + T} \left| \bra{\psi}  (\alpha - <\alpha>)  \ket{\tilde{\alpha}} \right|^2 d\alpha
\end{multline}
and from Schwartz's inequality: 

\begin{multline}
\frac{1}{T^2} \int_{\alpha_0}^{\alpha_0 + T}\left| \bra{\psi}  (\hat{H}_p - <E>) \ket{\tilde{\alpha}}\right|^2d\alpha \int_{\alpha_0}^{\alpha_0 + T} \left| \bra{\psi}  (\alpha - <\alpha>)  \ket{\tilde{\alpha}} \right|^2 d\alpha
\\ \ge \frac{1}{T^2} \left| \int_{\alpha_0}^{\alpha_0 + T}  \bra{\psi}(\hat{H}_p - <E>)\ket{\tilde{\alpha}} \bra{\tilde{\alpha}} (\alpha - <\alpha>) \ket{\psi} \right|^2.
\end{multline}
Looking at the product of the mean square deviations we have:

\begin{equation}
\Delta E \Delta\alpha \ge  \left|\bra{\psi} (\hat{H}_p - <E>)  (\hat{A} - <\alpha>)\ket{\psi}\right|
\end{equation}
and, being $\alpha$ a real value and $\hat{A}$ an Hermitian operator, we can write \cite{pegg}:
\begin{equation}\label{3principio1}
\left|\bra{\psi} (\hat{H}_p - <E>)  (\hat{A} - <\alpha>)\ket{\psi}\right| =   \frac{1}{2} \left| \bra{\psi} [ \hat{H}_p,\hat{A} ]     \ket{\psi}  \right| .
\end{equation} 
Then, using the fact that 

\begin{equation}\label{3principio2}
\begin{split}
\left[ \hat{H}_p , \hat{A} \right] & = i \left(\sum_{i \ne j} \frac{E_i}{E_i - E_j} e^{-i  (E_i - E_j) \alpha_0} \ket{E_i}\bra{E_j}  -  \sum_{i \ne j} \frac{E_j}{E_i - E_j} e^{-i  (E_i - E_j) \alpha_0} \ket{E_i}\bra{E_j}\right) = \\& = i \left( \sum_{i} \sum_{j}e^{-i (E_i - E_j) \alpha_0}e^{i (E_i - E_j) \alpha_0}  \ket{E_i}\bra{E_j} - \mathbb{1}\right) = i \left(\ket{\tilde{\alpha_0}}\bra{\tilde{\alpha_0}} - \mathbb{1}\right),
\end{split}
\end{equation} 
we finally obtain

\begin{equation}\label{3principio3}
\Delta E \Delta\alpha \ge  \frac{1}{2} \left| \bra{\psi} [ \hat{H}_p,\hat{A} ]     \ket{\psi}  \right| =  \frac{1}{2} \left|1- \braket{\psi | \tilde{\alpha_0}}\braket{\tilde{\alpha_0} | \psi}        \right|    .
\end{equation} 
Pegg notices here that, <<when the system is in an energy eigenstate the uncertainty in energy must be zero, even for a finite $\Delta \alpha$. In this case the right-hand side of equation (\ref{3principio3}) vanishies, ensuring consinstency>>. Furthermore, the term $\braket{\psi|\alpha_0}\braket{\alpha_0|\psi}$ is just $TP(\alpha_0)$ and therefore, <<if $\ket{\psi}$ is orthogonal to $\ket{\alpha_0}$ or at least if the probability distribution $P(\alpha)$ for the state $\ket{\psi}$ of the system is sufficiently narrow for this last term to be negligible, the uncertainty product takes the more usual form>>.

\section{Equally-Spaced Energy Levels}
In this Section we slightly deviate from Pegg's work and we consider an Hamiltonian with equally-spaced eigenvalues. We will see that for this kind of systems we can find an Hermitian operator to represent the $\alpha$ quantity. 

\begin{figure} [!t]
	\centering
	\includegraphics [height=13cm, angle=-90]{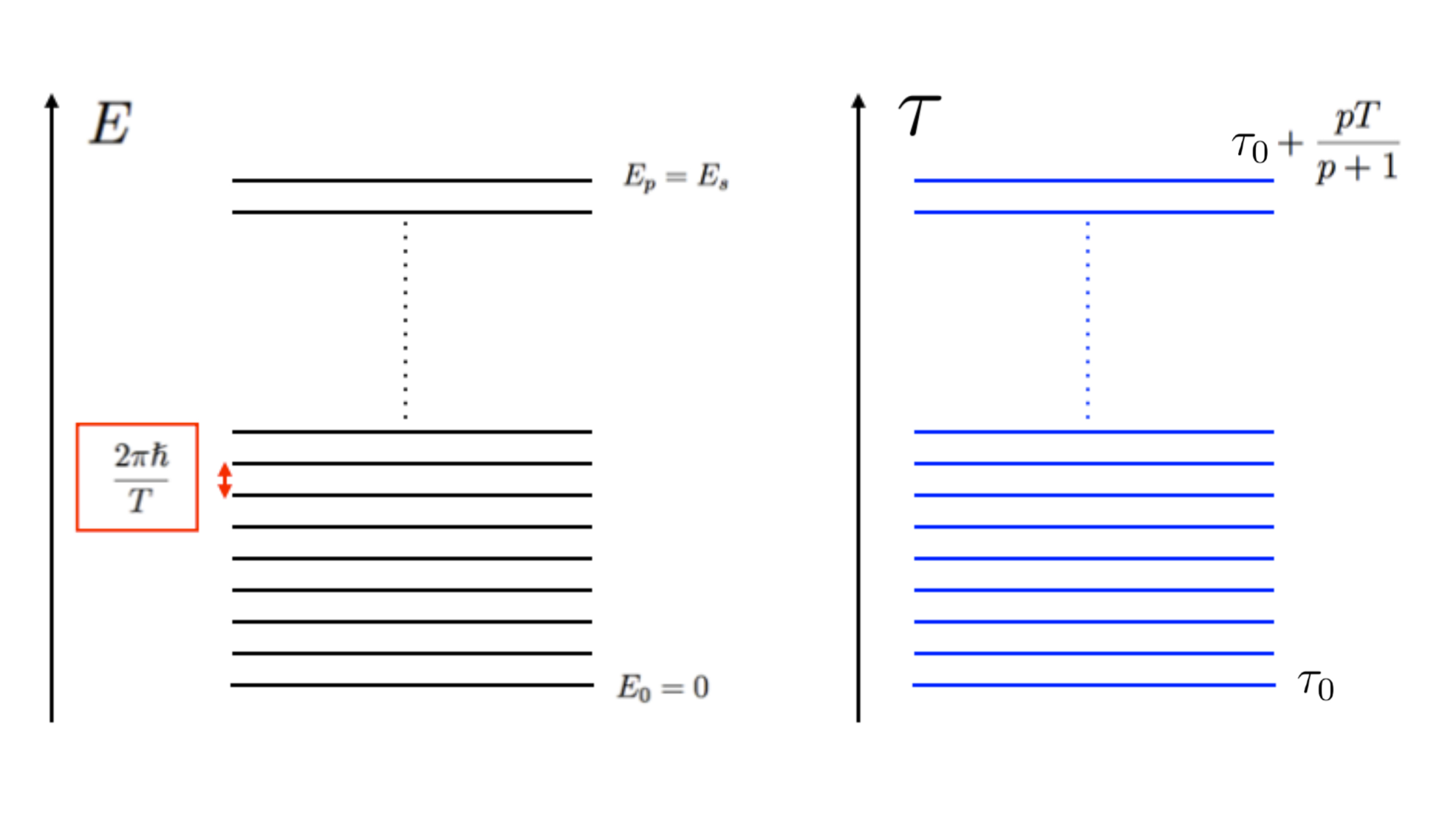} 
	\caption{Schematic representation of the energy spectrum (on the left) and the $p+1$ values of $\tau_{m}$ (on the right) in a system with equally-spaced energy levels.} 
	\label{3peggequally} 
\end{figure}

We have already seen that the operator we are searching can not be the operator $\hat{A}$ represented in Equation (\ref{3A}). Let's consider a system with $p+1$ non-degenerate equally-spaced energy levels. For simplicity we consider $E_0 = 0$. In this framework we notice that we are allowed to choose $s=p$. Indeed, in the case of equally-spaced energy levels we have $r_i = i$ which implies $r_p=p$. The constraint we have imposed on the numeber of the $\alpha$ states ($s+1>r_p$) then becomes 
\begin{equation}\label{3jjjjj}
\begin{split}
   s+1>r_p = p \:\: \Rightarrow \:\: s \ge p.
\end{split}
\end{equation}
Thus the key point of this Section is the choice of $s=p$. Following the path outlined in the previous Section, the energy eigenvalues can be written 

\begin{equation}\label{3jjjj}
E_n =   r_n \frac{2\pi }{T} =  n \frac{2\pi }{T}     
\end{equation}
(with $n=0,1,2,...,p$) and we can define the states $\ket{\tau_m}$ as

\begin{equation}\label{3taumket}
\ket{\tau_m}  = \frac{1}{\sqrt{p+1}} \sum_{n=0}^{p} e^{-i E_n \tau_m} \ket{E_n} 
\end{equation}
where

\begin{equation}\label{3jj}
\tau_m = \tau_0 + m\frac{T}{s+1} = \tau_0 + m\frac{T}{p+1}.
\end{equation}
As the $\alpha_{m}$ values of the previous Section, the $\tau_m$ values are spread on the range $\left[\tau_0, \tau_0 + T\right]$ but they are equal in number to the energy levels ($s=p$). For this reason, we can now consider that all the summations run between $0$ and $p$. The states $\ket{\tau_m}$ are orthogonal, namely (see Appendix 3.5.A at the end of this Chapter)

\begin{equation}
	\braket{\tau_{m}|\tau_{m'}} = \delta_{m,m'} 
\end{equation}
and they form a complete basis on the space $\mathcal{H}_p$. Indeed we have

\begin{equation}\label{3j}
\begin{split}
\sum_{m=0}^{p} \ket{\tau_m}\bra{\tau_m} & =\frac{1}{p+1} \sum_{m=0}^{p} \sum_{n=0}^{p} \sum_{n'=0}^{p} e^{-i  (E_n -E_{n'}) \tau_m} \ket{E_n} \bra{E_{n'}} = \\& 
= \frac{1}{p+1} \sum_{n=0}^{p} \sum_{n'=0}^{p}  e^{-i(E_n -E_{n'}) \tau_0} \sum_{m=0}^{p} e^{-i \frac{2\pi m}{p+1}(n-n')} \ket{E_n} \bra{E_{n'}} = \\& 
= \sum_{n=0}^{p} \sum_{n'=0}^{p} \delta_{n,n'} \ket{E_n} \bra{E_{n'}} = \sum_{n=0}^{p} \ket{E_n} \bra{E_{n'}} = \mathbb{1}_p, 
\end{split}
\end{equation}
where $\mathbb{1}_p$ is the Identity operator on the $p+1$ dimensional Hilbert space, and where we have used $\sum_{m=0}^{p} e^{-i (E_n -E_{n'}) \tau_m} = e^{-i(E_n -E_{n'}) \tau_0} \sum_{m=0}^{p} e^{-i \frac{2\pi m}{p+1}(n-n')}=(p+1)\delta_{n,n'}$. Furthermore the states are now the eigenstates of the Hermitian operator

\begin{equation}\label{3tauop}
\hat{\tau}  =  \sum_{m=0}^{p} \tau_m \ket{\tau_m}\bra{\tau_m} 
\end{equation}
with eigenvalues $\tau_m$ (see Appendix 3.5.B at the end of this Chapter). The term $\left| \braket{\tau_m|\psi} \right|^2$ is now the probability of projecting the state $\ket{\psi}$ on the state $\ket{\tau_m}$, that is the probability of obtaining the result $\tau_m$ by a measurement of the operator $\hat{\tau}$ given by equation (\ref{3tauop}). This is in accord with the result of the previous Section (\ref{3povm}), but now we can relate the measurement to the Hermitian operator $\hat{\tau}$. 

We now show that $\hat{\tau}$ is the conjugate, or complement, of $\hat{H}_p= \sum_{n=0}^{p}E_n\ket{E_n}\bra{E_n}$ in the sense that $\hat{H}_p$ is the generator of shifts in the values $\tau_m$ (eigenvalues of $\hat{\tau}$) and $\hat{\tau}$ is the generator of energy shifts. The first property, that follows from the definition of the states $\tau_m$ (\ref{3taumket}), reads

\begin{equation}
\begin{split}
e^{-i  \hat{H}_p (\tau - \tau_{0}) }\ket{\tau_0} & = \frac{1}{\sqrt{p+1}} \sum_{n=0}^{p} e^{-i E_n (\tau - \tau_{0}) } e^{-i E_n \tau_0} \ket{E_n} = \\& 
= \frac{1}{\sqrt{p+1}} \sum_{n=0}^{p} e^{-i E_n \tau} \ket{E_n} = \ket{\tau_{m}} 
\end{split}
\end{equation}  
where $\tau - \tau_{0}$ is an integer multiple of $\frac{T}{p+1}$. Clearly the cyclical repetition we found in the previous Section for the values $\alpha_m$ and the states $\ket{\alpha_{m}}$ still applies here for the values $\tau_{m}$ and the states $\ket{\tau_m}$ and we have $e^{-i  \hat{H}_p \frac{T}{p+1}} \ket{\tau_{m=p}} = \ket{\tau_0 + T} = \ket{\tau_0}$. The second property follows from the relation

\begin{equation}
\begin{split}
e^{i  \hat{\tau} (E_n - E_0) }\ket{E_0} & = \frac{1}{\sqrt{p+1}} \sum_{m=0}^{p} e^{i \tau_m (E_n - E_0)} e^{i \tau_m E_0} \ket{\tau_m} = \\& = \frac{1}{\sqrt{p+1}} \sum_{m=0}^{p}  e^{i \tau_m E_n} \ket{\tau_m} = \ket{E_n}
\end{split}
\end{equation} 
where $E_n - E_0$ is a multiple of $\frac{2\pi}{T}$ and where we used the fact that $\ket{E_n}$ can be expanded in terms of the \lq\lq time\rq\rq states basis $\left\{ \ket{\tau_m} \right\}$ as

\begin{equation}
\begin{split}
\ket{E_n} &= \sum_{m=0}^{p} \ket{\tau_m}\braket{\tau_m|E_n} =\\&= \frac{1}{\sqrt{p+1}}\sum_{m=0}^{p} e^{i  E_n \tau_m} \ket{\tau_m}.
\end{split}
\end{equation} 

Finally, to conclude this discussion we want to give the expression of the commutator between the $\hat{\tau}$ operator and the Hamiltonian operator, as this will be useful in the following. Using the definition of the Hamiltonian $\hat{H}_p$ and considering equations (\ref{3taumket}) and (\ref{3tauop}), we can find:

\begin{equation}\label{3comm}
\left[ \hat{\tau}, \hat{H}_p\right] =  \frac{T}{p+1} \sum_{n \ne n'} \frac{(E_n - E_{n'}) e^{i (E_n - E_{n'}) \tau_0}}{e^{i(E_n - E_{n'})T/(p+1)} - 1} \ket{E_{n'}}\bra{E_n}.
\end{equation} 
This is the commutator between $\hat{\tau}$ and the Hamiltonian $\hat{H}_p$ which can also be used to calculate the uncertainty relation in this particular case of Hamiltonian with equally-spaced energy levels. We do not elaborate further as it does not serve our purposes. For additional discussions we refer to \cite{pegg} (see also \cite{opfase}).

\section{Conclusions}
The questions Pegg has addressed in his work is basically whether or not there is a quantity connected to time that is complemetary to the Hamiltonian of a quantum system in the particular case of a finite system with discrete energy levels. Through Pegg's work we have seen that this quantity (the $\alpha$ quantity) can be represented by a POVM generated by the operators $\frac{p+1}{s+1} \ket{\alpha_m}\bra{\alpha_m}$, or $\ket{\tilde{\alpha}}\bra{\tilde{\alpha}} d\alpha$ in the limit $s \longrightarrow \infty$. In the special case of equally-spaced energy levels, we defined the Hermitian operator $\hat{\tau}$ complement of the Hamiltonian in the sense that $\hat{H}_p$ is the generator of shifts in the values $\tau_m$ (eigenvalues of $\hat{\tau}$) and $\hat{\tau}$ is the generator of energy shifts. 

Therefore, in conclusion, Pegg's $\alpha$ quantity has dimensions of time but it is not time. As we mentioned at the beginning of this Chapter, the $\alpha$ quantity is property of the system itself. It represent an observable of the quantum system and not time in the coordinate sense, as read on an external clock. For some states of the system, the rate of change of the mean value $<\alpha>$ in time can approach unity, but in other situation (i.e. the system is in an energy eigenstate), $<\alpha>$ does not change with time. This can be seen simply by noting that, considering the more general situation in which we have unequally-spaced energy levels and considering the $s \longrightarrow \infty$ limit, we have \cite{pegg}:

\begin{equation}
\begin{split}
\frac{d<\alpha >}{dt} & = i \bra{\psi} \left[ \hat{H}_p , \hat{A}\right] \ket{\psi} = \\&= 1 - \braket{\psi|\tilde{\alpha_0}} \braket{\tilde{\alpha_0}|\psi}
\end{split}
\end{equation}
where we used the property (\ref{3jkugv}) for $< \alpha >$ and equation (\ref{3principio2}) to evaluate the commutator between $\hat{H}_p$ and $\hat{A}$. So Pegg suggests calling this quantity with the name \lq\lq Age\rq\rq. We will see in the following how Age can be useful and find a revised physical justification within Page and Wooters theory.


\section{Appendices}

\subsection*{3.5.A \: Orthogonality of states $\ket{\tau_m}$}
We can easily see that the states $\ket{\tau_m}$ are orthogonal. Indeed we can easily calculate the scalar product $\braket{\tau_{m'} | \tau_m}$, where 

\begin{equation}
\ket{\tau_m} = \frac{1}{\sqrt{p+1}}\sum_{n=0}^{p} e^{-i E_n \tau_m} \ket{E_n} .
\end{equation} 
We have:
\begin{equation}
	\begin{split}
	\braket{\tau_{m'} | \tau_m} & = \frac{1}{p+1} \sum_{n=0}^{p} \sum_{n'=0}^{p} e^{-i (E_n \tau_m - E_n'\tau_{m'})} \braket{E_n'|E_n}	= 
	\\& = \frac{1}{p+1}  \sum_{n=0}^{p} e^{-i E_n (\tau_m - \tau_{m'})} =\\&=\frac{1}{p+1}  \sum_{n=0}^{p} e^{-i n \frac{2\pi }{T}\left(\tau_{m} - \tau_{m'}\right)}= 
	\\& = \frac{1}{p+1}  \sum_{n=0}^{p} e^{-i n \frac{2\pi}{p+1}\left(m - m' \right)}= \:\: \delta_{m,m'}.
	\end{split}		
\end{equation}
In this calculation we used the fact that $\sum_{n=0}^{p} e^{-i  \frac{2\pi n}{p+1} (m-m')} = (p+1)\delta_{m,m'}$. 

\subsection*{3.5.B \: $\ket{\tau_{m}}$ as eigenstates of $\hat{\tau}$}
We show here that the states $\ket{\tau_{m}}$ are eigenstates of the operator $\hat{\tau}$. Through the definitions (\ref{3taumket}) and (\ref{3tauop}) we have indeed:

\begin{equation} 
	\begin{split}
	\hat{\tau}\ket{\tau_m} & = \frac{1}{\sqrt{p+1}}  \sum_{m'=0}^{p}  \sum_{n=0}^{p} \tau_m'  e^{-iE_n\tau_m}  \ket{\tau_m'}\braket{\tau_m'|E_n} =
	\\& =  \frac{1}{p+1} \sum_{m'=0}^{p}  \sum_{n=0}^{p}  \tau_m'  e^{-iE_n\tau_m}  \ket{\tau_m'} \sum_{n'=0}^{p} e^{i\tau_m' E_n'}\braket{E_n'|E_n} =
	 \\& =   \frac{1}{p+1} \sum_{m'=0}^{p}  \sum_{n=0}^{p} \tau_m'  e^{-iE_n\tau_m}  \ket{\tau_m'} \sum_{n'=0}^{p} e^{i\tau_m' E_n'} \delta_{E_n,E_n'} =
	  \\& =  \frac{1}{p+1}  \sum_{m'=0}^{p}  \sum_{n=0}^{p} \tau_m'  e^{-iE_n(\tau_m - \tau_m')}  \ket{\tau_m'}  =
	   \\&=  \sum_{m'=0}^{p}   \tau_m'  \delta_{m,m'}  \ket{\tau_m'}   = \tau_m \ket{\tau_m} 
	\end{split}
	\end{equation}
	where we used again $\sum_{n=0}^{p} e^{-iE_n(\tau_m - \tau_m')} = \sum_{n=0}^{p}e^{-i\frac{2\pi n}{p+1}(m - m')} = (p+1)\delta_{m,m'}$. 

	\chapter{Time Observables within a Timeless Universe}
\label{timeobservables}

In this Chapter\footnote{Some parts of this Chapter are reprinted with permission from T. Favalli and A. Smerzi, \textit{Time Observables in a Timeless Universe}, Quantum \textbf{4}, 354 (2020).} we incorporate Pegg's formalism \cite{pegg} into Page and Wootters (PaW) theory\cite{pagewootters,wootters}, following \cite{nostro}. We start considering a closed quantum system that in the following we call the \lq\lq Universe\rq\rq. The Hilbert space of the Universe is composed by  a \lq\lq clock-subspace\rq\rq $C$ that keeps track of time and the \lq\lq system\rq\rq $S$ of the rest of the Universe. In absence of an external temporal reference frame we write the Schrödinger equation of the Universe as ($\hslash=1$): 

\begin{equation}\label{4sc}
(\hat{H}_S - i \frac{\partial}{\partial t_C} )\ket{\Psi} = 0 
\end{equation}
where the first term $\hat{H}_S$ is the Hamiltonian of $S$. We interpret the second term as a (possibly approximate) time representation of the clock-subspace Hamiltonian:

\begin{equation}\label{4ch}
- i  \frac{\partial}{\partial t_C} \to \hat{H}_C .
\end{equation}
Under the implicit assumption that the two subsystems $C$ and $S$ are not interacting, equation (\ref{4sc}) becomes\footnote{Equation (\ref{4scm}) represents the global constraint to which the whole Universe is subjected.}: 

\begin{equation} \label{4scm}
(\hat{H}_S + \hat{H}_C)\ket{\Psi} = 0 .
\end{equation}
The time representation of the clock Hamiltonian (\ref{4ch}) would be correct, $- i \frac{\partial}{\partial t_C} = \hat{H}_C$, only if $\hat{H}_C$ has a continuous, unbounded spectrum (as we see in Chapter 2 when we discussed the GLM proposal for the clock observable \cite{lloydmaccone}). In this case we could write the Hermitian time operator in the energy representation as $\hat T = - i \frac{\partial}{\partial E_C}$. Since we consider here an isolated physical system of finite size, the introduction of unbounded Hamiltonians with a continuous spectrum would not be possible. As a consequence, an Hermitian time operator written in differential form as in equation (\ref{4sc}) can not be introduced within standard approaches \cite{pauli}. 

As we saw in the previous Chapter, a way out was considered by Pegg \cite{pegg} who introduced the quantity \lq\lq Age\rq\rq. In the case of equally-spaced energy eigenvalues Age is described by an Hermitian operator complement of a bounded Hamiltonian. Pegg also considered a larger set of Hamiltonians having unequally-spaced energy eigenvalues. In this latter case we saw that the quantity Age is described by a POVM. The central result of this Chapter is to show that the Pegg formalism finds a sound physical interpretation when incorporated in the PaW framework. We initially introduce, as \lq\lq good\rq\rq clock, a device described by an Hamiltonian having equally-spaced eigenvalues. In this case time is described by an Hermitian operator. A device governed by an Hamiltonian having a generic spectrum can still provide a good clock mathematically described by a POVM. We show that in both cases we recover the Schr\"odinger dynamical evolution of the system $S$.

\section{Equally-Spaced Energy Levels for the Clock Hamiltonian}
\label{Time from Entanglement}

\subsection{The Clock Subspace}
\label{TheClockSubspace}
The first problem to deal with is the introduction of a good clock. We define as good clock a physical system governed by a lower-bounded Hamiltonian having discrete, equally-spaced, energy levels (a generalisation to unequally-spaced levels will be discussed in the following): 

\begin{equation}
\hat{H}_C= \sum_{n=0}^{s} E_n \ket{E_n}\bra{E_n}
\end{equation}
where $p+1$ is the dimension of the clock space (see also \cite{simile,simile2}). We now search for an Hermitian observable $\hat{\tau}$ in the clock space that is conjugated to the clock Hamiltonian $\hat{H}_C$. We define the time states 

\begin{equation}\label{4taumautovet}
\ket{\tau_m}_C = \frac{1}{\sqrt{p+1}} \sum_{n=0}^{p} e^{-i E_n \tau_m} \ket{E_n}_C 
\end{equation}
with 

\begin{equation}
	\tau_m = \tau_0 + m\frac{T}{p+1},
\end{equation}
where $E_n = E_0 + n\frac{2\pi}{T}$ and $m,n=0,1,...,p$. Equation (\ref{4taumautovet}) provides an orthonormal and complete basis since 

\begin{equation}\label{4ort}
\braket{\tau_m|\tau_{m'}}=\delta_{m,m'}
\end{equation}
and

\begin{equation}\label{4identity}
\sum_{m=0}^{p} \ket{\tau_m}\bra{\tau_m} = \mathbb{1}_C.
\end{equation}
With the states (\ref{4taumautovet}) we can define the Hermitian operator

\begin{equation}\label{4tauop}
\hat{\tau} = \sum_{m=0}^{p} \tau_m \ket{\tau_m}\bra{\tau_m}
\end{equation}
that is conjugated to the Hamiltonian $\hat{H}_C$. It is indeed easy to show\footnote{Proofs of (\ref{4ort}), (\ref{4identity}), (\ref{4property11}) and (\ref{4aaa}) are given in the Section 3.3 of the previous Chapter.} that $\hat{H}_C$ is the generator of shifts in $\tau_m$ values and, viceversa, $\hat{\tau}$ is the generator of energy shifts:

\begin{equation}\label{4property11}
\ket{\tau_m}_C = e^{-i \hat{H}_C(\tau_m - \tau_0)}\ket{\tau_0}_C
\end{equation}
and

\begin{equation}\label{4aaa}
\ket{E_n}_C = e^{i  \hat{\tau} (E_n - E_0)} \ket{E_0}_C	.
\end{equation}
A second important property of the clock states is their ciclic condition: $\ket{\tau_{m=p+1}}=\ket{\tau_{m=0}}$. The time taken by the system to return to its initial state is $T=\frac{2\pi}{\delta E}$ with $\delta E$ being the spacing between two neighbouring energy eigenvalues. Conversely, the smallest time interval is

\begin{equation}\label{42c}
\delta\tau= \tau_{m+1} -\tau_{m} = \frac{2\pi}{\delta E \left(p+1\right)} .
\end{equation}
To summarise: the greater is the spectrum of the clock Hamiltonian, the smaller is the spacing $\delta \tau$ between two eigenvalues of the clock. 
The smaller is the distance between two eigenvalues of the clock energy, the larger the range $T$ of the eigenvalues $\tau_m$. We conclude that a \lq\lq good clock\rq\rq is a system with a very small spacing between energy levels and a very large number of eigenvalues.

\subsection{Dynamics}
\label{EvolutionSingleState}
We consider the total Hilbert space of the Universe $\mathcal{H} = \mathcal{H}_C \otimes \mathcal{H}_S$, with $\mathcal{H}_C$ and $\mathcal{H}_S$ 
having dimension $d_C = p+1$ and $d_S$ respectively. We require that our \lq\lq good clock\rq\rq has $d_C \gg d_S$. A general bipartite state of the Universe can be written as

\begin{equation}
\ket{\Psi} = \sum_{n=0}^{d_C-1} \sum_{k=0}^{d_S-1} c_{n,k} \ket{E_n}_C \otimes \ket{E_k}_S.
\end{equation}
We impose the constraint (\ref{4scm}) $\hat{H}\ket{\Psi}=0$ and, under the assumption that the spectrum of the clock Hamiltonian is sufficiently dense (namely, that to each energy state of the system $S$ there is a state of the clock for which equation (\ref{4scm}) is satisfied), we obtain for the state of the Universe

\begin{equation}
\ket{\Psi} = \sum_{k=0}^{d_S-1} \tilde{c}_{k} \ket{E=-E_k}_C \otimes \ket{E_k}_S
\end{equation}
with $\sum_{k} \left|\tilde{c}_{k} \right|^2=1$. With the resolution of the identity (\ref{4identity}), we write

\begin{equation}\label{4stato1}
\begin{split}
\ket{\Psi} &= 
\sum_{m=0}^{d_C - 1} \ket{\tau_{m}}\braket{\tau_m|\Psi} = 
\\&= \frac{1}{\sqrt{d_C}} \sum_{m=0}^{d_C-1} \ket{\tau_m}_c \otimes  \sum_{k=0}^{d_S-1}\tilde{c}_{k} e^{-iE_k\tau_m}\ket{E_k}_S.
\end{split}
\end{equation}
By writing a generic state of the system as
$\ket{\phi(\tau_{m})}_S= \sum_{k=0}^{d_S-1}\tilde{c}_{k} e^{-iE_k\tau_m}\ket{E_k}_S$, the state (\ref{4stato1}) becomes

\begin{equation}\label{4stato2}
\ket{\Psi}  =\frac{1}{\sqrt{d_c}} \sum_{m=0}^{d_C-1} \ket{\tau_m}_C \otimes \ket{\phi(\tau_{m})}_S .  
\end{equation}
It is interesting to note, and we emphasize, that the state $\ket{\phi(\tau_{m})}_S$ is related to the the global $\ket{\Psi}$ of the Universe by 

\begin{equation}\label{4definition}
\ket{\phi(\tau_{m})}_S = \frac{\braket{\tau_m|\Psi}}{1/\sqrt{d_C}} 
\end{equation}
that is the Everett \textit{relative state} definition of the subsystem $S$ with respect to the clock system $C$ \cite{everett}. As
already mentioned, this kind of projection has nothing to do with a measurement. Rather, $\ket{\phi(\tau_{m})}_S$ are a  1-parameter family of states, each describing the state of $S$ conditioned to the clock $C$ in the state $\ket{\tau_{m}}_C$.

Now, following the PaW framework and using equation (\ref{4definition}), the constraint (\ref{4scm}) and 
equation (\ref{4property11}), we have:
\begin{equation}\label{4ev}
\begin{split}
\ket{\phi(\tau_{m})}_S &= \frac{\braket{\tau_{m}|\Psi}}{1/\sqrt{d_C}} = \sqrt{d_C}\bra{\tau_0}e^{i\hat{H}_C(\tau_m-\tau_0)}\ket{\Psi}=
 \\&= \sqrt{d_C}\bra{\tau_0}e^{i(\hat{H}-\hat{H}_S)(\tau_m-\tau_0)}\ket{\Psi}=\\&=\sqrt{d_C} ~ e^{-i\hat{H}_S(\tau_m-\tau_0)}\braket{\tau_0|\Psi}=
 \\&= e^{-i\hat{H}_S (\tau_{m}-\tau_0)}\ket{\phi(\tau_{0})}_S  
\end{split}
\end{equation}
where $\ket{\phi(\tau_{0})}_S =\frac{\braket{\tau_0|\Psi}}{1/\sqrt{d_C}}  = \sum_{k=0}^{d_S-1} \tilde{c}_k  e^{- i E_k \tau_0} \ket{E_k}_S$. Equation (\ref{4ev}) provides the Schrödinger evolution of $S$ with respect to the clock time. 

Now we can also consider the global state  written in the form (\ref{4stato2}) and, through (\ref{4ev}), we can consider the unitary operator $\hat{U}_{s} (\tau_{m}-\tau_0) = e^{-i \hat{H}_S (\tau_{m}-\tau_0)}$ \cite{lloydmaccone}. With this choice the state of the global system can be written as 

\begin{equation}\label{4lalla}
\begin{split}
\ket{\Psi}  =\frac{1}{\sqrt{d_C}} \sum_{m=0}^{d_C-1} \ket{\tau_m}_C \otimes \hat{U}_{S} (\tau_{m}-\tau_0) \ket{\phi_0}_S
\end{split}
\end{equation}
where the entire history of the Universe is again condensed and where, through the operator $\hat{U}_{s}(\tau_{m}-\tau_0)$, the form of the temporal evolution of the state $\ket{\phi(\tau_{m})}$ with respect to the values of $\tau_{m}$ is made explicit. Through (\ref{4lalla}) we can see that the conditional probability of obtaininig the outcome $a$ for the system $S$ when measuring the observable $A$ at a certain time $\tau_{m}$ is given, as expected, by the Born rule:

\begin{equation}
\begin{split}
P(a\: on\: S \: |\: \tau_m \: on \:C) =\frac{P(a \: on \: S , \: \tau_m \: on \: C)}{P( \tau_m \: on \: C)}  =  \left|	\bra{a}\hat{U}_{s} (\tau_{m}-\tau_0) \ket{\phi_0}\right|^2  .
\end{split}
\end{equation}

\section{The Hermitian Time Operator}

Here we show that within the PaW framework the operator Age  (called as $\hat{\tau}$ in the previous Section) has the expected properties of a Hermitian time observable. It is well known that Pauli objected about the existence of a time Hermitian operator because time is continuous and 
unbounded in the past and in the future while general Hamiltonians have a lower bounded (continuous or discrete) spectrum \cite{leonmaccone}.
Pegg's Age operator overcome the energy objection \cite{pauli} since $\hat{\tau}$ has a discrete spectrum and cyclical boundary conditions.
The question we address here is why $\hat{\tau}$ can not be considered as a proper time operator outside the PaW mechanism.
As clearly pointed out by Pegg, $\hat{\tau}$ has dimensions of time but it is a property of the quantum system, and it strongly depends on the state of the system. With a quantum system with Hamiltonian $\hat{H}$ we would be forced to consider $\hat{\tau}$ defined 
on the space of the system itself. So the evolution of the mean value of $\hat{\tau}$ operator with respect to an external time has to be constant or at least not zero, otherwise the dynamics would freeze:
\begin{equation}\label{4mammamia}
\begin{split}
\frac{d \langle \hat{\tau} \rangle}{dt} &= - i \bra{\psi} \left[ \hat{\tau}, \hat{H} \right]\ket{\psi} = \\& = -i \frac{T}{d} \sum_{n \ne n'} \frac{(E_n-E_{n'}) e^{i(E_n-E_{n'}) \tau_0}}{e^{i(E_n-E_{n'}) T/d)} - 1} \braket{\psi|E_{n'}}\braket{E_n|\psi}
\end{split}
\end{equation}
where $\ket{\psi}$ is a generic state of the system. If we consider the system in an energy eigenstate (that is $\ket{\psi}=\ket{E_i}$), we obtain
\begin{equation}\label{4mammamiamiamia}
\frac{d \langle \hat{\tau} \rangle}{dt} = 0 
\end{equation}
which means that the $\tau_{m}$ values stops running over time. Using Pegg's words we can say that {\it a system in a stationary state would not \lq\lq age\rq\rq as time goes on} \cite{pegg}. So, outside the PaW framework, the $\hat{\tau}$ operator can not be considered as a time observable, but as a property of the system that has dimension of time.

Conversely, within the PaW framework, we have a global stationary state that includes the whole time history of $S$ with respect to $C$.
So we consider the state:

\begin{equation}
\begin{split}
\ket{\Psi} &= \frac{1}{\sqrt{d_C}} \sum_{m=0}^{d_C-1} \ket{\tau_m}_C \otimes \ket{\phi(\tau_{m})}_S=\\&= \frac{1}{\sqrt{d_C}} \sum_{m=0}^{d_C-1} \ket{\tau_m}_C \otimes \sum_{k=0}^{d_S-1} \tilde{c}_k e^{-i  E_k \tau_{m}}\ket{E_k}_S
\end{split}
\end{equation}
that is the global state that satisfies the Wheeler-DeWitt constraint. By taking the state $\ket{\phi(\tau_{m})}_S$ of the system $S$ in an energy eigenstate, we obtain

\begin{equation}
\ket{\Psi} = \frac{1}{\sqrt{d_C}} \sum_{m=0}^{d_C-1} \ket{\tau_m}_C \otimes  e^{-i  E_k \tau_{m}}\ket{E_k}_S .
\end{equation}
So, the Everett relative state $\ket{\phi(\tau_{m})}_S$ is stationary since it only evolves with an unobservable global phase: 

\begin{equation}\label{4dausareperordinaryqm}
\ket{\phi(\tau_{m})}_S=e^{-i  E_k \tau_{m}}\ket{E_k}_S.
\end{equation}
However, this does not mean that the Universe stops. Indeed, from the fact that in the clock space $\hat{\tau}$ and $\hat{H}_C$ are conjugated operators, it follows that, even if taking the system $S$ in an energy eigenstate $\ket{E_k}$ forces the clock in an eigenstate of $\hat{H}_C$, all time states exist (indeed, thanks to the fact that $\hat{\tau}$ and $\hat{H}_c$ are incompatible observables, for construction we have $\ket{E_k} \propto \sum_{m} e^{-iE_k \tau_{m}} \ket{\tau_m}$). 
In this mechanism, thanks to this new conception of time, this means that all \lq\lq times\rq\rq exist.
The evolution expressed in equation (\ref{4dausareperordinaryqm}) describes what also happens in ordinary quantum mechanics when a system is in an energy eigenstate. Indeed, applying the Schrödinger equation to an energy eigenstate, we obtain exactly the same evolution. This is because a stationary state is not a state that does not evolve, but a state in which the probability distribution of any observable is constant over time. 

So, Age operator $\hat{\tau}$ that Pegg's defined as complement of the Hamiltonian becomes a proper time operator when included
in the PaW framework. This happens in general with any choice of the clock Hamiltonian, as discussed by Leon and Maccone in \cite{leonmaccone},
because in the Page and Wootter theory the concept of external time is eliminated (or in any case becomes irrelevant), and clock time as emerging from entanglement is not a property of the system $S$ but is a property of the clock subsystem $C$, which then enters the system $S$ through the Wheeler-DeWitt constraint.

\section{Unequally-Spaced Energy Levels for the Clock Hamiltonian}
\label{unequally}
With the perspective to extend the set of Hamiltonians useful to describe a clock, we now consider the case in which the clock Hamiltonian does not have equally-spaced energy levels, but non-degenerate eigenstates having rational energy ratios\footnote{As already mentioned, we will see that the limitation to spectra with rational energy ratios can be relaxed and the mechanism will work for any generic bounded Hamiltonian with discrete spectrum. We will discuss this point in paragraph 4.3.3.}. In this case we cannot define an Hermitian operator but we can still introduce Pegg's POVM, complement of such Hamiltonian \cite{pegg}.

\subsection{Discrete Flow of Time}
\label{discreteflow}
We introduce here again the framework discussed in Section 3.2 of Chapter 3, assuming now $E_0 \ne 0$. We consider a quantum system described by $p+1$ energy states $\ket{E_i}$ and $E_i$ energy levels with $i=0,1,2,...,p$ such that $\frac{E_i - E_0}{E_1 - E_0} = \frac{C_i}{B_i}$, where $C_i$ and $B_i$ are integers with no common factors. We can write

\begin{equation}\label{4ei}
E_i = E_0 + r_i \frac{2\pi}{T}
\end{equation}
where $T=\frac{2\pi r_1}{E_1 - E_0}$, $r_i = r_1\frac{C_i}{B_i}$ for $i>1$ (with $r_0=0$) and $r_1$ equal to the lowest common multiple of the values of $B_i$. In this space we define the states 

\begin{equation}\label{4defstatialpha}
\ket{\alpha_m}_C  = \frac{1}{\sqrt{d_C}}\sum_{i=0}^{d_C-1}e^{-i E_i \alpha_m}\ket{E_i}_C
\end{equation}
where $d_C = p+1$ and

\begin{equation}\label{4alpha1}
\alpha_m = \alpha_0 + m \frac{T}{s+1} = \alpha_0 + m \frac{T}{D}
\end{equation} 
with $m=0,1,2,...,s$ and $s+1=D \ge r_p$. The number of $\ket{\alpha_m}$ states is therefore greater than the number of energy states in $\mathcal{H}_C$ and the $s+1$ values of $\alpha_m$ are uniformly distributed over $T$. The resolution of the identity (\ref{4identity}) is now replaced by (see Appendix 4.8.A at the end of this Chapter):

\begin{equation}\label{4pomidentity}
\mathbb{1}_C = \frac{p+1}{s+1} \sum_{m=0}^{s} \ket{\alpha_{m}}\bra{\alpha_{m}} = \frac{d_C}{D} \sum_{m=0}^{s} \ket{\alpha_{m}}\bra{\alpha_{m}}.
\end{equation}
As in the previous discussion, we can now consider a general state in the space $\mathcal{H}=\mathcal{H}_c \otimes \mathcal{H}_s$ and require that it satisfies the PaW constraint. By writing 

\begin{equation}
\ket{\Psi} = \sum_{n=0}^{d_C-1} \sum_{k=0}^{d_S-1} c_{n,k} \ket{E_n}_C \otimes \ket{E_k}_S
\end{equation}
and imposing $\hat{H}\ket{\Psi} = 0$ (considering $d_C \gg d_S$), we obtain

\begin{equation}\label{4statoglobaleunequally}
\ket{\Psi} = \sum_{k=0}^{d_S-1} \tilde{c}_k \ket{E=- E_k}_C \otimes \ket{E_k}_S
\end{equation}
with $\sum_{k=0}^{d_S-1} \left| \tilde{c}_k \right|^{2} = 1$. We can now apply the resolution of the identity (\ref{4pomidentity}) to the state $\ket{\Psi}$ and obtain $(D=s+1)$:

\begin{equation}\label{4applicazionerisoluzioneidentita}
\begin{split}
\ket{\Psi} & = \frac{d_C}{D} \sum_{m=0}^{D-1} \ket{\alpha_{m}}\braket{\alpha_{m}|\Psi} = \\&= \frac{\sqrt{d_C}}{D} \sum_{m=0}^{D-1} \ket{\alpha_{m}}_C \otimes \sum_{k=0}^{d_S-1} \tilde{c}_k e^{-i E_k \alpha_{m}} \ket{E_k}_S . 
\end{split}
\end{equation}
We notice here that the states $\ket{\alpha_{m}}_C$ are not orthogonal. This introduces a possible conceptual warning that needs to be discussed. It is clear that by considering time states that are not orthogonal implies that these are partially indistinguishable with a single measurement, the probability of indistinguishability being proportional to $| \braket{\alpha_{m'}|\alpha_m}|^2$. The partial indistinguishability of the states $\ket{\alpha_{m}}_C$ implies an overlap between different times. We will show, however, that also in this case the time evolution is described by the Schr\"odinger equation. Considering

\begin{equation}\label{4hfhfhfhfhfhfh}
\ket{\phi(\alpha_{m})}_S =  \frac{\braket{\alpha_m|\Psi}}{1/\sqrt{d_C}} ,
\end{equation}
we obtain again $\ket{\phi(\alpha_{m})}_S = \sum_{k=0}^{d_S-1} \tilde{c}_k e^{-i \alpha_{m} E_k} \ket{E_k}_S$ for the state of the system $S$ (see Appendix 4.8.B at the end of this Chapter). Therefore, even if time states are partially indistinguishable, the state of the system $S$, conditioned on a given $\ket{\alpha_{m}}_C$,
evolves with $\alpha_m$. Thanks to the fact that 

\begin{equation}
\ket{\alpha_{m}}_c = e^{-i\hat{H}_c (\alpha_{m}-\alpha_0)} \ket{\alpha_{0}}_c,
\end{equation}
using once again the constraint (\ref{4scm}) and equation (\ref{4hfhfhfhfhfhfh}), we obtain

\begin{equation}\label{4evoluzionealpha}
\begin{split}
\ket{\phi(\alpha_{m})}_S &= \frac{\braket{\alpha_{m}|\Psi}}{1/\sqrt{d_C}} = \sqrt{d_C}\bra{\alpha_0}e^{i\hat{H}_C(\alpha_m-\alpha_0)}\ket{\Psi}=
\\&= \sqrt{d_C}\bra{\alpha_0}e^{i(\hat{H}-\hat{H}_S)(\alpha_m-\alpha_0)}\ket{\Psi}=\\&=\sqrt{d_C} ~ e^{-i\hat{H}_S(\alpha_m-\alpha_0)}\braket{\alpha_0|\Psi}=
\\&= e^{-i\hat{H}_S (\alpha_{m}-\alpha_0)}\ket{\phi(\alpha_{0})}_S  
\end{split}
\end{equation}
that is the Schrödinger evolution for the state $\ket{\phi(\alpha_{m})}_S$ with the Hamiltonian $\hat{H}_S$ and with respect to the $\alpha_{m}$ values. We notice here that POVMs generalizing the one introduced by Pegg have been discussed in \cite{review}. The consequence of using POVMs is that time states are not fully indistinguishable and this suggests a possible extension of the definition of the Everett relative states where is still possible to have a consistent dynamical evolution of the system $S$. 

To conclude this paragraph we notice that, through equation (\ref{4evoluzionealpha}), we can again define  the unitary operator $\hat{U}_{S} (\alpha_{m}-\alpha_0) = e^{-i \hat{H}_S (\alpha_{m}-\alpha_0)}$. With this choice the state of the global system can be written as 
\begin{equation}
\begin{split}
\ket{\Psi}  =\frac{\sqrt{d_C}}{D} \sum_{m=0}^{D-1} \ket{\alpha_m}_C \otimes \hat{U}_{S} (\alpha_{m}-\alpha_0) \ket{\phi(\alpha_{0})}_S
\end{split}
\end{equation}
and the conditional probability of obtaining the outcome $a$ for the system $S$ when measuring the observable $A$ at a certain time $\alpha_{m}$ is given again by the Born rule (see Appendix 4.8.C at the end of this Chapter):
\begin{equation}\label{4born}
\begin{split}
P(a\: on\: S \: |\: \alpha_m \: on \:C) =\frac{P(a \: on \: S, \: \alpha_m \: on \: C)}{P( \alpha_m \: on \: C)}   = \left|	\bra{a}\hat{U}_{S} (\alpha_{m}-\alpha_0) \ket{\phi(\alpha_{0})}\right|^2  .
\end{split}
\end{equation}
Equation (\ref{4born}) shows that the conditioned state of $S$ to a certain clock value $\alpha_{m}$ has no contributions from different times $\alpha_{m'} \ne \alpha_{m}$, and so interference phenomena are not present even if the time states are not orthogonal.

\subsection{Continuous Flow of Time}
\label{continuo}
So far we have considered a discrete flow of time. A continuous flow of time can be obtained in the limit $s \longrightarrow \infty$ \cite{pegg}. 
We define

\begin{equation}
\ket{\tilde{\alpha}} = \sum_{i=0}^{p} e^{- i E_i \alpha}\ket{E_i}_C
\end{equation}
where again $p+1$ is the number of energy eigenstates and $\alpha$ can now take any real value from $\alpha_0$ to $\alpha_0 + T$. In this framework the resolution of the identity (\ref{4pomidentity}) becomes

\begin{equation}\label{4newresolution}
\mathbb{1}_C = \frac{1}{T} \int_{\alpha_0}^{\alpha_0+T} d \alpha \ket{\tilde{\alpha}} \bra{\tilde{\alpha}}  .
\end{equation}
and the complement of the Hamiltonian is represented by the POVM generated by the infinitesimal operators $\ket{\tilde{\alpha}}\bra{\tilde{\alpha}}\frac{d \alpha}{T}$. The global state reads now

\begin{equation}\label{4applicazioneidentitacontinuo}
\begin{split}
\ket{\Psi}  =  \frac{1}{T} \int_{\alpha_0}^{\alpha_0 + T} d \alpha   \ket{\tilde{\alpha}} \braket{\tilde{\alpha}|\Psi} = 
 \frac{1}{T} \int_{\alpha_0}^{\alpha_0 + T} d \alpha   \ket{\tilde{\alpha}}_C \otimes \ket{\phi(\alpha)}_S 
\end{split}
\end{equation}
and, since $\ket{\phi (\alpha)}_s = \braket{\tilde{\alpha}|\Psi}$,
we derive the Schrödinger equation for the state $\ket{\phi(\alpha)}_S$ 

\begin{equation}
\begin{split}
i \frac{\partial}{\partial \alpha} \ket{\phi(\alpha) }_S &= i \frac{\partial}{\partial \alpha}\braket{\tilde{\alpha}|\Psi}= i \frac{\partial}{\partial \alpha} \sum_{k=0}^{d_C-1}\bra{E_k}e^{iE_k \alpha}\ket{\Psi} =\\&= - \sum_{k=0}^{d_C-1}\bra{E_k}E_ke^{i E_k \alpha}\ket{\Psi} =-\bra{\tilde{\alpha}}\hat{H}_C\ket{\Psi}= \hat{H}_S \ket{\phi(\alpha)}_S .
\end{split}
\end{equation}
We emphasize here that this framework allows us to recover a continuous flow of time for the subsytem $S$ maintaining a bounded and discrete clock Hamiltonian.

\subsection{Clock Hamiltonian with generic Spectrum}
To conclude the Section we briefly discuss the case of a clock Hamiltonian with a discrete spectrum and arbitrary (not rational) energy level ratios. Also in this scenario a Schr\"odinger evolution is recovered for the state of the system $S$ with respect to the clock. A caveat is that, in this case, the resolutions of the identity (\ref{4pomidentity}) and (\ref{4newresolution}) are no longer exact and the time states do not provide an overcomplete basis in $C$. Nevertheless, since any real number can be approximated with arbitrary precision by a ratio between two rational numbers, the residual terms in the resolution of the identity and consequent small corrections can be arbitrarily reduced.

\section{On the Quantum Speed Limit Time}
We consider in this Section the framework of paragraph 4.3.2, namely we work in the limit of continuous flow of time. An important question is the time scale over which the conditioned state $\ket{\phi(\alpha)}_{S}$ evolve into an othogonal configuration, thus becoming fully distinguishable. The state of the Universe satisfying the global constraint $\hat{H}\ket{\Psi} =0$ reads (\ref{4statoglobaleunequally}):
\begin{equation}\label{4STATOGLOBALE}
\ket{\Psi} = \sum_{k=0}^{d_S-1} \tilde{c}_k \ket{E=- E_k}_C \otimes \ket{E_k}_S .
\end{equation}
The conditioned state of $S$ can be obtained from (\ref{4STATOGLOBALE}), calculating $\ket{\phi(\alpha)}_{S} = \braket{\tilde{\alpha}|\Psi}$. We have ($\hslash \ne 1$):
\begin{equation}\label{4equazione}
\ket{\phi(\alpha)}_{S} = \sum_{k=0}^{d_S-1} \tilde{c}_k e^{-i\hslash^{-1} E_k \alpha} \ket{E_k}_S .
\end{equation} 
As outlined in \cite{leonmaccone} the crucial point for our argument here is to understand that in the PaW framework, since the clock $C$ and the system $S$ are entangled in the global state $\ket{\Psi}$, the coefficients in expansion (\ref{4STATOGLOBALE}) refers to the clock and the system together (notice that $\tilde{c}_k$ appears also in the conditioned state of $S$ (\ref{4equazione})). For this reason we will find that a limited spread in the expansion (\ref{4STATOGLOBALE}) will reduce the speed of the evolution for $S$ over time $\alpha$. Starting from equation (\ref{4equazione}) we can calculate in the space $S$:
\begin{equation}\label{4ultima1}
\begin{split}
\braket{\phi (\alpha_0) | \phi (\alpha)}  = \sum_{k=0}^{d_S - 1} |\tilde{c}_k|^2 e^{-i \hslash^{-1} E_k (\alpha - \alpha_0)} .
\end{split}
\end{equation} 
Looking at equation (\ref{4ultima1}), we can now consider the quantum speed limit time $\delta \alpha$ which gives us the minimum time needed for $S$ to evolve to an orthogonal configuration. We have \cite{speedlimit,margolus}:
\begin{equation}\label{4dalpha}
\delta \alpha  \geq  max \left( \frac{\pi\hslash}{2 (E_{S} - E_0)}  , \frac{\pi\hslash}{2 \Delta E} \right)
\end{equation}
where $E_{S} = \bra{\phi(\alpha_0)} \hat{H}_S\ket{\phi(\alpha_0)} $ and $\Delta E$ is the spread in energy related to the coefficients $\tilde{c}_k$ through $\Delta E = \sqrt{\bra{\phi(t_0)} (\hat{H}_S - E_{S} )^2\ket{\phi(t_0)}}$.


The aspect we emphasize is that the function $f(\alpha - \alpha_0) = \braket{\phi (\alpha_0) | \phi (\alpha)}$, and consequently the time scale on which $S$ varies significantly, is not related to the overlap of the states of the clock. This can be seen considering that in $f(\alpha - \alpha_0)$ do not enter time values different from $\alpha, \alpha_0$ and $f(\alpha - \alpha_0)$ takes the form expressed in \cite{speedlimit}. So the fact that our time states are not orthogonal does not have a consequence on the speed at which the state $\ket{\phi(\alpha)}_{S}$ evolves with respect to $\alpha$. Rather $f(\alpha - \alpha_0)$ is related to the spread of the coefficients $\tilde{c}_k$ appearing in (\ref{4STATOGLOBALE}) and (\ref{4equazione}). These considerations, together with equation (\ref{4dalpha}), indicate that a large spread in the coefficients $\tilde{c}_k$ within state $\ket{\phi(\alpha)}_{S}$, and so in the global state (\ref{4STATOGLOBALE}), is needed to make the time evolution of the subsystem $S$ faster \cite{asimmetry}.

\section{Interacting Clock and System}
\subsection{General Framework}
We consider here the case of interacting clock and system. We follow \cite{interacting} and we show that the same results can be found in our case of non-orthogonal time states. We also assume continuous time values $\alpha$, namely also here we use the framework of paragraph 4.3.2. 
The global Hamiltonian describing the whole Universe reads now:
\begin{equation}
	\hat{H} = \hat{H}_C + \hat{H}_S + \hat{H}_{int}
\end{equation}
where $\hat{H}_{int}$ is the term which describe the interaction present between the clock and the system $S$. For the relative state $\ket{\phi(\alpha)}_S=\braket{\tilde{\alpha}|\Psi}$ we can write ($\hslash =1$):

\begin{equation}
\begin{split}
i \frac{\partial}{\partial \alpha}\ket{\phi(\alpha)}_S & = i \frac{\partial}{\partial \alpha}\braket{\tilde{\alpha}|\Psi} = i \frac{\partial}{\partial \alpha} \sum_{k=0}^{d_C - 1}\bra{E_k} e^{i E_k \alpha} \ket{\Psi}=
\\&= - \sum_{k=0}^{d_C - 1}\bra{E_k} E_k e^{i E_k \alpha} \ket{\Psi}= - \bra{\tilde{\alpha}}\hat{H}_C\ket{\Psi} =
\\& = - \bra{\tilde{\alpha}}\hat{H} - \hat{H}_S - \hat{H}_{int}\ket{\Psi}
\end{split}
\end{equation}
from which we easily obtain

\begin{equation}\label{4int1}
i \frac{\partial}{\partial \alpha}\ket{\phi(\alpha)}_S = \hat{H}_S \ket{\phi(\alpha)}_S + \bra{\tilde{\alpha}}\hat{H}_{int}\ket{\Psi} .
\end{equation}
By inserting the resolution of the identity on $\mathcal{H}_C$ in terms of the clock states ($\mathbb{1}_C = \frac{1}{T} \int_{\alpha_0}^{\alpha_0+T} d \alpha \ket{\tilde{\alpha}} \bra{\tilde{\alpha}}$) between $\hat{H}_{int}$ and $\ket{\Psi}$ in the second term on the right side of equation (\ref{4int1}), we obtain

\begin{equation}\label{4int2}
i \frac{\partial}{\partial \alpha}\ket{\phi(\alpha)}_S = \hat{H}_S \ket{\phi(\alpha)}_S + \frac{1}{T} \int_{\alpha_0}^{\alpha_{0} + T} d\alpha' \hat{K}(\alpha,\alpha')\ket{\phi(\alpha')}_S
\end{equation}
where 
\begin{equation}
\hat{K}(\alpha,\alpha')= \bra{\tilde{\alpha}}\hat{H}_{int}\ket{\tilde{\alpha}'}
\end{equation}
is an operator acting on $\mathcal{H}_S$. It is easy to see that, when the interaction term vanishes, equation (\ref{4int2}) reduces to the usual Schrödinger equation. In equation (\ref{4int2}), the second term on the right side is an integral operator on $\mathcal{H}_S$. Thus the equation can be rewritten as

\begin{equation}\label{4int3}
i \frac{\partial}{\partial \alpha}\ket{\phi(\alpha)}_S = \left(\hat{H}_S + \hat{H}_K \right)\ket{\phi(\alpha)}_S
\end{equation} 
where $\hat{H}_K$ is defined through its action:

\begin{equation}
	\hat{H}_K \ket{\phi(\alpha)}_S =  \frac{1}{T} \int_{\alpha_0}^{\alpha_{0} + T} d\alpha' \hat{K}(\alpha,\alpha')\ket{\phi(\alpha')}_S .
\end{equation}
Note that $\hat{H}_K$ is a self-adjoint operator. This can be seen considering that 
\begin{equation}
\hat{K}^{\dagger}(\alpha,\alpha') = \left( \bra{\tilde{\alpha}} \hat{H}_{int}\ket{\tilde{\alpha}'}  \right)^{\dagger} = \left( \bra{\tilde{\alpha}'} \hat{H}_{int}\ket{\tilde{\alpha}}  \right) = \hat{K} (\alpha',\alpha).
\end{equation}
So equation (\ref{4int3}) can be considered as the ordinary Schrödinger equation with the Hamiltonian $\hat{H}_S$ replaced by the self-adjoint integral operator $\hat{H}_S + \hat{H}_K$.

As a last consideration we notice that equations (\ref{4int2}) and (\ref{4int3}) are non-local in time $\alpha$: this means that, in order to know whether $\ket{\phi(\alpha)}_S$ is a solution, we need the knowledge of $\ket{\phi(\alpha)}_S$ at all times $\alpha \in \left[\alpha_{0},\alpha_{0}+T\right]$. We will encounter again the Schrödinger equation corrected by a non-local term in Chapter 5.

\subsection{Gravitationally interacting Clock and System}
We can now discuss an example in order to examine how the dynamics of system S is modified by the interaction term present in the global Hamiltonian. We choose as an example the case in which the clock and the system S are coupled through Newtonian gravity. In this case the interaction Hamiltonian is:
\begin{equation}\label{4sto}
	\hat{H}_{int} = - \frac{G}{c^4 d} \hat{H}_C \otimes \hat{H}_S
\end{equation}
where distance $d$ between $C$ and $S$ is treated as a number and we have promoted the masses of $C$ and $S$ to operators through the mass-energy equivalence\footnote{In doing this we assume that the static mass is negligibly small compared with the dynamical one and we focus only on the effect due to the internal degrees of freedom of $C$ and $S$ \cite{entclockgravity,zych1}.}: 
\begin{equation}
m_C \to \frac{\hat{H}_C}{c^2} \:\:\: and \:\:\: m_S \to \frac{\hat{H}_S}{c^2} .
\end{equation}
With (\ref{4sto}), we can now calculate:
\begin{equation}\label{4trat}
	\begin{split}
	\hat{K}(\alpha,\alpha') &= \bra{\tilde{\alpha}} \hat{H}_{int}\ket{\tilde{\alpha}'} = - \frac{G}{c^4 d} \bra{\tilde{\alpha}} \hat{H}_C\otimes\hat{H}_S\ket{\tilde{\alpha}'} =
	\\&= - \frac{G}{c^4 d} \sum_{k=0}^{d_C -1} e^{iE_k \alpha} \bra{E_k} \sum_{k'=0}^{d_C-1}e^{-iE_{k'}\alpha'}E_{k'}\ket{E_{k'}} \hat{H}_S =
	\\&= - \frac{G}{c^4 d} \sum_{k=0}^{d_C -1}e^{iE_k (\alpha - \alpha')} E_k \hat{H}_S .
	\end{split}
\end{equation} 
Through equations (\ref{4trat}) and (\ref{4int2}) we can explicitly see how the dynamics of the system $S$ is modified by the gravitational coupling with the clock:
\begin{equation}\label{4definitiva}
	\begin{split}
i \frac{\partial}{\partial \alpha}\ket{\phi(\alpha)}_S &= \hat{H}_S \ket{\phi(\alpha)}_S - \frac{1}{T}\frac{G}{c^4 d} \int_{\alpha_0}^{\alpha_{0} + T} d\alpha' \sum_{k=0}^{d_C-1}  e^{iE_k (\alpha - \alpha')} E_k \hat{H}_S    \ket{\phi(\alpha')}_S=
\\&= \hat{H}_S \ket{\phi(\alpha)}_S - \frac{1}{T}\frac{G}{c^4 d} \int_{\alpha_0}^{\alpha_{0} + T} d\alpha' \sum_{k=0}^{d_C-1}  e^{iE_k (\alpha - \alpha')} E_k \hat{H}_S    \braket{\tilde{\alpha}'|\Psi}=
\\&= \hat{H}_S \ket{\phi(\alpha)}_S - \frac{1}{T}\frac{G}{c^4 d} \int_{\alpha_0}^{\alpha_{0} + T} d\alpha' \sum_{k=0}^{d_C-1}\sum_{k' =0}^{d_C-1} e^{iE_k\alpha} e^{-i\alpha'(E_k - iE_{k'} )}  E_k \hat{H}_S\ket{\Psi} 
\\&=  \hat{H}_S \ket{\phi(\alpha)}_S - \frac{G}{c^4 d} \hat{H}_S \sum_{k=0}^{d_C-1}  e^{iE_k \alpha} E_k \braket{E_k|\Psi}=
\\&= \hat{H}_S \ket{\phi(\alpha)}_S + \frac{G}{c^4 d} i  \hat{H}_S \frac{\partial}{\partial \alpha}\ket{\phi(\alpha)}_S
	\end{split}
\end{equation}
where we used (see Appendix 4.8.D at the end of this Chapter)
\begin{equation}\label{ultimaappendice}
\int_{\alpha_{0}}^{\alpha_{0} + T} d\alpha' e^{-i\alpha'(E_k - E_{k'})}= T \delta_{E_k,E_{k'}} . 
\end{equation}
After rearranging, equation (\ref{4definitiva}) becomes:
\begin{equation}\label{4ultimadef}
i \frac{\partial}{\partial \alpha}\ket{\phi(\alpha)}_S = \left[ \hat{H}_S + \frac{G}{c^4 d}   \hat{H}^{2}_S + \mathcal{O}\left( \frac{G^2}{c^8 d^2}\right)\right]\ket{\phi(\alpha)}_S
\end{equation}
in agreement with \cite{interacting}. From equation (\ref{4ultimadef}) we see that the gravitational interaction leads to a correction, within the Schrödinger equation, of the order $\sim G/c^4$ and inversely proportional to the distance $d$ between the clock $C$ and the system $S$.  
For a discussion on other examples we directly refer to \cite{interacting}.

\section{Considerations on the Arrow of Time}
\label{discussion}
Is there an arrow of time in the PaW formalism? The answer in clearly negative. However, it is possible to introduce an {\it emergent} arrow of time. Following \cite{vedral} one can consider for simplicity that the system $S$ consists only of two subsystems, \lq\lq the observer\rq\rq (${\Sigma}_1$) and \lq\lq the observed\rq\rq ($\Sigma_2$) which are initially in a product state. The arrow of time (with respect to clock time) can be provided by the increase in entanglement between the two subsystems within $S$, as the observer learns more and more about the observed. We can ask now where to find a good clock for the Universe. We have considered as a \lq\lq good\rq\rq clock a device defined in a Hilbert space larger than the Hilbert space of the system $S$, that is $d_C \gg d_S$. Indeed, if $d_c \le d_s$, it would not be possible to connect every energy state of the system $S$ to an energy state of $C$ satisfying the constraint (\ref{4scm}), and some states of $S$ would be excluded from the dynamics. Therefore, the clock introduced here has essentially two properties: it has to be larger than the system $S$ and it has to interact only weakly with $S$ or, in the ideal case, it should not interact at all. Does such a clock exist? Considering our closed quantum system as the whole Universe, a possible choice is to consider the non-observable Universe (namely, the Universe laying outside the light cone centred in the Earth) as a clock for the observable Universe. Indeed, in this case, the clock and the observable Universe $S$ are not interacting but can still be fully entangled, with the Hilbert space of the clock that can be quite larger that the Hilbert space of $S$. With this choice (which of course is just one among several equally speculative choices) the two requirements for a good clock are satisfied. We can then consider that during the evolution with respect to such a clock, inside the observable Universe (that is inside the interacting subsystems ${\Sigma}_1$ and ${\Sigma}_2$ of $S$) there is an increasing entanglement generated by $\hat{H}_S$ and, therefore, an increasing relative entropy and the emergence of a thermodynamic arrow of time. We return on this point in Chapter 5 where, again, we will speculate about possibility of applying our generalized version of the PaW mechanism to the whole Universe.

\section{Conclusions}
\label{Conclusions}
In this Chapter we incorporated Pegg's formalism into PaW theory. Thus, in the case of equally-spaced energy levels for the clock Hamiltonian, we introduced a clock observable (the Hermitian operator $\hat{\tau}$) complement of a bounded clock Hamiltonian with discrete spectrum. $\hat{\tau}$ is complement to the clock Hamiltonian in the sense that it is the generator of energy shifts while the clock Hamiltonian is the generator of translations of the eigenvalues of $\hat{\tau}$. In addition we showed that it is possible to extend this framework to any generic Hamiltonian with a discrete spectrum. In this case the clock observable is described by a POVM. We have demonstrated that, even if the time states are not fully distinguishable, the system $S$ still evolves with respect to the clock time according to the Schr\"odinger equation. Moreover the state of the system $\ket{\phi(\alpha_{m})}_s$ at certain time $\alpha_{m}$ is well defined, only depending on the $\alpha_m $ value and, when considering the conditional probabilities, the Born's rule is still obtained for the system $S$. In this framework we have also recovered a continuous flow of time for the subsystem $S$, still maintaining a bounded and discrete clock Hamiltonian. As mentioned in the introduction of this Chapter, when considering a closed physical system of finite size, the introduction of unbounded Hamiltonians with a continuous spectrum (as in the case of GLM) would not be possible. Our framework naturally solves this problem and, through our work, any generic quantum system of finite size can be used as a clock observable. Finally, we showed that the time scale over which the relative state of $S$ evolve into an othogonal configuration is not related to the overlap of the states of the clock. Rather it is related to the spread of the coefficients $\tilde{c}_k$ appearing in the global state of the Universe. The case of interacting clock and system is also considered and discussed.


\section{Appendices}

\subsection*{4.8.A \: Proof of Equation (\ref{4pomidentity})}
\label{appendiceb}
We prove that $\sum_{m=0}^{s}\ket{\alpha_{m}}\bra{\alpha_{m}}$ is equal to the identity:

\begin{equation}\label{4dada}
\begin{split}
\sum_{m=0}^{s}\ket{\alpha_{m}}\bra{\alpha_{m}} &= \frac{1}{p+1} \sum_{m=0}^{s}\sum_{i}\sum_{k}e^{-i\alpha_{m}E_{i}}e^{i\alpha_{m}E_k}\ket{E_{i}}\bra{E_k}=\\&= \frac{1}{p+1} \left[   \sum_{m=0}^{s}\sum_{k}\ket{E_{k}}\bra{E_k} +   \sum_{k \ne i}\sum_{m=0}^{s}  e^{i \alpha_{m}(r_k - r_{i})2\pi/T}\ket{E_i} \bra{E_k}  \right] .
\end{split}
\end{equation}
For $(E_k-E_0)/(E_1-E_0)$ rational, and thus $r_k - r_{i}$ integer, the second term of the right side of equation (\ref{4dada}) will be zero and then we have $\frac{p+1}{s+1} \sum_{m=0}^{s} \ket{\alpha_{m}}\bra{\alpha_{m}} = \mathbb{1}_C$.

\subsection*{4.8.B \: Relative State Definition for $S$ in case of non-orthogonal Time States}
\label{appendicec}
We start considering the global state $\ket{\Psi}$ written as $\ket{\Psi} = \sum_{k=0}^{d_S-1} \tilde{c}_k \ket{E=- E_k}_C \otimes \ket{E_k}_S$
and we apply in sequence the resolutions of the identity on the clock subspace 
\begin{equation}\label{4kkk}
\frac{d_C}{D} \sum_{m=0}^{D-1} \ket{\alpha_{m}}\bra{\alpha_{m}} = \mathbb{1}_C
\end{equation}
and 
\begin{equation}\label{4kk2}
\sum_{n=0}^{d_C-1}\ket{E_n}\bra{E_n}=\mathbb{1}_C.
\end{equation}
Applying (\ref{4kkk}) we have:
\begin{equation}\label{4speroultima}
\begin{split}
\ket{\Psi} &= \frac{d_C}{D}\sum_{m=0}^{D-1}\ket{\alpha_{m}}\braket{\alpha_{m}|\Psi}=
 \\&=\frac{d_C}{D}\sum_{m=0}^{D-1}\ket{\alpha_{m}}_C \otimes \sum_{k=0}^{d_S-1}\tilde{c}_k\braket{\alpha_{m}|E=-E_k} \ket{E_k}_S =
  \\&= \frac{\sqrt{d_C}}{D}\sum_{m=0}^{D-1}\ket{\alpha_{m}}_C \otimes \sum_{k=0}^{d_S-1}\tilde{c}_k e^{-i\alpha_{m}E_k} \ket{E_k}_S .
\end{split}
\end{equation}
Now we apply (\ref{4kk2}) to (\ref{4speroultima}) and we obtain
\begin{equation}
\begin{split}
\ket{\Psi} & = \sum_{n=0}^{d_C-1} \ket{E_n}\bra{E_n} \frac{\sqrt{d_C}}{D}\sum_{m=0}^{D-1}\ket{\alpha_{m}}_C \otimes \sum_{k=0}^{d_S-1}\tilde{c}_k e^{-i\alpha_{m}E_k} \ket{E_k}_S=
\\&= \frac{\sqrt{d_C}}{D} \sum_{n=0}^{d_C-1} \ket{E_n}_C \otimes \sum_{m=0}^{D-1} \braket{E_n|\alpha_{m}} \sum_{k=0}^{d_S-1} \tilde{c}_k e^{-i\alpha_{m}E_k} \ket{E_k}_S =
\\&= \frac{1}{D} \sum_{n=0}^{d_C-1} \ket{E_n}_C \otimes \sum_{m=0}^{D-1} e^{-i\alpha_{m}E_n} \sum_{k=0}^{d_S-1} \tilde{c}_k e^{-i\alpha_{m}E_k} \ket{E_k}_S =
\\&= \sum_{n=0}^{d_C-1} \ket{E_n}_C \otimes \sum_{k=0}^{d_S-1}\tilde{c}_k \frac{1}{D} \sum_{m=0}^{D-1} e^{-i \alpha_{m} \left(E_n + E_k\right) } \ket{E_k}_S
\end{split}
\end{equation}
from which we have

\begin{equation}\label{4delta}
\sum_{m'=0}^{D-1} e^{-i \alpha_{m'} \left(E_n + E_k\right) } = D\delta_{E_n,-E_k} .
\end{equation}
Considering now the definition of the state $\ket{\phi(\alpha_{m})}_S$ as 

\begin{equation}\label{4hfhfhfhfhfhfh1}
\ket{\phi(\alpha_{m})}_S = \sqrt{d_C} \braket{\alpha_m|\Psi} ,
\end{equation}
we have:

\begin{equation}
\begin{split}
\braket{\alpha_{m}|\Psi} & = \bra{\alpha_{m}} \frac{\sqrt{d_C}}{D} \sum_{m'=0}^{D-1} \ket{\alpha_{m'}}_C \otimes \sum_{k=0}^{d_S-1} \tilde{c}_k e^{-i\alpha_{m'} E_k} \ket{E_k}_S =
 \\&= \frac{\sqrt{d_C}}{D} \sum_{m'=0}^{D-1} \sum_{n,n'=0}^{d_c-1} \sum_{k=0}^{d_S-1} \frac{1}{d_C} e^{i E_n \alpha_{m}} e^{-i E_{n'} \alpha_{m'}} \braket{E_n|E_{n'}} \tilde{c}_k e^{-i E_k \alpha_{m'}} \ket{E_k}_S =
   \\&= \frac{1}{D\sqrt{d_C}} \sum_{n=0}^{d_C-1} e^{iE_n \alpha_{m}} \sum_{k=0}^{d_S-1} \tilde{c}_k \sum_{m'=0}^{D-1} e^{-i \alpha_{m'} \left(E_n + E_k\right)} \ket{E_k}_S 
\end{split}
\end{equation}
and, considering (\ref{4delta}), we obtain

\begin{equation}\label{4lolol}
\braket{\alpha_{m}|\Psi} = \frac{1}{\sqrt{d_C}} \sum_{k=0}^{d_S-1} \tilde{c}_k e^{-i\alpha_{m} E_k} \ket{E_k}_S.
\end{equation}
So,  for the relative state of the subsystem $S$ we have: 

\begin{equation}
\ket{\phi(\alpha_{m})}_S =\frac{\braket{\alpha_m|\Psi}}{1/\sqrt{d_C}} = \sum_{k=0}^{d_S-1} \tilde{c}_k e^{-i \alpha_{m} E_k} \ket{E_k}_S.
\end{equation}

\subsection*{4.8.C \: Proof of Equation (\ref{4born})}
\label{appendiced}
We start considering the global state written as 

\begin{equation}
\ket{\Psi} = \frac{\sqrt{d_C}}{D} \sum_{m=0}^{D-1} \ket{\alpha_{m}}_C \otimes \ket{\phi(\alpha_{m})}_S 
\end{equation}
where 

\begin{equation}
\ket{\phi(\alpha_{m})}_S = \sum_{k=0}^{d_S-1} \tilde{c}_k e^{-i E_k \alpha_{m}} \ket{E_k}_S .
\end{equation}
We can now calculate the conditional probability as follows
\begin{equation}\label{4ultimadim}
\begin{split}
P(a\: on\: S \: | \: \alpha_m \: on \:C) &=\frac{P(a \: on \: S , \: \alpha_m \: on \: C)}{P( \alpha_m \: on \: C)}  =
 \\&= \frac{\left| (\bra{\alpha_m}\bra{a})\ket{\Psi}\right|^2}{\sum_{a}\left| (\bra{\alpha_m}\bra{a})\ket{\Psi}\right|^2} =
 \\&= \frac{\left| (\bra{\alpha_m}\bra{a})\frac{\sqrt{d_C}}{D} \sum_{m'=0}^{D-1} \ket{\alpha_{m'}}_C\ket{\phi(\alpha_{m'})}_s\right|^2}{\sum_{a}\left| (\bra{\alpha_m}\bra{a})\frac{\sqrt{d_C}}{D} \sum_{m'=0}^{D-1} \ket{\alpha_{m'}}_C \ket{\phi(\alpha_{m'})}_S\right|^2} =
  \\&= \frac{\left| (\bra{\alpha_m}\bra{a}) \sum_{m'=0}^{D-1} \ket{\alpha_{m'}}_C \sum_{k=0}^{d_S-1} \tilde{c}_k e^{-i E_k \alpha_{m'}} \ket{E_k}_s\right|^2}{\sum_{a}\left| (\bra{\alpha_m}\bra{a}) \sum_{m'=0}^{D-1} \ket{\alpha_{m'}}_C \sum_{k=0}^{d_S -1} \tilde{c}_k e^{-i E_k \alpha_{m'}} \ket{E_k}_S \right|^2}  =
   \\&=  \frac{\left| \sum_{m'=0}^{D-1} \sum_{n=0}^{d_C-1} e^{iE_n(\alpha_{m}-\alpha_{m'})}\bra{a}   \sum_{k=0}^{d_S-1} \tilde{c}_k e^{-i E_k \alpha_{m'}} \ket{E_k}_S   \right|^2}{\sum_{a}\left| \sum_{m'=0}^{D-1} \sum_{n=0}^{d_C-1} e^{iE_n(\alpha_{m}-\alpha_{m'})} \bra{a}    \sum_{k=0}^{d_S-1} \tilde{c}_k e^{-i E_k \alpha_{m'}} \ket{E_k}_S   \right|^2} =
    \\&=  \frac{\left|  \sum_{n=0}^{d_C-1} e^{iE_n\alpha_{m}} \bra{a} \sum_{k=0}^{d_S-1} \tilde{c}_k \sum_{m'=0}^{D-1} e^{-i (E_k+E_n) \alpha_{m'}} \ket{E_k}_S   \right|^2}{\sum_{a}\left| \sum_{n=0}^{d_C-1} e^{iE_n\alpha_{m}} \bra{a} \sum_{k=0}^{d_S-1} \tilde{c}_k \sum_{m'=0}^{D-1} e^{-i (E_k+E_n) \alpha_{m'}} \ket{E_k}_S   \right|^2} .
\end{split}
\end{equation}
Thanks to equation (\ref{4delta}), that is $\sum_{m'=0}^{D-1} e^{-i \alpha_{m'} \left(E_n + E_k\right) } = D\delta_{E_n,-E_k}$, we have

\begin{equation}
\begin{split}
P(a\: on\: S \: | \: \alpha_m \: on \:C) &=  \frac{\left| \bra{a}\sum_{k=0}^{d_S-1} \tilde{c}_k e^{-iE_k \alpha_{m}} \ket{E_k}_S   \right|^2}{\sum_{a}\left|  \bra{a}\sum_{k=0}^{d_S-1} \tilde{c}_k e^{-iE_k \alpha_{m}} \ket{E_k}_S  \right|^2} =
\\&= \frac{\left| \braket{a|\phi(\alpha_{m})}   \right|^2}{\sum_{a}\left|  \braket{a|\phi(\alpha_{m})}   \right|^2} = \left| \braket{a|\phi(\alpha_{m})}   \right|^2 
\end{split}
\end{equation}
and, considering that 
\begin{equation}
\ket{\phi(\alpha_{m})}_S = \hat{U}_{S} (\alpha_{m}-\alpha_0) \ket{\phi(\alpha_{0})} ,
\end{equation}
we obtain

\begin{equation}
P(a\: on\: S \: |\: \alpha_m \: on \:C) = \left| \bra{a}\hat{U}_{S} (\alpha_{m}-\alpha_0) \ket{\phi(\alpha_{0})}\right|^2 .
\end{equation}

\subsection*{4.8.D \: Proof of Equation (\ref{ultimaappendice})}
We start here considering a generic state $\ket{\psi} \in \mathcal{H}_C$, that is
\begin{equation}
	\ket{\psi}= \sum_{k=0}^{d_C - 1} c_k \ket{E_k}_C .
\end{equation} 
We now apply on this state in sequence the resolutions of the identity 
\begin{equation}
\mathbb{1}_{C} = \frac{1}{T} \int_{\alpha_0}^{\alpha_0+T} d\alpha \ket{\tilde{\alpha}} \bra{\tilde{\alpha}}
\end{equation}
and 
\begin{equation}
\mathbb{1}_{C} = \sum_{n=0}^{d_C - 1} \ket{E_n}\bra{E_n} .
\end{equation}
In this way we obtain:
\begin{equation}\label{4hhh}
\begin{split}
\ket{\psi} & = \frac{1}{T} \int_{\alpha_0}^{\alpha_0+T} d\alpha \ket{\tilde{\alpha}}\braket{\tilde{\alpha}|\psi} = \frac{1}{T} \int_{\alpha_0}^{\alpha_0+T} d\alpha \sum_{k=0}^{d_C - 1} c_k e^{i \alpha E_k} \ket{\tilde{\alpha}}_C= 
\\&= \sum_{n=0}^{d_C-1}\ket{E_n}\bra{E_n}\frac{1}{T} \int_{\alpha_0}^{\alpha_0+T} d\alpha \sum_{k=0}^{d_C - 1} c_ke^{i\alpha E_k} \ket{\tilde{\alpha}}_C = 
\\&= \sum_{n=0}^{d_C - 1} \sum_{k=0}^{d_C-1} c_k \frac{1}{T} \int_{\alpha_0}^{\alpha_0+T} d\alpha~ e^{i\alpha (E_k - E_n)}\ket{E_n}_C .
\end{split}
\end{equation}
Since the state $\ket{\Psi}$ in equation (\ref{4hhh}) has to be equal to the initial state $\ket{\psi}= \sum_{k=0}^{d_C -1} c_k\ket{E_k}_C$, we have: 
\begin{equation}
\int_{\alpha_0}^{\alpha_0+T} d\alpha ~ e^{i \alpha (E_k - E_n)}= T \delta_{k,n} .
\end{equation}
	
\begin{quote}
\label{bianca2}

\end{quote}
	
	\chapter{Thermal Equilibrium and Emergence of Time}
\label{canonical}

Considering a closed quantum system (the ``Universe") where a small susbsystem $S$ is weakly coupled with a large environment $B$, it is well known that the canonical ensemble is obtained for $S$ when the global system $S+B$ is described by the microcanonical ensemble. So, assuming the subsystem $S$ in thermal equilibrium at inverse temperature $\beta$, we have:

\begin{equation}\label{6rhobeta}
\hat{\Omega}_S = \frac{1}{Z} e^{-\beta \hat{H}_S}
\end{equation} 
that is the canonical density matrix, where $\hat{H}_S$ is the Hamiltonian of the subsystem $S$ and $Z = Tr \left[ e^{-\beta \hat{H}_S}\right]$. Equation (\ref{6rhobeta}) is simply obtained by calculating the reduced density matrix of the subsystem $S$, considering that the global system $S+B$ is described by the microcanonical density matrix at an appropriate total energy $E$ and assuming that $S$ is weakly coupled with the environment $B$. 

Despite this the foundations of statistical mechanics are still a subject of debate. One controversial issue is the validity of the postulate of \textit{equal a priori probability}, which is always assumed but cannot be proven. This is why a new approach to quantum statistical mechanics has recently been proposed. In \cite{canonical} (see also \cite{popescu1,popescu2,zurek}) the authors show how in quantum systems, without invoking the principle of \textit{equal a priori probability} and under a suitable constraint on the total energy, can still be obtained the canonical distribution for the system $S$ when weakly coupled with a large environment $B$. Therefore these authors prove that, \textit{even if the global state of the system $S+B$ corresponds to a pure wave function instead of a mixture, for the overwhelming majority of randomly chosen pure wave-functions of the Universe the reduced density matrix of the system $S$ is canonical}. They call this property \textit{Canonical Typicality} in \cite{canonical} or \textit{General Canonical Principle} in \cite{popescu1}. 

The goal of this Chapter\footnote{Some parts of this Chapter are reprinted with permission from T. Favalli and A. Smerzi, \textit{Peaceful coexistence of thermal equilibrium and the emergence of time}, Phys. Rev. D \textbf{105}, 023525 (2022). Copyright by the American Physical Society.} (following \cite{nostro2}) is to show that canonical thermal equilibrium and dynamics can coexist in a quantum Universe, with the environment providing the clock for the evolution of the system $S$. The dynamics is governed by the Schr\"odinger equation corrected by a non-local term which vanishes in the limit of fixed total energy of the Universe. The paradox of the coexistence of thermal equilibrium and non-trivial evolution of $S$ is solved because the trace over the degrees of freedom of the environment coincides with a temporal average over the entire life of $S$. The temporal dynamics of the system $S$ emerges by considering the relative states of $S$ (in Everett sense \cite{everett}) with respect to the states of the environment.

\section{Canonical Equilibrium Distribution}

\subsection{Canonical Typicality: an Overview}
We consider our \lq\lq Universe\rq\rq consisting of a small subsystem $S$ and a large environment $B$ with $d_S \ll d_B$, where $d_S$ and $d_B$ are the dimension of $S$ and $B$ respectively. We also assume the Universe obeying a global constraint $R$ which identifies a subspace of the global Hilbert space: $\mathcal{H}_R \subseteq \mathcal{H}_B \otimes\mathcal{H}_S$.
The Universe is governed by the global Hamiltonian:
\begin{equation}
\hat{H} = \hat{H}_B + \hat{H}_S
\end{equation}  
where $\hat{H}_B$ and $\hat{H}_S$ are the Hamiltoninian of the subsystems acting on $\mathcal{H}_B$ and $\mathcal{H}_S$ respectively. The relatively small interaction between the environment and the system $S$ is neglected \cite{canonical,popescu1}. In analogy with the standard derivation of the canonical distribution of a subsystem given a global microcanonical distribution, we impose, as constraint, the total energy in the interval $\left[E,E+\delta\right]$, 
where $\delta \ll E, \Delta E^{(S)}$ (being $\Delta E^{(S)}$ the typical spacing between energy levels of the system $S$) but large enough to contain many eigenvalues of $\hat{H}$. In this framework the equiprobable state of the Universe in $\mathcal{H}_R$ is
\begin{equation}
\hat{\Omega}_R = d^{-1}_R \hat{\Pi}_R
\end{equation}
where $d^{-1}_R$ is the dimension of $\mathcal{H}_R$ and $\hat{\Pi}_R$ is the projection on such a space. As pointed out in \cite{cantyp}, in this case usually equal probabilities and random phases are assigned to all the states of the Universe which are consistent with the constraint $R$. The canonical state of $S$ is then defined as the trace over the environment $B$:
\begin{equation}
\hat{\Omega}_S = Tr_B \left[ \hat{\Omega}_R \right] \approx \frac{1}{Z} e^{-\beta \hat{H}_S} .
\end{equation}
that is the same of equation (\ref{6rhobeta}) we found in the introduction of this Chapter. 

In order to illustrate the Canonical Typicality we do not cosnsider the Universe in the equiprobable state $\hat{\Omega}_R$, which describes our ignorance about the global state, but we assume the Universe in the pure state $\ket{\Psi} \in \mathcal{H}_R$. In this case the state of the subsystem $S$ is described by the reduced density matrix
\begin{equation}
\hat{\rho}_S = Tr_B\left[\ket{\Psi}\bra{\Psi}\right].
\end{equation}
The crucial point is now to focus on the distance between the state $\hat{\rho}_S$ and the the canonical state $\hat{\Omega}_S$, namely we can ask how much $\hat{\rho}_S$ differs from $\hat{\Omega}_S$. The Principle of Canonical Typicality states that: \textit{given a sufficiently small subsystem $S$ of the Universe, for almost every pure states $\ket{\Psi} \in \mathcal{H}_R$, the subsystem $S$ is (at least approximately) in the canonical state $\hat{\Omega}_S$}, that is
\begin{equation}\label{6dadimostrare}
\hat{\rho}_S \approx \hat{\Omega}_S .
\end{equation}
As emphasised in \cite{popescu1}, this means that for almost every pure state $\ket{\Psi} \in \mathcal{H}_R$ of the Universe, the subsystem $S$ behaves as if the Universe was in the equiprobable state $\hat{\Omega}_R$. Equation (\ref{6dadimostrare}) implies that the thermal state of $S$ can be derived either from a (randomly chosen) pure state $\ket{\Psi} \in \mathcal{H}_R$ or from the maximally mixed state $\hat{\Omega}_R$.

\subsection{Derivation of Canonical Typicality}
We give a brief proof of Canonical Typicality following \cite{canonical}. As in the previous paragraph, we consider the Universe divided into a small subsystem $S$ and a large environment $B$ (with $d_S \ll d_B$) assuming the energy of the Universe within the interval $\left[E,E+\delta\right]$, where $\delta$ is small on the macroscopic scale ($\delta \ll E,\Delta E^{(S)}$ being $\Delta E^{(S)}$ the typical spacing between energy levels of the system $S$) but large enough for the interval to contain many eigenvalues of the global Hamiltonian $\hat{H}$. Neglecting the relatively small interaction between $B$ and $S$, the global Hamiltonian can be written $\hat{H} = \hat{H}_B + \hat{H}_S$.
What we want to demonstrate is the validity of equation (\ref{6dadimostrare}), namely we want to prove that  
\begin{equation}
\hat{\rho}_S = Tr_{B} \left[\ket{\Psi}\bra{\Psi}\right] \approx \hat{\Omega}_S \approx \frac{1}{Z} e^{-\beta \hat{H}_S}
\end{equation}
for typical (randomly chosen) pure wave-functions of the whole Universe $ \ket{\Psi} \in \mathcal{H}_R$.

As authors do in \cite{canonical} we begin by recalling the standard derivation of the canonical ensemble from the microcanonical. We start considering the global system described by the microcanonical density matrix as $\hat{\Omega}_R = d^{-1}_R \hat{\Pi}_R$. So, tracing on the degrees of freedom of $B$, we obtain the reduced density matrix $\hat{\Omega}_S$ of the system $S$:
\begin{equation}\label{6formuladadimostrare}
\hat{\Omega}_S = d^{-1}_R \sum_{i} d^{(i)}_B \ket{E^{(S)}_i}\bra{E^{(S)}_{i}} 
\end{equation}
where $d^{(i)}_B$ is the dimension of the spectral subspace of $\mathcal{H}_B$ associated with energies in $\left[   E - E^{(S)}_i, E - E^{(S)}_i + \delta    \right]$ and $\left\{ \ket{E^{(S)}_{i} }\right\}$ is the energy eigenbasis for the subsystem $S$ with corresponding eigenvalues $E^{(S)}_{i}$. At this point can be rigorously proven (under suitable conditions) that, when the environment $B$ is sufficiently large, we have $\hat{\Omega}_S \approx \frac{1}{Z} e^{-\beta \hat{H}_S}$. This follows from the fact that the entropy of the environment can be written $S(E) \approx \log\left[ d^{(R)}_B \right]$, leading to
\begin{equation}\label{6miserveora}
d^{(i)}_B \approx e^{S(E-E^{(S)}_i)} \approx e^{S(E) - \beta E^{(S)}_i} \sim e^{-\beta E^{(S)}_i} 
\end{equation}
where $\beta$ is defined through $\beta = dS(E)/dE$. 
Thus, in order to demonstrate Canonical Typicality, it is only necessary to show that (\ref{6formuladadimostrare}) holds (at least approximately) when $\hat{\Omega}_S$ is replaced by $\hat{\rho}_S = Tr_{B} \left[\ket{\Psi}\bra{\Psi}\right]$ for typical pure states $ \ket{\Psi} \in \mathcal{H}_R$.

As outlined in \cite{canonical} we consider that a random vector $ \ket{\Psi} \in \mathcal{H}_R$ <<can always be regarded as arising by normalization $\ket{\Psi}=\ket{\Phi}/||\ket{\Phi}||$, 
where $\ket{\Phi} \in \mathcal{H}_R$ is a Gaussian random vector with mean zero and covariance given by the identity operator on $\mathcal{H}_R$. This means that in the decomposition
\begin{equation}\label{6decomposition}
\ket{\Phi} = \sum_{i} \sum_{j} C_{ij} \ket{E^{(B)}_j} \otimes \ket{E^{(S)}_i}
\end{equation}
the real and the imaginary parts of the coefficients are indipendent real Gaussian random variables with mean zero and variance $1/2$ for those $i$ and $j$ for which $E^{(B)}_j + E^{(S)}_i \in \left[E,E+\delta\right]$ (and $C_{ij} =0$ otherwise)>>. 
We can now rewrite (\ref{6decomposition}) as:
\begin{equation}
\ket{\Psi} = \sum_{i} \ket{\phi_i} \otimes \ket{E^{(S)}_i}
\end{equation}
with 
\begin{equation}\label{613articolo}
\ket{\phi_i} = \sum_{j \in I_i} C_{ij} \ket{E^{(B)}_j}
\end{equation}
where $I_i$ is the set of the levels $j$ in $B$ for which $E^{(B)}_j \in \left[   E - E^{(S)}_i, E - E^{(S)}_i + \delta    \right]$. The reduced density matrix $\hat{\rho}_S$ for the subsystem $S$ is obtained by

\begin{equation}\label{6red}
\begin{split}
\hat{\rho}_S = Tr_{B} \left[\ket{\Psi}\bra{\Psi}\right] = \frac{1}{\Vert \Phi \Vert ^2} Tr_{B}  \left[\ket{\Phi}\bra{\Phi}\right]= \frac{1}{\Vert \Phi \Vert ^2} \sum_{i,i'} \braket{\phi_i|\phi_{i'}} \ket{E^{(S)}_i}\bra{E^{(S)}_{i'}} .
\end{split}
\end{equation}
Here the assumptions on the energy interval $\left[E,E+\delta\right]$ turn to be fundamental. Remembering that $\delta$ is so small that the system's energy spacings $\Delta E^{(S)}_i = E^{(S)}_{i+1} - E^{(S)}_{i}$ are all grater than $\delta$, we have that the relevant energy intervals $I_i$ for $B$ are pairwise disjoint and the states $\ket{\phi_i}$ pairwise orthogonal, that is $\braket{\phi_i|\phi_{i'}} = \delta_{i,i'} \Vert \phi_i \Vert ^2$.
Now, as indicated again in \cite{canonical}, from equation (\ref{613articolo}) <<we have that
\begin{equation}
\Vert \phi_i \Vert ^2 = \sum_{j} \left| C_{ij} \right|^2 ,
\end{equation}
so that $\Vert \phi_i \Vert ^2$ is the sum of the $N_i = d^{(i)}_B$ indipendent, identically distribuited random variables $\left| C_{ij} \right|^2$ with mean $1$>>. 
The last step is now to consider the law of large numbers, from which it follows that typically $\Vert \phi_i \Vert ^2  \approx d^{(i)}_B$.
This implies, for the reduced density matrix (\ref{6red}), the following expression:
\begin{equation}\label{6ultima}
\hat{\rho}_S \approx d^{-1}_R \sum_{i} d^{(i)}_B \ket{E^{(S)}_i}\bra{E^{(S)}_{i}} .
\end{equation} 
Equation (\ref{6ultima}) is exactly what was needed to show, indeed we have obtained for $\hat{\rho}_S$, at least approximately, the same expression of the right side of equation (\ref{6formuladadimostrare}).

\section{Environment as a Clock}

\subsection{General Framework}

We are now going to merge Canonical Typicality and PaW theory \cite{pagewootters,wootters}. The key point is to recognise the environment as a clock:
\begin{equation}
\hat{H}_B \equiv \hat{H}_C.
\end{equation}
Notice that in the PaW framework a good clock has to have a Hilbert space dimension larger than the dimension of the system $S$, otherwise it would no longer be possible to relate each energy eigenstate of the system to an energy eigenstate of the clock (see Chapter 4 and \cite{nostro}). Furthermore a good clock has to interact weakly with the system $S$ or, in the ideal case, it should not interact at all. These conditions coincides with those required for the environment by Canonical Typicality.

In the original PaW theory the global state of the Universe is an eigenstate of the total Hamiltonian with zero eigenvalue, i.e. $\hat{H}\ket{\Psi}=0$. 
Here we follow a slightly different path and we weakly relax the PaW constraint considering the total energy of the Universe
within the energy shell $\left[E,E+\delta\right]$ where $\delta \ll E,\Delta E^{(S)}$ but large enough to contain many energy eigenvalues of the clock $\equiv$ environment $C$. In this framework we will find a non-local Schrödinger-like evolution for the relative state of $S$ that reduces to the usual Schrödinger dynamics for times $t-t_0 \ll 1/\delta$ (where $t_0$ is the initial time) or for all times in the limit $\delta \rightarrow 0$.

We define the time states in the Hilbert space of the clock $\equiv$ environment using the approach developed in Chapter 3 and 4, when we considered Pegg's POVM complement of an Hamiltonian with unequally-spaced energy eigenvalues. Therefore we assume that the Hamiltonian of the environment has non-degenerate energy eigenstates with rational energy differences
\begin{equation}\label{6numraz}
\frac{E^{(C)}_i }{E^{(C)}_1} = \frac{A_i}{D_i},
\end{equation}
where $A_i$ and $D_i$ are integers with no common factors and $E^{(C)}_0 =0$. This implies that all energy values are integer multiples of a (arbitrarily small) step ($\hslash=1$):
\begin{equation}\label{6ei}
E^{(C)}_i = r_i \frac{2\pi}{T}
\end{equation}
where $T=\frac{2\pi r_1}{E_1}$, $r_i = r_1\frac{A_i}{D_i}$ for $i>1$ (with $r_0=0$) and $r_1$ is 
equal to the lowest common multiple of the values of $D_i$. In this space we define the states 
\begin{equation}
\ket{t_m}  = \frac{1}{\sqrt{p+1}}\sum_{i=0}^{p}e^{-i E^{(C)}_i t_m}\ket{E^{(C)}_i}
\end{equation}
with $p+1 = d_{C}$,
$t_m = t_0 + mT /(s+1)$, 
$m=0,1,2,...,s$ and $s+1 \ge r_p$. The number of states $\ket{t_m}$ is therefore greater than the number of energy states 
in $\mathcal{H}_{C}$ and the $s+1$ values of $t_m$ are uniformly distributed over $T$. These states are not othogonal but provide an overcomplete basis in $\mathcal{H}_{C}$ with the resolution of the identity
\begin{equation}\label{6pomidentity}
\frac{p+1}{s+1} \sum_{m=0}^{s} \ket{t_{m}}\bra{t_{m}} = \mathbb{1}_{C}.
\end{equation}
In order to obtain a continuous flow of time we can now consider the limit $s \rightarrow \infty $ and define
\begin{equation}\label{6alphastateinf}
\ket{{t}} = \sum_{i=0}^{p} e^{- i E^{(C)}_i t}\ket{E^{(C)}_i}
\end{equation}
where $t$ can now take any real value from $t_0$ to $t_0 + T $. In this limit the resolution of the identity (\ref{6pomidentity}) becomes
\begin{equation}\label{6newresolution}
\frac{1}{T} \int_{t_0}^{t_0+T} d t \ket{t} \bra{t} = \mathbb{1}_{C} 
\end{equation}
and the states $\ket{t}$ provide again an overcomplete basis in $\mathcal{H}_C$. 
As mentioned in Section 4.3.3 of Chapter 4, if we would consider non-rational ratios of energy levels, the resolutions of the identity (\ref{6pomidentity}) and (\ref{6newresolution}) are no longer exact and the time states $\ket{t_m}$ and $\ket{t}$ do not provide an overcomplete basis in $C$. However, since any real number can be approximated with arbitrary precision by a ratio between two rational numbers, the residual terms in the resolutions of the identity can be arbitrarily reduced. This allow us to consider any generic bounded Hamiltonian with discrete spectrum in describing the clock $\equiv$ environment $C$.

\subsection{Random Universe and Dynamics}

Here we show that a Universe in a (randomly chosen) pure state is compatible with the emergence of time and a non-trivial dynamical evolution of the system $S$. 
The global state of the Universe is
\begin{equation}\label{6statoglobale}
\ket{\Psi} = \sum_{j}\sum_{i} c_{ij} \ket{E^{(C)}_i} \otimes \ket{E^{(S)}_j}
\end{equation}
where we take the coefficients $c_{ij}$ distribuited as in \cite{canonical} (and already treated in Section 5.1.2).
With $\ket{\Psi}=\ket{\Phi}/||\ket{\Phi}||$ and $\ket{\Phi} =\sum_{j}\sum_{i} C_{ij} \ket{E^{(C)}_i} \otimes \ket{E^{(S)}_j}$,
the real and imaginary parts of the coefficients $C_{ij}=c_{ij}||\ket{\Phi}||$ are chosen as
independent real Gaussian random variables with mean zero and variance $1/2$ for the 
values of $i, j$ such that
$E^{(C)}_i + E^{(S)}_j \in \left[E,E+\delta\right]$ and $C_{ij}=0$ otherwise. 
Since
$\delta  \ll \Delta E^{(S)}_j = E^{(S)}_{j+1} - E^{(S)}_{j}~ \forall j$
and considering that the spectrum of $\hat{H}_{C}$ is much denser than the spectrum of $\hat{H}_{S}$, 
the constraint on the total energy implies that each level of the system is coupled with several neighbour levels of the clock. 
The random choice of the coefficients provides Canonical Typicality \cite{canonical}: here we show that is also sufficient
to provide the temporal dynamics for the relative state (in Everett sense \cite{everett}) of the subsystem $S$.
The action of the global Hamiltonian $\hat{H}$ on the global state of the Universe $\ket{\Psi}$ gives
\begin{equation}\label{6applicazionehtot}
\begin{split}
\hat{H}\ket{\Psi} & = 
\left( \hat{H}_{C} + \hat{H}_S  \right)\sum_{j}\sum_{i \in I_j} c_{ij} \ket{E^{(C)}_i} \otimes\ket{E^{(S)}_j} =
\\&= \sum_{j}\sum_{i \in I_j} c_{ij} \left( E^{(C)}_i + E^{(S)}_j  \right) \ket{E^{(C)}_i} \otimes\ket{E^{(S)}_j} =
\\&= E \ket{\Psi} + \sum_{j}\sum_{i \in I_j} c_{ij} \Delta_{ij} \ket{E^{(C)}_i}\otimes \ket{E^{(S)}_j}
\end{split} 
\end{equation} 
where $I_j$ is the set of the environment levels such that $E^{(C)}_i \in \left[E- E^{(S)}_j , E- E^{(S)}_j + \delta \right]$ 
and where we have written $E^{(C)}_i + E^{(S)}_j  = E + \Delta_{ij}$ with $\Delta_{ij} \in \left[0,\delta\right]$. The relative state of the system $S$ at a certain ``time" $t$ is defined by
\begin{equation}\label{6defstatorelativo}
\ket{\phi (t)}_S = \braket{t|\Psi}
\end{equation}
(notice that (\ref{6defstatorelativo}) is still a pure state) and its time evolution can be easily calculated:

\begin{equation}\label{6eqschroedinger}
\begin{split}
i \frac{\partial}{\partial t} \ket{\phi(t) }_S &= \frac{\partial}{\partial t} \sum_{k}^{d_{C}}\bra{E_k}e^{iE_k t}\ket{\Psi} = - \sum_{k}^{d_{C}}\bra{E_k}E_ke^{i E_k t}\ket{\Psi}= \\&
=-\bra{t}\hat{H}_{C}\ket{\Psi}=
\hat{H}_S \braket{t|\Psi} - \bra{t}\hat{H}\ket{\Psi}  =\\ \\&
= \left(\hat{H}_S - E\right) \ket{\phi(t)}_S - \bra{t}\sum_{j}\sum_{i \in I_j} c_{ij} \Delta_{ij} \ket{E^{(C)}_i}\otimes \ket{E^{(S)}_j} 
\end{split}
\end{equation}
where we have used (\ref{6alphastateinf}), $\hat{H} = \hat{H}_{C} + \hat{H}_S$, (\ref{6defstatorelativo}) and (\ref{6applicazionehtot}). By defining the operator $\hat{\Delta} =  \sum_{j} \sum_{i \in I_j} \Delta_{ij} \ket{E^{(C)}_i} \ket{E^{(S)}_j} \bra{E^{(C)}_i}  \bra{E^{(S)}_j}$
and removing the term related to $E$ which gives an irrelevant phase factor in the evolution of $S$, (\ref{6eqschroedinger}) becomes the time non-local Schrödinger-like equation:

\begin{equation}\label{6eqschrodinger2}
i \frac{\partial}{\partial t} \ket{\phi(t) }_S =  \hat{H}_S  \ket{\phi(t)}_S - \frac{1}{T} \int_{t_0}^{t_0+T} dt' \hat{\Delta}(t,t')\ket{\phi(t')}_S
\end{equation}
where $\hat{\Delta}(t,t') = \bra{t}\hat{\Delta}\ket{t'}$. 
The last term in the right-hand side of the equation is an integral operator acting on $S$. We already found the same kind of evolution for the state of $S$ when we addressed the case of interacting clock and system in Chapter 4. 


For times $t-t_0 \ll 1/\delta$ (and so $t-t_0 \ll 1/\Delta_{ij}$ for typical $\Delta_{ij}$) 
(\ref{6eqschrodinger2}) reduces to the ordinary Schr\"odinger equation. Indeed we have:

\begin{equation}\label{617new}
\begin{split}
\ket{\phi(t)}_S &= \braket{t|\Psi} = \bra{t_0}e^{i\hat{H}_C(t-t_0)}\ket{\Psi}= e^{-i\hat{H}_S(t-t_0)}\bra{t_0}e^{i\hat{H}(t-t_0)}\ket{\Psi} =\\&
= e^{-i\hat{H}_S(t-t_0)}\bra{t_0}\sum_{j}\sum_{i \in I_j}c_{ij} e^{i(E^{(C)}_i + E^{(C)}_j)(t-t_0)}\ket{E^{(C)}_i}\otimes\ket{E^{(S)}_j}=\\&
= e^{-i\hat{H}_S(t-t_0)}\bra{t_0}\sum_{j}\sum_{i \in I_j}c_{ij} e^{iE(t-t_0)} e^{i\Delta_{ij}(t-t_0)} \ket{E^{(C)}_i}\otimes\ket{E^{(S)}_j}
\end{split}
\end{equation}
where we used (\ref{6defstatorelativo}), $\hat{H} = \hat{H}_{C} + \hat{H}_S$ and $E^{(C)}_i + E^{(S)}_j  = E + \Delta_{ij}$. For $t-t_0 \ll 1/\delta$, considering $e^{i\Delta_{ij}(t-t_0)}\simeq 1$ and removing the irrelevant global phase factor $e^{iE(t-t_0)}$, (\ref{617new}) becomes

\begin{equation}
\ket{\phi(t)}_S \simeq e^{-i \hat{H}_S(t-t_0)}\ket{\phi(t_0)}_S 
\end{equation}
which provides the Schr\"odinger evolution for the system $S$. In Section 5.3 we'll briefly discuss the effect of the non-local term in equation (\ref{6eqschrodinger2}) for times $t- t_0 \ge 1/\delta$.

Equation (\ref{6eqschrodinger2}) can be explicitly solved: with $E^{(C)}_i + E^{(S)}_j = \Delta_{ij}$ (the term related to $E$ has been removed) we obtain (see Appendix 5.6.A):

\begin{equation}\label{6metodo2}
\begin{split}
\ket{\phi (t)}_S =  \sum_{j} \alpha_j(t)  e^{-i  E^{(S)}_j t}  \ket{E^{(S)}_j}
\end{split}
\end{equation}
with
\begin{equation}\label{6alphat}
\alpha_j (t) = \sum_{i \in I_j} c_{ij}e^{i \Delta_{ij} t} .
\end{equation}
Equation (\ref{6metodo2}) provides an additional time dependence to the Schr\"odinger evolution through the coefficients $\alpha_j(t)$. 
In the case $t-t_0 \ll 1/\delta$ we have
$\alpha(t) \simeq  \sum_{i \in I_j}c_{ij}e^{i\Delta_{ij}t_0} \equiv \alpha_j(t_0)$ and the state (\ref{6metodo2}) becomes 
\begin{equation}\label{6evscr}
\begin{split}
\ket{\phi (t)}_S \simeq 
\sum_{j} \alpha_j(t_0)  e^{-i E^{(S)}_j t} \ket{E^{(S)}_j} 
\end{split}
\end{equation}
where we recognize again the Schr\"odinger evolution for the state of the system $S$.

We emphasize that (\ref{6eqschrodinger2}) does not preserve the norm of the state $\ket{\phi(t)}_S$ over time. Indeed, calculating the scalar product $\braket{\phi(t)|\phi(t)}$ through (\ref{6metodo2}) we obtain
\begin{equation}\label{6evoluzionenorma}
\begin{split}
\braket{\phi(t)|\phi(t)} &= \sum_{j} \left|\alpha_j(t)\right|^2 = \sum_{j} \sum_{i \in I_j} \sum_{k \in I_j} c_{ij}c^{*}_{kj} e^{i(\Delta_{ij} - \Delta_{kj})t} =\\&
=  \sum_{j} \sum_{i \in I_j} \sum_{k \in I_j}|c_{ij}||c_{kj}|\cos((\Delta_{ij} - \Delta_{kj})t - \Delta\varphi^{(j)}_{ik})
\end{split} 
\end{equation}
where $c_{ij}=|c_{ij}|e^{i\varphi_{ij}}$, $c_{kj}=|c_{kj}|e^{i\varphi_{kj}}$ and $\Delta\varphi^{(j)}_{ik} = \varphi_{kj} - \varphi_{ij}$. However, the corrections remain small when $t-t_0 \ll 1/\delta$ (and vanish in the limit $\delta\rightarrow 0$) 
being	
$\braket{\phi(t)|\phi(t)} \simeq \sum_{j} \sum_{i,k \in I_j} c_{ij}c^{*}_{kj}e^{i(\Delta_{ij} - \Delta_{kj})t_0}=\sum_{j} |\alpha_j(t_0)|^2$ that is different from the unity but (approximately) constant over time. We notice that a similar problem arose, for different reasons, in \cite{interacting} (where the case of interacting clock and system is treated) and it was 
handled by introducing a new definition for the inner product. 



To summarise, the environment can provide the clock for the dynamical evolution of the system $S$. 
The state of $S$ conditioned to a certain value of the clock through (\ref{6defstatorelativo}) consists of a pure state obeying a non-local
dynamical Schrödinger-like equation (that reduces to the standard Schrödinger equation for $t-t_0 \ll 1/\delta$). Nevertheless, after tracing out the degrees of freedom of the clock $\equiv$ environment, we find the system in a state of thermal equilibrium provided by the canonical distribution. This compatibility is simply explained by the fact that 
the trace over the environment degrees of freedom is equivalent to the trace over all times. Indeed we have (see Appendix 5.6.B at the end of this Chapter):
\begin{equation}\label{6mediatemporale}
\hat{\rho}_S = Tr_C\left[\ket{\Psi}\bra{\Psi}\right] = \frac{1}{T} \int_{t_0}^{t_0+T} dt \braket{t|\Psi}\braket{\Psi|t}   \approx \frac{1}{Z} e^{-\beta \hat{H}_S}
\end{equation}
where again $\beta = dS(E)/dE$ is the inverse temperature with $S(E)$ the clock $\equiv$ environment entropy and $Z = Tr \left[ e^{-\beta \hat{H}_S}\right]$ as in equation (\ref{6rhobeta}). 

\subsection{Initial Conditions for $S$}
\indent

The merging of Canonical Typicality and PaW imposes a constraint on the allowed initial conditions of the 
subsystem $S$. 
The initial condition for the solution state (\ref{6metodo2}) is 
\begin{equation}\label{6defstatoiniziale}
\ket{\phi (t_0)}_S =   \sum_{j}\alpha_j(t_0) e^{-iE^{(S)}_jt_0 } \ket{E^{(S)}_j}
\end{equation}
with
\begin{equation}\label{6alpha}
\alpha_j(t_0) = \sum_{i \in I_j} c_{ij} e^{i\Delta_{ij}t_0}
\end{equation}
The reduced density matrix $\hat{\rho}_S$ of the subsystem $S$
is
\begin{equation}\label{6mdr}
\begin{split}
\hat{\rho}_S = Tr_C\left[\ket{\Psi}\bra{\Psi}\right] 
= \sum_{j} \sum_{i\in I_j}\left|c_{ij} \right|^2 \ket{E^{(S)}_j}\bra{E^{(S)}_j}
\end{split}
\end{equation}
where we have again considered that the relevant energy intervals for the clock $\equiv$ environment coupled with each level $E^{(S)}_j$ of $S$ are pairwise disjoint (which is his is a consequence of $\delta  \ll \Delta E^{(S)}_j = E^{(S)}_{j+1} - E^{(S)}_{j}~ \forall j$). Canonical Typicality implies
\begin{equation}\label{6condcoeff}
\sum_{i \in I_j}\left|  c_{ij} \right|^2 \approx \frac{1}{Z} e^{-\beta E^{(S)}_j}
\end{equation}
which constraints the initial conditions by selecting a set of allowed $\alpha_j(t_0)$ through (\ref{6alpha}). It is crucial to note that 
the condition (\ref{6condcoeff}) contraints the sum of the absolute values of the coefficients $c_{ij}$ and then leaves a large margin of freedom 
on the possible values of $\alpha_j(t_0)$. 	
In conclusion, Canonical Typicality states that the reduced density matrix $\hat{\rho}_S=Tr_{C}\left[\ket{\Psi}\bra{\Psi}\right]$ of the \textit{overwhelming majority} of the pure wave-functions $\ket{\Psi} \in \mathcal{H}_R$ is canonical. This means that the overwhelming majority of the randomly chosen coefficients $c_{ij}$ satisfies the condition (\ref{6condcoeff}) which, in our framework, is
consistent with a non-trivial dynamical evolution of $S$. 

\section{A Toy Model}

We look now at a simple example assuming that the subsystem $S$ consists of a one-dimensional harmonic oscillator 
with Hamiltonian $\hat{H}_S = \frac{\hat{P}^2}{2m} + \frac{1}{2}m\omega^2\hat{X}^2$. 
We assume that the dynamics is confined among the two lowest energy levels of the oscillator and therefore the global state of the Universe (\ref{6statoglobale}), satisfying the constraint on the total energy, is 
	\begin{equation}
	\ket{\Psi} = \sum_{i \in I_0} c_{i0} \ket{E^{(C)}_i}\otimes\ket{0^{(S)}} + \sum_{i \in I_1} c_{i1} \ket{E^{(C)}_i}\otimes\ket{1^{(S)}} .
	\end{equation}

With (\ref{6defstatoiniziale}) and $t_0 = 0$, we look at the (pure) initial state of the subsystem $S$: $\ket{\phi(0)}_S = \alpha_0(0) \ket{0^{(S)}} + \alpha_1(0) \ket{1^{(S)}}$, 	
where we set the initial values $\alpha_0(0)$ and $\alpha_1(0)$ according to (\ref{6alpha}). Thanks to (\ref{6metodo2}) and (\ref{6alphat}), for the conditioned state of $S$ at a generic time $t$, we have

\begin{equation}\label{6statotoymodel}
\begin{split}
\ket{\phi(t)}_S &= \alpha_0(t) e^{-i E^{(S)}_0t} \ket{0^{(S)}} +  \alpha_1(t) e^{-i E^{(S)}_1 t} \ket{1^{(S)}} =\\&
= \sum_{i \in I_0} c_{i0} e^{-i (E^{(S)}_0 - \Delta_{i0})t} \ket{0^{(S)}} + \sum_{i \in I_1} c_{i1} e^{-i (E^{(S)}_1 - \Delta_{i1})t} \ket{1^{(S)}}.
\end{split}
\end{equation}
We recall here that the state (\ref{6statotoymodel}) is not normalized. To restore the normalization (for $t\ll 1/\delta$ where the norm is approximately preserved) we should divide (\ref{6statotoymodel}) by $\sqrt{\alpha^2_0(0) + \alpha^2_1(0)}$. However, for the sake of simplicity, we proceed  with the calculation without considering the normalization of the state.

We look now at the time dependence of the expectation value of the position operator $\hat{X} = \sqrt{\frac{1}{2m\omega}} (\hat{a} + \hat{a}^{\dagger})$ where $\hat{a}$ and $\hat{a}^{\dagger}$ are the usual lowering and rising operators. We calculate $\langle \hat{X} \rangle_{t} = \bra{\phi(t)} \hat{X} \ket{\phi(t)}$ and we obtain (see Appendix 5.6.C at the end of this Chapter):

\begin{equation}\label{6evestgrande}
\begin{split}
\langle \hat{X} \rangle_{t} &=\sqrt{\frac{2}{m\omega}} |\alpha_{0}(t)||\alpha_{1}(t)|\cos(\omega t - \Delta\phi(t)) =\\&
=  \sqrt{\frac{2}{m\omega}} \sum_{i \in I_0} \sum_{k \in I_1}  |c_{i0}||c_{k1}|\cos((\omega  + (\Delta_{i0} - \Delta_{k1}))t - \Delta\varphi^{(0,1)}_{ik})
\end{split}
\end{equation}
where we have considered $\alpha_{0}(t)= |\alpha_{0}(t)|e^{i\phi_{0}(t)}$, $\alpha_{1}(t)= |\alpha_{1}(t)|e^{i\phi_{1}(t)}$, $\Delta\phi (t)= \phi_{1}(t) - \phi_{0}(t)$, $c_{i0}= |c_{i0}|e^{i\varphi_{i0}}$, $c_{k1}= |c_{k1}|e^{i\varphi_{k1}}$, $\Delta\varphi^{(0,1)}_{ik} = \varphi_{k1} - \varphi_{i0}$ and $E^{(S)}_1 - E^{(S)}_0 = \omega$. 

For times $t \ll 1/|\Delta_{i0} - \Delta_{k1}|$, up to first order of approximation in $t( \Delta_{i0} - \Delta_{k1})$, (\ref{6evestgrande}) reduces to (see Appendix 5.6.D at the end of this Chapter):

\begin{multline}\label{6dimappe}
\langle \hat{X} \rangle_{t} \simeq  \sqrt{\frac{2}{m\omega}}  |\alpha_0(0)||\alpha_1(0)|\cos(\omega t - \Delta \phi(0)) + \\ -  \sqrt{\frac{2}{m\omega}} \sum_{i \in I_0} \sum_{k \in I_1}  |c_{i0}||c_{k1}|t( \Delta_{i0} - \Delta_{k1} )\sin(\omega t  - \Delta\varphi^{(0,1)}_{ik})  
\end{multline}
where $\alpha_0(0) = |\alpha_0(0)|e^{i \phi_0(0)}$, $\alpha_1(0) = |\alpha_1(0)|e^{i \phi_1(0)}$, $\Delta \phi(0) = \phi_1(0) - \phi_0(0)$ and where we used again (\ref{6alpha}). Equation (\ref{6dimappe}) indicates, as expected, that the expectation value $\langle \hat{X} \rangle_{t}$ oscillates between $\pm \sqrt{\frac{2}{m\omega}} |\alpha_0(0)||\alpha_1(0)|$ with frequency $\omega$ (apart from small corrections) and this is not surprising since we know that for $t\ll 1/\delta$ the system $S$ exhibits a Scr\"odinger-like evolution. 

Conversely (\ref{6evestgrande}) shows the behaviour of $\langle \hat{X} \rangle_{t}$ for times $t \ge 1/\delta$, which results a little more tricky because of the impossibility, in this case, of neglecting the effects due to the non-local term in (\ref{6eqschrodinger2}). We can ask here how long is the period of time in which the evolution of $S$ is Scr\"odinger-like. The answer is very simple: such a duration strongly depends on the magnitude of $\delta$. Clearly a smaller $\delta$ will increase the time interval in which $S$ exhibits a Scr\"odinger-like evolution and in the limit $\delta \rightarrow 0$, as already metioned, we recover the Scr\"odinger dynamics for all times. 	

Instead of exploring what happens at each instant of time through the conditional state $\ket{\phi(t)}_S$, we can trace over the clock $\equiv$ environment degrees of freedom which correponds, in our framework, to a time average. Through equations (\ref{6mediatemporale}), (\ref{6mdr}) and (\ref{6condcoeff}), for the overwhelming majority of the randomly chosen $c_{ij}$ we have:

\begin{equation}\label{ultimaes}
\begin{split}
\hat{\rho}_S & = \frac{1}{T} \int_{0}^{T} dt \braket{t|\Psi}\braket{\Psi|t} = \sum_{i \in I_0}|c_{i0}|^2 \ket{0^{(S)}}\bra{0^{(S)}} + \sum_{i \in I_1}|c_{i1}|^2 \ket{1^{(S)}}\bra{1^{(S)}} \approx  \\&
\approx \frac{1}{Z}\left( e^{-\beta E^{(S)}_0} \ket{0^{(S)}}\bra{0^{(S)}} + e^{-\beta E^{(S)}_1} \ket{1^{(S)}}\bra{1^{(S)}} \right) .
\end{split}
\end{equation}
This is the canonical mixed density matrix for the subsystem $S$ that, as mentioned, describes the equilibrium state resulting from tracing over all times.

\section{Non-Observable Universe as Clock}

A condition to merge Canonical Typicality and PaW time is to have an environment that is negligeably interacting with $S$.
This is certainly not the case in our everyday life where decoherence due to the
surrounding environment is often non-negligeable. So we return here to the point addressed in Section 4.6 of Chapter 4. Considering indeed our closed quantum global system as the whole Universe where $S$ is the observable universe, a possible choice of a good clock is the non-observable Universe.
In this respect, it is intriguing to notice that the recent observations on the cosmic microwave background \cite{planck} toghether with the inflationary paradigm indicate that at the beginning of cosmic inflation the Universe was in a pure state with highly-correlated quantum fluctuations \cite{bianchi}. Furthermore, it has been suggested that the assumption that the observable and the non-observable Universe might be entangled provides an argument in support of inflation \cite{vedraltemperature}, \cite{vedralinflation}. It is therefore somehow natural to speculate that the non-observable Universe acts as a clock for the observable Universe. Indeed in this framework the two requirements for a ``good clock" are satisfied: the clock and the system $S$ are non-interacting 
(or weakly interacting) and, in addition, the dimension of the clock is presumably larger than the dimension of $S$ (i.e. the the non-observable Universe is bigger than the observable Universe). 
A very simple estimate supports the consistency of the scenario.  
The spacing between the energy levels in the clock space is: 

\begin{equation}\label{6spacingenergy}
\delta E^{(C)}_i = E^{(C)}_{i+1}  + E^{(C)}_i = \frac{2\pi \hslash}{T} \left( r_{i+1} - r_i \right)  
\end{equation}
where $r_{i+1} - r_i$ is an integer. Notice that $\delta E^{(C)}_{min} = 2\pi \hslash/T$ is the minimum energy step value, so that all other energy values can be considered as integer multiples of this minimum step. The $\delta E^{(C)}_{min}$ is inversely proportional to the value of $T$, that is the time taken by the clock to return to its initial state.
So, considering the global system as the whole Universe, 
we can relate $T$ to the current age of the Universe $T_U$ by assuming 

\begin{equation}
T \ge T_{U} \sim 13.8 \times 10^{9} y \simeq 4.35 \times 10^{17} s
\end{equation}
and $t_0 = 0$ the instant of the Big Bang. This leads to 

\begin{equation}\label{633}
\delta E^{(C)}_{min} = \frac{2\pi \hslash}{T} \le \frac{2\pi \hslash}{T_{U}} \simeq 1.5 \times 10^{-51} J .
\end{equation}
This upper limit for $\delta E^{(C)}_{min}$ is very small compared to other energies on subatomic scales and fits into our framework of constructing the spectrum of the Hamiltonian $\hat{H}_{C}$ with integer multiples of a minimum energy step. Equation (\ref{633}) constitutes however an upper bound for the value of $\delta E^{(C)}_{min}$ that can be further reduced. Indeed, considering a cosmological model for which $T\gg T_U$ (i.e. the entire life of the Universe is assumed much greater than the current age), we have $\delta E^{(C)}_{min} \ll \frac{2\pi \hslash}{T_{U}}$. 


As a last consideration on this point we note that the fact of considering the non-observable Universe as clock for the observable Universe implies that from inside the system $S$ we have no no access to the clock. This is not in contradiction with the PaW mechanism where indeed the clock and the system $S$ are not interacting, but are entangled. It is thanks to entanglement that clock values $t$ enter in the evolution of system $S$ which is therefore able to perceive the passage of time. On this issue we have already made our considerations in Section 2.5 of Chapter 2.

\section{Conclusions}

In this Chapter we have merged Canonical Typicality and the PaW quantum time following \cite{nostro2}.
We considered a quantum Universe made by a small system $S$ and a large environment which serves as a clock for $S$. Thanks to Canonical Typicality we know that for almost all pure states in which the whole Universe can be, after tracing over the environment the system $S$ is in a state of equilibrium described by the canonical distribution. In the same scenario we find a Schrödinger-like evolution corrected by a non-local term for the relative state of $S$ with respect to the clock $\equiv$ environment. Canonical typicality and dynamics can coexist because in our protocol the action of tracing out the environment is equivalent to tracing over all times: the average over the environment coincides with a temporal average.


\section{Appendices}

\subsection*{5.6.A \: Proof of Equation (\ref{6metodo2})}

Here we show that the state 

\begin{equation}\label{6soluzione}
\ket{\phi (t)}_S =   \sum_{j} \sum_{i \in I_j} c_{ij}  e^{-i ( E^{(S)}_j - \Delta_{ij} )t}  \ket{E^{(S)}_j}
\end{equation}
is a solution for the equation 

\begin{equation}\label{6eqdimezzo}
i \frac{\partial}{\partial t} \ket{\phi(t) }_S = \hat{H}_S  \ket{\phi(t)}_S - \bra{t}\hat{\Delta}\ket{\Psi} 
\end{equation}
(where $\hat{\Delta} =  \sum_{j} \sum_{i \in I_j} \Delta_{ij} \ket{E^{(C)}_i} \ket{E^{(S)}_j} \bra{E^{(C)}_i}  \bra{E^{(S)}_j}$) and for equation (\ref{6eqschrodinger2}). In the first case we substitute (\ref{6soluzione}) in (\ref{6eqdimezzo}) thus obtaining

\begin{multline}
i \left( -i \sum_{j}\sum_{i\in I_j} c_{ij} E^{(S)}_j e^{-i(E^{(S)}_j - \Delta_{ij})t}\ket{E^{(S)}_j} + i  \sum_{j}\sum_{i\in I_j} c_{ij} \Delta_{ij} e^{-i(E^{(S)}_j - \Delta_{ij})t}\ket{E^{(S)}_j} \right) = \\= \hat{H}_S \ket{\phi (t)}_S - \bra{t}\sum_{j} \sum_{i\in I_j} c_{ij} \Delta_{ij} \ket{E^{(C)}_i} \otimes\ket{E^{(S)}_j}
\end{multline}
\begin{multline}
\Rightarrow \hat{H}_S \ket{\phi (t)}_S - \sum_{j}\sum_{i\in I_j} c_{ij} \Delta_{ij} e^{-i(E^{(S)}_j - \Delta_{ij})t}\ket{E^{(S)}_j} =\\=  \hat{H}_S \ket{\phi (t)}_S -  \bra{t}\sum_{j} \sum_{i\in I_j} c_{ij} \Delta_{ij} \ket{E^{(C)}_i} \otimes\ket{E^{(S)}_j} .
\end{multline}
So we have:

\begin{equation}\label{644}
\sum_{j}\sum_{i\in I_j} c_{ij} \Delta_{ij} e^{-i(E^{(S)}_j - \Delta_{ij})t}\ket{E^{(S)}_j} = \bra{t}\sum_{j} \sum_{i\in I_j} c_{ij} \Delta_{ij} \ket{E^{(C)}_i}\otimes \ket{E^{(S)}_j}
\end{equation}
that is an identity, considering that in the right-hand side of the equation (\ref{644}) $\braket{t|E^{(C)}_i} = e^{iE^{(C)}_i t}$ and the global constraint on total energy gives $E^{(C)}_i = - E^{(S)}_j + \Delta_{ij}$ (remember that we have removed the term related to the total energy $E$).

To verify that the (\ref{6soluzione}) is a solution for (\ref{6eqschrodinger2}) we have first to see how $\hat{\Delta}(t,t')$ acts on $\ket{\phi (t')}_S$. We have:
\begin{equation}\label{641}
\hat{\Delta}(t,t') \ket{\phi (t')}_S = \sum_{j}\sum_{k \in I_j} \sum_{i \in I_j} e^{iE^{(C)}_k t} \Delta_{kj} e^{-iE^{(C)}_k t'}c_{ij} e^{iE^{(C)}_it'}\ket{E^{(S)}_j}.  
\end{equation}
So, by substituting the state (\ref{6soluzione}) in equation (\ref{6eqschrodinger2}) and using (\ref{641}), we obtain:
\begin{multline}
\sum_{j}\sum_{i\in I_j} c_{ij} \Delta_{ij} e^{-i(E^{(S)}_j - \Delta_{ij})t}\ket{E^{(S)}_j} =\\= \frac{1}{T} \int_{0}^{T} dt' \sum_{j}\sum_{k\in I_j} \sum_{i\in I_j} e^{iE^{(C)}_k t} \Delta_{kj} e^{-iE^{(C)}_k t'} c_{ij} e^{iE^{(C)}_it'}\ket{E^{(S)}_j}
\end{multline}
\begin{equation}
\begin{split}
\Rightarrow \sum_{j}\sum_{i\in I_j} c_{ij} \Delta_{ij} e^{-i(E^{(S)}_j - \Delta_{ij})t}\ket{E^{(S)}_j} = \sum_{j}\sum_{i\in I_j} c_{ij} \Delta_{ij} e^{-i(E^{(S)}_j - \Delta_{ij})t}\ket{E^{(S)}_j}
\end{split}
\end{equation}
where in the last step we used again the constraint on the energy $E^{(C)}_i = - E^{(S)}_j + \Delta_{ij}$ and (see Appendix 4.8.D in Chapter 4 for the proof): 
\begin{equation}
\int_{t_0}^{t_0+T} dt'  e^{i (E^{(C)}_i - E^{(C)}_k)t'} =T \delta_{i,k} .
\end{equation}


\subsection*{5.6.B \: Proof of Equation (\ref{6mediatemporale})}

Here we prove equation (\ref{6mediatemporale}), namely we show that, although the states $\ket{t}$ are not orthogonal, we have

\begin{equation}\label{641b}
\hat{\rho}_S =Tr_C\left[\ket{\Psi}\bra{\Psi}\right]= \frac{1}{T} \int_{t_0}^{t_0+T} dt \braket{t|\Psi}\braket{\Psi|t}.
\end{equation}
Then, thanks to Canonical Typicality, we know that $\hat{\rho}_S \approx \frac{1}{Z} e^{-\beta \hat{H}_S}$ where $\beta = dS(E)/dE$ is the inverse temperature (with $S(E)$ the entropy of $C$) and $Z = Tr \left[ e^{-\beta \hat{H}_S}\right]$ \cite{canonical}. We start calculating the partial trace of the global state $Tr_C\left[\ket{\Psi}\bra{\Psi}\right]$ through the energy basis in the subspace $C$:

\begin{equation}\label{651}
\begin{split}
\hat{\rho}_S &= Tr_C\left[ \ket{\Psi}\bra{\Psi}\right] = \sum_{n} \braket{E^{(C)}_n|\Psi}\braket{\Psi|E^{(C)}_n} = \\&
=\sum_{n}  \sum_{j}\sum_{i\in I_j}  \sum_{l} \sum_{k \in I_l} c_{ij}c^{*}_{kl} \braket{E^{(C)}_n|E^{(C)}_i}\braket{E^{(C)}_k|E^{(C)}_n}\ket{E^{(S)}_j}\bra{E^{(S)}_l}
\end{split}
\end{equation}  
where $I_j$ is the set of the environment levels such that 
\begin{equation}
E^{(C)}_i \in \left[E- E^{(S)}_j , E- E^{(S)}_j + \delta \right] .
\end{equation}
Now, being $\delta  \ll \Delta E^{(S)}_j ~ \forall j$, then the energy intervals for the clock $\equiv$ environment coupled with each level $E^{(S)}_j$ of $S$ are pairwise disjoint. So equation (\ref{651}) becomes

\begin{equation}\label{6asterisco}
\hat{\rho}_S = Tr_C\left[ \ket{\Psi}\bra{\Psi}\right] 
=  \sum_{j} \sum_{i\in I_j}\left|c_{ij} \right|^2 \ket{E^{(S)}_j}\bra{E^{(S)}_j} .
\end{equation}
Going instead to calculate the right-hand side of equation (\ref{641b}) we have

\begin{equation}
\begin{split}
\frac{1}{T} \int_{t_0}^{t_0+T} dt \braket{t|\Psi}\braket{\Psi|t} &= \frac{1}{T} \int_{t_0}^{t_0+T} dt \sum_{j}\sum_{i\in I_j} \sum_{l}  \sum_{k \in I_l} c_{ij}c^{*}_{kl} \braket{t|E^{(C)}_i}\braket{E^{(C)}_k|t} \ket{E^{(S)}_j}\bra{E^{(S)}_l}= \\&
= \sum_{j}\sum_{i\in I_j} \sum_{l}  \sum_{k \in I_l} c_{ij}c^{*}_{kl} \frac{1}{T} \int_{t_0}^{t_0+T} dt e^{-it(E^{(C)}_i - E^{(C)}_k)} \ket{E^{(S)}_j}\bra{E^{(S)}_l}
\end{split}
\end{equation}
and, considering $\int_{t_0}^{t_0+T} dt e^{-it(E^{(C)}_i - E^{(C)}_k)} = T\delta_{i,k}$ (see Appendix 4.8.D in Chapter 4), we obtain

\begin{equation}\label{654}
\frac{1}{T} \int_{t_0}^{t_0+T} dt \braket{t|\Psi}\braket{\Psi|t} 
= \sum_{j} \sum_{i\in I_j}\left|c_{ij} \right|^2 \ket{E^{(S)}_j}\bra{E^{(S)}_j}
\end{equation}
where we used again the fact that the relevant energy intervals for the clock $\equiv$ environment coupled with each level $E^{(S)}_j$ of $S$ are pairwise disjoint. Equation (\ref{654}) that is the same of (\ref{6asterisco}), so 

\begin{equation}
\frac{1}{T} \int_{t_0}^{t_0+T} dt \braket{t|\Psi}\braket{\Psi|t} = Tr_C\left[\ket{\Psi}\bra{\Psi}\right] = \hat{\rho}_S .
\end{equation}

\subsection*{5.6.C \: Proof of Equation (\ref{6evestgrande})}

We calculate here the expectation value $\langle \hat{X} \rangle_{t} = \bra{\phi(t)} \hat{X} \ket{\phi(t)}$, where $\hat{X}=\sqrt{\frac{1}{2m\omega}} (\hat{a} + \hat{a}^{\dagger})$, considering the relative state $\ket{\phi(t)}_S$ written as 

\begin{equation}
\ket{\phi(t)}_S = \sum_{i \in I_0} c_{i0} e^{-i (E^{(S)}_0 - \Delta_{i0})t} \ket{0^{(S)}} + \sum_{i \in I_1} c_{i1} e^{-i (E^{(S)}_1 - \Delta_{i1})t} \ket{1^{(S)}}.
\end{equation}
We have:

\begin{multline}
\langle \hat{X} \rangle_{t} = \left( \sum_{i \in I_0} c^{*}_{i0}e^{i(E^{(S)}_0 - \Delta_{i0})t}\bra{0^{(S)}} + \sum_{i \in I_1} c^{*}_{i1}e^{i(E^{(S)}_1 - \Delta_{i1})t}\bra{1^{(S)}} \right) \times \\
\times \sqrt{\frac{1}{2m\omega}} (\hat{a} + \hat{a}^{\dagger}) \left(  \sum_{i \in I_0} c_{i0}e^{-i(E^{(S)}_0 - \Delta_{i0})t}\ket{0^{(S)}} + \sum_{i \in I_1} c_{i1}e^{-i(E^{(S)}_1 - \Delta_{i1})t}\ket{1^{(S)}} \right)
\end{multline}
\begin{multline}\label{6contox1}
\Rightarrow  \langle \hat{X} \rangle_{t} = \sqrt{\frac{1}{2m\omega}} \left( \sum_{i \in I_0} c^{*}_{i0}e^{i(E^{(S)}_0 - \Delta_{i0})t}\bra{0^{(S)}} + \sum_{i \in I_1} c^{*}_{i1}e^{i(E^{(S)}_1 - \Delta_{i1})t}\bra{1^{(S)}} \right) \times \\
\times \left( \sum_{i \in I_1}c_{i1} e^{-i(E^{(S)}_1 - \Delta_{i1})t}\ket{0^{(S)}} + \sum_{i \in I_0}c_{i0} e^{-i(E^{(S)}_0 - \Delta_{i0})t}\ket{1^{(S)}} + \sqrt{2} \sum_{i \in I_1}c_{i1} e^{-i(E^{(S)}_1 - \Delta_{i1})t}\ket{2^{(S)}}\right) .
\end{multline}
From equation (\ref{6contox1}) we have:

\begin{multline}\label{6fff}
\langle \hat{X} \rangle_{t} = \sqrt{\frac{1}{2m\omega}}  \sum_{i \in I_0} \sum_{k \in I_1} c^{*}_{i0} c_{k1} e^{i(E^{(S)}_0 - E^{(S)}_1 -\Delta_{i0} +\Delta_{k1})t } +\\+ \sqrt{\frac{1}{2m\omega}}  \sum_{i \in I_0} \sum_{k \in I_1} c_{i0} c^{*}_{k1} e^{-i(E^{(S)}_0 - E^{(S)}_1 -\Delta_{i0} +\Delta_{k1})t } .
\end{multline}
Now, writing $c_{i0}= |c_{i0}|e^{i\varphi_{i0}}$, $c_{k1}= |c_{k1}|e^{i\varphi_{k1}}$, $\Delta\varphi^{(0,1)}_{ik} = \varphi_{k1} - \varphi_{i0}$ and considering that $E^{(S)}_1 - E^{(S)}_0 = \omega$, we obtain:

\begin{equation}\label{6contox2}
\langle \hat{X} \rangle_{t} = \sqrt{\frac{2}{m\omega}} \sum_{i \in I_0} \sum_{k \in I_1}  |c_{i0}||c_{k1}|\cos((\omega  + (\Delta_{i0} - \Delta_{k1}))t - \Delta\varphi^{(0,1)}_{ik}) 
\end{equation}
which is what we had to prove, since equation (\ref{6contox2}) is the same of the second part of equation (\ref{6evestgrande}). 

If we want instead to consider the expectation value $\langle \hat{X} \rangle_{t}$ expressed in terms of the $\alpha_j(t)$, through the definition (\ref{6alphat}): 
\begin{equation}
\alpha_j (t) = \sum_{i \in I_j} c_{ij}e^{i \Delta_{ij} t} ,
\end{equation}
we can rewrite equation (\ref{6fff}) as

\begin{equation}\label{6ffff}
\langle \hat{X} \rangle_{t} = \sqrt{\frac{1}{2m\omega}} \left(  \alpha^{*}_{0}(t) \alpha_{1}(t) e^{i(E^{(S)}_0 - E^{(S)}_1)t } +  \alpha_{0}(t) \alpha^{*}_{1}(t) e^{-i(E^{(S)}_0 - E^{(S)}_1)t } \right).
\end{equation}
Writing $\alpha_{0}(t)= |\alpha_{0}(t)|e^{i\phi_{0}(t)}$, $\alpha_{1}(t)= |\alpha_{1}(t)|e^{i\phi_{1}(t)}$, $\Delta\phi (t)= \phi_{1}(t) - \phi_{0}(t)$ and considering again $E^{(S)}_1 - E^{(S)}_0 = \omega$, (\ref{6ffff}) becomes:

\begin{equation}\label{6ggg}
\langle \hat{X} \rangle_{t} =\sqrt{\frac{2}{m\omega}} |\alpha_{0}(t)||\alpha_{1}(t)|\cos(\omega t - \Delta\phi(t)).
\end{equation}
Equation (\ref{6ggg}) is the same of the first part of equation (\ref{6evestgrande}) and shows the expectation value $\langle \hat{X} \rangle_{t}$ expressed in terms of the time-dependent coefficients $\alpha_j(t)$.

\subsection*{5.6.D \: Proof of Equation (\ref{6dimappe})}

We prove here that equation (\ref{6evestgrande}) reduces to (\ref{6dimappe}) in the case of $t \ll 1/|\Delta_{i0} - \Delta_{k1}|$. We start considering the second part of equation (\ref{6evestgrande}), that is:
\begin{equation}\label{6evestgrandeapp}
\langle \hat{X} \rangle_{t} = \sqrt{\frac{2}{m\omega}} \sum_{i \in I_0} \sum_{k \in I_1}  |c_{i0}||c_{k1}|\cos(\omega t  + (\Delta_{i0} - \Delta_{k1})t - \Delta\varphi^{(0,1)}_{ik}) .
\end{equation}
This can be rewritten as

\begin{multline}
\langle \hat{X} \rangle_{t} = \sqrt{\frac{2}{m\omega}} \left( \sum_{i \in I_0} \sum_{k \in I_1}|c_{i0}||c_{k1}|\cos(\omega t  - \Delta\varphi^{(0,1)}_{ik})\cos( (\Delta_{i0} - \Delta_{k1})t) \right) + \\
-       \sqrt{\frac{2}{m\omega}}  \left( \sum_{i \in I_0} \sum_{k \in I_1}|c_{i0}||c_{k1}| \sin(\omega t  - \Delta\varphi^{(0,1)}_{ik})\sin((\Delta_{i0} - \Delta_{k1})t)  \right) 
\end{multline}
We impose now the condition $t \ll 1/|\Delta_{i0} - \Delta_{k1}|$, so we can consider the Taylor expansions of $\cos((\Delta_{i0} - \Delta_{k1})t)$ and $\sin((\Delta_{i0} - \Delta_{k1})t)$ as

\begin{equation}
\begin{split}
&\cos((\Delta_{i0} - \Delta_{k1})t) = 1 - \frac{1}{2}(\Delta_{i0} - \Delta_{k1})^2 t^2 + o((\Delta_{i0} - \Delta_{k1})^2 t^2)
\\& \sin((\Delta_{i0} - \Delta_{k1})t) =  (\Delta_{i0} - \Delta_{k1})t - \frac{1}{6} (\Delta_{i0} - \Delta_{k1})^3 t^3 + o((\Delta_{i0} - \Delta_{k1})^3 t^3) .
\end{split}
 \end{equation}
Thus we obtain:

\begin{equation}\label{6hp}
\begin{split}
\langle \hat{X} \rangle_{t} &= \sqrt{\frac{2}{m\omega}} \sum_{i \in I_0}\sum_{k \in I_1}|c_{i0}||c_{k1}|\cos(\omega t - \Delta\varphi^{(0,1)}_{ik}) + \\& - \sqrt{\frac{2}{m\omega}} \sum_{i \in I_0}\sum_{k \in I_1}|c_{i0}||c_{k1}| t(\Delta_{i0} - \Delta_{k1})\sin(\omega t  - \Delta\varphi^{(0,1)}_{ik}) +  \\&
- \sqrt{\frac{1}{2m\omega}}   \sum_{i \in I_0}\sum_{k \in I_1}|c_{i0}||c_{k1}|t^2(\Delta_{i0} - \Delta_{k1})^2\cos(\omega t  - \Delta\varphi^{(0,1)}_{ik}) + 
\\& + \frac{1}{3} \sqrt{\frac{1}{2m\omega}}  \sum_{i \in I_0}\sum_{k \in I_1}|c_{i0}||c_{k1}| t^3(\Delta_{i0} - \Delta_{k1})^3 \sin(\omega t  - \Delta\varphi^{(0,1)}_{ik})  + ... \:\: . 
\end{split}
\end{equation}
With $\alpha_0(0) = |\alpha_0(0)|e^{i \phi_0(0)}$, $\alpha_1(0) = |\alpha_1(0)|e^{i \phi_1(0)}$, $\Delta \phi(0) = \phi_1(0) - \phi_0(0)$ and $\alpha_j(0) = \sum_{i \in I_j} c_{ij}$, we have 

\begin{equation}
\sum_{i \in I_0} \sum_{k \in I_1}  |c_{i0}||c_{k1}|\cos(\omega t  - \Delta\varphi^{(0,1)}_{ik}) =  |\alpha_0(0)||\alpha_1(0)|\cos(\omega t - \Delta \phi(0)) .
\end{equation}
So, considering up to the first order of approximation in $t( \Delta_{i0} - \Delta_{k1} )$, we have finally for the expectation value $\langle \hat{X} \rangle_{t}$:

\begin{multline}\label{658}
\langle \hat{X} \rangle_{t} \simeq  \sqrt{\frac{2}{m\omega}} |\alpha_0(0)||\alpha_1(0)|\cos(\omega t - \Delta \phi(0)) + \\ -  \sqrt{\frac{2}{m\omega}}\sum_{i \in I_0} \sum_{k \in I_1}  |c_{i0}||c_{k1}| t( \Delta_{i0} - \Delta_{k1} )\sin(\omega t  - \Delta\varphi^{(0,1)}_{ik}) 
\end{multline}
that is what we needed to show being (\ref{658}) the same of equation (\ref{6dimappe}).

	\chapter{Quantum Spacetime}
\label{spacetime}

In this Chapter\footnote{A large part of this Chapter is reprinted from T. Favalli and A. Smerzi, \textit{A model of quantum spacetime}, AVS Quantum Sci. \textbf{4}, 044403 (2022), with the permission of AIP Publishing.} we introduce the spatial degree of freedom into our discussion following \cite{nostro3}. The Page and Wootters (PaW) framework \cite{pagewootters,wootters} can be read as an internalization of the time reference frame, with the clock being an appropriately chosen physical system and time is considered as \lq\lq what is shown on a quantum clock\rq\rq. We explore here the possibility to extend this protocol in order to internalizing, together, the temporal and the spatial reference frames. In this approach also space becomes an emerging property of entangled subsystems 
and the concept of position is recovered relative to \lq\lq what is shown on a quantum rod\rq\rq. 

As mentioned in the Introduction, quantum reference frames for the spatial degree of freedom have been extensively studied in quantum information and quantum foundations (see for example \cite{QRF1,QRF2,QRF3,QRF4,QRF5,QRF6,QRF7,QRF8,QRF9,QRF10,QRF11,QRF12,QRF13,QRF14,QRF15,QRF16,QRF17,QRF18,QRF19,QRF20,QRF21}). In the quantum gravity literature, it has been suggested that quantum reference frames are needed to formulate a quantum theory of gravity \cite{dewitt,afundamental,QG1,QG2}. In \cite{change1} (see also \cite{change2,change3,change4,change5,change6,change7,change8,change9}) has also been introduced a formalism for describing transformations to change the description between different quantum reference frames in various contexts. Just as the PaW mechanism has been extensively studied in order to describe the temporal degree of freedom, all these works have dealt only with internalization of the spatial reference frame leaving time as an external parameter in the theory. Only recently, in \cite{giacomini}, has been introduced a fully relational formulation of a $1+1$ dimensional spacetime for the case of $N$ relativistic quantum particles in a weak gravitational field.

In this Chapter we first focus on space and we divide our global quantum system (the \lq\lq Universe\rq\rq) in two entangled subsystems, $R$ and $S$, where $R$ is the quatum rod that acts as a spatial reference frame for $S$. A generalization of the PaW mechanism for the spatial degree of freedom has already been addressed in \cite{hoehn1,hoehn2,hoehn3} (see also \cite{change2,change3}). Here we give our own version adopting and generalizing the approach outlined in Chapter 3 \cite{pegg,peggbar}, which we have already encountered several times throughout this work. We consider indeed discrete spectra for the momentum operators and we take the spatial degree of freedom described by generalized Pegg's POVMs. This choice allow us to recover continuous values for the spatial degrees of freedom even if the momenta have discrete spectra (the generalization to the case of a continuous spectrum is also disussed). We therefore assume the Universe satisfying a constraint on total momentum $\hat{P}_{tot} \ket{\Psi} = 0$. Even if the global position is completely undetermined, a well-defined relative position emerges from the entanglement between the two subsystems $R$ and $S$. 
Finally we introduce an additional subsystem $C$ acting as a clock and we consider the Universe satisfying a double constraint: both on total momentum and total energy, that is $\hat{P}_{tot} \ket{\Psi} = 0$ and $ \hat{H}_{tot}\ket{\Psi} = 0$. We show that this framework can be implemented consistently and we thus provide a model of non-relativistic quantum spacetime emerging from entanglement. 
In order to facilitate the reading and simplify the notation, we proceed as follows: in Sections 6.2, 6.3 and 6.4 we consider a single spatial degree of freedom for the subsystems $R$ and $S$. In Section 6.5 we generalize the results to the case of $3+1$ dimensional spacetime and we discuss some examples.

\section{Relative-Position Localization through Entanglement}
	Let's start considering a thought experiment proposed by A. V. Rau, J. A. Dunningham and K. Burnett (RDB) in \cite{burnett}. Their work shows how entanglement could be crucial in the localization process. Following \cite{burnett}, we consider two non-interacting particles with equal mass in a momentum product state, that is $\ket{\Psi_i} =\ket{p_1} \ket{p_2} $,
	where $\ket{\Psi_i}$ is the global state of the two particles and $\ket{p_1}$, $\ket{p_2}$ are the states of the individual particles in the basis of the momentum eigenstate. We immediately note that this system has relative and absolute positions which are totally unknown. So, the authors imagine to have a photon scattering on this system, and the result will be that the state of the two particles has now become a linear combination of the results of the photon hitting each particle. We will therefore have:

	\begin{equation}
	\ket{\Psi_i} =\ket{p_1}\ket{p_2} \longrightarrow \ket{\Psi_f}
	\end{equation}
	and

	\begin{equation}
	\ket{\Psi_f} =\alpha\ket{p_1 +\Delta p}\ket{p_2} +\beta\ket{p_1 } \ket{p_2 +\Delta p}
	\end{equation}
	where $\alpha$, $\beta$ is the probability amplitudes and $\Delta p$ is the moment transferred to the system by scattering with the photon. What has changed fundamentally is that the two particles are now entangled. The state $\ket{\Psi_f}$ of the two particles has still a well-defined momentum for its center of mass (CM), and it is $\ket{\psi_{CM}} =\ket{p_1 + p_2 +\Delta p}$. Therefore the momentum of the CM is shifted but still well-defined. For this reason, no particle in the system will have a definite absolute position.

	We look now at the relative coordinates. Let's take for simplicity $\left| \alpha\right| =\left| \beta \right|$ and write the state of the relative momentum (after the common phase has been removed). We have

	\begin{equation}
	\ket{p_{rel}} =\frac{1}{\sqrt{2}}\left[ \ket{p_i +\frac{\Delta p}{2}} + e^{i\phi} \ket{p_i - \frac{\Delta p}{2}} \right] .
	\end{equation}
	where
	\begin{equation}
	p_i =\frac{p_1 - p_2}{2} 
	\end{equation} 
	which is the relative momentum before scattering. As RDB observe: <<The relative phase between the two possible events it is related to any observed interference between the alternative path of the photon at the detector and will turn out to be critical to the emerging relative position of the two particles>>. What happens is that scattering broadens the relative-momentum wavefunction and reduces uncertainty in the space of the relative position. In this way, through multiple scattering, the particles become more and more entangled and their relative positions increasingly better localized. As pointed out by the authors, this type of framework suggests a prominent role for entanglement in the localization process.

	In our case this thought experiment is useful as an introduction, since to apply the PaW mechanism to space we will need a constraint on the global momentum. As previously mentioned, indeed, we will have to set the global momentum of the Universe equal to zero, and we will have to look at the emergence of relative position thanks to the entanglement between two subsystems within the Universe. When finally we will introduce a third subsystem that keeps track of time, we will see that there will be an evolution in time of the subsystems, and the relative position between the particles will also vary with the passage of time.

\section{Emergent Space from Entanglement}

\subsection{General Framework}
We divide the total Hilbert space (the \lq\lq Universe\rq\rq) in two subsystems $R$ and $S$ where
$R$ acts as quantum reference frame for $S$. We consider the two subsystems non-interacting and entangled with global Hamiltonian

\begin{equation}
\hat{H} = \hat{H}_R + \hat{H}_S
\end{equation}
where $\hat{H}_R$ and $\hat{H}_S$ act on $R$ and $S$ respectively. We consider the momenta of $R$ and $S$ having a discrete, bounded, non-degenerate spectra and introduce the spatial degrees of freedom as POVMs. The case of momenta with continuous, unbounded spectrum will be discussed in Section 6.3.5. We begin by first considering the subspace $R$, then we will extend the framework to $S$.

In introducing the POVMs of space we follow a generalization of the framework of Chapter 3, 
namely we assume the momentum operator $\hat{P}_R$ of $R$ with point-like spectrum, equally-spaced eigenvalues and non-degenerate eigenstates. It can be illustrated by taking  $d_R$ momentum states $\ket{p_k}$ and $p_k$ momentum levels with $k=0,1,2,...,d_R -1$ such that (we set $\hslash=1$):

\begin{equation}\label{7pk}
p_k = p^{(R)}_0 + \frac{2\pi}{L_R} k .
\end{equation}
In this space we can define the states 

\begin{equation}\label{7statixj}
\ket{x_j}_R  = \frac{1}{\sqrt{d_R}}\sum_{k=0}^{d_R -1}e^{- i p_k x_j}\ket{p_k}_R
\end{equation}
and the values $x_j = x_0 + j \frac{L_R}{D_R}$
with $j=0,1,2,...,z=D_R -1$ and with the constraint $z+1=D_R \ge d_R$. If we take $D_R=d_R$ the states (\ref{7statixj}) are orthogonal (with the $D_R=d_R$ values of $x_j$ uniformly distributed over $L_R$) and the spatial degree of freedom is described by the Hermitian operator $\hat{X}_R = \sum_{j=0}^{d_R -1} x_j \ket{x_j}\bra{x_j}$ complement of $\hat{P}_R$. If instead we take $D_R > d_R$, the number of the states $\ket{x_j}_R$ is greater than the number of momenta states in $R$ (with the $D_R$ values of $x_j$ again uniformly distributed over $L_R$). These states still satisfy the key property $\ket{x_j}_R = e^{- i\hat{P}_R(x_j - x_0) }\ket{x_0}_R$ and it can furthermore be used for writing the resolution of the identity:

\begin{equation}\label{7pomidentity}
\mathbb{1}_{R} = \frac{d_R}{D_R} \sum_{j=0}^{D_R -1} \ket{x_j}\bra{x_j}.
\end{equation}
Thanks to (\ref{7pomidentity}) the spatial degree of freedom is represented by a POVM, being $d_R D^{-1}_{R} \ket{x_j}\bra{x_j}$ the $D_R$ non-orthogonal elements. In order to obtain a continuous representation of the coordinate $x$ (maintaining a discrete momentum spectrum), we can now consider the limit $z \longrightarrow \infty$, defining 

\begin{equation}\label{7xstateinf}
\ket{x}_R = \sum_{k=0}^{d_R -1 } e^{- i p_k x}\ket{p_k}_R
\end{equation}
where $x$ can now take any real value from $x_0$ to $x_0 + L_R$. In this limiting case the resolution of the identity (\ref{7pomidentity}) becomes

\begin{equation}\label{7newresolution}
\mathbb{1}_{R} = \frac{1}{L_R} \int_{x_0}^{x_0+L_R} dx \ket{x} \bra{x} .
\end{equation}
The states $\ket{x_j}$ and $\ket{x}$ are not orthogonal, but we will see in the following that this will not constitute a problem in our derivation of emerging spacetime. 

As mentioned, also the subspace $S$ can be equipped with the POVMs of space considering $\hat{P}_S$ with discrete, bounded spectrum and applying the same formalsim adopted in the subspace $R$. So we assume that also in $S$ all the momentum eigenvalues can be written as multiples of a minimum step, that is $p_k = p^{(S)}_0 + \frac{2\pi}{L_S} k$ with $k=0,1,2,...,d_S -1$. We thus define the states $\ket{y_l}_S  = \frac{1}{\sqrt{d_S}}\sum_{k=0}^{d_S -1}e^{-i p_k y_l}\ket{p_k}_S$ and the $g+1=D_S$ values $y_l = y_0 + l \frac{L_S}{D_S}$ or, in the limiting case ($g \longrightarrow \infty$) in which $y$ take any real value from $y_0$ to $y_0 + L_S$, we consider the states $\ket{y}_S = \sum_{k=0}^{d_S -1 } e^{- i p_k y}\ket{p_k}_S$. Also in this case, when taking $D_S=d_S$ we can define the operator $\hat{Y}_S = \sum_{l}y_l\ket{y_l}\bra{y_l}$ (complement of $\hat{P}_S$) which is an Hermitian operator.

\subsection{Emergent Relative Distance}
In order to obtain the emergence of space from entanglement we consider now the following constraint on the total momentum:

\begin{equation}\label{7constmomentum}
\hat{P}\ket{\Psi} = ( \hat{P}_R + \hat{P}_S)\ket{\Psi} = 0
\end{equation}
where $\hat{P}_R $ and $\hat{P}_S$ act on $R$ and $S$ respectively. Assuming $d_R \gg d_S$ the global state $\ket{\Psi}$ satisfying (\ref{7constmomentum}) can be writen as

\begin{equation}\label{7miserveperlafine}
\ket{\Psi} = \sum_{k=0}^{d_S -1} c_k \ket{p=-p_k}_R\otimes\ket{p_k}_S .
\end{equation}
We can now expand $\ket{\Psi}$ in the $\left\{\ket{x_j}\right\}$ basis on $R$ through (\ref{7pomidentity}), thus obtaining

\begin{equation}\label{7stato1}
\begin{split}
\ket{\Psi} & =  \frac{d_R}{D_R} \sum_{j=0}^{D_R-1} \ket{x_j} \braket{x_j|\Psi} = \\& = \frac{\sqrt{d_R}}{D_R} \sum_{j=0}^{D_R-1} \ket{x_j}_R\otimes\sum_{k=0}^{d_S -1} c_k e^{- i p_k x_j}\ket{p_k}_S = \\&= \frac{\sqrt{d_R}}{D_R} \sum_{j=0}^{D_R-1} \ket{x_j}_R\otimes\ket{\phi(x_j)}_S 
\end{split}
\end{equation}
where in last step we have defined $\ket{\phi(x_j)}_S = \sum_{k=0}^{d_S -1} c_k e^{- i p_k x_j}\ket{p_k}_S$. This state can be obtained from the global state $\ket{\Psi}$ through the \textit{relative state} definition (in Everett sense \cite{everett}) of the subsystem $S$ with respect to $R$ \cite{nostro}:

\begin{equation}\label{7defstatorelativo}
\ket{\phi(x_j)}_S = \frac{\braket{x_j|\Psi}}{1/\sqrt{d_R}}.
\end{equation}
Now using the fact that $\ket{x_j }_R = e^{- i\hat{P}_R (x_j -x_0)}\ket{x_0}_R$ and considering equations (\ref{7constmomentum}) and (\ref{7defstatorelativo}), we obtain
\begin{equation}\label{7trasldiscreta}
\begin{split}
\ket{\phi(x_j)}_S & = \sqrt{d_R}\braket{x_j|\Psi} = \sqrt{d_R} \bra{x_0}e^{i\hat{P}_R (x_j -x_0)}\ket{\Psi} =  \\&
=        \sqrt{d_R} \bra{x_0}e^{i (\hat{P} - \hat{P}_S) (x_j -x_0)}\ket{\Psi}        = e^{- i \hat{P}_S (x_j -x_0)}\ket{\phi(x_0)}_S
\end{split}
\end{equation}
that is, we have that the operator $\hat{P}_S$ is the generator of spatial traslations in the coordinate $x_j$. In this framework it is evident that the translation moves the system with respect to the coordinate of the reference frame $R$ and therefore \lq\lq external\rq\rq to $S$. As in our previous discussion regarding the emergence of time in Chapters 4 and 5, we notice that having non-orthogonal position states does not constitute a problem in our framework. Indeed, even if the $\ket{x_j}_R$ are partially indistinguishable, the state of the system $S$ conditioned on a given $x_j$ does not dependes on $x_i \ne x_j$ (as we can clearly see in equations (\ref{7stato1}) and (\ref{7trasldiscreta})) and so interference phenomena are not present even if the coordinates in $R$ are not orthogonal.

These results can easily extended in the limiting case $z,g \longrightarrow \infty$. Indeed in this case the global state satisfying the constraint (\ref{7constmomentum}) can be written:
\begin{equation}\label{7statoglobalecontinuo}
\begin{split}
\ket{\Psi} & =  \frac{1}{L_R} \int_{x_0}^{x_0 + L_R} d x   \ket{x} \braket{x|\Psi} = \\&= \frac{1}{L_R} \int_{x_0}^{x_0 + L_R} d x   \ket{x}_R \otimes \sum_{k=0}^{d_S-1} c_k e^{- i p_k x} \ket{p_k}_S =\\&= \frac{1}{L_R} \int_{x_0}^{x_0 + L_R} d x   \ket{x}_R \otimes \ket{\phi(x)}_S 
\end{split}
\end{equation}
and, for the relative state $\ket{\phi(x)}_S = \braket{x|\Psi}$, can be easily obtained
\begin{equation}\label{7traslcontinua}
\begin{split}
\hat{P}_S\ket{\phi(x)}_S &= \bra{x}\hat{P}_S\ket{\Psi} = \bra{x}(\hat{P} - \hat{P}_R)\ket{\Psi}  = - \bra{x}\hat{P}_R\ket{\Psi} = \\& 
= - \sum_{k=0}^{d_R -1} \bra{p_k}p_k e^{i p_k x} \ket{\Psi} = i\frac{\partial}{\partial x} \braket{x|\Psi} =  i\frac{\partial}{\partial x} \ket{\phi(x)}_S
\end{split}
\end{equation}
showing again that the momentum $\hat{P}_S$ is the generator of translations in the coordinates $x$, but written through the differential expression\footnote{We notice here that we can write this differential equation for the state $\ket{\phi(x)}_S$ since the coordinate $x$ enters into the state of the system $S$ through entanglement.} $\hat{P}_S\ket{\phi(x)}_S = i\frac{\partial}{\partial x} \ket{\phi(x)}_S$. 

In this framework the absolute position of $R+S$ is totally indeterminate. However, considering discrete values for the coordinates in $R$ and $S$, we can look for the conditional probability of obtaining $y_l$ on $S$ conditioned of having $x_j$ on $R$, where $y_l=y_0 + l\frac{L_S}{D_S}$, $x_j=x_0 + j\frac{L_R}{D_R}$. We have (see Appendix 6.7.A):

\begin{equation}\label{7conditionalprobabilitydiscreta}
P(y_l \: on \: S \:|\: x_j \: on \: R) =\frac{d_S}{D_S} \left|\braket{y_l|\phi(x_j)} \right|^2 = \frac{1}{D_S} \left| \sum_{k=0}^{d_S-1} c_k e^{i p_k(y_l-x_j)} \right|^2
\end{equation} 
that is a well-defined probability distribution, where $\sum_{l=0}^{D_S -1} P(y_l \: on \: S \:|\: x_j \: on \: R) = 1$. Working instead in the limit $z,g \longrightarrow \infty$, the probability for a value of $y$ in the small range between $y$ and $y + dy$ is given by $P(y \: on \: S \:|\: x \: on \: R)dy$, where the probability density $P(y \: on \: S \:|\: x \: on \: R)$ is (the proof is given in Appendix 6.7.B):

\begin{equation}\label{7conditionalprobability}
\begin{split}
P(y \: on \: S \:|\: x \: on \: R) = \frac{1}{L_S} \left| \braket{y|\phi(x)} \right|^2  =\frac{1}{L_S} \left| \sum_{k=0}^{d_S-1} c_k e^{ i p_k(y-x)} \right|^2
\end{split}
\end{equation}
that is again a well-defined probability density distribution depending on the distance between $S$ and $R$. Indeed, also in this case, if we consider the integral over all possible values of $y$ we obtain 	
\begin{equation}
\int_{y_0}^{y_0 + L_S} dy P(y \: on \: S \:|\: x \: on \: R) =	\frac{1}{L_S} \int_{y_0}^{y_0 + L_S} dy \sum_{k}\sum_{n} c_k c^{*}_n e^{i(p_k - p_n)(y-x)} = 1
\end{equation}
where we have used (see Appendix 6.7.C at the end of this Chapter):

\begin{equation}\label{7delta} 
\int_{y_0}^{y_0 + L_S} dy e^{iy(p_k - p_n)} = L_S\delta_{p_k,p_n} .
\end{equation}
Equations (\ref{7conditionalprobabilitydiscreta}) and (\ref{7conditionalprobability}) display an essential feature of the complementarity between positions and momenta. Indeed, if the system $S$ is in an eigenstate of the momentum, in the right-hand side of (\ref{7conditionalprobabilitydiscreta}) and (\ref{7conditionalprobability}) there is only one term of modulus unity and we have $P(y_l \: on \: S \:|\: x_j \: on \: R) = 1/D_S$ and $P(y \: on \: S \:|\: x \: on \: R) = 1/L_S$. So, in this case, the probability $P(y_l \: on \: S \:|\: x_j \: on \: R)$ and the probability density $P(y \: on \: S \:|\: x \: on \: R)$ are constant across the whole interval $\left[y_0,y_0+L_S \right]$ indicating that, when the momentum of the system $S$ can be determined exactly, thus the position with respect to the reference frame $R$ is completely random. 

We have so considered a \lq\lq positionless\rq\rq Universe, satisfying the constraint (\ref{7constmomentum}) on the total momentum (where the absolute position is totally indeterminate) and we found the well-defined conditional probability $P(y_l \: on \: S \:|\: x_j \: on \: R)$ (for the case of discrete coordinates) and the probability density $P(y \: on \: S \:|\: x \: on \: R)$ (for the limiting case $z,g \longrightarrow \infty$ where $x$ and $y$ have continuous values) depending on the relative distance between the two entangled subsystems $R$ and $S$.

\subsection{On the Position-Momentum Uncertainty Relation}
In this paragraph we want to focus on the quantity $\delta x$, namely the minimum interval in the $x$ values of $R$ over which the state of the system $\ket{\phi (x)}_S$ varies significantly. We will show that, if the momentum spread in the expansion (\ref{7miserveperlafine}) is $\Delta p$, than $\delta x \ge \hslash /\Delta p$ (re-introducing $\hslash \ne 1$). This means that will be impossible to distinguish states of the system $S$ conditioned to $x$ values on $R$ which are close to each other less than $\simeq \hslash /\Delta p$, in accordance with position-momentum uncertainty relation. 

For the sake of simplicity we consider here $D_R=D_S=d_R=d_S = d$ and $L_R = L_S = L$. We assume also discrete values for the space, so we have 
\begin{equation}
\ket{x_j}_R = \frac{1}{\sqrt{d}} \sum_{k}e^{-ip_kx_j} \ket{p_k}_R
\end{equation}
and 
\begin{equation}
\ket{y_l}_S = \frac{1}{\sqrt{d}} \sum_{k}e^{- ip_k y_l} \ket{p_k}_S 
\end{equation}
with $x_j=x_0 + j\frac{L}{d}$ and $y_l=y_0 + l\frac{L}{d}$. As already mentioned, in this particular case we can define the operators $\hat{X}_R = \sum_{j=0}^{d-1} x_j\ket{x_j}\bra{x_j}$ and $\hat{Y}_S = \sum_{l=0}^{d-1} y_l\ket{y_l}\bra{y_l}$ (complement of $\hat{P}_R$ and $\hat{P}_S$ respectively) 
which are Hermitian operators.

The crucial point for our argument here is to understand that in this framework, since the reference frame $R$ and the system $S$ are entangled in the global state $\ket{\Psi}$, the $\Delta p$ related to the spread of the coefficients in the expansion (\ref{7miserveperlafine}), that is $\ket{\Psi} = \sum_{k=0}^{d_S -1} c_k \ket{p=-p_k}_R\otimes\ket{p_k}_S$, does not refer exclusively to $R$, but to $R$ and $S$ together. For this reason what we will find is that a limited spread in the expansion in the momentum eigenbasis will reduce the distinguishability of states of $S$ conditioned on proximal values of $R$. So, starting from equation  

\begin{equation}\label{7equazioneevoluzione2}
\ket{\phi(x_j)}_S = \frac{\braket{x_j|\Psi}}{1/\sqrt{d}} = \sum_{k=0}^{d -1} c_k e^{- i \hslash^{-1} p_k x_j}\ket{p_k}_S,
\end{equation}
we can calculate in the space $S$:

\begin{equation}\label{7ultima}
\begin{split}
\braket{\phi (x_i) | \phi (x_j)} 
= \sum_{k=0}^{d-1} \left| c_k \right|^2 e^{- i\hslash^{-1} p_k (x_j - x_i)}. 
\end{split}
\end{equation} 
Equations (\ref{7ultima}) indicates that, if $\left| c_k \right|^2$ has a spread $\simeq \Delta p$, than the scalar product $f(x_j - x_i) = \braket{\phi (x_i) | \phi (x_j)}$ will have a spread of the order $\simeq \hslash/ \Delta p$. This means that the state $\ket{\phi(x_j)}_S$ of the subsystem $S$ varies significantly for intervals 

\begin{equation}
\delta x  \geq \hslash/ \Delta p
\end{equation}
where $\Delta p$ is indeed the uncertainty in momentum related to the spread of the coefficients $c_k$. 
This means that it is impossible to distinguish states of the system $S$ conditioned to $x$ values on $R$ which are closer than $\simeq \hslash /\Delta p$ to each other, in accordance with the position-momentum uncertainty relation. 

Expression (\ref{7ultima}) for $f(x_j - x_i)$ holds also in the case of non-orthogonal space states where the spatial degrees of freedom are described by POVMs. So we notice here that the function $f(x_j - x_i)$, and consequently the scale on which the system $S$ varies significantly, is not related to the overlap of the states in $R$. Namely the fact of using coordinates states that are not orthogonal does not have an effect on the derivation of the function $f(x_j - x_i)$. As in the case of Section 4.4 in Chapter 4, this is because, calculating the conditioned state of $S$ to a certain value $x_j$ on $R$ through (\ref{7equazioneevoluzione2}), we find no contributions from different positions $x_i \ne x_j$, and so interference phenomena are not present even if the position states are not orthogonal. Rather $f(x_j - x_i)$ is related to the spread of the coefficients appearing in the global state $\ket{\Psi}$ and this fact shows us a condition for a good functioning of our framework: a large spread in the coefficients within the global state in the expansion (\ref{7miserveperlafine}) is needed in order to distinguish states of $S$ projected to closer values of $R$ \cite{asimmetry}.


\section{Spacetime from Entanglement}
	
\subsection{Introducing the $C$ Subspace}
In order to introduce the temporal degree of freedom we have to consider an additional Hilbert space and assign it to time. We so assume that the Universe is divided into three subsystems, namely we work in $\mathcal{H} = \mathcal{H}_C\otimes\mathcal{H}_R\otimes\mathcal{H}_S$, where $\mathcal{H}_C$ is the time Hilbert space. The three subsystems are non-interacting but still entangled and the global Hamiltonian reads: 

\begin{equation}
\hat{H} = \hat{H}_C + \hat{H}_R + \hat{H}_S
\end{equation}
where $\hat{H}_C$, $\hat{H}_R$ and $\hat{H}_S$ act on $C$, $R$ and $S$ respectively. To be as general as possible, we consider the clock Hamiltonian $\hat{H}_C$ with bounded, discrete spectrum with unequally-spaced energy levels and we introduce also the time observable described by a Pegg's POVM. Let us repeat again such a framework. 

We start assuming $\hat{H}_C$ with point-like spectrum, non-degenerate eigenstates and having rational energy ratios. 
The framewowrk can be illustrated by taking $p+1 = d_{C}$ energy states $\ket{E_i}_C$ and $E_i$ energy levels with $i=0,1,2,...,d_C -1$ such that $\frac{E_i -E_0 }{E_1 - E_0} = \frac{A_i}{B_i}$,	
where $A_i$ and $B_i$ are integers with no common factors. Doing this we obtain (we set again $\hslash=1$):
\begin{equation}\label{7ei}
E_i = E_0 + r_i \frac{2\pi}{T}
\end{equation}
where $T=\frac{2\pi r_1}{E_1}$, $r_i = r_1\frac{A_i}{B_i}$ for $i>1$ (with $r_0=0$) and $r_1$ equal to the lowest common multiple of the values of $B_i$. In this space we define the states 

\begin{equation}\label{7timestates}
\ket{t_m}_C  = \frac{1}{\sqrt{d_C}}\sum_{i=0}^{d_C -1 }e^{-i E_i  t_m}\ket{E_i}_C
\end{equation}
where $t_m = t_0 + m \frac{T}{D_C}$ with $m=0,1,2,...,s=D_C-1$ and $s+1 \ge r_p$\footnote{As we have already seen throughout our work, by assuming the energy spectrum with equally-spaced eigenvalues and taking $D_C=d_C$, the time states (\ref{7timestates}) are orthogonal and the temporal degree of freedom is described by the Hermitian operator $\hat{T}_C = \sum_{m}t_m \ket{t_m}\bra{t_m}$.}. The number of $\ket{t_m}_C$ states is therefore greater than the number of energy states in $\mathcal{H}_{C}$ and the $D_C$ values of $t_m$ are uniformly distributed over $T$. These states satisfy the key property $\ket{t_m}_C = e^{- i\hat{H}_C(t_m - t_0) }\ket{t_0}_C$ and furthermore can be used for writing the resolution of the identity in the $C$ subspace:
\begin{equation}\label{7pomidentity2}
\mathbb{1}_{C} = \frac{d_C}{D_C} \sum_{m=0}^{D_C -1} \ket{t_{m}}\bra{t_{m}}.
\end{equation}
Thanks to property (\ref{7pomidentity2}) also time is represented by a POVM, being $d_C D^{-1}_{C} \ket{t_m}\bra{t_m}$ the $D_C$ non-orthogonal elements. As we did for space, in order to obtain a continuous flow of time, we can now consider the limit $s \longrightarrow \infty$, defining 
\begin{equation}\label{7alphastateinf}
\ket{t}_C = \sum_{i=0}^{d_C-1} e^{- i E_i t}\ket{E_i}_C
\end{equation}
where $t$ can now take any real value from $t_0$ to $t_0 + T$. In this limiting case the resolution of the identity (\ref{7pomidentity2}) becomes
\begin{equation}\label{7newresolution2}
\mathbb{1}_{C} = \frac{1}{T} \int_{t_0}^{t_0+T} d t \ket{t} \bra{t} .
\end{equation}
We emphasize here again that this framework allows us to use a generic Hamiltonian as a clock Hamiltonian. Indeed, in the case of non-rational ratios of energy levels, the residual terms in the resolutions of the identity (\ref{7pomidentity2}) and (\ref{7newresolution2}) and consequent small corrections can be arbitrarily reduced.

\subsection{Emergent $1+1$ Dimensional Spacetime}
We want now to obtain a model of spacetime emerging from entanglement. So we consider the global state $\ket{\Psi} \in \mathcal{H}_C\otimes\mathcal{H}_R\otimes\mathcal{H}_S$ simultaneously satisfying
\begin{equation}\label{71}
\hat{H}\ket{\Psi} = (\hat{H}_C + \hat{H}_R + \hat{H}_S )\ket{\Psi}=0
\end{equation}
and
\begin{equation}\label{72}
\hat{P}\ket{\Psi} = (\hat{P}_R + \hat{P}_S )\ket{\Psi}=0
\end{equation}
where we have assumed $\hat{P}_C = 0$. The mechanism works also with $\hat{P}_C \ne 0$ but, in this case, there could be limitations in the allowed momenta to ensure that (\ref{71}) and (\ref{72}) are simultaneously satisfied (see Appendix 6.7.D). The framework with $\hat{P}_C = 0$ can be implemented for example by assuming the subspace $C$ describing the internal degree of freedom of the subsystem $R$.

If we want to look at the explicit form of the state $\ket{\Psi}$ we need to know the relation between the momenta and the energy of $R$ and $S$. Nevertheless, assuming again $d_C, d_R \gg d_S$, we can write in general:

\begin{equation}\label{7statoglobalespaziotempo}
\ket{\Psi} = \sum_{k=0}^{d_S -1} c_k \ket{E=-\epsilon_k}_C\otimes\ket{p=-p_k}_R\otimes\ket{p_k}_S
\end{equation} 
where $$\epsilon_k= E^{(R)}(-p_k) + E^{(S)}(p_k)$$ is the energy function related to the momenta of $R$ and $S$\footnote{For simplicity we consider here that the energy function depends only on the momenta and not on the coordinates. Clearly the model works also considering the presence of external potentials in $R$ and $S$ (equations (\ref{7evoluzioneRS}), (\ref{7evoluzioneRSc}) and (\ref{7m}) that we will find in the following would be indeed still valid) but, in this case, the state $\ket{\Psi}$ can not be written in the simple form (\ref{7statoglobalespaziotempo}) and we can not explicitly calculate the conditional probabilities (\ref{7probfinalediscreta}) and (\ref{7probfinale}). For this reason we prefer to simplify the model, as we believe this choice helps to capture the essence of the mechanism. In Section 6.3.5 we consider however the case where external potentials are present in $R$ and $S$.}. If, for example, we consider $R$ and $S$ as free particles (with mass $M$ and $m$ respectively) we have $\hat{H}_R = \frac{\hat{P}^{2}_R}{2M}$ and $\hat{H}_S = \frac{\hat{P}^{2}_S}{2m}$, which implies $\epsilon_k = \frac{p^{2}_k}{2M} + \frac{p^{2}_k}{2m}$.

Starting from the state $\ket{\Psi}$ satisfying (\ref{71}) and (\ref{72}), we can now expand it on the basis $\left\{\ket{t_m}_C\right\}$ in $C$ thanks to (\ref{7pomidentity2}), thus obtaining
\begin{equation}\label{7serveperGLM}
\begin{split}
\ket{\Psi} &= \frac{d_C}{D_C} \sum_{m=0}^{D_C -1} \ket{t_m} \braket{t_m|\Psi} =  \frac{\sqrt{d_C}}{D_C} \sum_{m=0}^{D_C -1} \ket{t_m}_C \otimes \ket{\phi(t_m)}_{R,S}
\end{split}
\end{equation}
where $\ket{\phi(t_m)}_{R,S} = \sqrt{d_C} \braket{t_m|\Psi}$ is state of the composite system $R+S$ at time $t_m$, namely it is the \textit{relative state} (in Everett sense \cite{everett}) of $R+S$ conditioned on having the value $t_m$ on $C$. For such a state, through (\ref{71}) and the relative state definition, it is easy to find the time evolution with respect to the clock $C$:

\begin{equation}\label{7evoluzioneRS}
\begin{split}
\ket{\phi(t_m)}_{R,S} &=\sqrt{d_C}\braket{t_m|\Psi} = \sqrt{d_C} \bra{t_0} e^{i\hat{H}_C(t_m - t_0)}\ket{\Psi}= \\&
= \sqrt{d_C} \bra{t_0} e^{-i (\hat{H}_R + \hat{H}_S - \hat{H})(t_m - t_0)}\ket{\Psi}= e^{-i (\hat{H}_R + \hat{H}_S)(t_m - t_0)}\ket{\phi(t_0)}_{R,S}
\end{split}
\end{equation} 
where $\ket{\phi(t_0)}_{R,S}= \sqrt{d_C} \braket{t_0|\Psi}$ is the state of $R+S$ conditioned on $t_0$ that is the value of the clock taken as initial time. Equation (\ref{7evoluzioneRS}) shows, as expected, the simultaneous evolution of $R$ and $S$ over time. Having indeed considered a quantum spatial reference frame, it was reasonable to expect that it also evolves in time together with the subsystem $S$. We can consider then the limiting case $s \longrightarrow \infty$ where $t$ takes all the real values between $t_0$ and $t_0+T$. The global state reads: 
\begin{equation}
\begin{split}
\ket{\Psi} = \frac{1}{T} \int_{t_0}^{t_0 +T} dt \ket{t} \braket{t|\Psi} = \frac{1}{T} \int_{t_0}^{t_0 +T} dt \ket{t}_C \otimes \ket{\phi(t)}_{R,S}
\end{split}
\end{equation}
and defining the relative state of $R+S$ as $\ket{\phi(t)}_{R,S} = \braket{t|\Psi}$ we obtain \cite{nostro}:
\begin{equation}\label{7evoluzioneRSc}
i \frac{\partial}{\partial t}\ket{\phi(t)}_{R,S} = \left(\hat{H}_R + \hat{H}_S\right)\ket{\phi(t)}_{R,S}
\end{equation}
that is the Schrödinger evolution for the state of $R+S$ with respect to the clock time $t$, written in the usual differential form.

We can therefore expand the state $\ket{\Psi}$ in the coordinates $\left\{\ket{x_j}_R\right\}$ in $R$, thus obtaining:
\begin{equation}
\ket{\Psi} = \frac{d_R}{D_R} \sum_{j=0}^{D_R -1} \ket{x_j} \braket{x_j|\Psi} =  \frac{\sqrt{d_R}}{D_R} \sum_{j=0}^{D_R -1} \ket{x_j}_R \otimes \ket{\varphi(x_j)}_{C,S}
\end{equation}
where $\ket{\varphi(x_j)}_{C,S}=\sqrt{d_R}\braket{x_j|\Psi}$ is the relative state of $C+S$ conditioned to the value $x_j$ on the reference frame $R$. All the results found in the previous Section apply to the state $\ket{\varphi(x_j)}_{C,S}$. Indeed we have also in this case
\begin{equation}\label{7m}
\ket{\varphi(x_j)}_{C,S} = e^{-i \hat{P}_S(x_j - x_0)} \ket{\varphi(x_0)}_{C,S}
\end{equation}
where the momentum of the clock $C$ does not appear since we have chosen $\hat{P}_C = 0$. Also here we consider the limit $z \longrightarrow \infty$, where again $x$ can take all the real values between $x_0$ and $x_0 + L_R$. In this case the global state can be written
\begin{equation}\label{7mm}
\begin{split}
\ket{\Psi} = \frac{1}{L_R} \int_{x_0}^{x_0 +L_R} dx \ket{x} \braket{x|\Psi} = \frac{1}{L_R} \int_{x_0}^{x_0 +L_R} dx \ket{x}_R \otimes \ket{\varphi(x)}_{C,S}
\end{split}
\end{equation}
and defining the relative state of $C+S$ as $\ket{\varphi(x)}_{C,S} = \braket{x|\Psi}$ we obtain $\hat{P}_S \ket{\varphi(x)}_{C,S} = i \frac{\partial}{\partial x} \ket{\varphi(x)}_{C,S}$. Through this latter equation and (\ref{7m}) we can see again that $\hat{P}_S$ is the generator of translations in the coordinate values $x$ for the relative state $\ket{\varphi(x)}_{C,S}$.

Finally we can expand the state $\ket{\Psi}$ simultaneously on the coordinates $\left\{\ket{x_j}_R\right\}$ in $R$ and on the time basis $\left\{\ket{t_m}_C\right\}$ in $C$. We have for the global state:

\begin{equation}\label{745}
\begin{split}
\ket{\Psi} &= \left(\frac{d_C}{D_C} \sum_{m=0}^{D_C -1} \ket{t_m}\bra{t_m} \otimes  \frac{d_R}{D_R} \sum_{j=0}^{D_R -1} \ket{x_j}\bra{x_j} \right)\ket{\Psi}=\\&
= \frac{\sqrt{d_C}}{D_C} \frac{\sqrt{d_R}}{D_R} \sum_{m=0}^{D_C -1}\sum_{j=0}^{D_R -1}\ket{t_m}_C\otimes\ket{x_j}_R\otimes\ket{\psi(t_m,x_j)}_S
\end{split}
\end{equation}
where $\ket{\psi(t_m,x_j)}_S =\sqrt{d_C} \sqrt{d_R}(\bra{t_m}\otimes\bra{x_j})\ket{\Psi}$ is the relative state of the system $S$ at time $t_m$ conditioned on the value $x_j$ for the reference frame $R$. With the state $\ket{\psi(t_m,x_j)}_S$ we have not yet defined the position of the system $S$. This state indeed gives us the value of the time that enters as a parameter thanks to the entanglement with the subspace $C$ and indicates the position of the reference frame $R$. What we can now search is the conditional probability of having a certain position $y_l$ in $S$ at time $t_m$ and knowing that the reference frame is in $x_j$, that is (see Appendix 6.7.E at the end of this Chapter):

\begin{multline}\label{7probfinalediscreta}
P(y_l \: on\: S\:|\:x_j\:on\:R, \: t_m \: on \:C) = \frac{d_S}{D_S} |\braket{y_l|\psi(t_m,x_j)}|^2 = \frac{1}{D_S} \left| \sum_{k=0}^{d_S -1} c_k e^{-i\epsilon_k t_m}e^{ip_k(y_l-x_j)} \right|^2
\end{multline}
where, we recall, $\epsilon_k$ is the energy function related to the momenta $p_k$ of $R$ and $S$ and where it is easy to verify that $\sum_{l=0}^{D_S -1} 	P(y_l \: on\: S\:|\:x_j\:on\:R, \: t_m \: on \:C) = 1$ given each $x_j$ and $t_m$. Clearly we can extend these results also to the limiting cases $z,g,s \longrightarrow \infty$. Indeed we can write the global state $\ket{\Psi}$ as

\begin{equation}\label{747}
\begin{split}
\ket{\Psi} &= \left( \frac{1}{T}\int_{t_0}^{t_0 + T} dt \ket{t}\bra{t} \otimes \frac{1}{L_R}\int_{x_0}^{x_0 + L_R} dx \ket{x}\bra{x} \right) \ket{\Psi} = \\&
= \frac{1}{T} \frac{1}{L_R} \int_{t_0}^{t_0 + T} dt \int_{x_0}^{x_0 + L_R} dx \ket{t}_C \otimes \ket{x}_R \otimes \ket{\psi(t,x)}_S
\end{split}
\end{equation}
where again $\ket{\psi(t,x)}_S = (\bra{t}\otimes\bra{x})\ket{\Psi}$ is the relative state of the system $S$ at time $t$ conditioned on the value $x$ for the reference frame $R$. The conditional probability density of having a certain position $y$ in $S$ at time $t$ and knowing that the reference frame is in $x$ is (see Appendix 6.7.F at the end of this Chapter):

\begin{equation}\label{7probfinale}
P(y \: on\: S\:|\:x\:on\:R,\: t \: on \:C) = \frac{1}{L_S}\left| \braket{y|\psi(t,x)} \right|^2= \frac{1}{L_S}\left| \sum_{k=0}^{d_S -1} c_k e^{-i\epsilon_k t}e^{ip_k(y-x)} \right|^2 .
\end{equation}
We notice that also the probability density (\ref{7probfinale}) is well-defined for each time (indeed it is easy to verify that $	\int_{y_0}^{y_0 + L_S} dy P(y \: on\: S\:|\:x\:on\:R,\: t \: on \:C) =1$ for all $x$ and $t$) and it depends on time $t$ as well as on the distance $y-x$ between $S$ and $R$.

So, through entanglement, we have found for the subsystem $S$ a conditional probability density that give us informations about the evolution of $S$ in time and space, where for time we consider the clock time and for space we consider the relative distance between $S$ and the quantum reference frame $R$. All these results are obtained within a globally static and \lq\lq positionless\rq\rq Universe.

To conclude this paragraph we notice that a good spatial reference frame is a reference that moves only slightly in time. If a good spatial reference frame is considered, one can look at the evolution of $S$ by itself. We show this point with an example: assuming $R$ and $S$ as free particles (with mass $M$ and $m$ respectively), we could take $M \gg m$ thus obtaining $\frac{\hat{P}^{2}_R}{2M} \ll \frac{\hat{P}^{2}_S}{2m}$. 
Starting from the state (\ref{747}) we can consider the relative state $\ket{\psi(t,x)}_S = (\bra{t}\otimes\bra{x})\ket{\Psi}$ and investigate its evolution. If the mass $M$ is sufficiently large, we have:

\begin{equation}\label{7evS2}
i \frac{\partial}{\partial t}\ket{\psi(t,x)}_{S} \simeq \hat{H}_S \ket{\psi(t,x)}_{S} .
\end{equation}
Equation (\ref{7evS2}) shows that, if $M$ is sufficiently large, the evolution of $S$ alone can be recovered with respect to time $t$ and with respect to a spatial reference frame that does not evolve (or that move negligibly in time). Furthermore, equation (\ref{7evS2}) together with the property $\hat{P}_S \ket{\psi(t,x)}_{S} = i \frac{\partial}{\partial x} \ket{\psi(t,x)}_{S}$ lead to: 

\begin{equation}\label{7evfinale31+1}
i \frac{\partial}{\partial t} \ket{\psi(t, x)}_S \simeq - \frac{1}{2m}\frac{\partial^{2}}{\partial x^{2}}  \ket{\psi(t,x)}_{S}
\end{equation}
which clearly describes the dynamics of the particle in $S$ with respect to the coordinates of the $1+1$ dimensional quantum reference frame. 
We emphasize here that, in this case, we can write the equation (\ref{7evfinale31+1}) for the state $\ket{\psi(t,x)}_{S}$ because the values of time and space of the subspaces $C$ and $R$ enter as parameters in $S$ thanks to the entanglement present in the global state $\ket{\Psi}$. We will return to this point later, in Section 6.5, when we also discuss the example of relativistic particles. 


\subsection{A simple Example}
We consider here a simple example assuming $R$ and $S$ as free particles with mass $M$ and $m$ respectively and $d_R=d_S =3$. We start assuming $D_R =D_S = d_R =d_S$, $L_R=L_S=L$ and discrete values of space and time. We have therefore: $p^{(R)}_k = p^{(S)}_k = p_0 + \frac{2\pi}{L} k$,	
with $p_0=-\frac{2\pi}{L}$ (that implies $p_1=0$, $p_2=\frac{2\pi}{L}$), $x^{(R)}_j = x_0+j\frac{L}{3} = 0,\frac{L}{3},\frac{2L}{3}$ and $y^{(S)}_l = y_0+l\frac{L}{3} = 0,\frac{L}{3},\frac{2L}{3}$. The global state satisfying the constraints on total energy and total momentum can be written as:
\begin{equation}
\ket{\Psi} = c_0 \ket{E_{2,0}}_C\ket{p_2}_R\ket{p_0}_S  + c_1 \ket{E_{1,1}}_C\ket{p_1}_R\ket{p_1}_S + c_2 \ket{E_{0,2}}_C\ket{p_0}_R\ket{p_2}_S
\end{equation}
where $E_{k,n} = -( \frac{p^{2}_k}{2M} + \frac{p^{2}_n}{2m} )$ and where we assume, for simplicity, the coefficients $c_i$ to be real. Furthermore we have $E_{2,0}= E_{0,2}=\epsilon= (\frac{2\pi}{L})^2( \frac{1}{2M} + \frac{1}{2m})$ and $E_{1,1}=0$. We can now expand the global state $\ket{\Psi}$ simultaneously on the coordinates $\left\{\ket{x_j}_R\right\}$ in $R$ and on the time basis $\left\{\ket{t_m}_C\right\}$ in $C$, and then we search the state $\ket{\psi(t_m,x_j)}_S = \sqrt{d_C}\sqrt{3}(\bra{t_m}\otimes\bra{x_j})\ket{\Psi}$, thus obtaining:
\begin{equation}
\begin{split}
\ket{\psi(t_m,x_j)}_S &= \sqrt{3} \left[c_0 e^{-i\epsilon t_m}\braket{x_j|p_2}\ket{p_0}_S  + c_1 \braket{x_j|p_1}\ket{p_1}_S + c_2  e^{-i\epsilon t_m}\braket{x_j|p_0}\ket{p_2}_S \right] =  \\&
=c_0  e^{-i\epsilon t_m}e^{i\frac{2\pi}{L}x_j}\ket{-\frac{2\pi}{L}}_S  +c_1 \ket{0}_S +c_2 e^{-i\epsilon t_m}e^{-i\frac{2\pi}{L}x_j}\ket{\frac{2\pi}{L}}_S.
\end{split}
\end{equation}
We can now calculate the conditional probability of obtaining $y_l$ on $S$ conditioned on having $x_j$ on $R$ and $t_m$ on $C$. Considering $d_S = D_S =3$, we have:
\begin{multline}\label{7esempio1}
P(y_l \: on\: S\:|\:x_j\:on\:R,\: t_m \: on \:C) = \left| \braket{y_l| \psi(t_m,x_j)}\right|^2 = \\
= \frac{1}{3}\left| c_0  e^{-i\epsilon t_m}e^{-i\frac{2\pi}{L}(y_l - x_j)}  + c_1 +  c_2 e^{-i\epsilon t_m}e^{i\frac{2\pi}{L}(y_l-x_j)} \right|^2 .
\end{multline}
Proceeding with the calculations from equation (\ref{7esempio1}) and remembering we have real coefficients, we obtain
\begin{multline}
P(y_l \: on\: S\:|\:x_j\:on\:R,\: t_m \: on \:C) = \frac{1}{3} + \frac{2}{3}c_0 c_1 \cos(\epsilon t_m + \frac{2\pi}{L}( y_l - x_j))+ \\ + \frac{2}{3}c_1 c_2 \cos(\epsilon t_m - \frac{2\pi}{L}(y_l -x_j))  + \frac{2}{3}c_0 c_2\left( 1- 2\sin^{2}( \frac{2\pi}{L}(y_l-x_j))  \right) 
\end{multline}
that is the expression of how the probability of having a certain relative distance between the particles $S$ and $R$ given $x_j$ for the reference frame $R$ at time $t_m$. 

If we consider the limiting cases $s,z\longrightarrow\infty$ (mantaining the assumption $L_R=L_S=L$), 
we obtain for the probability density:
\begin{multline}
P(y \: on\: S\:|\:x\:on\:R,\: t \: on \:C) = \frac{1}{L} + \frac{2}{L}c_0 c_1 \cos(\epsilon t + \frac{2\pi}{L}( y - x))+ \\ + \frac{2}{L}c_1 c_2 \cos(\epsilon t - \frac{2\pi}{L}(y -x))  + \frac{2}{L}c_0 c_2\left( 1- 2\sin^{2}( \frac{2\pi}{L}(y-x))  \right) .
\end{multline}

\subsection{On the Quantum Speed Limit Time}

We reiterate here the considerations made in Section 4.4 of Chapter 4, seeing how they apply to the present case of emerging spacetime.Namely we look at the time scale over which the conditioned state $\ket{\phi (t)}_{R,S}$ evolve into an othogonal configuration, thus becoming fully distinguishable. For the sake of simplicity we consider again the case of $d_R=D_R$, $d_S=D_S$ that leads to discrete values of space and orthogonal states. For the temporal degree of freedom we consider the limiting case $s \longrightarrow \infty$. The conditioned state of $R+S$ can be obtained from (\ref{7statoglobalespaziotempo}), calculating $\ket{\phi(t)}_{R,S} = \braket{t|\Psi}$. We have ($\hslash \ne 1$):

\begin{equation}\label{7equazione}
\ket{\phi(t)}_{R,S} = \sum_{k=0}^{d_S - 1} c_k e^{-i \hslash^{-1} \epsilon_k t}\ket{p=-p_k}_R\otimes\ket{p_k}_S
\end{equation} 
where $\epsilon_k= E^{(R)}(-p_k) + E^{(S)}(p_k)$ is the energy function related to the momenta $p_k$ of $R$ and $S$. Starting from equation (\ref{7equazione}) we can calculate in the space $R+S$:

\begin{equation}\label{7ultima1}
\begin{split}
\braket{\phi (t_0) | \phi (t)}  = \sum_{k=0}^{d_S - 1} |c_k|^2 e^{-i \hslash^{-1} \epsilon_k (t - t_0)} .
\end{split}
\end{equation} 
Looking at equation (\ref{7ultima1}), we can now consider the quantum speed limit time $\delta t$ which gives us the minimum time needed for $R+S$ to evolve to an orthogonal configuration. We have \cite{speedlimit,margolus}:

\begin{equation}\label{7dalpha}
\delta t  \geq  max \left( \frac{\pi\hslash}{2 [E_{R,S} - (E^{(R)}_0 + E^{(S)}_0)]}  , \frac{\pi\hslash}{2 \Delta E} \right)
\end{equation}
where $E_{R,S} = \bra{\phi(t_0)}\hat{H}_R + \hat{H}_S\ket{\phi(t_0)} $ and $\Delta E$ is the spread in energy related to the coefficients $c_k$ through 

\begin{equation}
\Delta E = \sqrt{\bra{\phi(t_0)} (\hat{H}_R + \hat{H}_S - E_{R,S} )^2\ket{\phi(t_0)}}. 
\end{equation}

We can see here how, again, the function $f(t - t_0) = \braket{\phi (t_0) | \phi (t)}$, and consequently the time scale on which $R+S$ varies significantly, is not related to the overlap of the states of the clock. This can be seen considering that in $f(t - t_0)$ do not enter time values different from $t, t_0$ and $f(t - t_0)$ takes the form expressed in \cite{speedlimit}. So the fact that our time states are not orthogonal does not have a consequence on the speed at which the state $\ket{\phi(t)}_{R,S}$ evolves with respect to $t$. Rather $f(t - t_0)$ is related to the spread of the coefficients $c_k$ appearing in the state (\ref{7equazione}). These considerations, together with equation (\ref{7dalpha}), indicate the key point in this framework: a large spread in the coefficients $c_k$ within state $\ket{\phi(t)}_{R,S}$, and so in the global state 
$\ket{\Psi} = \sum_{k=0}^{d_S -1} c_k \ket{E=-\epsilon_k}_C\otimes\ket{p=-p_k}_R\otimes\ket{p_k}_S$, is needed to make the time evolution of the subsystem $R+S$ faster \cite{asimmetry}.

\subsection{Introducing External Potentials in $R$ and $S$}
We illustrate here the case in which external potentials are present within $R$ and $S$ by considering these two subspaces consisting of two harmonic oscillators. We use this example also to extend our framework assuming momentum operators with continuous spectra. So far we have indeed always used momentum operators in $R$ and $S$ with discrete spectra, but the entire discussion can be easily generalized to the case of continuous values for momenta and coordinates, with orthogonal position states. Such a framework allows us to more easily address the problem of considering the presence of external potentials in $R$ and $S$. What we need is indeed to have Hermitian operators $\hat{X}$ (within $R$) and $\hat{Y}$ (within $S$) in order to describe the potentials of the harmonic oscillators. In the subspace $C$ we assume the framework described so far working in the limiting case $s \longrightarrow \infty$.

We therefore start considering the Hamiltonians of $R$ and $S$ written as:

\begin{equation}
\begin{split}
& \hat{H}_R= \frac{\hat{P}^{2}_R}{2M} + V_R(\hat{X}) =  \frac{\hat{P}^{(2)}_R}{2M} + \frac{1}{2}M\omega_R\hat{X}^{2}
\\&  \hat{H}_S= \frac{\hat{P}^{2}_S}{2m} + V_S(\hat{Y}) =  \frac{\hat{P}^{(2)}_S}{2m} + \frac{1}{2}m \omega_S \hat{Y}^{2} .
\end{split}
\end{equation} 
In this framework, the global state $\ket{\Psi}$ satisfying the constraint on total momentum $\left( \hat{P}_R + \hat{P}_S\right)\ket{\Psi}=0$ is:

\begin{equation}\label{7Gstato1}
\ket{\Psi} = \sum_{n=0}^{d_C - 1} \int dp ~ c_n \psi(p) \ket{E_n}_C \otimes \ket{-p}_R \otimes \ket{p}_S
\end{equation}
where $\sum_{n=0}^{d_C - 1} |c_n|^2 = \int dp |\psi(p)|^2=1$. The state (\ref{7Gstato1}) can be rewritten in the energy eigenbasis for $R$ and $S$ as

\begin{equation}\label{7Gstato2}
\ket{\Psi} = \sum_{n=0}^{d_C - 1} \sum_{k} \sum_{l} \int dp ~ c_n \psi(p) \beta(-p,E_k) \beta(p,E_l) \ket{E_n}_C \otimes \ket{E_k}_R \otimes \ket{E_l}_S
\end{equation}
where $\beta(a,b) = \braket{b|a}$. Through the state (\ref{7Gstato2}) we can now impose the constraint on total energy $\hat{H}\ket{\Psi} = \left( \hat{H}_C + \hat{H}_R + \hat{H}_S\right)\ket{\Psi}=0$ where the Hamiltonians of $R$ and $S$ can be rewritten ($\hslash=1$): $\hat{H}_R = \omega_R\left( \hat{a}^{\dagger}_R\hat{a}_R + \frac{1}{2} \right)$ and $\hat{H}_S = \omega_S\left( \hat{a}^{\dagger}_S \hat{a}_S + \frac{1}{2} \right)$, being $\hat{a}^{\dagger}_R$, $\hat{a}^{\dagger}_S$, $\hat{a}_R$ and $\hat{a}_S$ the usual rising and lowering operators for the subsystems $R$ and $S$. For the global state $\ket{\Psi}$ we thus find:

\begin{equation}\label{7Gstatoe}
\ket{\Psi} = \sum_{k} \sum_{l} \int dp ~ \tilde{c}_{kl} \psi(p) \beta(-p,E_k) \beta(p,E_l) \ket{E=-\epsilon_{kl}}_C \otimes \ket{E_k}_R \otimes \ket{E_l}_S
\end{equation}
with

\begin{equation}
\epsilon_{kl} = \omega_R\left( k + \frac{1}{2} \right) + \omega_S \left( l + \frac{1}{2} \right) .
\end{equation}
The state (\ref{7Gstatoe}) provides the explicit form of the global state of our quantum Universe simultaneously satisfying both the constraints: on total energy and total momentum. We can then expand again the subspaces $R$ and $S$ back on the momentum eigenbasis and rewriting (\ref{7Gstatoe}) as

\begin{multline}\label{7Gstatodef}
\ket{\Psi} = \int dp' \sum_{k} \sum_{l} \int dp ~ \tilde{c}_{kl} \psi(p) \beta(-p,E_k) \beta(p,E_l) \\
\times \beta^{*}(-p',E_k) \beta^{*}(p',E_l) \ket{E=-\epsilon_{kl}}_C \otimes \ket{-p'}_R \otimes \ket{p'}_S .
\end{multline} 

Through the states (\ref{7Gstatoe}) and (\ref{7Gstatodef}) we can find again all the results shown in Section 6.3.2. For the relative state $\ket{\phi(t)}_{R,S} = \braket{t|\Psi}$ we have: 

\begin{equation}
\begin{split}
i \frac{\partial}{\partial t}\ket{\phi(t)}_{R,S} & = \left(\hat{H}_R + \hat{H}_S\right)\ket{\phi(t)}_{R,S} = \\&
= \left(\frac{\hat{P}^{(2)}_R}{2M} + \frac{1}{2}M\omega_R\hat{X}^{2} + \frac{\hat{P}^{(2)}_S}{2m} + \frac{1}{2}m \omega_S \hat{Y}^{2}\right)\ket{\phi(t)}_{R,S} .
\end{split}
\end{equation}
which provides the Schrödinger evolution for the subsystem $R+S$. In the same way, for the relative state $\ket{\varphi(x)}_{C,S} = \sqrt{2\pi}\braket{x|\Psi}$ we have: $ \ket{\varphi(x+a)}_{C,S}= e^{-i a \hat{P}_S}\ket{\varphi(x)}_{C,S}$
which shows how the operator $\hat{P}_S$ is the generator of translations in the values of the reference frame $R$ for the relative state of the subsystem $C+S$. Finally we expand the state $\ket{\Psi}$ simultaneously on the coordinates in $R$ and on the time basis in $C$ thus finding:
\begin{equation}\label{7Gespansione}
\begin{split}
\ket{\Psi} 
= \frac{1}{T \sqrt{2\pi} } \int dt \int dx \ket{t}_C \otimes \ket{x}_R \otimes \ket{\psi(t,x)}_S
\end{split}
\end{equation}
where the integral on $dt$ is evaluated from $t_0$ and $t_0 +T$ and where now the relative state $\ket{\psi(t,x)}_S =  \sqrt{2\pi} \left( \bra{t}\otimes\bra{x}\right)\ket{\Psi}$ reads

\begin{multline}\label{7Gfin}
\ket{\psi(t,x)}_S = \int dp'	\sum_{k} \sum_{l}  \int dp ~ \tilde{c}_{kl} \psi(p) \beta(-p,E_k) \beta(p,E_l) \\ \times \beta^{*}(-p',E_k) \beta^{*}(p',E_l) e^{-i\epsilon_{kl}t} e^{-ixp'} \ket{p'}_S .     
\end{multline} 
Through the state (\ref{7Gfin}) 
we can search also in this case the probability density $P(y \: on\: S\:|\:x\:on\:R,\: t \: on \:C) = \left| \braket{y|\psi(t,x)} \right|^2$ obtaining:
\begin{multline}\label{7Gprobfinale}
P(y \: on\: S\:|\:x\:on\:R,\: t \: on \:C) = \frac{1}{2\pi} |\int dp' \sum_{k} \sum_{l}  \int dp ~ \tilde{c}_{kl} \psi(p) \beta(-p,E_k) \beta(p,E_l) \\ \times \beta^{*}(-p',E_k) \beta^{*}(p',E_l) e^{-i\epsilon_{kl}t} e^{ip'(y-x)}|^2
\end{multline}
which still depends on time $t$ and on the relative distance $y-x$ between $S$ and $R$.

\section{Multiple Time Measurements}
In this Section we show how Kuchar objection \cite{kuchar} to the PaW theory can be overcome in our framework of emerging spacetime. We recall here that Kuchar emphasized that the PaW mechanism is not able to provide the correct propagators when considering multiple measurements. Indeed measurements of the system at two times will give the wrong statistics because the first measurement \lq\lq collapses\rq\rq the time state and freezes the rest of the Universe (namely $R$ and $S$ in our framework). The two possible ways out of this problem we have addressed in Section 2.4 of Chapter 2 are the GPPT proposal in \cite{gppt} (see also \cite{esp1}) and the GLM theory in \cite{lloydmaccone}. 

We will now explore how these proposals can be applied to our framework of emerging spacetime. For the sake of simplicity we consider again a simple case: we start from the framework described in Sections 6.2.1 and 6.3.1, we work with discrete values of space and time and we choose $d_C=D_C$, $d_R=D_R$, $d_S=D_S$. This latter assumption implies that we are considering an equally-spaced spectrum for $\hat{H}_C$ and we emphasize that this choice leads to orthogonal states of time and position, namely  
    \begin{equation}
    \begin{split}
& \braket{t_m|t_{m'}}=\delta_{m,m'} \: on \: C 
\\& \braket{x_i|x_{j}}=\delta_{i,j} \: on \: R 
\\& \braket{y_l|y_{k}}=\delta_{l,k} \: on \: S \: 
    \end{split}
   \end{equation} 
In the case of GPPT theory, this simplified framework can then be readily generalized to the case of unequally-spaced levels for $\hat{H}_C$ and to the limiting cases $z,g,s \longrightarrow \infty$, where time and position are represented by POVMs. Conversely, for the GLM proposal the assumption of orthogonal time states in $C$ and orthogonal position states in $R$ and $S$ will be necessary.

\subsection{The GPPT proposal}
As pointed out in \cite{esp1} one of the main ingredients of the GPPT theroy is the averaging over the abstract coordinate time (the \lq\lq external time\rq\rq) in order to eliminate any external time dependence in the observables. With the perspective of calculating the probabilities for multiple time measurements, we look in this Section at the probability $P(y_l \: on \: S, \: x_j \: on \: R \:  | \: t_m \: on \: C)$ of having $y_l$ on $S$ and $x_j$ on $R$ conditioned to having $t_m$ on $C$. This probability essentially is not different from the probabilities we have calculated so far (apart from a numerical factor). We will see indeed that it will depend on the relative distance between $R$ and $S$ in the same way as we found in the previous Section. What clearly is different is the interpretation of this probability, where the value of the reference frame $R$ is not given and can vary. Following GPPT this probability is given by \cite{gppt}:

\begin{equation}\label{7cpGPPT}
\begin{split}
P(y_l \: on \: S, \: x_j \: on \: R \:  | \: t_m \: on \: C) &= \frac{\int d\theta \: Tr \left[ \hat{\Pi}_{t_m,x_j,y_l}(\theta) \hat{\rho} \right]}{\int d\theta \: Tr \left[ \hat{\Pi}_{t_m}(\theta) \hat{\rho} \right]} = \\&
= \frac{1}{d_R} \frac{1}{d_S} \left| \sum_{k=0}^{d_S - 1} c_k e^{-i\epsilon_k t_m } e^{i p_k (y_l - x_j)}\right|^2
\end{split}
\end{equation}
where $\hat{\rho}=\ket{\Psi}\bra{\Psi}= \sum_{k}\sum_{k'}c_k c^{*}_{k'}\ket{E=-\epsilon_k}\bra{E=-\epsilon_{k'}}\otimes\ket{p=-p_k}\bra{p=-p_{k'}}\otimes\ket{p_k}\bra{p_{k'}}$ is the global state of the Universe, $\theta$ is the external time, $\hat{\Pi}_{t_m}(\theta)=\hat{U}^{\dagger} (\theta) \hat{\Pi}_{t_m} \hat{U}(\theta)$ (with $\hat{U}(\theta) = e^{-i\hat{H}\theta}$) is the projector relative to the result $t_m$ for a clock measurement at external time $\theta$ and $\hat{\Pi}_{t_m,x_j,y_l}(\theta)=\hat{U}^{\dagger} (\theta) \hat{\Pi}_{t_m,x_j,y_l} \hat{U}(\theta)$ is the projector relative to the result $y_l$ for a measurement on $S$, $x_j$ for a measurement on $R$ and $t_m$ for a measurement on $C$ at external time $\theta$ (we are working here in the Heisenberg picture with respect to the external time $\theta$). Equation (\ref{7cpGPPT}) now takes the place of equations (\ref{7probfinalediscreta}) and (\ref{7probfinale}) and, as expected, depends on time $t_m$ and on the distance $y_l - x_j$ between the two subsystems $S$ and $R$. 

Equation (\ref{7cpGPPT}) can be generalized to the case of multiple time measurements. For two measurements at times $t_m$ and $t_{m'}$ (with $t_{m'} > t_m$) we have: 

\begin{multline}\label{7cpGPPT2}
P(y_{l'} \: on \: S, x_{j'} \: on \: R \: | \:t_{m'} \: on \: C,y_l, x_j, t_m) =\\
= \frac{\int d\theta \int d\theta' \: Tr \left[\hat{\Pi}_{t_{m'},x_{j'},y_{l'}}(\theta)  \hat{\Pi}_{t_m,x_j,y_l}(\theta') \hat{\rho} \hat{\Pi}_{t_m,x_j,y_l}(\theta') \right]}{\int d\theta \int d\theta' \: Tr \left[\hat{\Pi}_{t_{m'}}(\theta) \hat{\Pi}_{t_m,x_j,y_l}(\theta') \hat{\rho} \hat{\Pi}_{t_m,x_j,y_l}(\theta') \right]}
\end{multline}
which provides the conditional probability of obtaining $y_{l'}$ and $x_{j'}$ on $S$ and $R$ at clock time $t_{m'}$, given that a \lq\lq previous\rq\rq joint measurement of $S$, $R$ and $C$ returns $y_l$, $x_j$ and $t_m$. Proceeding with the calculations from the equation (\ref{7cpGPPT2}) we obtain:

\begin{equation}\label{7cpGPPT3}
\begin{split}
P(y_{l'} \: on \: S, x_{j'} \: on \: R \: | \:t_{m'} \: on \: C,y_l, x_j, t_m) 
= \frac{1}{d^{2}_R} \frac{1}{d^{2}_S} \left| \sum_{k=0}^{d_S-1} e^{-i\epsilon_k(t_{m'} - t_m)} e^{ip_k(\Delta_f - \Delta_i)}\right|^2
\end{split}
\end{equation}
where we have written $\Delta_i=y_l - x_j$ and $\Delta_f=y_{l'} - x_{j'}$ respectively the initial distance and the final distance (namely the distance at the first measurement at time $t_m$ and the distance at the second measurement at time $t_{m'}$) between the particle $S$ and $R$. This equation provides the the correct propagator: the same result can be indeed obtained calculating 

\begin{equation}
\left|\bra{x_{j'}} \bra{y_{l'}} e^{-i(\hat{H}_R + \hat{H}_S)(t_{m'} - t_m)}\ket{x_j}\ket{y_l}\right|^2 .
\end{equation}
Furthermore, we can easily see that probability (\ref{7cpGPPT3}) depends, as expected, on the initial and the final relative distances between $S$ and $R$. 

As previously mentioned, these results can be readily generalized to the case of unequally-spaced levels for $\hat{H}_C$ and to the limiting cases $z,s \longrightarrow \infty$. Indeed, in our framework, the fact of using POVMs in describing time and space does not constitute a problem in the application of the GPPT theory because when we calculate the probability $P(y_{l'} \: on \: S, x_{j'} \: on \: R \: | \:t_{m'} \: on \: C,y_l, x_j, t_m)$ through the (\ref{7cpGPPT2}) we do not find terms related to the overlap between the states and therefore interference phenomena are not present even if the states are not orthogonal \cite{nostro2}.

\subsection{GLM's Multiple Measurements}
We focus now on the GLM proposal applying it directly to our case of emergent spacetime. In this paragraph we consider the global state of the Universe written as in (\ref{7serveperGLM}), that in our particular case of orthogonal states of time and space ($d_C=D_C$, $d_R=D_R$, $d_S=D_S$) becomes

\begin{equation}\label{7statoGLM}
\begin{split}
\ket{\Psi} 
= \frac{1}{\sqrt{d_C}}\sum_{m=0}^{d_C -1} \ket{t_m}_C \otimes \hat{U}_{R,S}(t_m - t_0)\ket{\phi(t_0)}_{R,S}
\end{split}
\end{equation}
where $\ket{\phi(t_0)}_{R,S}$ is the state of $R+S$ conditioned on $t_0$ that is the value of the clock taken as initial time and where, thanks to the (\ref{7evoluzioneRS}), we have defined the unitary operator $\hat{U}_{R,S}(t_m - t_0)=e^{-i(\hat{H}_R + \hat{H}_S)(t_{m} - t_0)}$.

Again we start considering to perform a single measurement within $R+S$ at time $t_{m'}$. We divide the spaces $R$ and $S$ into two subsystems respectively: $\mathcal{H}_R = \mathcal{H}_{Q_R}\otimes\mathcal{H}_{M_R}$ and $\mathcal{H}_S = \mathcal{H}_{Q_S}\otimes\mathcal{H}_{M_S}$ where $Q_R$, $Q_S$ are the systems to be measured (the \textit{observed}) and $M_R$, $M_S$ are the ancillary memory systems (the \textit{observers}). In this framework GLM use von Neumann's prescription for measurements \cite{vonneumann}, where a measurement apparatus essentially consists in an (ideally instantaneous) interaction between the observed and the observers. The interaction correlates $Q_R$ with $M_R$ and $Q_S$ with $M_S$ along the eigenbasis $\{ \ket{x_j , y_l}\}$ of the observables $\hat{X}=\sum_{j}x_j\ket{x_j}\bra{x_j}$ and $\hat{Y}= \sum_{l}y_l\ket{y_l}\bra{y_l}$ to be measured, that is  

\begin{equation}\label{7mapping}
\ket{\phi(t_m)}_{Q_R,Q_S}\otimes\ket{r,r}_{M_R,M_S} \rightarrow \sum_{j=0}^{d_R -1}\sum_{l=0}^{d_S -1}\left( \braket{x_j,y_l|\phi(t_m)} \right)\ket{x_j,y_l}_{Q_R,Q_S}\otimes\ket{x_j,y_l}_{M_R,M_S}
\end{equation}
where $\ket{r,r}_{M_R,M_S}$ is the stete of the memories before the interaction and $\braket{x_j,y_l|\phi(t_m)}$ is the probabiliy amplitude of obtaining $x_j$ and $y_l$ when measuring the observables $\hat{X}$ and $\hat{Y}$. In this framework the Hamiltonian of $R+S$ can be written as

\begin{equation}\label{7hamiltonianaGLM}
\hat{H}_R(t_m) + \hat{H}_S(t_m)= \hat{H}_{Q_R} + \hat{H}_{Q_S} + \delta_{m,m'}\left( \hat{h}_{Q_R,M_R} + \hat{h}_{Q_S,M_S} \right)
\end{equation} 
where $\hat{h}_{Q_i,M_i}$ are responsible for the mapping equation (\ref{7mapping}). So we can write the global state (\ref{7statoGLM}) including the measurement thus obtaining \cite{lloydmaccone,esp2}

\begin{multline}\label{7singolamisuraGLM}
\ket{\Psi} = \frac{1}{\sqrt{d_C}}\sum_{m < m'} \ket{t_m}_C \otimes \hat{U}_{R,S}(t_m - t_0)\ket{\phi(t_0)}_{Q_R,Q_S}\otimes\ket{r,r}_{M_R,M_S} + \\
+ \frac{1}{\sqrt{d_C}}\sum_{m \ge m'} \ket{t_m}_C \otimes \sum_{j=0}^{d_R -1}\sum_{l=0}^{d_S -1}\left( \braket{x_j,y_l|\phi(t_m)} \right) \hat{U}_{Q_R,Q_S}(t_m - t_{m'})\ket{x_j,y_l}_{Q_R,Q_S}\otimes\ket{x_j,y_l}_{M_R,M_S} 
\end{multline}
where the first summation describes the evolution of $R+S$ prior to the measurement, when the memories are in the state $\ket{r,r}_{M_R,M_S}$, whereas the second summation describes the evolution after the measurement, when the memories are correlated with the subsystems $Q_R$ and $Q_S$. Now the probability that, at a given time $t_{m'}$, the values $x_{j'}$ and $y_{l'}$ will be registered by the memories $M_R$ and $M_S$ respectively can be expressed as \cite{lloydmaccone}:

\begin{equation}\label{7probGLM}
P(x_{j'},y_{l'} \:  | \: t_{m'} ) = \frac{|| \left( _{C}\bra{t_{m'}}\otimes _{M_R}\bra{x_{j'}}\otimes _{M_S}\bra{y_{l'}}\right)\ket{\Psi}||^2}{1/d_C}
\end{equation}
where we use the norm of a vector as $||\ket{v}||^2 = \braket{v|v}$. Equation (\ref{7probGLM}) returns the correct result 

\begin{equation}
P(x_{j'},y_{l'} \:  | \: t_{m'} ) =  \left| \braket{x_{j'},y_{l'}|\phi(t_{m'})} \right|^2 = \frac{1}{d_R} \frac{1}{d_S} \left| \sum_{k=0}^{d_S - 1} c_k e^{-i\epsilon_k t_{m'} } e^{i p_k (y_{l'} - x_{j'})}\right|^2  . 
\end{equation}
The GLM formalism allows us easily to calculate also the probability $P(y_{l'} \:  | \:x_{j'}, t_{m'} )$ simply by dividing equation (\ref{7probGLM}) by $1/d_R$. In this case we obtain: $P(y_{l'} \:  | \:x_{j'}, t_{m'} ) = \frac{1}{d_S} \left| \sum_{k=0}^{d_S - 1} c_k e^{-i\epsilon_k t_{m'} } e^{i p_k (y_{l'} - x_{j'})}\right|^2$
in accordance with our previous results.

It is now possible to extend equations (\ref{7hamiltonianaGLM}), (\ref{7singolamisuraGLM}) and (\ref{7probGLM}) in order to perform multiple measurements. The framework allows an arbitrary number of measurements but we consider here only the simple case of a double measurement at times $t_{m'}$ and $t_{m''}$ (with $t_{m''} > t_{m'}$). What we have to do is simply consider a larger set of memories $M^{(1)}_R$, $M^{(1)}_S$, $M^{(2)}_R$ and $M^{(2)}_S$ (where $M^{(1)}_R$, $M^{(1)}_S$ refer to the first measurement and $M^{(2)}_R$, $M^{(2)}_S$ refer to the second measurement) which couple with $Q_R$ and $Q_S$ through the time-dependent Hamiltonian
\begin{multline}
\hat{H}_R(t_m) + \hat{H}_S(t_m)= \hat{H}_{Q_R} + \hat{H}_{Q_S} +\\+ \delta_{m,m'}\left( \hat{h}_{Q_R,M^{(1)}_R} + \hat{h}_{Q_S,M^{(1)}_S} \right) 
+ \delta_{m,m''}\left( \hat{h}_{Q_R,M^{(2)}_R} + \hat{h}_{Q_S,M^{(2)}_S} \right) .
\end{multline}
The global state (\ref{7statoGLM}) including the double measurement can then be written \cite{lloydmaccone}:
\begin{multline}\label{7doppiamisuraGLM}
\ket{\Psi} = \frac{1}{\sqrt{d_C}}\sum_{m < m'} \ket{t_m}_C \otimes \hat{U}_{R,S}(t_m - t_0)\ket{\phi(t_0)}_{Q_R,Q_S}\otimes\ket{r,r}_{M^{(1)}_R,M^{(1)}_S}\otimes \ket{r,r}_{M^{(2)}_R,M^{(2)}_S} + \\
+ \frac{1}{\sqrt{d_C}} \sum_{m' \le m < m''} \ket{t_m}_C \otimes \sum_{j=0}^{d_R -1}\sum_{l=0}^{d_S -1}\left( \braket{x_j,y_l|\phi(t_m)} \right)\\ 
\times \hat{U}_{Q_R,Q_S}(t_m - t_{m'})\ket{x_j,y_l}_{Q_R,Q_S}  \otimes\ket{x_j,y_l}_{M^{(1)}_R,M^{(1)}_S}\otimes\ket{r,r}_{M^{(2)}_R,M^{(2)}_S} + \\
+ \frac{1}{\sqrt{d_C}} \sum_{m \ge m''} \ket{t_m}_C \otimes  \sum_{i=0}^{d_R -1}\sum_{k=0}^{d_S -1}  \sum_{j=0}^{d_R -1}\sum_{l=0}^{d_S -1} \left( \bra{x_i,y_k}\hat{U}_{Q_R,Q_S}(t_{m''} -t_{m'})\ket{x_j,y_l}\right) \left( \braket{x_j,y_l|\phi(t_m)}\right) \\
\times \hat{U}_{Q_R,Q_S}(t_m - t_{m''})\ket{x_i,y_k}_{Q_R,Q_S}\otimes\ket{x_j,y_l}_{M^{(1)}_R,M^{(1)}_S}\otimes\ket{x_i,y_k}_{M^{(2)}_R,M^{(2)}_S} .
\end{multline}
Through the state (\ref{7doppiamisuraGLM}) we search now the probability of obtaining $x_{j''}$ and $y_{l''}$ on $R$ and $S$ at time $t_{m''}$, given that a \lq\lq previous\rq\rq measurement at time $t_{m'}$ returns $x_{j'}$ and $y_{l'}$. This can be formally expressed as follows \cite{lloydmaccone}
\begin{multline}\label{7probdoppioGLM}
P\left(    \left( x_{j''},y_{l''} \:  | \: t_{m''} \right)\:|\:\left( x_{j'},y_{l'} \:  | \: t_{m'} \right)  \right) = \frac{P(x_{j''},y_{l''},x_{j'},y_{l'}\:|\:t_{m''})}{P(x_{j'},y_{l'} \:  | \: t_{m'})}=\\
= \frac{||(_{C}\bra{t_{m''}}\otimes _{M^{(1)}_R}\bra{x_{j'}}\otimes _{M^{(1)}_S}\bra{y_{l'}} \otimes _{M^{(2)}_R}\bra{x_{j''}}\otimes _{M^{(2)}_S}\bra{y_{l''}})\ket{\Psi} ||^2}{||  (_{C}\bra{t_{m'}}\otimes _{M^{(1)}_R}\bra{x_{j'}}\otimes _{M^{(1)}_S}\bra{y_{l'}})\ket{\Psi} ||^2} 
\end{multline}
which returns the correct result for a two-times measurement:
\begin{equation}\label{7pf} 
\begin{split}
P\left(    \left( x_{j''},y_{l''} \:  | \: t_{m''} \right)\:|\:\left( x_{j'},y_{l'} \:  | \: t_{m'} \right)  \right) &= \left| \bra{x_{j''},y_{l''}}\hat{U}_{Q_R,Q_S}(t_{m''} -t_{m'})\ket{x_{j'},y_{l'}}\right|^2 =\\&
= \frac{1}{d^{2}_R} \frac{1}{d^{2}_S} \left| \sum_{k=0}^{d_S-1} e^{-i\epsilon_k(t_{m''} - t_{m'})} e^{ip_k(\Delta_f - \Delta_i)}\right|^2
\end{split}
\end{equation}
where $\Delta_i=y_{l'} - x_{j'}$ and $\Delta_f=y_{l''} - x_{j''}$ are again the distances between $R$ and $S$ at times $t_{m'}$ and $t_{m''}$ respectively.
The result (\ref{7pf}) follows from the fact that $P(x_{j''},y_{l''},x_{j'},y_{l'}\:|\:t_{m''}) = \left| \left( \bra{x_{j''},y_{l''}}\hat{U}_{Q_R,Q_S}(t_{m''} -t_{m'})\ket{x_{j'},y_{l'}}\right) \left( \braket{x_{j'},y_{l'}|\phi(t_{m'})}\right) \right|^2$ obtained from the third summation in (\ref{7doppiamisuraGLM}) and $P(x_{j'},y_{l'} \:  | \: t_{m'}) = \left| \braket{x_{j'},y_{l'}|\phi(t_{m'})} \right|^2$ obtained from the second summation in (\ref{7doppiamisuraGLM}). It is finally easy to verify that the probability $P\left(    \left( y_{l''} \:  | \: x_{j''}, t_{m''} \right)\:|\:\left( y_{l'} \:  | \:x_{j'}, t_{m'} \right)  \right)$ is obtained dividing (\ref{7pf}) by $1/d^2_{R}$. So we find: $P\left(    \left( y_{l''} \:  | \: x_{j''}, t_{m''} \right)\:|\:\left( y_{l'} \:  | \:x_{j'}, t_{m'} \right)  \right)=\frac{1}{d^{2}_S} \left| \sum_{k=0}^{d_S-1} e^{-i\epsilon_k(t_{m''} - t_{m'})} e^{ip_k(\Delta_f - \Delta_i)}\right|^2$.

As we mentioned previously the GLM proposal can not be generalized, in our framework, to the case of non-orthogonal states of time and space and consequently in applying the GLM theory we are forced to assume $D_R=d_R$, $D_S=d_S$ and equally-spaced energy levels in $\hat{H}_C$.


\section{Generalization to $3+1$ Dimensional Spacetime}

We generalize here the results of Section 6.3.2 to the case of $3+1$ dimensional spacetime, meaning that we have now three degrees of freedom within the subspaces $R$ and $S$. The constraint on total energy reads again:

\begin{equation}\label{74constraint1}
\hat{H}\ket{\Psi} = ( \hat{H}_C + \hat{H}_R +\hat{H}_S)\ket{\Psi}=0 .
\end{equation}
The Hamiltonians $\hat{H}_R$ and $\hat{H}_S$ in (\ref{74constraint1}) depend respectively on the operators $\hat{P}^{(1)}_R$, $\hat{P}^{(2)}_R$, $\hat{P}^{(3)}_R$ and $\hat{P}^{(1)}_S$, $\hat{P}^{(2)}_S$, $\hat{P}^{(3)}_S$ where $1,2,3$ are the three degrees of freedom identifying three orthogonal directions in space\footnote{Also in this case, for simplicity, we assume that no external potentials are present in $R$ and $S$. The case of 3+1 dimensional spacetime with the presence of external potentials in $R$ and $S$ is a straightforward generalization of the argument we addressed in Section 6.3.5 of this Chapter.}. The constraint on the total momentum reads now $\vec{P}\ket{\Psi} = (\vec{P}_R + \vec{P}_S)\ket{\Psi}=0$, which we rewrite as

\begin{equation}\label{74constraint2}
\begin{split}
&   (\hat{P}^{(1)}_R + \hat{P}^{(1)}_S)\ket{\Psi}=0
\\& (\hat{P}^{(2)}_R + \hat{P}^{(2)}_S)\ket{\Psi}=0 
\\& (\hat{P}^{(3)}_R + \hat{P}^{(3)}_S)\ket{\Psi}=0
\end{split}
\end{equation} 
where, also in this case, for simplicity we have chosen $\vec{P}_C = 0$ (with $\hat{P}^{(1)}_C = \hat{P}^{(2)}_C=\hat{P}^{(3)}_C=0$). So, assuming $d_C, d^{(1)}_R,d^{(2)}_R,d^{(3)}_R \gg d^{(1)}_S,d^{(2)}_S,d^{(3)}_S$ and simplifying the notation as much as possible, the global state $\ket{\Psi}$ satisfying the constraints (\ref{74constraint1}) and (\ref{74constraint2}) can be written as

\begin{equation}\label{74statoglobale}
\ket{\Psi} = \sum_{i=0}^{d^{(1)}_S - 1} \sum_{j=0}^{d^{(2)}_S - 1} \sum_{k=0}^{d^{(3)}_S - 1} c_{ijk}\ket{-\epsilon_{ijk}}_C \otimes \ket{-p^{(1)}_i , -p^{(2)}_j, -p^{(3)}_k }_R \otimes \ket{p^{(1)}_i , p^{(2)}_j, p^{(3)}_k }_S 
\end{equation}
where $d^{(1)}_R,d^{(2)}_R,d^{(3)}_R$ and $d^{(1)}_S,d^{(2)}_S,d^{(3)}_S$ are the dimensions of the subspaces of $R$ and $S$ (associated with the three spatial directions 1,2,3) and $\epsilon_{ijk}$ is the energy function related with the momenta of $R$ and $S$. In the next two paragraphs we will study separately the case of discrete values for space and time and then the case of continuous values for space and time. 

\subsection{Discrete Values for Space and Time}
Starting from the state $\ket{\Psi}$ satisfying (\ref{74constraint1}) and (\ref{74constraint2}), we can now expand it on the basis $\left\{\ket{t_a}_C\right\}$ in $C$ thanks to (\ref{7pomidentity2}), thus obtaining

\begin{equation}
\begin{split}
\ket{\Psi} = \frac{d_C}{D_C} \sum_{a=0}^{D_C -1}\ket{t_a}\braket{t_a|\Psi} = \frac{\sqrt{d_C}}{D_C}\sum_{a=0}^{D_C -1}\ket{t_a}_C \otimes \ket{\phi(t_a)}_{R,S}
\end{split}
\end{equation}
where the relative state $\ket{\phi(t_a)}_{R,S} = \sqrt{d_C}\braket{t_a|\Psi}$ takes now the form

\begin{equation}
\ket{\phi(t_a)}_{R,S} = \sum_{i=0}^{d^{(1)}_S - 1} \sum_{j=0}^{d^{(2)}_S - 1} \sum_{k=0}^{d^{(3)}_S - 1} c_{ijk} e^{-it_a \epsilon_{ijk}} \ket{-p^{(1)}_i , -p^{(2)}_j, -p^{(3)}_k }_R \otimes \ket{p^{(1)}_i , p^{(2)}_j, p^{(3)}_k }_S .
\end{equation}
For this relative state we again easily find the Schrödiger evolution with respect to the clock values, namely

\begin{equation}
\ket{\phi(t_a)}_{R,S} = e^{-i (\hat{H}_R + \hat{H}_S)(t_a - t_0)}\ket{\phi(t_0)}_{R,S}
\end{equation}
where $\ket{\phi(t_0)}_{R,S}= \sqrt{d_C} \braket{t_0|\Psi}$ is the state of $R+S$ conditioned on $t_0$ that is the value of the clock taken as initial time. All the consideration made in Section 6.3.2 are clearly still valid in this case and we do not repeat them.

We emphasize here that in each subspace of $R$, related to the three degrees of freedom, we apply the formalism described in Section 6.2. This means considering the states $\ket{x^{(J)}_n} = 1/\sqrt{d^{(J)}_R}\sum_{k=0}^{d^{(J)}_R -1}e^{-i p^{(J)}_k x^{(J)}_n}\ket{p^{(J)}_k}$ and the values $x^{(J)}_n = x^{(J)}_0 + n L^{(J)}_R/D^{(J)}_R$ with $n=0,1,2,...,z=D^{(J)}_R -1$ for $J=1,2,3$. The same clearly holds for the subspaces within the system $S$ that we equip with the states $\ket{y^{(J)}_q} = 1/\sqrt{d^{(J)}_S}\sum_{k=0}^{d^{(J)}_S -1}e^{-i p^{(J)}_k y^{(J)}_q}\ket{p^{(J)}_k}$ and with the values $y^{(J)}_q = y^{(J)}_0 + q L^{(J)}_S/D^{(J)}_S$ with $q=0,1,2,...,z=D^{(J)}_S -1$ (for semplicity we choose $D^{(J)}_R = D^{(J)}_S =z+1 ~ \forall J$). We can now expand the state $\ket{\Psi}$ in the coordinates $\left\{ \ket{x^{(1)}_l , x^{(2)}_m, x^{(3)}_n}_R\right\}$ in $R$:

\begin{equation}
\begin{split}
\ket{\Psi} &=\frac{d^{(1)}_R}{D^{(1)}_R} \frac{d^{(2)}_R}{D^{(2)}_R} \frac{d^{(3)}_R}{D^{(3)}_R} \sum_{l=0}^{D^{(1)}_R - 1} \sum_{m=0}^{D^{(2)}_R - 1} \sum_{n=0}^{D^{(3)}_R - 1}\ket{x^{(1)}_l , x^{(2)}_m, x^{(3)}_n}\braket{x^{(1)}_l , x^{(2)}_m, x^{(3)}_n|\Psi} =\\& 
= \frac{\sqrt{d^{(1)}_R}}{D^{(1)}_R} \frac{\sqrt{d^{(2)}_R}}{D^{(2)}_R} \frac{\sqrt{d^{(3)}_R}}{D^{(3)}_R} \sum_{l=0}^{D^{(1)}_R - 1} \sum_{m=0}^{D^{(2)}_R - 1} \sum_{n=0}^{D^{(3)}_R - 1}\ket{x^{(1)}_l , x^{(2)}_m, x^{(3)}_n}_R \otimes \ket{\varphi(x^{(1)}_l , x^{(2)}_m, x^{(3)}_n)}_{C,S} 
\end{split}
\end{equation}
where $\ket{\varphi(x^{(1)}_l , x^{(2)}_m, x^{(3)}_n)}_{C,S}= \sqrt{d^{(1)}_R d^{(2)}_R d^{(3)}_R} \braket{x^{(1)}_l , x^{(2)}_m, x^{(3)}_n|\Psi}$ is the relative state of $C+S$ conditioned to the value $(x^{(1)}_l , x^{(2)}_m, x^{(3)}_n)$ of the reference frame $R$. For this relative state we find:

\begin{multline}\label{74trasl}
\ket{\varphi(x^{(1)}_l , x^{(2)}_m, x^{(3)}_n)}_{C,S} = \sqrt{d^{(1)}_R d^{(2)}_R d^{(3)}_R} \braket{x^{(1)}_l , x^{(2)}_m, x^{(3)}_n|\Psi}=\\
= \sqrt{d^{(1)}_R d^{(2)}_R d^{(3)}_R} \bra{x^{(1)}_l , x^{(2)}_m, x^{(3)}_n}e^{i\hat{P}^{(1)}_R(x^{(1)}_l - x^{(1)}_0)}e^{i\hat{P}^{(2)}_R(x^{(2)}_m - x^{(2)}_0)}e^{i\hat{P}^{(3)}_R(x^{(3)}_n - x^{(3)}_0)}\ket{\Psi}=\\
=e^{-i\hat{P}^{(1)}_S(x^{(1)}_l - x^{(1)}_0)}e^{-i\hat{P}^{(2)}_S(x^{(2)}_m - x^{(2)}_0)}e^{-i\hat{P}^{(3)}_S(x^{(3)}_n - x^{(3)}_0)} \ket{\varphi(x^{(1)}_0 , x^{(2)}_0, x^{(3)}_0)}_{C,S}
\end{multline}
where we used the relative state definition and the constraint (\ref{74constraint2}). Equation (\ref{74trasl}) can be rewritten in a more compact form as

\begin{equation}
\ket{\varphi(\vec{x_0} + \vec{a})}_{C,S} = e^{-i\vec{a} \cdot \vec{P}_S} \ket{\varphi(\vec{x_0})}_{C,S}
\end{equation}
where $\vec{P}_S = (\hat{P}^{(1)}_S,\hat{P}^{(2)}_S,\hat{P}^{(3)}_S)$, $\vec{x_0} = (x^{(1)}_0 , x^{(2)}_0, x^{(3)}_0)$ is the initial position of the reference frame and the translation vector is $\vec{a} = (x^{(1)}_l - x^{(1)}_0, x^{(2)}_m - x^{(2)}_0, x^{(3)}_n - x^{(3)}_0)$.

Finally we can expand the state $\ket{\Psi}$ simultaneously on time $\left\{\ket{t_a}_C\right\}$ and on the coordinates $\left\{ \ket{x^{(1)}_l , x^{(2)}_m, x^{(3)}_n}_R\right\}$, thus obtaining:
\begin{equation}
\begin{split}
\ket{\Psi} &= \left( \frac{d_C}{D_C} \sum_{a}
\ket{t_a}\bra{t_a} \otimes \frac{d^{(1)}_R}{D^{(1)}_R} \frac{d^{(2)}_R}{D^{(2)}_R} \frac{d^{(3)}_R}{D^{(3)}_R} \sum_{l,m,n} 
\ket{x^{(1)}_l , x^{(2)}_m, x^{(3)}_n}\bra{x^{(1)}_l , x^{(2)}_m, x^{(3)}_n} \right)\ket{\Psi} = \\&
= \frac{\sqrt{d_C}}{D_C}\frac{\sqrt{d^{(1)}_R}}{D^{(1)}_R} \frac{\sqrt{d^{(2)}_R}}{D^{(2)}_R} \frac{\sqrt{d^{(3)}_R}}{D^{(3)}_R} \sum_{a}\sum_{l,m,n} 
\ket{t_a}_C \otimes \ket{x^{(1)}_l , x^{(2)}_m, x^{(3)}_n}_R \otimes \ket{\psi(t_a,x^{(1)}_l , x^{(2)}_m, x^{(3)}_n)}_S 
\end{split}
\end{equation}
where the summation on time runs from $0$ to $D_C -1$, the summations on $l,m,n$ run from $0$ to $D^{(1)}_R -1$, $D^{(2)}_R -1$, $D^{(3)}_R -1$ respectively and where $\ket{\psi(t_a,x^{(1)}_l , x^{(2)}_m, x^{(3)}_n)}_S = \sqrt{d_C}\sqrt{d^{(1)}_R d^{(2)}_R d^{(3)}_R}(\bra{t_a}\otimes\bra{x^{(1)}_l , x^{(2)}_m, x^{(3)}_n})\ket{\Psi}$ is the relative state of the system $S$ at time $t_a$ and conditioned on the value $(x^{(1)}_l , x^{(2)}_m, x^{(3)}_n)$ for the spatial reference frame $R$. Through this state, extending the formalism of Section 6.3.2, we can search the conditional probability:
\begin{multline}\label{74probdiscreta}
P(y^{(1)}_p, y^{(2)}_q,y^{(3)}_r\:|\:x^{(1)}_l , x^{(2)}_m, x^{(3)}_n,t_a) = \\
=  \frac{d^{(1)}_S}{D^{(1)}_S} \frac{d^{(2)}_S}{D^{(2)}_S} \frac{d^{(3)}_S}{D^{(3)}_S} 
\left|\braket{y^{(1)}_p, y^{(2)}_q,y^{(3)}_r|\psi(t_a,x^{(1)}_l , x^{(2)}_m, x^{(3)}_n)}\right|^2=\\
=\frac{1}{D^{(1)}_S}  \frac{1}{D^{(2)}_S} \frac{1}{D^{(3)}_S} \left| \sum_{i,j,k}
c_{ijk} e^{-it_a\epsilon_{ijk}} e^{ip^{(1)}_i(y^{(1)}_p  -  x^{(1)}_l )} e^{ip^{(2)}_j (y^{(2)}_q  -  x^{(2)}_m )} e^{ip^{(3)}_k(y^{(3)}_r  -  x^{(3)}_n )}\right|^2
\end{multline}  
where the summations on $i,j,k$ run from $0$ to $d^{(1)}_S - 1$, $d^{(2)}_S - 1$ and $d^{(3)}_S - 1$ respectively. Equation (\ref{74probdiscreta}) provides the probability of having a certain position $(y^{(1)}_p, y^{(2)}_q,y^{(3)}_r)$ on $S$ at time $t_a$ and knowing that the spatial reference frame $R$ is in $(x^{(1)}_l , x^{(2)}_m, x^{(3)}_n)$. This conditional probability is well-defined (indeed it is easy to verify that we have $\sum_{p=0}^{D^{(1)}_S - 1} \sum_{q=0}^{D^{(2)}_S - 1} \sum_{r=0}^{D^{(3)}_S - 1} P(y^{(1)}_p, y^{(2)}_q,y^{(3)}_r\:|\:x^{(1)}_l , x^{(2)}_m, x^{(3)}_n,t_a)=1$) and, as expected depends on the relative distance between $R$ and $S$.

\subsection{Continuous Values for Space and Time}
We consider now the case of continuous values for space and time. We therefore assume $s\longrightarrow \infty$ for the space $C$ and $z\longrightarrow \infty$ within the subspaces of $R$ and $S$. This implies that we have the states $\ket{x^{(J)}} = \sum_{k=0}^{d^{(J)}_R -1}e^{-i p^{(J)}_k x^{(J)}}\ket{p^{(J)}_k}$ with $x^{(J)} \in \left[ x^{(J)}_0 , x^{(J)}_0 + L^{(J)}_R \right]$ within $R$ and the states $\ket{y^{(J)}} = \sum_{k=0}^{d^{(J)}_S -1}e^{-i p^{(J)}_k y^{(J)}}\ket{p^{(J)}_k}$ with $y^{(J)} \in \left[ y^{(J)}_0 , y^{(J)}_0 + L^{(J)}_S \right]$ within $S$ for $J=1,2,3$. Starting from (\ref{74statoglobale}) we can write:

\begin{equation}
\ket{\Psi} = \frac{1}{T} \int_{t_0}^{t_0 + T} dt \ket{t}\braket{t|\Psi} = \frac{1}{T} \int_{t_0}^{t_0 + T} dt \ket{t}_C\otimes\ket{\phi(t)}_{R,S}
\end{equation}
where the relative state $\ket{\phi(t)}_{R,S} = \braket{t|\Psi}$ takes now the form

\begin{equation}\label{786}
\ket{\phi(t)}_{R,S} = \sum_{i=0}^{d^{(1)}_S - 1} \sum_{j=0}^{d^{(2)}_S - 1} \sum_{k=0}^{d^{(3)}_S - 1} c_{ijk} e^{-it \epsilon_{ijk}} \ket{-p^{(1)}_i , -p^{(2)}_j, -p^{(3)}_k}_R \otimes \ket{p^{(1)}_i , p^{(2)}_j, p^{(3)}_k }_S .
\end{equation}
As in the $1+1$ dimensional case, using (\ref{74constraint1}) and (\ref{786}), for this state we find:

\begin{equation}\label{74evcont}
i \frac{\partial}{\partial t}\ket{\phi(t)}_{R,S} = (\hat{H}_R + \hat{H}_S)\ket{\phi(t)}_{R,S}
\end{equation}
which provides the Scrhödinger evolution of $R+S$ with respect to the clock time $t$.   

We can now expand the global state $\ket{\Psi}$ in the coordinate basis $\left\{ \ket{x^{(1)} , x^{(2)}, x^{(3)} }_R\right\}$ in $R$, that is

\begin{equation}\label{787}
\begin{split}
\ket{\Psi} &= \frac{1}{L^{(1)}_R} \frac{1}{L^{(2)}_R} \frac{1}{L^{(3)}_R} \int dx^{(1)} \int dx^{(2)} \int dx^{(3)} \ket{x^{(1)} , x^{(2)}, x^{(3)}} \braket{x^{(1)} , x^{(2)}, x^{(3)}|\Psi}=\\&
=\frac{1}{L^{(1)}_R} \frac{1}{L^{(2)}_R} \frac{1}{L^{(3)}_R} \int dx^{(1)} \int dx^{(2)} \int dx^{(3)} \ket{x^{(1)} , x^{(2)}, x^{(3)}}_R \otimes \ket{\varphi(x^{(1)} , x^{(2)}, x^{(3)})}_{C,S}
\end{split}
\end{equation}
where each integral on $dx^{(J)}$ is evaluated from $x^{(J)}_0$ to $x^{(J)}_0 + L^{(J)}_R$ and where the state $\ket{\varphi(x^{(1)} , x^{(2)}, x^{(3)})}_{C,S}$ takes the form:

\begin{equation}\label{74statoCS}
\ket{\varphi(x^{(1)} , x^{(2)}, x^{(3)})}_{C,S} = \sum_{i,j,k} 
c_{ijk} e^{-ip^{(1)}_i x^{(1)}} e^{-ip^{(2)}_j x^{(2)}} e^{-ip^{(3)}_k x^{(3)}} \ket{-\epsilon_{ijk}}_C \otimes \ket{p^{(1)}_i , p^{(2)}_j, p^{(3)}_k }_S 
\end{equation}
where the summations on $i,j,k$ run from $0$ to $d^{(1)}_S - 1$, $d^{(2)}_S - 1$ and $d^{(3)}_S - 1$ respectively. Using the definition (\ref{74statoCS}) and the constraint (\ref{74constraint2}), for the relative state $\ket{\varphi(x^{(1)} , x^{(2)}, x^{(3)})}_{C,S} = \braket{x^{(1)} , x^{(2)}, x^{(3)}|\Psi}$ we can now obtain:

\begin{multline}
i\left(\frac{\partial}{\partial x^{(1)}} + \frac{\partial}{\partial x^{(2)}} + \frac{\partial}{\partial x^{(3)}} \right)\ket{\varphi(x^{(1)} , x^{(2)}, x^{(3)})}_{C,S}= \\
= \sum_{i,j,k} c_{ijk} p^{(1)}_i e^{ip^{(1)}_i x^{(1)}} e^{ip^{(2)}_j x^{(2)}} e^{ip^{(3)}_k x^{(3)}} \ket{-\epsilon_{ijk}}_C \otimes \ket{p^{(1)}_i , p^{(2)}_j, p^{(3)}_k }_S  + \\
+ \sum_{i,j,k} c_{ijk} p^{(2)}_j e^{ip^{(1)}_i x^{(1)}} e^{ip^{(2)}_j x^{(2)}} e^{ip^{(3)}_k x^{(3)}} \ket{-\epsilon_{ijk}}_C \otimes \ket{p^{(1)}_i , p^{(2)}_j, p^{(3)}_k }_S + \\
+ \sum_{i,j,k} c_{ijk} p^{(3)}_k e^{ip^{(1)}_i x^{(1)}} e^{ip^{(2)}_j x^{(2)}} e^{ip^{(3)}_k x^{(3)}} \ket{-\epsilon_{ijk}}_C \otimes \ket{p^{(1)}_i , p^{(2)}_j, p^{(3)}_k }_S = \\ = \left(\hat{P}^{(1)}_S + \hat{P}^{(2)}_S +\hat{P}^{(3)}_S \right) \ket{\varphi(x^{(1)} , x^{(2)}, x^{(3)})}_{C,S}
\end{multline}
which shows $\vec{P}_S=(\hat{P}^{(1)}_S,\hat{P}^{(2)}_S,\hat{P}^{(3)}_S)$ to be the generator of translations for the states of $C+S$ acting on the coordinates of the spatial reference frame $\vec{x} = (x^{(1)} , x^{(2)}, x^{(3)})$.

Expanding the state $\ket{\Psi}$ simultaneously on the coordinates basis $\left\{ \ket{x^{(1)} , x^{(2)}, x^{(3)} }_R\right\}$ and on the time basis $\left\{\ket{t}_C\right\}$ we find:

\begin{multline}\label{74esp}
\ket{\Psi} = A \left( \int dt \ket{t}\bra{t} \otimes   \int dx^{(1)} \int dx^{(2)} \int dx^{(3)} \ket{x^{(1)} , x^{(2)}, x^{(3)}} \bra{x^{(1)} , x^{(2)}, x^{(3)}} \right) \ket{\Psi} =\\
= A \int dt  \int dx^{(1)} \int dx^{(2)} \int dx^{(3)} \ket{t}_C\otimes \ket{x^{(1)} , x^{(2)}, x^{(3)}}_R \otimes \ket{\psi(t,x^{(1)} , x^{(2)}, x^{(3)})}_S
\end{multline}
where $A= \frac{1}{T}\frac{1}{L^{(1)}_R} \frac{1}{L^{(2)}_R} \frac{1}{L^{(3)}_R}$. In equation (\ref{74esp}) the integral on time is evaluated from $t_0$ to $t_0 + T$ and each integral on $dx^{(J)}$ is evaluated from $x^{(J)}_0$ to $x^{(J)}_0 + L^{(J)}_R$. The state $\ket{\psi(t,x^{(1)} , x^{(2)}, x^{(3)})}_S = (\bra{t}\otimes\bra{x^{(1)} , x^{(2)}, x^{(3)}})\ket{\Psi}$ 
takes now the form 

\begin{equation}
\ket{\psi(t,x^{(1)} , x^{(2)}, x^{(3)})}_S = \sum_{i,j,k} c_{ijk}  e^{-it\epsilon_{ijk}} e^{-ip^{(1)}_i x^{(1)}} e^{-ip^{(2)}_j x^{(2)}} e^{-ip^{(3)}_k x^{(3)}} \ket{p^{(1)}_i , p^{(2)}_j, p^{(3)}_k }_S 
\end{equation}
where the summations on $i,j,k$ run again from $0$ to $d^{(1)}_S - 1$, $d^{(2)}_S - 1$ and $d^{(3)}_S - 1$ respectively.
Through this state we can calculate the conditional probability density of having the position $(y^{(1)},y^{(2)},y^{(3)})$ on $S$ at time $t$ and knowing that the spatial reference frame $R$ is in $(x^{(1)} , x^{(2)}, x^{(3)})$. We have: 
\begin{multline}
P(y^{(1)},y^{(2)},y^{(3)}\:|\:x^{(1)} , x^{(2)}, x^{(3)},t) =\\
=  	\frac{1}{L^{(1)}_S} \frac{1}{L^{(2)}_S} \frac{1}{L^{(3)}_S} \left| \braket{y^{(1)},y^{(2)},y^{(3)}|\psi(t,x^{(1)} , x^{(2)}, x^{(3)})} \right|^2 = \\
= 	\frac{1}{L^{(1)}_S} \frac{1}{L^{(2)}_S} \frac{1}{L^{(3)}_S} \left| \sum_{i,j,k}
c_{ijk} e^{-it \epsilon_{ijk}} e^{ip^{(1)}_i(y^{(1)}  -  x^{(1)} )} e^{ip^{(2)}_j (y^{(2)}  -  x^{(2)} )} e^{ip^{(3)}_k(y^{(3)}  -  x^{(3)} )} \right|^2 .
\end{multline}
As in the previous cases, this probability density is well-defined (indeed we have $\int dy^{(1)} \int dy^{(2)} \int dy^{(3)} P(y^{(1)},y^{(2)},y^{(3)}\:|\:x^{(1)} , x^{(2)}, x^{(3)},t)=1$ with each integral on $dy^{(J)}$ evaluated from $y^{(J)}_0$ to $y^{(J)}_0 + L^{(J)}_S$) and it depends on time and on the relative distance between $R$ and $S$ in a 3-dimensional continuous space.

\subsection{Free Particles (with $M\gg m$) in $3+1$ Spacetime}
We give here a simple example of an emerging $3+1$ dimensional spacetime using continuous values of space and time, by starting from the framework described in Section 6.5.2. In doing so we adopt a slightly different formalism, which allows us to emphasize how space and time are treated here on equal footing. 

We consider two systems that we call $\mathfrak{R}$ and $S$. The system $\mathfrak{R}$ acts as spacetime reference frame for $S$ and it is composed of a free particle of mass $M$ together with an additional degree of freedom (with zero momentum) that acts as a clock. We choose also $S$ as free particle of mass $m$ and we assume $M \gg m$ (as mentioned in Section 6.3.2 this choice implies that we will be able to neglect the kinetic energy term of $\mathfrak{R}$ with respect to the kinetic energy of $S$, namely $\mathfrak{R}$ is a good reference frame and moves very slightly in time). 
The global Hamiltonian can be written:

\begin{equation}
\hat{H}=\hat{H}^{(0)}_{\mathfrak{R}} +\hat{H}^{(1)}_{\mathfrak{R}}+ \hat{H}^{(2)}_{\mathfrak{R}} + \hat{H}^{(3)}_{\mathfrak{R}} +\hat{H}_{S}
\end{equation}
where $\hat{H}^{(0)}_{\mathfrak{R}}$ is the Hamiltonian of the temporal reference frame (which takes the place of what was $\hat{H}_C$ in the previous discussion), $\hat{H}^{(1)}_{\mathfrak{R}}$, $\hat{H}^{(2)}_{\mathfrak{R}}$, $\hat{H}^{(3)}_{\mathfrak{R}}$ are the Hamiltonians depending on the momenta of the reference frame in the three spatial directions through $\hat{H}^{(1)}_{\mathfrak{R}} = \left( \hat{P}^{(1)}_{\mathfrak{R}} \right)^2/2M$, $\hat{H}^{(2)}_{\mathfrak{R}} = \left( \hat{P}^{(2)}_{\mathfrak{R}} \right)^2/2M$, $\hat{H}^{(3)}_{\mathfrak{R}} = \left( \hat{P}^{(3)}_{\mathfrak{R}} \right)^2/2M$ and $\hat{H}_S = \hat{H}^{(1)}_S + \hat{H}^{(2)}_S+ \hat{H}^{(3)}_S$ is the Hamiltonian of $S$ depending on the momenta $\hat{P}^{(1)}_S$, $\hat{P}^{(2)}_S$, $\hat{P}^{(3)}_S$ through $\hat{H}^{(1)}_{S} = \left( \hat{P}^{(1)}_{S} \right)^2/2m$, $\hat{H}^{(2)}_{S} = \left( \hat{P}^{(2)}_{S} \right)^2/2m$, $\hat{H}^{(3)}_{S} = \left( \hat{P}^{(3)}_{S} \right)^2/2m$. 
With the intent of treating space and time on equal footing, within the time subspace we rewrite the time states as $\ket{x^{(0)}} = \sum_{k=0}^{d^{(0)}_{\mathfrak{R}} -1} e^{-ix^{(0)} p^{(0)}_k} \ket{p^{(0)}_k}$ and the time values as $x^{(0)} \in \left[ x^{(0)}_0, x^{(0)}_0 + L^{(0)}_{\mathfrak{R}} \right]$, where $d^{(0)}_{\mathfrak{R}}$ is the dimension of the time subspace, $p^{(0)}_k$ are the eigenvalues of $\hat{H}^{(0)}_{\mathfrak{R}}$ and where $L^{(0)}_{\mathfrak{R}}$ takes now the place of what was $T$ in the previous discussion. Furthermore we redefine the position states as $\ket{x^{(J)}} = \sum_{k=0}^{d^{(J)}_{\mathfrak{R}} -1} e^{i x^{(J)} p^{(J)}_k} \ket{p^{(J)}_k}$ in $\mathfrak{R}$ and $\ket{y^{(J)}} = \sum_{k=0}^{d^{(J)}_{S} -1} e^{i y^{(J)} p^{(J)}_k} \ket{p^{(J)}_k}$ in $S$ for $J=1,2,3$.     

The constraints (\ref{74constraint1}) and (\ref{74constraint2}) read now:

\begin{equation}\label{75constraint1}
\hat{H}\ket{\Psi} = \left( \hat{H}^{(0)}_{\mathfrak{R}} +\hat{H}^{(1)}_{\mathfrak{R}}+ \hat{H}^{(2)}_{\mathfrak{R}} + \hat{H}^{(3)}_{\mathfrak{R}} +\hat{H}_{S} \right)\ket{\Psi}=0
\end{equation}
and
\begin{equation}\label{75constraint2}
\begin{split}
\vec{P}\ket{\Psi}	= \left( \vec{P}_{\mathfrak{R}} + \vec{P}_S \right)\ket{\Psi} = 0
\end{split}
\end{equation} 
with $(\hat{P}^{(1)}_{\mathfrak{R}} + \hat{P}^{(1)}_S)\ket{\Psi}=(\hat{P}^{(2)}_{\mathfrak{R}} + \hat{P}^{(2)}_S)\ket{\Psi}= (\hat{P}^{(3)}_{\mathfrak{R}} + \hat{P}^{(3)}_S)\ket{\Psi}=0$. 
Assuming again $d^{(0)}_{\mathfrak{R}}, d^{(1)}_{\mathfrak{R}},d^{(2)}_{\mathfrak{R}},d^{(3)}_{\mathfrak{R}} \gg d^{(1)}_S,d^{(2)}_S,d^{(3)}_S$, the global state $\ket{\Psi}$ simultaneously satisfying (\ref{75constraint1}) and (\ref{75constraint2}) can be written as

\begin{equation}
\ket{\Psi} = \sum_{i=0}^{d^{(1)}_S - 1} \sum_{j=0}^{d^{(2)}_S - 1} \sum_{k=0}^{d^{(3)}_S - 1} c_{ijk} \ket{p^{(0)}=- \epsilon_{ijk}, -p^{(1)}_i , -p^{(2)}_j, -p^{(3)}_k}_{\mathfrak{R}} \otimes \ket{p^{(1)}_i , p^{(2)}_j, p^{(3)}_k }_S
\end{equation}
where $p^{(0)}$ is the value for the energy of the time reference and the energy function $\epsilon_{ijk}$ is: $\epsilon_{ijk} = \left(\frac{1}{2M} + \frac{1}{2m}\right) \left( \left(p^{(1)}_i\right)^{2} + \left(p^{(2)}_j\right)^{2} + \left(p^{(3)}_k\right)^{2}\right) \simeq \frac{1}{2m}\left( \left(p^{(1)}_i\right)^{2} + \left(p^{(2)}_j\right)^{2} + \left(p^{(3)}_k\right)^{2}\right)$. 

We can now expand the global state $\ket{\Psi}$ in the basis $\left\{\ket{x^{(0)},x^{(1)},x^{(2)},x^{(3)}}_{\mathfrak{R}} \right\}$ in the space $\mathfrak{R}$, thus obtaining:

\begin{multline}
\ket{\Psi} = A \int dx^{(0)} \int dx^{(1)} \int dx^{(2)} \int dx^{(3)} \ket{x^{(0)},x^{(1)},x^{(2)},x^{(3)}}\braket{x^{(0)},x^{(1)},x^{(2)},x^{(3)}|\Psi} =\\
= A\int dx^{(0)} \int dx^{(1)} \int dx^{(2)} \int dx^{(3)} \ket{x^{(0)},x^{(1)},x^{(2)},x^{(3)}}_{\mathfrak{R}} \otimes \ket{\psi(x^{(0)},x^{(1)},x^{(2)},x^{(3)})}_S
\end{multline}
where $A= \frac{1}{L^{(0)}_{\mathfrak{R}}} \frac{1}{L^{(1)}_{\mathfrak{R}}} \frac{1}{L^{(2)}_{\mathfrak{R}}} \frac{1}{L^{(3)}_{\mathfrak{R}}}$ and the integrals on $dx^{(J)}$ are evaluated from $x^{(J)}_0$ to $x^{(J)}_0 + L^{(J)}_{\mathfrak{R}}$ for $J=0,1,2,3$. The state $ \ket{\psi(x^{(0)},x^{(1)},x^{(2)},x^{(3)})}_S = \braket{x^{(0)},x^{(1)},x^{(2)},x^{(3)}|\Psi}$ is the relative state of the system $S$ conditioned on the value $(x^{(0)},x^{(1)},x^{(2)},x^{(3)})$ for the spacetime reference frame $\mathfrak{R}$ and it takes the form:
\begin{multline}\label{75statoS}
\ket{\psi(x^{(0)},x^{(1)} , x^{(2)}, x^{(3)})}_S \simeq \sum_{i=0}^{d^{(1)}_S - 1} \sum_{j=0}^{d^{(2)}_S -1} \sum_{k=0}^{d^{(3)}_S - 1} c_{ijk}  e^{-i \frac{1}{2m}\left( (p^{(1)}_i)^{2} + (p^{(2)}_j)^{2} + (p^{(3)}_k)^{2}\right) x^{(0)}} \\ \times e^{ip^{(1)}_i x^{(1)}} e^{ip^{(2)}_j x^{(2)}} e^{ip^{(3)}_k x^{(3)}} \ket{p^{(1)}_i , p^{(2)}_j, p^{(3)}_k }_S .
\end{multline}
For the relative state (\ref{75statoS}), through (\ref{75constraint1}) and (\ref{75constraint2}), we easily find:
\begin{equation}\label{7evfinale1}
i \frac{\partial}{\partial x^{(0)}} \ket{\psi(x^{(0)}, x^{(1)} , x^{(2)}, x^{(3)})}_S \simeq \hat{H}_S \ket{\psi(x^{(0)},x^{(1)} , x^{(2)}, x^{(3)})}_S
\end{equation}
and 
\begin{equation}\label{7evfinale2}
- i\frac{\partial}{\partial x^{(J)}} \ket{\psi(x^{(0)},x^{(1)} , x^{(2)}, x^{(3)})}_{S} = \hat{P}^{(J)}_S \ket{\psi(x^{(0)},x^{(1)} , x^{(2)}, x^{(3)})}_{S} 
\end{equation}
for $J=1,2,3$. Equations (\ref{75statoS}), (\ref{7evfinale1}) and (\ref{7evfinale2}) lead to ($\vec{x}=(x^{(1)} , x^{(2)}, x^{(3)})$): 

\begin{equation}\label{7evfinale3}
i \frac{\partial}{\partial x^{(0)}} \ket{\psi(x^{(0)}, \vec{x})}_S \simeq - \frac{1}{2m}\left(\frac{\partial^{2}}{\left(\partial x^{(1)}\right)^2} + \frac{\partial^{2}}{\left(\partial x^{(2)}\right)^2} + \frac{\partial^{2}}{\left(\partial x^{(3)}\right)^2} \right)\ket{\psi(x^{(0)},\vec{x})}_{S}
\end{equation}
which describes the dynamics of the particle in $S$ with respect to the coordinates of the $3+1$ dimensional quantum reference frame. 
We emphasize here that the formalism adopted allows us to write the equation (\ref{7evfinale3}) for $\ket{\psi(x^{(0)},x^{(1)} , x^{(2)}, x^{(3)})}_S$ because the values of time and space of the subspace $\mathfrak{R}$ enter as parameters in $S$ thanks to the entanglement present in the global state $\ket{\Psi}$.

\subsection{System $S$ as a Relativistic Particle}
The formalism adopted in the previous paragraph is particularly well suited in describing the behavior of a relativistic particle. Considering indeed the system $S$ to be a relativistic particle with no internal degree of freedom (namely with spin $0$), we have for the energy function ($c=1$):

\begin{equation}\label{7KGenergy}
\epsilon_{ijk} \simeq \pm \sqrt{(p^{(1)}_i)^2 + (p^{(2)}_j)^2 +(p^{(3)}_k)^2 + m^2} 
\end{equation}
which can be obtained from the energy constraint\footnote{The constraint (\ref{7cos}) derives from assuming the Hamiltonian of $S$ as $\hat{H}_S = \sqrt{|\vec{P}_{S}|^2 + m^2}$.} 

\begin{equation}\label{7cos}
\left( \left(\hat{H}^{(0)}_{\mathfrak{R}} \right)^2 - \left|\vec{P}_{S}\right|^2 -m^2 \right)\ket{\Psi}\simeq 0
\end{equation}
where $\vec{P}_{S}=\left(\hat{P}^{(1)}_{S},\hat{P}^{(2)}_{S},\hat{P}^{(3)}_{S} \right)$ and the kinetic energy term of $\mathfrak{R}$ has been neglected.
The relative state (\ref{75statoS}) of the system $S$ conditioned on the value $(x^{(0)},x^{(1)},x^{(2)},x^{(3)})$ of the spacetime reference frame $\mathfrak{R}$ reads now:

\begin{multline}\label{75statoS2}
\ket{\psi_{\pm}(x^{(0)},x^{(1)} , x^{(2)}, x^{(3)})}_S \simeq \sum_{i=0}^{d^{(1)}_S - 1} \sum_{j=0}^{d^{(2)}_S -1} \sum_{k=0}^{d^{(3)}_S - 1} c_{ijk}  e^{\mp i x^{(0)} \sqrt{(p^{(1)}_i)^2 + (p^{(2)}_j)^2 +(p^{(3)}_k)^2 + m^2} } \\ \times e^{ip^{(1)}_i x^{(1)}} e^{ip^{(2)}_j x^{(2)}} e^{ip^{(3)}_k x^{(3)}} \ket{p^{(1)}_i , p^{(2)}_j, p^{(3)}_k }_S 
\end{multline}
where we have considered together both the results for the energy function (\ref{7KGenergy}). For the relative state (\ref{75statoS2}) we find:
\begin{multline}\label{75.1}
\frac{\partial^2}{\partial (x^{(0)})^2}  \ket{\psi_{\pm}(x^{(0)},x^{(1)} , x^{(2)}, x^{(3)})}_S \simeq  \\ \simeq - \left( \left(\hat{P}^{(1)}_S\right)^2 + \left(\hat{P}^{(2)}_S\right)^2 + \left(\hat{P}^{(3)}_S\right)^2 + m^2 \right) \ket{\psi_{\pm}(x^{(0)},x^{(1)} , x^{(2)}, x^{(3)})}_S
\end{multline}
and
\begin{equation}\label{75.2}
\begin{split}
\frac{\partial^2}{\partial (x^{(J)})^2} \ket{\psi_{\pm}(x^{(0)},x^{(1)} , x^{(2)}, x^{(3)})}_S = - \left(\hat{P}^{(J)}_S\right)^2 \ket{\psi_{\pm}(x^{(0)},x^{(1)} , x^{(2)}, x^{(3)})}_S
\end{split}
\end{equation}
with $J=1,2,3$. Through equations (\ref{75.1}) and (\ref{75.2}) we easily obtain: 

\begin{equation}\label{7KG}
\left(\frac{\partial^2}{\partial (x^{(0)})^2} -  \frac{\partial^2}{\partial (x^{(1)})^2} -  \frac{\partial^2}{\partial (x^{(2)})^2} - \frac{\partial^2}{\partial (x^{(3)})^2} +m^2 \right) \ket{\psi_{\pm}(x^{(0)}, \vec{x})}_S \simeq 0
\end{equation}
which describes the dynamics of the particle in $S$ with respect to the coordinates of the $3+1$ dimensional quantum reference frame. Equation (\ref{7KG}) has the form of the Klein-Gordon equation but differs from it being the derivatives applied to the state $\ket{\psi_{\pm}(x^{(0)},x^{(1)} , x^{(2)}, x^{(3)})}_S$ and not to the wave function. Also in this case, this is possible since the time and space values of $\mathfrak{R}$ enter as parameters in the state of $S$ through the entanglement present in the global state $\ket{\Psi}$.

A similar result can be obtained considering the system $S$ as a relativistic particle with spin $1/2$. In this case the global state of the Universe can be written:

\begin{equation}
\ket{\Psi} = \sum_{\sigma = 0}^{3} \sum_{n=0}^{d^{(0)}_{\mathfrak{R}} - 1} \sum_{i=0}^{d^{(1)}_S - 1} \sum_{j=0}^{d^{(2)}_S - 1} \sum_{k=0}^{d^{(3)}_S - 1} c^{(\sigma)}_{nijk} \ket{p^{(0)}_n , -p^{(1)}_i , -p^{(2)}_j, -p^{(3)}_k}_{\mathfrak{R}} \otimes \ket{p^{(1)}_i , p^{(2)}_j, p^{(3)}_k,\sigma}_S
\end{equation} 
where we have introduced the spin degree of freedom within the subspace $S$ in accordance to \cite{dirac,librodirac} and where the value of $p^{(0)}_n$ is constrained through 

\begin{equation}\label{7dirac1}
\left( \hat{H}^{(0)}_{\mathfrak{R}} + \vec{\alpha} \cdot \vec{P}_S + \beta m \right)\ket{\Psi} \simeq 0 .
\end{equation}
In equation (\ref{7dirac1}) we have written for the system $S$ the free Dirac Hamiltonian as $\hat{H}_S = \vec{\alpha}\cdot \vec{P}_S + \beta m$ \cite{dirac} and we have again neglected the kinetic energy term of $\mathfrak{R}$. The state $\ket{\psi(x^{(0)},x^{(1)} , x^{(2)}, x^{(3)})}_S = \braket{x^{(0)},x^{(1)} , x^{(2)}, x^{(3)}|\Psi}$ reads now:

\begin{multline}\label{7diracrel}
\ket{\psi (x^{(0)},x^{(1)} , x^{(2)}, x^{(3)})}_S = \sum_{\sigma = 0}^{3} \sum_{n=0}^{d^{(0)}_{\mathfrak{R}} - 1} \sum_{i=0}^{d^{(1)}_S - 1} \sum_{j=0}^{d^{(2)}_S -1} \sum_{k=0}^{d^{(3)}_S - 1} c^{(\sigma)}_{nijk} e^{- ip^{(0)}_n x^{(0)}} \\ \times e^{ip^{(1)}_i x^{(1)}} e^{ip^{(2)}_j x^{(2)}} e^{ip^{(3)}_k x^{(3)}} \ket{p^{(1)}_i , p^{(2)}_j, p^{(3)}_k,\sigma}_S.
\end{multline}
For the relative state (\ref{7diracrel}) still holds

\begin{equation}\label{7dirac2}
i \frac{\partial}{\partial x^{(0)}} \ket{\psi(x^{(0)},x^{(1)} , x^{(2)}, x^{(3)})}_S \simeq \hat{H}_S \ket{\psi(x^{(0)},x^{(1)} , x^{(2)}, x^{(3)})}_S
\end{equation}
and
\begin{equation}\label{7dirac3}
- i \frac{\partial}{\partial x^{(J)}} \ket{\psi(x^{(0)},x^{(1)} , x^{(2)}, x^{(3)})}_{S} = \hat{P}^{(J)}_S   \ket{\psi(x^{(0)},x^{(1)} , x^{(2)}, x^{(3)})}_{S} .
\end{equation}
So, starting from equations (\ref{7dirac2}) and (\ref{7dirac3}), writing $\vec{\alpha}=(\alpha^{(1)}, \alpha^{(2)}, \alpha^{(3)})$ and remembering that $\hat{H}_S = \vec{\alpha}\cdot \vec{P}_S + \beta m$, we obtain:

\begin{multline}\label{7mmm}
i \frac{\partial}{\partial x^{(0)}} \ket{\psi(x^{(0)},\vec{x})}_S \simeq  \left(-i \alpha^{(1)}\frac{\partial}{\partial x^{(1)}} -i \alpha^{(2)}\frac{\partial}{\partial x^{(2)}} -i \alpha^{(3)}\frac{\partial}{\partial x^{(3)}} + \beta m   \right)  \ket{\psi(x^{(0)},\vec{x})}_S
\end{multline}
which has the form of the Dirac equation and again describes the dynamics of the particle in $S$ with respect to the coordinates of the $3+1$ dimensional quantum reference frame. All the considerations made for equation (\ref{7KG}) still apply in this case. Namely (\ref{7mmm}) differs from the Dirac equation being the derivatives applied to the state $\ket{\psi_{\pm}(x^{(0)},x^{(1)} , x^{(2)}, x^{(3)})}_S$ and not to the wave function. Also in this case, this is possible since the time and space values of $\mathfrak{R}$ enter as parameters in the state of $S$ through the entanglement present in the global state $\ket{\Psi}$. 

Clearly, in order to give a complete relativistic generalization of the model, in addition to this discussion, we need to consider relativistic reference frames and a protocol that allows to change the point of view between different observers in different reference frames (so that dilation of times and contraction of lengths can be derived), but this is beyond the scope of the present work.


\section{Conclusions}

The PaW mechanism was originally introduced in order to describe the emergence of time from entanglement.
In this Chapter (following \cite{nostro3}) we first extended the PaW mechanism at the spatial degree of freedom and then we provide a description of a model of non-relativistic quantum spacetime. In doing this we started focusing on space and we showed that, in a closed quantum system satisfying a global constraint on total momentum (and therefore with the absolute position totally indeterminate), a well-defined relative position emerge between the subsystems $S$ and $R$, where the latter is taken as quantum spatial reference frame. In the spaces $R$ and $S$, generalizing the approach outlined in Chapter 3 \cite{pegg}, we considered non-degenerate, discrete spectra for the momentum operators and we introduce POVMs in describing the spatial degrees of freedom. In this way we recovered continuous values of space in $S$ and $R$ also for a discrete momentum spectra (the case of momentum with continuous spectrum was then also treated in Section 6.3.5). Finally we introduced in the Universe an additional subsystem $C$ acting as a clock and we considered the Universe satisfying a double constraint: both on total momentum and total energy. We showed how this framework can be implemented without contradiction in the simple case of one spatial degree of freedom (considering also the case of multiple time measurements) and in the \lq\lq more physical\rq\rq case of three spatial degrees of freedom thus providing a $3+1$ dimensional quantum spacetime emerging from entanglement within a closed quantum system.



\section{Appendices}

\subsection*{6.7.A \: Proof of Equation (\ref{7conditionalprobabilitydiscreta})}
We start considering the global state written as

\begin{equation}
\ket{\Psi} =  \frac{\sqrt{d_R}}{D_R} \sum_{j=0}^{D_R-1} \ket{x_j}_R\otimes\ket{\phi(x_j)}_S 
\end{equation}
where $\ket{\phi(x_j)}_S = \sum_{k=0}^{d_S-1} c_k e^{-i p_k x_j} \ket{p_k}_S$. We can now calculate the conditional probability as follows

\begin{equation}
\begin{split}
& P(y_l \: on \: S \:|\: x_j \: on \: R) =\frac{d_S}{D_S} \frac{\braket{\Psi|x_j}\bra{x_j}\otimes\ket{y_l}\braket{y_l|\Psi}}{\braket{\Psi|x_j}\braket{x_j|\Psi}} = \\&
= \frac{d_S}{D_S} \frac{\frac{d_R}{D^{2}_R} \sum_{n}\sum_{m} \braket{x_n|x_j}\braket{x_j|x_m} \braket{\phi(x_n)|y_l} \braket{y_l|\phi(x_m)}}{\frac{d_R}{D^{2}_R} \sum_{n'}\sum_{m'} \braket{x_{n'}|x_j}\braket{x_j|x_{m'}} \braket{\phi(x_{n'})|\phi(x_{m'})}}= \\&
= \frac{d_S}{D_S} \frac{\sum_{n}\sum_{m} \braket{x_n|x_j}\braket{x_j|x_m} \sum_{k}\sum_{a}c^{*}_kc_a e^{ip_k x_n} e^{-i p_a x_m} \braket{p_k|y_l} \braket{y_l|p_a}}{\sum_{n'}\sum_{m'} \braket{x_{n'}|x_j}\braket{x_j|x_{m'}} \sum_{b}\sum_{k'}c^{*}_b c_{k'} e^{ip_b x_{n'}} e^{-i p_{k'}x_{m'}} \braket{p_b|p_{k'}}}=  \\&
=\frac{1}{D_S} \frac{\sum_{n}\sum_{m}  \sum_{i}\sum_{g}e^{- p_i(x_j - x_n)} e^{ip_g(x_j - x_m)}      \sum_{k}\sum_{a}c^{*}_kc_a e^{ip_k x_n} e^{-i p_a x_m} e^{-i y_l p_k} e^{i y_l p_a}}{\sum_{n'}\sum_{m'}    \sum_{i'}\sum_{g'}e^{- p_{i'}(x_j - x_{n'})} e^{ip_{g'}(x_j - x_{m'})}      \sum_{k'} |c_{k'}|^2 e^{ip_{k'} (x_{n'} -x_{m'})}     }
\end{split}
\end{equation}
We use now the fact that (see Appendix 6.7.C for the proof):

\begin{equation}\label{7delta2}
\sum_{j=0}^{D_R -1} e^{-i x_j(p_k - p_n)} = D_R\delta_{p_k,p_n}
\end{equation}
and so we obtain

\begin{equation}
\begin{split}
P(y_l \: on \: S \:|\: x_j \: on \: R) &= \frac{1}{D_S} \frac{\sum_{k}\sum_{a} c^{*}_k c_a e^{i(p_a - p_k)(y_l-x_j)}}{\sum_{k'} |c_{k'}|^2} = \\& = \frac{d_S}{D_S}\left| \braket{y_l|\phi(x_j)} \right|^2 = \frac{1}{D_S} \left| \sum_{k=0}^{d_S-1} c_k e^{i p_k(y_l-x_j)} \right|^2
\end{split}
\end{equation}
where we have considered that $\sum_{k'} |c_{k'}|^2 = 1$.

\subsection*{6.7.B \:Proof of Equation (\ref{7conditionalprobability})}
We start considering the global state written as $\ket{\Psi} = \frac{1}{L_R} \int_{x_0}^{x_0 + L_R} d x   \ket{x}_R \otimes \ket{\phi(x)}_S $
where $\ket{\phi(x)}_S = \sum_{k=0}^{d_S-1} c_k e^{-i p_k x} \ket{p_k}_S$. We can now calculate the conditional probability density as follows (all the integrals are evaluated between $x_0$ and $x_0+L_R$):

\begin{equation}
\begin{split}
& P(y \: on \: S \:|\: x \: on \: R)  = \frac{1}{L_S} \frac{\braket{\Psi|x}\bra{x}\otimes\ket{y}\braket{y|\Psi}}{\braket{\Psi|x}\braket{x|\Psi}} = \\ \\&
= \frac{1}{L_S} \frac{\frac{1}{L^{2}_R} \int dx' \int dx^{''} \braket{x'|x}\braket{x|x^{''}} \braket{\phi(x')|y}\braket{y|\phi(x^{''})}}{\frac{1}{L^{2}_R} \int dx' \int dx^{''} \braket{x'|x}\braket{x|x^{''}} \braket{\phi(x')|\phi(x^{''})}} = \\ \\&
= \frac{1}{L_S} \frac{  \int dx' \int dx^{''} \sum_{m}\sum_{l}e^{-ip_m(x-x')}e^{ip_l(x-x^{''})} \sum_{k}\sum_{n} c^{*}_k c_n e^{ip_k x' }e^{-ip_n x^{''}} \braket{p_k|y}\braket{y|p_n}}{ \int dx' \int dx^{''} \sum_{m'}\sum_{l'}e^{-ip_{m'} (x-x')}e^{ip_{l'}(x-x^{''})} \sum_{k'}\sum_{n'} c^{*}_{k'} c_{n'} e^{ip_{k'} x' }e^{-ip_{n'} x^{''}}\braket{p_{k'}|p_{n'}}} = \\ \\&
= \frac{1}{L_S} \frac{  \int dx' \int dx^{''} \sum_{m}\sum_{l}e^{-ip_m(x-x')}e^{ip_l(x-x^{''})} \sum_{k}\sum_{n} c^{*}_k c_n e^{ip_k x' }e^{-ip_n x^{''}}  e^{-ip_k y} e^{ip_n y}          }{   \int dx' \int dx^{''} \sum_{m'}\sum_{l'}e^{-ip_{m'} (x-x')}e^{ip_{l'}(x-x^{''})} \sum_{k'} |c_{k'}|^2 e^{ip_{k'} (x' - x^{''})}} .
\end{split}
\end{equation}
We use now the fact that (see Appendix 6.7.C for the proof): $\int_{x_0}^{x_0 + L_R} d x' e^{-i x' (p_k - p_n)} = L_R \delta_{p_k,p_n}$,
thus we obtain:

\begin{equation}
\begin{split}
P(y \: on \: S \:|\: x \: on \: R) &= \frac{1}{L_S} \frac{\sum_{k}\sum_{n} c^{*}_k c_n e^{i(p_n - p_k)(y-x)}}{\sum_{k'} |c_{k'}|^2} =  \\& = \frac{1}{L_S}\left| \braket{y|\phi(x)} \right|^2 = \frac{1}{L_S} \left| \sum_{k=0}^{d_S-1} c_k e^{i p_k(y-x)} \right|^2
\end{split}
\end{equation}
where again we have considered that $\sum_{k'} |c_{k'}|^2 = 1$.

\subsection*{6.7.C \: Proof of Equations (\ref{7delta}) and (\ref{7delta2})}
We first approach the case with discrete space values (namely we first demonstrate equation (\ref{7delta2})). We start considering the global state $\ket{\Psi}$ written as
\begin{equation}\label{7eee}
\ket{\Psi} = \sum_{k=0}^{d_S -1} c_k \ket{p=-p_k}_R\otimes\ket{p_k}_S
\end{equation}
and we apply in sequence the resolutions of the identity
\begin{equation}
\mathbb{1}_{R} = \frac{d_R}{D_R} \sum_{j=0}^{D_R -1} \ket{x_j}\bra{x_j}  \:\:\: and \:\:\: \mathbb{1}_{R} = \sum_{n=0}^{d_R -1} \ket{p_n}\bra{p_n} .
\end{equation}
We obtain
\begin{equation}
\begin{split}
\ket{\Psi} & =	\frac{d_R}{D_R} \sum_{j=0}^{D_R -1} \ket{x_{j}}\braket{x_{j}|\Psi}=  \\&
=	\frac{\sqrt{d_R}}{D_R} \sum_{j=0}^{D_R -1} \ket{x_{j}}_R\otimes \sum_{k=0}^{d_S -1} c_ke^{-ip_k x_{j}}\ket{p_k}_S = \\&
= \sum_{n=0}^{d_R -1} \ket{p_n}\bra{p_n}\frac{\sqrt{d_R}}{D_R} \sum_{j=0}^{D_R -1} \ket{x_{j}}_R\otimes \sum_{k=0}^{d_S -1} c_ke^{-ip_k x_{j}}\ket{p_k}_S = \\&
= \sum_{n=0}^{d_R -1}\sum_{k=0}^{d_S -1}c_k \frac{1}{D_R} \sum_{j=0}^{D_R -1}   e^{-ix_j (p_n + p_k)} \ket{p_n}_R \otimes \ket{p_k}_S
\end{split}
\end{equation}
from which we have
\begin{equation}
\sum_{j=0}^{D_R -1} e^{-i x_j(p_k + p_n)} = D_R\delta_{p_n,-p_k} .
\end{equation}

Let us now consider the limiting case $z \longrightarrow \infty$, i.e. we assume the continuous representation of the coordinate $x$ which can now take any real value from $x_0$ to $x_0 + L_R$. We start again from the global state written as (\ref{7eee}) and we apply in sequence the resolutions of the identity
\begin{equation}
\mathbb{1}_{R} = \frac{1}{L_R} \int_{x_0}^{x_0+L_R} dx \ket{x} \bra{x}  \:\:\: and \:\:\: \mathbb{1}_{R} = \sum_{n=0}^{d_R -1} \ket{p_n}\bra{p_n} .
\end{equation}
We have:
\begin{equation}
\begin{split}
\ket{\Psi} & =	\frac{1}{L_R} \int_{x_0}^{x_0 + L_R} dx \ket{x}\braket{x|\Psi}=  \\&
=\frac{1}{L_R} \int_{x_0}^{x_0 + L_R} dx \ket{x}_R\otimes \sum_{k=0}^{d_S -1} c_ke^{-ip_kx}\ket{p_k}_S = \\&
= \sum_{n=0}^{d_R -1} \ket{p_n}\bra{p_n} \frac{1}{L_R} \int_{x_0}^{x_0 + L_R} dx \ket{x}_R\otimes \sum_{k=0}^{d_S -1} c_k e^{-ip_kx}\ket{p_k}_S= \\&
= \sum_{n=0}^{d_R -1}\sum_{k=0}^{d_S -1}c_k \frac{1}{L_R} \int_{x_0}^{x_0 + L_R} dx e^{-ix(p_n + p_k)} \ket{p_n}_R \otimes \ket{p_k}_S
\end{split}
\end{equation}
from which we can easily see that $\int_{x_0}^{x_0 + L_R} dx e^{- i x(p_n + p_k)} = L_R \delta_{p_n, -p_k}$.

\subsection*{6.7.D \:Emergent Spacetime with $\hat{P}_C \ne 0$}
We discuss here our model of $1+1$ dimensional spacetime in the case of $\hat{P}_C \ne 0$. The contraints for the global state $\ket{\Psi} \in \mathcal{H}_C\otimes\mathcal{H}_R\otimes\mathcal{H}_S$ are now:
\begin{equation}\label{1A}
\hat{H}\ket{\Psi} = (\hat{H}_C + \hat{H}_R + \hat{H}_S )\ket{\Psi}=0
\end{equation}
and
\begin{equation}\label{2A}
\hat{P}\ket{\Psi} = (\hat{P}_C + \hat{P}_R + \hat{P}_S )\ket{\Psi}=0 .
\end{equation}
As we mentioned in the main text, in this case, there could be limitations in the allowed momenta to ensure that (\ref{1A}) and (\ref{2A}) are together satisfied and the global state $\ket{\Psi}$ can not be written here in the simple form (\ref{7statoglobalespaziotempo}). The limitations could arise from the fact that energies and momenta of $C$, $R$ and $S$ simultaneously have to sum to zero and the difficulty of the problem depends on the type of dispersion relations in the three subspaces. 
However, assuming $\ket{\Psi}$ satisfying the constraints, the discussion follows as in the case of $\hat{P}_C = 0$. Namely, we expand also here the global state on the basis $\left\{\ket{t_m}_C\right\}$ in $C$ thus obtaining
\begin{equation}\label{serveperGLMA}
\begin{split}
\ket{\Psi} = \frac{d_C}{D_C} \sum_{m=0}^{D_C -1} \ket{t_m} \braket{t_m|\Psi} =  \frac{\sqrt{d_C}}{D_C} \sum_{m=0}^{D_C -1} \ket{t_m}_C \otimes \ket{\phi(t_m)}_{R,S}
\end{split}
\end{equation}
where $\ket{\phi(t_m)}_{R,S} = \sqrt{d_C} \braket{t_m|\Psi}$ is state of the composite system $R+S$ at time $t_m$. For such a state, through (\ref{1A}) and the relative state definition, it is easy to find again the time evolution with respect to the clock $C$:
\begin{equation}\label{evoluzioneRSA}
\begin{split}
\ket{\phi(t_m)}_{R,S} 
= e^{-i (\hat{H}_R + \hat{H}_S)(t_m - t_0)}\ket{\phi(t_0)}_{R,S}
\end{split}
\end{equation} 
where $\ket{\phi(t_0)}_{R,S}= \sqrt{d_C} \braket{t_0|\Psi}$ is the state of $R+S$ conditioned on $t_0$ that is the value of the clock taken as initial time. Equation (\ref{evoluzioneRSA}) shows, as expected, the simultaneous evolution of $R$ and $S$ over time. In the limiting case $s \longrightarrow \infty$ where $t$ takes all the real values between $t_0$ and $t_0+T$ the global state can again be written 
\begin{equation}
\ket{\Psi} = \frac{1}{T} \int_{t_0}^{t_0 +T} dt \ket{t} \braket{t|\Psi} = \frac{1}{T} \int_{t_0}^{t_0 +T} dt \ket{t}_C \otimes \ket{\phi(t)}_{R,S}
\end{equation}
and defining the relative state of $R+S$ as $\ket{\phi(t)}_{R,S} = \braket{t|\Psi}$ we obtain:
\begin{equation}\label{evoluzioneRScA}
i \frac{\partial}{\partial t}\ket{\phi(t)}_{R,S} = \left(\hat{H}_R + \hat{H}_S\right)\ket{\phi(t)}_{R,S}
\end{equation}
that is the Schrödinger evolution for $R+S$ with respect to the clock time $t$, written in the usual differential form. 

We can therefore expand the state $\ket{\Psi}$ in the coordinates $\left\{\ket{x_j}_R\right\}$ in $R$, thus obtaining:
\begin{equation}
\begin{split}
\ket{\Psi} = \frac{d_R}{D_R} \sum_{j=0}^{D_R -1} \ket{x_j} \braket{x_j|\Psi} =  \frac{\sqrt{d_R}}{D_R} \sum_{j=0}^{D_R -1} \ket{x_j}_R \otimes \ket{\varphi(x_j)}_{C,S}
\end{split}
\end{equation}
where $\ket{\varphi(x_j)}_{C,S}=\sqrt{d_R}\braket{x_j|\Psi}$ is the relative state of $C+S$ conditioned to the value $x_j$ on the reference frame $R$. For the state $\ket{\varphi(x_j)}_{C,S}$ we find now:
\begin{equation}\label{mA}
\begin{split}
\ket{\varphi(x_j)}_{C,S} & = \sqrt{d_R}\braket{x_j|\Psi} = \sqrt{d_R} \bra{x_0}e^{i\hat{P}_R (x_j -x_0)}\ket{\Psi} =  \\&
=        \sqrt{d_R} \bra{x_0}e^{i (\hat{P} - \hat{P}_C - \hat{P}_S )(x_j -x_0)}\ket{\Psi}  = e^{- i ( \hat{P}_C + \hat{P}_S) (x_j -x_0)}\ket{\phi(x_0)}_{C,S}
\end{split}
\end{equation}
where the momentum of the clock $C$ appear in the equation since we have $\hat{P}_C \ne 0$. Also here we consider the limit $z \longrightarrow \infty$, where again $x$ can take all the real values between $x_0$ and $x_0 + L_R$. In this case the global state can be written

\begin{equation}\label{mmA}
\ket{\Psi} = \frac{1}{L_R} \int_{x_0}^{x_0 +L_R} dx \ket{x} \braket{x|\Psi} = \frac{1}{L_R} \int_{x_0}^{x_0 +L_R} dx \ket{x}_R \otimes \ket{\varphi(x)}_{C,S}
\end{equation}
and, defining the relative state of $C+S$ as $\ket{\varphi(x)}_{C,S} = \braket{x|\Psi}$, we obtain 

\begin{equation}
\begin{split}
\left(\hat{P}_C + \hat{P}_S\right) \ket{\varphi(x)}_{C,S} &= \bra{x}\left( \hat{P}_C + \hat{P}_S \right)\ket{\Psi}
= \bra{x}\left( \hat{P} - \hat{P}_R \right)\ket{\Psi}
\\&= -  \left( \sum_{k=0}^{d_R - 1} p_k e^{i p_k x} \bra{p_k} \right) \ket{\Psi}
= i \frac{\partial}{\partial x} \ket{\varphi(x)}_{C,S} .
\end{split}
\end{equation} 
Through this latter equation and (\ref{mA}) we can see again that now the generator of translations in the values $x$ for the state $\ket{\varphi(x)}_{C,S}$ is the operator $\hat{P}_C + \hat{P}_S$.

Finally we can expand the state $\ket{\Psi}$ simultaneously on the coordinates $\left\{\ket{x_j}_R\right\}$ in $R$ and on the time basis $\left\{\ket{t_m}_C\right\}$ in $C$. We have for the global state: $\ket{\Psi} = \frac{\sqrt{d_C}}{D_C} \frac{\sqrt{d_R}}{D_R} \sum_{m=0}^{D_C -1}\sum_{j=0}^{D_R -1}\ket{t_m}_C\otimes\ket{x_j}_R\otimes\ket{\psi(t_m,x_j)}_S$
where $\ket{\psi(t_m,x_j)}_S$ (obtained from $\ket{\psi(t_m,x_j)}_S=\sqrt{d_C} \sqrt{d_R}(\bra{t_m}\otimes\bra{x_j})\ket{\Psi}$) is the relative state of the system $S$ at time $t_m$ conditioned on the value $x_j$ for the reference frame $R$. Also here we can search the conditional probability of having a certain position $y_l$ in $S$ at time $t_m$ and knowing that the reference frame is in $x_j$, that is:

\begin{equation}\label{probfinalediscretaA}
P(y_l \: on\: S\:|\:x_j\:on\:R, \: t_m \: on \:C) = \frac{d_S}{D_S} |\braket{y_l|\psi(t_m,x_j)}|^2 
\end{equation}
that will depend on time $t_m$ and on the values $x_j$ and $y_l$ of $R$ and $S$ respectively.
Clearly we can extend these results also to the limiting cases $z,g,s \longrightarrow \infty$. Indeed we can write the global state $\ket{\Psi}$ as 
\begin{equation}
\ket{\Psi}= \frac{1}{T} \frac{1}{L_R} \int_{t_0}^{t_0 + T} dt \int_{x_0}^{x_0 + L_R} dx \ket{t}_C \otimes \ket{x}_R \otimes \ket{\psi(t,x)}_S
\end{equation}
where again $\ket{\psi(t,x)}_S = (\bra{t}\otimes\bra{x})\ket{\Psi}$ is the relative state of the system $S$ at time $t$ conditioned on the value $x$ for the reference frame $R$. The conditional probability density of having a certain position $y$ in $S$ at time $t$ and knowing that the reference frame is in $x$ is:
\begin{equation}\label{probfinaleA}
P(y \: on\: S\:|\:x\:on\:R,\: t \: on \:C) = \frac{1}{L_S}\left| \braket{y|\psi(t,x)} \right|^2 . 
\end{equation}
Also this probability density 
will depend on time $t$ and on the values of $x$ and $y$. 
We can not explicitly calculate the probabilities (\ref{probfinalediscretaA}) and (\ref{probfinaleA}) here since we can not write the state $\ket{\Psi}$ explicitly in the form (\ref{7statoglobalespaziotempo}). 

In summary we assume as a \lq\lq good clock\rq\rq a system with $\hat{P}_C = 0$, a framework that can be easily implemented using an internal degree of freedom in describing the clock. If we want to use $\hat{P}_C \ne 0$ we have to find a state $\ket{\Psi}$ that satisfies (\ref{1A}) and (\ref{2A}), undergoing the limitations that this choice imposes and the discussion then follows as shown in this Appendix. A different (less elegant) framework might be to take $(\hat{P}_R + \hat{P}_S)\ket{\Psi} = 0$ as a constraint for the theory even in the case of a clock with non-zero momentum, but we believe that this choice (even if functional) does not have a good physical justification. In this latter case indeed the Universe would not be in an eigenstate of the global momentum and thus the symmetry with respect to the temporal degree of freedom 
would be lost.

\subsection*{6.7.E \:Proof of Equation (\ref{7probfinalediscreta})}
We start here considering the global state $\ket{\Psi}$ written as in equation \ref{7statoglobalespaziotempo}, that is 
\begin{equation}
	\ket{\Psi} = \sum_{k=0}^{d_S -1} c_k \ket{E=-\epsilon_k}_C\otimes\ket{p=-p_k}_R\otimes\ket{p_k}_S ,
\end{equation}
where 
\begin{equation}
\epsilon_k= E^{(R)}(-p_k) + E^{(S)}(p_k)
\end{equation}
is the energy function related to the momenta $p_k$ of the subsystems $R$ and $S$. Now we write the conditional probability as:
\begin{equation}\label{775}
\begin{split}
& P(y_l \: on\: S\:|\:x_j\:on\:R,\: t_m \: on \:C) = \frac{d_S}{D_S} \frac{\braket{\Psi|t_m}\bra{t_m}\otimes\ket{x_j}\bra{x_j}\otimes\ket{y_l}\braket{y_l|\Psi}}{\braket{\Psi|t_m}\bra{t_m}\otimes\ket{x_j}\braket{x_j|\Psi}} = \\&
= \frac{d_S}{D_S} \frac{\sum_{k}\sum_{n} c_k c^{*}_n \braket{E=-\epsilon_n|t_m}\braket{t_m|E=-\epsilon_k}\braket{p=-p_n|x_j}\braket{x_j|p=-p_k}\braket{p_n|y_l}\braket{y_l|p_k}}{\sum_{k'}\sum_{n'}c_{k'} c^{*}_{n'} \braket{E=-\epsilon_{n'}|t_m}\braket{t_m|E=-\epsilon_{k'}}\braket{p=-p_{n'}|x_j}\braket{x_j|p=-p_{k'}}\braket{p_{n'}|p_{k'}}} = \\&
=\frac{1}{D_S}  \frac{\sum_{k}\sum_{n} c_k c^{*}_n e^{-it_m(\epsilon_k -\epsilon_n)} e^{i (p_k - p_n)(y_l-x_j)}}{\sum_{k'} |c_{k'}|^2} = \frac{1}{D_S} \left| \sum_{k=0}^{d_S -1} c_k e^{-i\epsilon_k t_m}e^{ip_k(y_l-x_j)} \right|^2
\end{split}
\end{equation}
where, in the last step, we have considered $\sum_{k'=0}^{d_S-1}|c_{k'}|^2 = 1$. From equation (\ref{775}) and considering $\ket{\psi(t_m,x_j)}_S= \sum_{k=0}^{d_S -1}c_k e^{-i\epsilon_k t_m} e^{-ip_k x_j}\ket{p_k}$, 
we can also see that the probability $P(y_l \: on\: S\:|\:x_j\:on\:R,\: t_m \: on \:C)$ can be written as $\frac{d_S}{D_S}|\braket{y_l|\psi(t_m,x_j)}|^2$.

\subsection*{6.7.F \: Proof of Equation (\ref{7probfinale})}
We start also here considering the global state written as in (\ref{7statoglobalespaziotempo}), that is
\begin{equation}
\ket{\Psi} = \sum_{k=0}^{d_S -1} c_k \ket{E=-\epsilon_k}_C\otimes\ket{p=-p_k}_R\otimes\ket{p_k}_S
\end{equation}
where $\epsilon_k= E^{(R)}(- p_k) + E^{(S)}(p_k)$ 
is the energy function related to the momenta $p_k$ of $R$ and $S$. Now we write the probability density as:

\begin{equation}\label{776}
\begin{split}
& P(y \: on\: S\:|\:x\:on\:R,\: t \: on \:C) = \frac{1}{L_S} \frac{\braket{\Psi|t}\bra{t}\otimes\ket{x}\bra{x}\otimes\ket{y}\braket{y|\Psi}}{\braket{\Psi|t}\bra{t}\otimes\ket{x}\braket{x|\Psi}} = \\&
=  \frac{1}{L_S} \frac{\sum_{k}\sum_{n} c_k c^{*}_n \braket{E=-\epsilon_n|t}\braket{t|E=-\epsilon_k}\braket{p=-p_n|x}\braket{x|p=-p_k}\braket{p_n|y}\braket{y|p_k}}{\sum_{m}\sum_{l}c_m c^{*}_l \braket{E=-\epsilon_l|t}\braket{t|E=-\epsilon_m}\braket{p=-p_l|x}\braket{x|p=-p_m}\braket{p_l|p_m}} = \\&
= \frac{1}{L_S} \frac{\sum_{k}\sum_{n} c_k c^{*}_n e^{-it(\epsilon_k -\epsilon_n)} e^{i (p_k - p_n)(y-x)}}{\sum_{m} |c_m|^2} =  \frac{1}{L_S}\left| \sum_{k=0}^{d_S -1} c_k e^{-i\epsilon_k t}e^{ip_k(y-x)} \right|^2
\end{split}
\end{equation}
where, in the last step, we have considered again $\sum_{m=0}^{d_S-1}|c_m|^2 = 1$. Also in this case, from equation (\ref{776}) and considering $\ket{\psi(t,x)}_S= \sum_{k=0}^{d_S -1}c_k e^{-i\epsilon_k} e^{-ip_k}\ket{p_k}$, we can see that $P(y \: on\: S\:|\:x\:on\:R,\: t \: on \:C) =  \frac{1}{L_S}\left| \braket{y|\psi(t,x)} \right|^2$.

\subsection*{6.7.G \: On the Emergence of Spatial Rotations}
We consider here a short digression regarding the emergence of spatial rotations within our quantum Universe. We start from the formalism introduced in Section 6.3.5 and we work with a generalization of such a framework to the case of $3+1$ dimensional spacetime. We therefore assume continuous spectra for the momenta and continuous values for the coordinates in $R$ and $S$ with orthogonal position states. We need indeed Hermitian $\vec{X}$, $\vec{P}_R$ and $\vec{Y}$, $\vec{P}_S$ (within $R$ and $S$) with continuous spectra in order to obtain well-defined angular momentum operators $\vec{L}_R$ and $\vec{L}_S$. 

We consider therefore the global state of the Universe $\ket{\Psi}$ satisfying the constraints on total energy and total momentum, namely we assume: 
\begin{equation}
	\hat{H}\ket{\Psi} = (\hat{H}_C + \hat{H}_R +\hat{H}_S)\ket{\Psi}=0 .
\end{equation}
and 
\begin{equation}
\vec{P}\ket{\Psi} = (\vec{P}_R + \vec{P}_S)\ket{\Psi}=0
\end{equation}
with $(\hat{P}^{(1)}_R + \hat{P}^{(1)}_S)\ket{\Psi} = (\hat{P}^{(2)}_R + \hat{P}^{(2)}_S)\ket{\Psi} = (\hat{P}^{(3)}_R + \hat{P}^{(3)}_S)\ket{\Psi}= 0$.
Also here we have chosen $\vec{P}_C = 0$ (with $\hat{P}^{(1)}_C = \hat{P}^{(2)}_C=\hat{P}^{(3)}_C=0$). For the state $\ket{\Psi}$ we now add an additional constraint: $\vec{L}\ket{\Psi} = \left( \vec{L}_R + \vec{L}_S \right)\ket{\Psi}=0$, 
that we rewrite as
\begin{equation}\label{74constraintL2}
	\begin{split}
		&   (\hat{L}^{(1)}_R + \hat{L}^{(1)}_S)\ket{\Psi}=0
		\\& (\hat{L}^{(2)}_R + \hat{L}^{(2)}_S)\ket{\Psi}=0
		\\& (\hat{L}^{(3)}_R + \hat{L}^{(3)}_S)\ket{\Psi}=0 
	\end{split}
\end{equation}
where $\vec{L}_R = (\hat{L}^{(1)}_R, \hat{L}^{(2)}_R, \hat{L}^{(3)}_R)$ and $\vec{L}_S = (\hat{L}^{(1)}_S, \hat{L}^{(2)}_S, \hat{L}^{(3)}_S)$ are the angular momentum operators for the systems $R$ and $S$. Our choice of having $\hat{P}^{(1)}_C = \hat{P}^{(2)}_C=\hat{P}^{(3)}_C=0$ leads to $\vec{L}_C = 0$ (with $\hat{L}^{(1)}_C = \hat{L}^{(2)}_C = \hat{L}^{(3)}_C = 0$) that does not enter in (\ref{74constraintL2}). 

Having found (if it exists!) the global state $\ket{\Psi}$ that simultaneously satisfies the constraints (\ref{74constraintL2}) on the angular momentum and both on total energy and total momentum, we can expand it into the coordinate basis in $R$, thus obtaining
\begin{multline}
	\ket{\Psi} = \int dx^{(1)} \int dx^{(2)} \int dx^{(3)} \ket{x^{(1)} , x^{(2)}, x^{(3)}} \braket{x^{(1)} , x^{(2)}, x^{(3)}|\Psi}=\\
	= \frac{1}{(2\pi)^{\frac{3}{2}}}\int dx^{(1)} \int dx^{(2)} \int dx^{(3)} \ket{x^{(1)} , x^{(2)}, x^{(3)}}_R \otimes \ket{\varphi(x^{(1)} , x^{(2)}, x^{(3)})}_{C,S} .
\end{multline}
which differs from the (\ref{787}) for the fact that now the position states are orthogonal.
So, for the relative state $\ket{\varphi(x^{(1)}_f , x^{(2)}_f, x^{(3)}_f)}_{C,S} = (2\pi)^{ \frac{3}{2}} \braket{x^{(1)}_f , x^{(2)}_f, x^{(3)}_f|\Psi}$, we have:
\begin{equation}
	\begin{split}
		\ket{\varphi(x^{(1)}_f , x^{(2)}_f, x^{(3)}_f)}_{C,S} &= (2\pi)^{ \frac{3}{2}} \bra{x^{(1)}_0, x^{(2)}_0, x^{(3)}_0}e^{i\vec{\theta}\cdot\vec{L}_R}\ket{\Psi}  = \\&
		= (2\pi)^{ \frac{3}{2}} \bra{x^{(1)}_0, x^{(2)}_0, x^{(3)}_0} e^{i (\theta^{(1)}\hat{L}^{(1)}_R + \theta^{(2)}\hat{L}^{(2)}_R + \theta^{(3)}\hat{L}^{(3)}_R)}\ket{\Psi} = \\&
		= e^{-i(\theta^{(1)}\hat{L}^{(1)}_S + \theta^{(2)}\hat{L}^{(2)}_S + \theta^{(3)}\hat{L}^{(3)}_S)} (2\pi)^{ \frac{3}{2}} \braket{x^{(1)}_0, x^{(2)}_0, x^{(3)}_0|\Psi} = \\&
		= e^{-i\vec{\theta}\cdot\vec{L}_S}\ket{\varphi(x^{(1)}_0 , x^{(2)}_0, x^{(3)}_0)}_{C,S}
	\end{split}
\end{equation}
that can be rewritten in the more compact form
\begin{equation}\label{7ultimaL}
	\ket{\varphi(\vec{x}_f)}_{C,S} = e^{-i\vec{\theta}\cdot\vec{L}_S}\ket{\varphi(\vec{x}_0)}_{C,S}
\end{equation}
where $\ket{\varphi(\vec{x}_0)}_{C,S}$ is the initial state and $\ket{\varphi(\vec{x}_f)}_{C,S}$ the final state for a given angular vector $\vec{\theta}= (\theta^{(1)},\theta^{(2)},\theta^{(3)})$. In analogy to the case of the spatial translations, equation (\ref{7ultimaL}) shows how the operator $\vec{L}_S$ is responsible for rotations of the relative state $\ket{\varphi(\vec{x})}_{C,S}$ in $S$ with respect to the coordinates of the reference frame $R$. 

If therefore exists a global state $\ket{\Psi}$ simultaneously satisfying the three constraints (\ref{74constraint1}), (\ref{74constraint2}) and (\ref{74constraintL2}), for the case of continuous spectra of momenta and coordinates in $R$ and $S$ (with Hermitian position and momentum operators), this framework is able to describe the emergence of time and space, with translations and rotations, all arising from entanglement in a constrained quantum Universe.
	
\begin{quote}
\label{bianca2}

\end{quote}
	
	\chapter{Quantum Clocks in a Gravitational Field}
\label{gravity}

Time is the physical quantity that is measured by a clock. On the other hand, we can think a clock as a physical system that evolves over time. 
In this Chapter we investigate, through the Page and Wootters (PaW) framework \cite{pagewootters,wootters}, the time evolution of two quantum clocks $A$ and $B$ interacting with gravitational field produced by a spherical mass $M$. The model of clocks we use has been introduced in Chapter 4 (see also \cite{nostro}) where the time states were chosen to belongs to the complement of an Hamiltonian proposed by Pegg in \cite{pegg} and described in Chapter 3.

We start showing that free clocks (namely not perturbed by the gravitational field) evolve synchronously over time. Then we place clocks $A$ and $B$ at different distances from the center of the spherical mass $M$ (source of the field) and we find that, as time (read by a far-away observer) goes on, the time states of $A$ and $B$ suffer a different delay in the phases which translates into a different ticking rate for the clocks. When considering a Newtonian gravitational field, the time dilation effect we find is in agreement with the first order expansion of the gravitational time dilation as derived from the Schwarzschild metric.

Initially we do not assume relativistic corrections to the gravitational energy, but we only consider the interaction between the clock and the Newtonian gravitational field. Nevertheless, in calculating such interaction, we promote the masses of the clocks to operators using the mass-energy equivalence $m \rightarrow m + \hat{H}_{clock}/c^{2}$ \cite{entclockgravity,zych1} (see also \cite{zych2,zych3,zych4}). In this sense, we can say that our results originate from a non-relativistic quantum framework with the only exception of using the mass-energy equivalence. 
Finally, as a last point, we introduce the \textit{relativistic gravitational potential} \cite{gravpot1,gravpot2} within our framework and, with this choice, the agreement between our time dilation effect and the exact Schwarzschild solution is obtained.

As already mentioned, we perform our analysis using the PaW framework, where time is a quantum degree of freedom which belongs to an ancillary Hilbert space.
In describing the time evolution of a system $S$, PaW theory assume a global space $\mathcal{H}=\mathcal{H}_C \otimes \mathcal{H}_S$, where $\mathcal{H}_C$ is the Hilbert space assigned to time. The flow of time then emerges thanks to the entanglement between $S$ and the temporal degree of freedom. In this Chapter, the system $S$ consists of the two clocks $A$ and $B$, and their evolution is investigated with respect to the third clock $C$. This latter is placed, for convenience, at infinite distance from the mass $M$ (source of the gravitational field) and it will thus play the role of a far-away, not perturbed, reference clock. In Section 7.2 we consider the clocks $A$ and $B$ described by Pegg's Hermitian operators complement of Hamiltonians with bounded, discrete spectra and equally-spaced energy levels. In Section 7.3 we generalize to the case where the clocks $A$ and $B$ have continuous time values, namely described by Pegg's POVMs.

\section{General Framework}
As previously mentioned, we work adopting the PaW theory and so we assume the Universe divided into two subsystems $C$ and $S$ where the first acts as time reference for the second. The global Hamiltonian reads $\hat{H} = \hat{H}_C + \hat{H}_S$ and the global state of the Universe $\ket{\Psi} \in \mathcal{H}_C \otimes \mathcal{H}_S$ satisfies the constraint:
\begin{equation}\label{7wdw}
	\hat{H}\ket{\Psi} = \left(\hat{H}_C + \hat{H}_S\right)\ket{\Psi} = 0 .
\end{equation}
We do not give here a summary of PaW theory, 
rather we focus on the $C$ subspace. 

In the $C$ subspace we still adopt the framework of paragraph 4.3.2 in Chapter 4, considering the Hamiltonian in $C$ with a generic spectrum\footnote{Having an Hamiltonian with generic spectrum in $C$ allows us to choose any Hamiltonian for $S$ being sure that every energy state of $S$ is connected with an energy state of $C$ satisfying (\ref{7wdw}).}. 
We repeat such a framework for the last time. We assume $\hat{H}_C$ with point-like spectrum, non-degenerate eigenstates and having rational energy ratios. Namely, we take $d_{C}$ energy states $\ket{E_i}_C$ and $E_i$ energy levels with $i=0,1,2,...,d_C -1$ such that $\frac{E_i -E_0 }{E_1 - E_0} = \frac{A_i}{B_i}$,	
where $A_i$ and $B_i$ are integers with no common factors. Doing this we obtain ($\hslash=1$):
\begin{equation}\label{5ei}
E_i = E_0 + r_i \frac{2\pi}{T_C}
\end{equation}
where $T_C =\frac{2\pi r_1}{E_1}$, $r_i = r_1\frac{A_i}{B_i}$ for $i>1$ (with $r_0=0$) and $r_1$ equal to the lowest common multiple of the values of $B_i$. In this space we define the states 
\begin{equation}\label{5alphastateinf}
\ket{t}_C = \sum_{i=0}^{d_C-1} e^{- i E_i t}\ket{E_i}_C
\end{equation}
where $t$ can take any real value from $t_0$ to $t_0 + T_C$. These states satisfy the key property $\ket{t}_C = e^{- i\hat{H}_C(t - t_0) }\ket{t_0}_C$ and furthermore can be used for writing the resolution of the identity in the $C$ subspace:
\begin{equation}\label{5newresolution2}
\mathbb{1}_{C} = \frac{1}{T_C} \int_{t_0}^{t_0+T_C} d t \ket{t} \bra{t} .
\end{equation}
As we have seen several times, thanks to property (\ref{5newresolution2}) the clock in $C$ is represented by a Pegg's POVM generated by the infinitesimal operators $\frac{1}{T_C} \ket{t}\bra{t} dt$.

This framework for the subspace $C$ allow us to consider any generic Hamiltonian as Hamiltonian for the $C$ subspace. We recall here that, in the case of non-rational ratios of energy levels, (\ref{5newresolution2}) is not exact but, since any real number can be approximated with arbitrary precision by a ratio between two rational numbers, the residual terms and consequent small corrections can be arbitrarily reduced. Consequently we can use the PaW mechanism with any Hamiltonian $\hat{H}_S$ being sure that no energy state of $S$ is excluded from the dynamics, simply assuming $d_C\gg d_S$ and a sufficiently large $T_C$. Furthermore, considering continuous time values for the clock in $C$, allow us to recover a continuous flow of time in the subsystem $S$. 

In the following we will consider also the system $S$ consisting of two quantum clocks ($A$ and $B$) and we will examine their evolution with respect to the clock in the $C$ subspace which play the role of a far-away time reference.

\section{Clocks $A$ and $B$ with Discrete Time Values}
We assume here the subsystem $S$ consisting of two clocks, $A$ and $B$, with discrete time values and we study their evolution with respect to the clock in the $C$ subspace. In Section 7.2.1 we consider two free clocks (namely not perturbed by the gravitational field), we define operators $\hat{\tau}$ and $\hat{\theta}$ complement of $\hat{H}_A$ and $\hat{H}_B$ and we show, as expected, that the clocks evolve synchronously. In Section 7.2.2 we place clocks $A$ and $B$ within the gravitational potential, at distance $x+h$ and $x$ respectively from the center of a non-rotating, spherical mass $M$. In this new case operators $\hat{\tau}$ and $\hat{\theta}$ are complement of Hamiltonians $\hat{H'}_A$ and $\hat{H'}_B$ modified by the gravitational field and we will show how clock $B$ ticks at a lower rate than $A$. Finally we bring $A$ to an infinite distance from the mass in Section 7.2.3.

\subsection{Evolution of Free Clocks}
As mentioned, we consider here the system $S$ consisting of two clocks $A$ and $B$. The global space reads now $\mathcal{H} = \mathcal{H}_C \otimes \mathcal{H}_S = \mathcal{H}_C \otimes \mathcal{H}_A  \otimes \mathcal{H}_B$ where we assume $d_A = d_B = d$ and $d_C \gg d$. The global Hamiltonian reads:
\begin{equation}
\hat{H} = \hat{H}_C + \hat{H}_A + \hat{H}_B
\end{equation}  
where the clocks $A$ and $B$ are governed by Hamiltonians with bounded, discrete spectra and equally-spaced energy levels. Being the clocks $A$ and $B$ identical, we have: 
\begin{equation}
\hat{H}_A = \hat{H}_B = \sum_{k=0}^{d-1}  \frac{2\pi}{T} k \ket{k} \bra{k}
\end{equation}
where
\begin{equation}
\frac{2\pi}{T}k= E^{(A)}_k = E^{(B)}_k 
\end{equation} 
are the equally-spaced energy eigenvalues. We simultaneously introduce the operators $\hat{\tau} = \sum_{m=0}^{d-1} \tau_m \ket{\tau_m}\bra{\tau_m}$ and $\hat{\theta} = \sum_{l=0}^{d-1} \theta_l \ket{\theta_l}\bra{\theta_l}$ complement of $\hat{H}_A$ and $\hat{H}_B$ respectively, with 
\begin{equation}\label{nuovo1}
\ket{\tau_m}_A = \frac{1}{\sqrt{d}} \sum_{k=0}^{d-1} e^{-i \frac{2\pi}{T}k \tau_m} \ket{k}_A = \frac{1}{\sqrt{d}} \sum_{k=0}^{d-1} e^{-i \frac{2\pi}{d}k m} \ket{k}_A
\end{equation} 
and

\begin{equation}\label{nuovo2}
\ket{\theta_l}_B = \frac{1}{\sqrt{d}} \sum_{k=0}^{d-1} e^{-i \frac{2\pi}{T}k \theta_l} \ket{k}_B = \frac{1}{\sqrt{d}} \sum_{k=0}^{d-1} e^{-i \frac{2\pi}{d}k l} \ket{k}_B 
\end{equation} 
where we have defined: $\tau_m = m\frac{T}{d}$ and $\theta_l = l\frac{T}{d}$ ($m,l=0,1,2,...,d-1$). 
The Hermitian operator $\hat{\tau}$ is complement of the Hamilitonian $\hat{H}_A$ in the sense that $\hat{H}_A$ is generator of shifts in eigenvalues of $\hat{\tau}$ and, viceversa, $\hat{\tau}$ is the generator of energy shifts. The same holds for $\hat{\theta}$ with respect to $\hat{H}_B$. The quantity $T$ is the time taken by the clocks to return to their initial state, indeed we have $\ket{\tau_m + T}_A = \ket{\tau_m}_A$ and $\ket{\theta_l + T}_B = \ket{\theta_l}_B$. We want the clocks $A$ and $B$ to be independent, so we assume them in a product state in $S$. The global state $\ket{\Psi}$ satisfying the constraint (\ref{7wdw}) can be written:
\begin{equation}\label{statoglobaleD}
\begin{split}
\ket{\Psi} &= \frac{1}{T_C} \int_{0}^{T_C} dt \ket{t}_C \otimes \ket{\psi(t)}_S =
\\& = \frac{1}{T_C} \int_{0}^{T_C} dt \ket{t}_C \otimes \ket{\varphi(t)}_A\otimes \ket{\phi(t)}_B
\end{split}
\end{equation}
where $\ket{\psi(t)}_S=\braket{t|\Psi}= \ket{\varphi(t)}_A\otimes \ket{\phi(t)}_B$ is the relative state of $S=A+B$ at time $t$, namely the product state of $A$ and $B$ conditioned on having $t$ in $C$.

\begin{figure}[t!]
	\centering 
	\includegraphics [height=3.5cm]{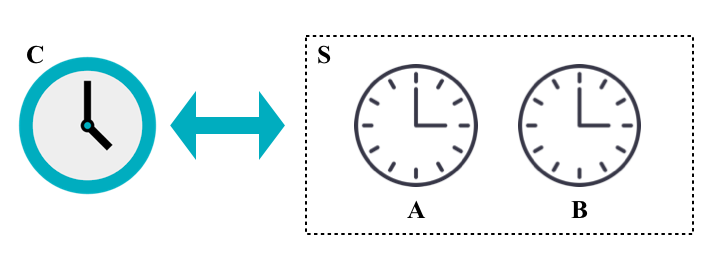} 
	\caption{The subsystem $S$ consists of two clocks, $A$ and $B$, and we study their evolution with respect to the clock $C$.} 
	\label{immagine1} 
\end{figure}

Through the state $\ket{\psi(t)}_S$ we can investigate the time evolution of the clocks $A$ and $B$. For the initial state of the clocks we choose the state 
\begin{equation}\label{5iniziale}
\ket{\psi(0)}_S = \ket{\varphi(0)}_A\otimes  \ket{\phi(0)}_B = \frac{1}{\sqrt{d}}\sum_{k=0}^{d-1} \ket{k}_A \otimes \frac{1}{\sqrt{d}}\sum_{n=0}^{d-1} \ket{n}_B
\end{equation}
namely we consider the clocks $A$ and $B$ at time $t=0$ to be in the time states $\ket{\tau_0=0}_A$ and $\ket{\theta_0=0}_B$. This implies that the state of $S$ at generic time $t$ reads:
\begin{equation}\label{5evoluzioneD}
\begin{split}
\ket{\psi(t)}_S &= \ket{\varphi(t)}_A\otimes \ket{\phi(t)}_B =
\\& = \frac{1}{d}\sum_{k=0}^{d-1} e^{-i \frac{2\pi}{T} k t} \ket{k}_A \otimes \sum_{n=0}^{d-1} e^{-i \frac{2\pi}{T} n t} \ket{n}_B
\end{split}
\end{equation}
indicating the clocks evolving with the Schrödinger evolution. As times $t$ goes on, the clocks \lq\lq click\rq\rq all the time states $\ket{\tau_m}_A$ and $\ket{\theta_l}_B$ until they reaches $t=T$, thus completing one cycle and returning to their initial state. As expected, the clocks evolve synchronously over time: considering indeed to be at time $t=m\frac{T}{d}$, we have
\begin{equation}
\begin{split}
\ket{\psi(t=m\frac{T}{d})}_S &= \ket{\varphi(t=m\frac{T}{d})}_A\otimes \ket{\phi(t=m\frac{T}{d})}_B = \\&
 = \frac{1}{d}\sum_{k=0}^{d-1} e^{-i \frac{2\pi}{d} k m} \ket{k}_A \otimes \sum_{n=0}^{d-1} e^{-i \frac{2\pi}{d} n m} \ket{n}_B
\end{split}
\end{equation}
where $A$ clicks the $m-th$ state $\ket{\tau_m}_A$ and also $B$ clicks its $m-th$ state $\ket{\theta_m}_B$.

\subsection{$A$ and $B$ interacting with the Gravitational Field}
\begin{figure}[t!] 
	\centering 
	\includegraphics [height=5cm]{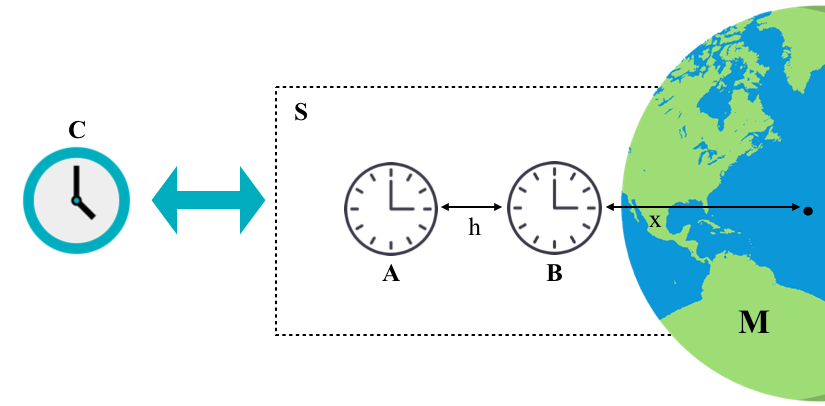} 
	\caption{The subsystem $S$ consists of two clocks $A$ and $B$ both placed within the gravitational potential. $B$ is at distance $x$ from the center of the mass $M$, while $A$ is at distance $x+h$.} 
	\label{nuova1} 
\end{figure}

We consider now the case in which clocks $A$ and $B$ are placed within a gravitational potential. In doing this we assume $B$ at a distance $x$ from the center of a spherical mass $M$ and $A$ placed at a distance $x+h$ (see Fig. \ref{nuova1}). In the interaction between the clocks and the gravitational potential we consider the Newtonian potential 
\begin{equation}
\phi(x) = - \frac{GM}{x} 
\end{equation}
where $G$ is the universal gravitational constant and  the coordinate $x$ is treated as a number. In calculating such interaction we promote the mass of the clocks to operator using the mass–energy equivalence: $m_A \rightarrow m_A +  \hat{H}_A/c^{2}$ and $m_B \rightarrow m_B  +  \hat{H}_B/c^{2}$. We assume the static mass of the clocks negligibly small when compared with the dynamical one and we focus only on the effects due to the internal degrees of freedom. However, considering the contributions of the static masses we would only obtain an unobservable global phase factor in the evolution of the clocks.

For the Hamiltonians $\hat{H}_A$ and $\hat{H}_B$ we have again $\hat{H}_A = \hat{H}_B = \sum_{k=0}^{d-1}\frac{2\pi}{T}k\ket{k}\bra{k}$, and so the global Hamiltonian now reads:
\begin{equation}
\begin{split}
\hat{H} &= \hat{H}_C + \hat{H}_A + \hat{H}_B + \frac{\hat{H}_A}{c^2}\phi(x+h) + \frac{\hat{H}_B}{c^2}\phi(x) =\\&=  \hat{H}_C + \hat{H'}_A + \hat{H'}_B
\end{split}
\end{equation}
where 
\begin{equation}\label{HAprimo}
\begin{split}
\hat{H'}_A = \hat{H}_A \left(1+ \frac{\phi(x + h)}{c^2}\right)=\hat{H}_A \left(1- \frac{GM}{(x+h)c^2}\right) 
\end{split}
\end{equation}
and
\begin{equation}\label{Hprimo}
\hat{H'}_B = \hat{H}_B \left(1+ \frac{\phi(x)}{c^2}\right)=\hat{H}_B \left(1- \frac{GM}{xc^2}\right) .
\end{equation}
The Hamiltonians $\hat{H'}_A$ and $\hat{H'}_B$ can be written:

\begin{equation}\label{HAprimo2}
\hat{H'}_A = \sum_{k=0}^{d-1} \frac{2\pi}{T^{\prime\prime}} k \ket{k}\bra{k}
\end{equation}
and
\begin{equation}\label{Hprimo2}
\hat{H'}_B = \sum_{k=0}^{d-1} \frac{2\pi}{T'} k \ket{k}\bra{k} ,
\end{equation}
where we have defined 
\begin{equation}
T^{\prime\prime}= \frac{T}{1- \frac{GM}{(x+h)c^2}} 
\end{equation}
and
\begin{equation}
T'= \frac{T}{1- \frac{GM}{xc^2}} .
\end{equation}

In the subspace $S$ we introduce again the operators $\hat{\tau}$ and $\hat{\theta}$ where now the first is complement of $ \hat{H'}_A$ and the second is complement of $\hat{H'}_B$. The time states of the clocks read here:
\begin{equation}\label{nuovo3A}
\ket{\tau_m}_A = \frac{1}{\sqrt{d}}\sum_{k=0}^{d-1} e^{-i \frac{2\pi}{T^{\prime\prime}} k \tau_m} \ket{k}_A 
\end{equation}
with
\begin{equation}\label{gggA}
\tau_m = m\frac{T^{\prime\prime}}{d} = m\frac{T}{d(1- \frac{GM}{(x+h)c^2})} 
\end{equation} 
and

\begin{equation}\label{nuovo3}
\ket{\theta_l}_B = \frac{1}{\sqrt{d}}\sum_{k=0}^{d-1} e^{-i \frac{2\pi}{T'} k \theta_l} \ket{k}_B 
\end{equation}
with
\begin{equation}\label{77ggg}
\theta_l = l\frac{T'}{d} = l\frac{T}{d(1- \frac{GM}{xc^2})} . 
\end{equation} 
We notice that the presence of the gravitational field does not change the form of the time states. Indeed, through (\ref{gggA}) and (\ref{77ggg}), we can rewrite the states of the clocks (\ref{nuovo3A}) and (\ref{nuovo3}) as

\begin{equation}
\ket{\tau_m}_A = \frac{1}{\sqrt{d}}\sum_{k=0}^{d-1} e^{-i \frac{2\pi}{d} k m} \ket{k}_A
\end{equation}
and
\begin{equation}
\ket{\theta_l}_B = \frac{1}{\sqrt{d}}\sum_{k=0}^{d-1} e^{-i \frac{2\pi}{d} k l} \ket{k}_B
\end{equation}
which are the same of (\ref{nuovo1}) and (\ref{nuovo2}). The global state satisfying the constraint (\ref{7wdw}) can again be written as in (\ref{statoglobaleD}) and we consider also here the clocks starting in the product state of the time states $\ket{\tau_0=0}_A$ and $\ket{\theta_0=0}_B$.

We can now look at the state $\ket{\psi(t)}_S$ investigating the time evolution of the two clocks $A$ and $B$. At generic time $t$ we have:

\begin{equation}\label{5evoluzioneD3}
\begin{split}
\ket{\psi(t)}_S &= \ket{\varphi(t)}_A\otimes \ket{\phi(t)}_B =\\&= \frac{1}{d}\sum_{k=0}^{d-1} e^{-i \frac{2\pi}{T} k t (1- \frac{GM}{(x+h)c^2})} \ket{k}_A \otimes \sum_{n=0}^{d-1} e^{-i \frac{2\pi}{T} n t(1- \frac{GM}{xc^2})} \ket{n}_B .
\end{split}
\end{equation}
Equation (\ref{5evoluzioneD3}) provides again the Schrödinger evolution for $A$ and $B$ where we can notice a different delay in the phases of the two clocks. This different delay indicates the two clocks ticking different time states at a given time $t$ as read by the clock $C$. To better see this point we can consider to be at time $t=m \frac{T}{d}$ as read by $C$. Equation (\ref{5evoluzioneD3}) becomes:

\begin{equation}
\begin{split}
\ket{\psi(t=m \frac{T}{d})}_S &= \ket{\varphi(t=m \frac{T}{d})}_A\otimes \ket{\phi(t=m \frac{T}{d})}_B =\\&= \frac{1}{d}\sum_{k=0}^{d-1} e^{-i \frac{2\pi}{d} k m(1- \frac{GM}{(x+h)c^2})} \ket{k}_A \otimes \sum_{n=0}^{d-1} e^{-i \frac{2\pi}{d} n m(1- \frac{GM}{xc^2})} \ket{n}_B 
\end{split}
\end{equation}
showing that no one, between $A$ and $B$, has clicked the $m-th$ state. Clock $B$ has indeed clicked a number of states $m'=m\left( 1- \frac{GM}{xc^2} \right)$ and $A$ has clicked a number of states $m^{\prime\prime}=m\left( 1- \frac{GM}{(x+h)c^2} \right)$ ($m',m^{\prime\prime} \in \mathbb{R}$). From this we easily derive: 

\begin{equation}\label{gen}
\begin{split}
m' 	=  m^{\prime\prime} \left( 1- \frac{GM}{xc^2} \right) \left( 1- \frac{GM}{(x+h)c^2} \right)^{-1} 
\end{split}
\end{equation}
which is in agreement with the time dilation between two clocks at a (radial) distance $h$ from each other as obtained in the first order expansion of the Schwarzschild metric. We notice that the \lq\lq agreement\rq\rq only regards the time dilation effect:
the $x$ and $x+h$ coordinates represent here simply the distance between the clocks and the center of the spherical mass $M$. We do not define the Schwarzschild radial coordinate, as we do not define coordinate time and proper time. We are indeed working in a purely quantum non-relativistic framework, with the only exception of having promoted the masses of the clocks to operators using the mass-energy equivalence.
Expanding the second term in the right-hand side of (\ref{gen}) and neglecting terms of the order $\sim \left(\frac{GM}{xc^2}\right)^2$, the equation reads
\begin{equation}\label{2clock1}
\begin{split}
m' & \simeq m^{\prime\prime} \left( 1- \frac{GM}{xc^2} \right) \left( 1 +  \frac{GM}{(x+h)c^2}\right)  
\\& \simeq  m^{\prime\prime}\left( 1 - \frac{GMh}{x(x+h)c^2} \right) 
\end{split}
\end{equation}
which, for $h\ll x$, becomes:
\begin{equation}
m'  \simeq m^{\prime\prime} \left( 1 - \frac{ah}{c^2} \right) 
\end{equation}
where we have defined the gravitational acceleration as $a=\frac{GM}{x^2}$.
This time dilation effect is in agreement with the relativistic result in the case, for example, of two clocks placed on the earth and separated by a vertical distance $h$, sufficiently small relative to the radius of the planet. We notice here that our results follows from having a finite $T$ and, if we consider clocks with unbounded time states, we would not find the same effect. Taking indeed the limit of infinite $T$ of our framework, the spacing between the energy levels goes to zero (namely, the energy spectrum turns to be continuous) and $d$ must approach infinity as well. The spacing between two neighbors time eigenvalues becomes however indeterminate and we would not be able to find the time dilation for the clock states.

\subsection{Bringing $A$ to an Infinite Distance}
As a last point, we consider the clock $B$ again placed within the gravitational potential at a distance $x$ from origin of the gravitational field, but we bring clock $A$ at an infinite distance from $M$ (see Fig. (\ref{immagine2})). In this limiting case the Hamiltonian of clock $A$ becomes: 
\begin{equation}
\hat{H'}_A \longrightarrow \hat{H}_A = \sum_{k=0}^{d-1}  \frac{2\pi}{T} k \ket{k} \bra{k} 
\end{equation}
and so the global Hamiltonian reads
\begin{equation}
\begin{split}
\hat{H} &= \hat{H}_C + \hat{H}_A + \hat{H}_B + \frac{\hat{H}_B}{c^2}\phi(x) =\\&=  \hat{H}_C + \hat{H}_A + \hat{H'}_B
\end{split}
\end{equation}
where $\hat{H'}_B$ is again defined as in (\ref{Hprimo}) and (\ref{Hprimo2}). 
In the subspace $S$ we introduce again the operators $\hat{\tau}$ and $\hat{\theta}$ where the first is complement of $ \hat{H}_A$ and the second is complement of $\hat{H'}_B$. The time states of clock $A$ are provided here by (\ref{nuovo1}) ($A$ is now not perturbed by the gravitational field), while time states and time values of the clock $B$ are again defined by (\ref{nuovo3}) and (\ref{77ggg}).
In this limiting case equation (\ref{5evoluzioneD3}) becomes:
\begin{figure}[t!] 
	\centering 
	\includegraphics [height=5cm]{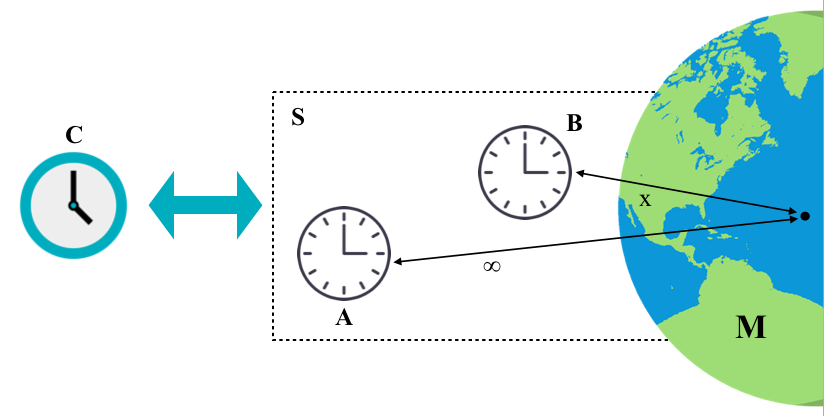} 
	\caption{The subsystem $S$ consists of two clocks $A$ and $B$. $B$ is placed within the gravitational potential at distance $x$ from the center of the spherical mass $M$, while $A$ is non-perturbed at an infinite distance from the mass.} 
	\label{immagine2} 
\end{figure}
\begin{equation}\label{5evoluzioneD2}
\begin{split}
\ket{\psi(t)}_S &= \ket{\varphi(t)}_A\otimes \ket{\phi(t)}_B =\\&= \frac{1}{d}\sum_{k=0}^{d-1} e^{-i \frac{2\pi}{T} k t} \ket{k}_A \otimes \sum_{n=0}^{d-1} e^{-i \frac{2\pi}{T} n t(1- \frac{GM}{xc^2})} \ket{n}_B .
\end{split}
\end{equation}
The delay in the phase is present now only in the state of clock $B$, proportional to $\left(1- \frac{GM}{xc^2}\right)$. As times $t$ goes on, the clocks $A$ and $B$ click again all the time states $\ket{\tau_m}_A$ and $\ket{\theta_l}_B$ but, during the time in which the clock $A$ reaches state $\ket{\tau_{m^{\prime\prime}}}_A$, clock $B$ does not reach the state $\ket{\theta_{m^{\prime\prime}}}_B$: while $A$ clicks the $m^{\prime\prime}-th$ state, the clock $B$ has clicked a number of states 
\begin{equation}\label{5key}
m'=m^{\prime\prime}\left(1- \frac{GM}{xc^2}\right) .
\end{equation}  
Equation (\ref{5key}) can also be obtained by taking the limit $h\longrightarrow \infty$ in equation (\ref{gen}) and it is again in agreement with the first order expansion of
the gravitational time dilation as derived from the Schwarzschild metric
where we have indeed: 

\begin{equation}
d \tau = \left(1 - \frac{2GM}{rc^2} \right)^{\frac{1}{2}}dt \simeq \left(1- \frac{GM}{rc^2}\right)dt ,
\end{equation}
with $d\tau$ the infinitesimal proper time measured by the clock placed within the gravitational field, $dt$ the time measured by a far-away observer and $r$ the Schwarzschild radial coordinate.

\section{Clocks $A$ and $B$ with Continuous Time Values}
We replicate here the discussion made in Section 7.2, but considering the clocks $A$ and $B$ represented by Pegg's POVM with continuous time values. As in the previous Section, we first study the time evolution of free clocks (namely not perturbed by the gravitational field), then we will introduce the gravitational field and we study two cases: we will start by considering $A$ and $B$ within the gravitational potential at distances $x+h$ and $x$ respectively from the origin of the field, then we bring $A$ at infinite distance from the large, spherical mass $M$.

\subsection{Evolution of Free Clocks}
The system $S$ consists here of two free clocks $A$ and $B$ with continuos time values. The global Hilbert space reads $\mathcal{H} = \mathcal{H}_C \otimes \mathcal{H}_S = \mathcal{H}_C \otimes \mathcal{H}_A  \otimes \mathcal{H}_B$ where we assume $d_A = d_B = d$ and $d_C \gg d$. The global Hamiltonian reads:

\begin{equation}
\hat{H} = \hat{H}_C + \hat{H}_A + \hat{H}_B 
\end{equation} 
and the clocks are governed again by Hamiltonians with bounded, discrete spectra and equally-spaced energy levels. Namely we have:

\begin{equation}
\hat{H}_A = \hat{H}_B = \sum_{k=0}^{d-1}  \frac{2\pi}{T} k \ket{k} \bra{k}
\end{equation}
where $\frac{2\pi}{T}k= E^{(A)}_k = E^{(B)}_k$ are the equally-spaced energy eigenvalues. In this paragraph we don not consider the Hermitian operators $\hat{\tau}$ and $\hat{\theta}$ complement of $\hat{H}_A$ and $\hat{H}_B$. Rather we introduce the time states: 

\begin{equation}\label{statitaufree3}
\ket{\tau_f}_A = \sum_{k=0}^{d-1} e^{-i \frac{2\pi}{T}k \tau_f} \ket{k}_A = \sum_{k=0}^{d-1} e^{-i 2\pi k f} \ket{k}_A
\end{equation} 
and

\begin{equation}\label{statithetafree3}
\ket{\theta_g}_B = \sum_{k=0}^{d-1} e^{-i \frac{2\pi}{T}k \theta_g} \ket{k}_B = \sum_{k=0}^{d-1} e^{-i 2\pi k g} \ket{k}_B 
\end{equation} 
where we have defined $\tau_f=fT$ and $\theta_g=gT$, with $f$ and $g$ taking any real values in the interval $\left[0,1\right]$. 
Assuming the clocks $A$ and $B$ in a product state in $S$, the global state $\ket{\Psi}$ satisfying the constraint (\ref{7wdw}) can again be written:
\begin{equation}\label{statoglobaleC}
\begin{split}
\ket{\Psi} &= \frac{1}{T_C} \int_{0}^{T_C} dt \ket{t}_C \otimes \ket{\psi(t)}_S =
\\& = \frac{1}{T_C} \int_{0}^{T_C} dt \ket{t}_C \otimes \ket{\varphi(t)}_A\otimes \ket{\phi(t)}_B
\end{split}
\end{equation}
where $\ket{\psi(t)}_S=\braket{t|\Psi}= \ket{\varphi(t)}_A\otimes \ket{\phi(t)}_B$ is the relative state of $S=A+B$ at time $t$, namely the product state of $A$ and $B$ conditioned on having $t$ in $C$.

Through the state $\ket{\psi(t)}_S$ we investigate again the time evolution of the clocks $A$ and $B$. For the initial state of the clocks we choose the state 
\begin{equation}\label{5iniziale2}
\ket{\psi(0)}_S = \ket{\varphi(0)}_A\otimes  \ket{\phi(0)}_B = \frac{1}{d}\sum_{k=0}^{d-1} \ket{k}_A \otimes \sum_{n=0}^{d-1} \ket{n}_B
\end{equation}
namely we consider the clocks $A$ and $B$ at time $t=0$ to be in the time states $\ket{\tau_{f=0}}_A$ and $\ket{\theta_{g=0}}_B$ (apart from a normalization constant). This implies that the state of $S$ at time $t=fT$ reads:
\begin{equation}\label{5evoluzioneC}
\begin{split}
\ket{\psi(t=fT)}_S &= \ket{\varphi(t=fT)}_A\otimes \ket{\phi(t=fT)}_B =
\\& = \frac{1}{d}\sum_{k=0}^{d-1} e^{-i 2\pi k f} \ket{k}_A \otimes \sum_{n=0}^{d-1} e^{-i 2\pi n f} \ket{n}_B
\end{split}
\end{equation}
indicating the clocks evolving together. Also in this case we can indeed see that $A$ and $B$ are clicking simultaneously the time states $\ket{\tau_f}_A$ and $\ket{\theta_{f}}_B$. 

\subsection{$A$ and $B$ interacting with the Gravitational Field}
We consider now the case of clocks $A$ and $B$ with continuous time values both interacting with the Newtonian gravitaional field. Clock $B$ is placed at a distance $x$ from the center of the mass $M$, while $A$ is at a distance $x+h$. We have again $\hat{H}_A = \hat{H}_B = \sum_{k=0}^{d-1}\frac{2\pi}{T}k\ket{k}\bra{k}$, and so the global Hamiltonian reads:
\begin{equation}
\begin{split}
\hat{H} &= \hat{H}_C + \hat{H}_A + \frac{\hat{H}_A}{c^2}\phi(x+h) + \hat{H}_B + \frac{\hat{H}_B}{c^2}\phi(x) =\\&=  \hat{H}_C + \hat{H'}_A + \hat{H'}_B
\end{split}
\end{equation}
where
\begin{equation}
\hat{H'}_A 
=\hat{H}_A \left(1- \frac{GM}{(x+h)c^2}\right) = \sum_{k=0}^{d-1} \frac{2\pi}{T^{\prime\prime}} k \ket{k}\bra{k}
\end{equation}
and
\begin{equation}\label{hBgrav2}
\hat{H'}_B = \hat{H}_B \left(1- \frac{GM}{xc^2}\right)= \sum_{k=0}^{d-1} \frac{2\pi}{T'} k \ket{k}\bra{k}
\end{equation}
with $T^{\prime\prime}=T/(1- \frac{GM}{(x+h)c^2})$ and $T'=T/(1- \frac{GM}{xc^2})$. 
As in the case of previous Section, we notice that the presence of the gravitational field does not change the form of the time states for the clocks. Indeed we have:
\begin{equation}\label{taugrav}
\ket{\tau_f}_A = \sum_{k=0}^{d-1} e^{-i\frac{2\pi}{T^{\prime\prime}}k \tau_f}\ket{k}_A = \sum_{k=0}^{d-1} e^{-i 2\pi k f}\ket{k}_A 
\end{equation}
and
\begin{equation}\label{tetagrav}
\ket{\theta_g}_B = \sum_{k=0}^{d-1} e^{-i\frac{2\pi}{T'}k \theta_g}\ket{k}_B = \sum_{k=0}^{d-1} e^{-i 2\pi k g}\ket{k}_B 
\end{equation}
where we now consider $\tau_f = fT^{\prime\prime}$ and $\theta_g = gT'$ with $f,g \in \left[0,1\right]$. Definitions (\ref{taugrav}) and (\ref{tetagrav}) for the time states are therefore unchanged from (\ref{statitaufree3}) and (\ref{statithetafree3}). 

Choosing the product state (\ref{5iniziale2}) as initial condition for the clocks, we can look at the evolution of $A$ and $B$. Assuming clock $C$ to be in $t=fT$, equation (\ref{5evoluzioneC}) becomes: 
\begin{equation}\label{5evoluzioneC2.2}
\begin{split}
\ket{\psi(t= fT)}_S &= \ket{\varphi(t=fT)}_A\otimes \ket{\phi(t=fT)}_B =\\&= \frac{1}{d}\sum_{k=0}^{d-1} e^{-i 2\pi k f(1- \frac{GM}{(x+h)c^2})} \ket{k}_A \otimes \sum_{n=0}^{d-1} e^{-i 2\pi n f(1- \frac{GM}{xc^2})} \ket{n}_B .
\end{split}
\end{equation}
Equation (\ref{5evoluzioneC2.2}) shows that, when the clock $C$ reads time $t=fT$, clock $A$ has not reached the state $\ket{\tau_f}_A$ and $B$ has not reached the state $\ket{\theta_{f}}_B$. Rather $A$ is clicking the state $\ket{\tau_{f^{\prime\prime}}}_A$ with $f^{\prime\prime}=f\left( 1- \frac{GM}{(x+h)c^2} \right)$ and $B$ is clicking the state $\ket{\theta_{f'}}_B$ with $f'=f\left( 1- \frac{GM}{xc^2} \right)$. This implies
\begin{equation}
f' = f^{\prime\prime} \left( 1- \frac{GM}{xc^2} \right) \left( 1- \frac{GM}{(x+h)c^2} \right)^{-1} 
\end{equation}
which is the same of equation (\ref{gen}) in this new case of clocks with continuous time values. As we did in Section 7.2.2, neglecting terms of the order $\sim \left(\frac{GM}{xc^2}\right)^2$, writing the gravitational acceleration $a=\frac{GM}{x^2}$ and for $h \ll x$, we can derive:
\begin{equation}
f' \simeq f^{\prime\prime} \left( 1 - \frac{ah}{c^2} \right) 
\end{equation}
where we can see $B$ ticking slower then $A$ in agreement with the relativistic result.

\subsection{Bringing $A$ to an Infinite Distance}
Bringing now clock $A$ at infinite distance from $M$, we have again $\hat{H'}_A \longrightarrow \hat{H}_A = \sum_{k=0}^{d-1}  \frac{2\pi}{T} k \ket{k} \bra{k}$ 
and so the global Hamiltonian reads:
\begin{equation}
\begin{split}
\hat{H} &= \hat{H}_C + \hat{H}_A + \hat{H}_B + \frac{\hat{H}_B}{c^2}\phi(x) =\\&=  \hat{H}_C + \hat{H}_A + \hat{H'}_B
\end{split}
\end{equation}
with $ \hat{H'}_B$ again defined as in (\ref{hBgrav2}). 
The time states of clocks $A$ and $B$ are provided by (\ref{statitaufree3}) and (\ref{tetagrav}) respectively, while for the time values we have $\tau_f = fT$ and $\theta_g=gT'$ with $f,g \in \left[0,1\right]$. We can so look at  the time evolution of the clocks $A$ and $B$. Assuming clock $C$ to be in $t=f^{\prime\prime}T$, we get: 

\begin{equation}\label{evoluzioneC22}
\begin{split}
\ket{\psi(t=f^{\prime\prime}T)}_S &= \ket{\varphi(t=f^{\prime\prime}T)}_A\otimes \ket{\phi(t=f^{\prime\prime}T)}_B =\\&= \frac{1}{d}\sum_{k=0}^{d-1} e^{-i 2\pi k f^{\prime\prime}} \ket{k}_A \otimes \sum_{n=0}^{d-1} e^{-i 2\pi n f^{\prime\prime} (1- \frac{GM}{xc^2})} \ket{n}_B .
\end{split}
\end{equation}
Equation (\ref{evoluzioneC22}) shows that, when the clock $A$ reaches the state $\ket{\tau_{f^{\prime\prime}}}_A$, $B$ has not reached the state $\ket{\theta_{f^{\prime\prime}}}_B$ but rather $B$ is in the state $\ket{\theta_{f'}}_B$ with 
\begin{equation}\label{s2}
f'=f^{\prime\prime}\left(1- \frac{GM}{xc^2}\right) .
\end{equation}
Again we found clock $B$ ticking slower in agreement with the first order expansion of the gravitational time dilation as derived from the Schwarzschild metric.

\section{Gravitational Redshift}
In this Section we calculate the gravitational redshift as produced in our framework. The present analysis applies to both cases discussed, namely clocks with discrete time values and clocks with continuous time values. We consider the scenario in which clocks $A$ and $B$ are both placed within the gravitational field at distance $x+h$ and $x$ respectively from the center of the spherical mass $M$. We are therefore working in the framework described in paragraphs 7.2.2 and 7.3.2.

For the Hamiltonians $\hat{H}_A$ and $\hat{H}_B$ we have: $\hat{H}_A = \hat{H}_B = \sum_{k=0}^{d-1}\frac{2\pi}{T}k\ket{k}\bra{k}$, with $E^{(A)}_k=E^{(B)}_k = \frac{2\pi}{T}k$ the energy eigenvalues. As already seen, such Hamiltonians are modified by the interaction with the gravitational field in the following way:

\begin{equation}
	\hat{H}_A \longrightarrow	\hat{H'}_A = \hat{H}_A \left(1- \frac{GM}{(x+h)c^2}\right) 
\end{equation}
and

\begin{equation}
\hat{H}_B \longrightarrow	\hat{H'}_B = \hat{H}_B \left(1- \frac{GM}{xc^2}\right) . 
\end{equation}
The two modified Hamiltonians $\hat{H'}_A$ and $\hat{H'}_B$ can be written:
\begin{equation}
	\hat{H'}_A = \sum_{k=0}^{d-1} \frac{2\pi}{T^{\prime\prime}} k \ket{k}\bra{k}
\end{equation}
and
\begin{equation}
	\hat{H'}_B = \sum_{k=0}^{d-1} \frac{2\pi}{T'} k \ket{k}\bra{k} ,
\end{equation}
where we have again defined 
\begin{equation}
	T^{\prime\prime}= \frac{T}{1- \frac{GM}{(x+h)c^2}} 
\end{equation}
and
\begin{equation}
	T'= \frac{T}{1- \frac{GM}{xc^2}} .
\end{equation} 
 We assume an observer in $A$ receiving a light signal emitted in $B$. We consider the frequency of the light signal as proportional to the spacing between two neighboring energy levels of the clocks, namely $1/T$ for a free clock. The observer $A$ can read the frequency $\nu_O$ coming from $B$ and compare it with its own spectrum, thus obtaining:
\begin{equation}
	\delta \nu = \nu_O - \nu = \frac{1}{T^{'}} - \frac{1}{ T^{\prime\prime}}
\end{equation}
that, for $h \ll x$, becomes
\begin{equation}
	\begin{split}
		\delta \nu & = \frac{1}{T} \left[  \left(1 - \frac{GM}{xc^2}\right) - \left(1 - \frac{GM}{(x+h)c^2}\right)\right]   
		\\&
		=\frac{1}{T}  \frac{GM}{c^2} \left( \frac{1}{x+h} - \frac{1}{x} \right)
		\simeq - \frac{1}{T} \frac{GMh}{x^2c^2}  . 
	\end{split}
\end{equation}
Writing now again the gravitational acceleration $a=\frac{GM}{x^2}$ and neglecting terms of the order $\sim \left(\frac{GM}{xc^2}\right)^2$, we obtain
\begin{equation}\label{frequenzaD}
	\frac{\delta \nu}{\nu} \simeq - \frac{ah}{c^2} 
\end{equation}
that is in agreement, at the first order of approximation, with the relativistic result \cite{misuragravita}. Equation (\ref{frequenzaD}) clearly holds even when considering the spacing between any two energy levels and not two neighbors.

\section{Relativistic Gravitational Potential}
The agreement between our results and the relativistic solution clearly holds up to the first order of approximation in the Taylor expansion of the Schwarzschild metric. 
This does not include, for example, the case of clocks placed close to the event horizon of a black hole.
Nevertheless, we can extend our analysis considering the expression of the relativistic gravitational potential within the global Hamiltonian. Indeed, in this case, the potential energy $V$ of a clock placed at a distance $x$ from the center of the spherical mass $M$ reads \cite{gravpot1,gravpot2}:
\begin{equation}
V = m_{clock}c^2 \left[\left(1 - \frac{2GM}{xc^2}\right)^{\frac{1}{2}} - 1  \right] .
\end{equation}
Therefore, considering the case of $A$ and $B$ in the gravitational field (at distances $x+h$ and $x$ respectively from the center of the spherical mass $M$) and promoting the mass of the clocks to operator using the mass-energy equivalence, we have the global Hamiltonian:
	\begin{equation}
	\begin{split}
	\hat{H} &= \hat{H}_C + \hat{H}_A + \hat{H}_{A}\left[\left(1 - \frac{2GM}{(x+h)c^2}\right)^{\frac{1}{2}} - 1  \right] + \hat{H}_B + \hat{H}_{B}\left[\left(1 - \frac{2GM}{xc^2}\right)^{\frac{1}{2}} - 1  \right] = \\&
	= \hat{H}_C +  \hat{H}_{A}\left(1 - \frac{2GM}{(x+h)c^2}\right)^{\frac{1}{2}} + \hat{H}_{B} \left(1 - \frac{2GM}{xc^2}\right)^{\frac{1}{2}} 
	=\hat{H}_C +  \hat{H'}_{A} +  \hat{H'}_{B}
	\end{split}
	\end{equation}
where now

\begin{equation}
\hat{H'}_{A} = \hat{H}_{A}\left(1 - \frac{2GM}{(x+h)c^2}\right)^{\frac{1}{2}} = \sum_{k=0}^{d-1} \frac{2\pi}{T^{\prime\prime}} k \ket{k}\bra{k}
\end{equation}
and

\begin{equation}
\hat{H'}_{B} = \hat{H}_{B}\left(1 - \frac{2GM}{xc^2}\right)^{\frac{1}{2}} = \sum_{k=0}^{d-1} \frac{2\pi}{T'} k \ket{k}\bra{k}
\end{equation}
with $T^{\prime\prime}= T \left(1- \frac{2GM}{(x+h)c^2}\right)^{ - \frac{1}{2}}$ and $T'= T \left(1- \frac{2GM}{xc^2}\right)^{ - \frac{1}{2}}$.
We can now investigate the time evolution of $A$ and $B$. At generic time $t$ (as read by the clock $C$) we have:

\begin{equation}\label{evrelpot}
\begin{split}
\ket{\psi(t)}_S &= \ket{\varphi(t)}_A\otimes \ket{\phi(t)}_B =\\& = \frac{1}{d}
\sum_{k=0}^{d-1} e^{-i \frac{2\pi}{T} k t (1- \frac{2GM}{(x+h)c^2})^{\frac{1}{2}}} \ket{k}_A \otimes \sum_{n=0}^{d-1} e^{-i \frac{2\pi}{T} n t(1- \frac{2GM}{xc^2})^{\frac{1}{2}}} \ket{n}_B .
\end{split}
\end{equation}
In the case of clocks with discrete time values equation (\ref{evrelpot}) implies that, when $A$ clicks its $m^{\prime\prime}-th$ time state, clock $B$ has clicked a number of states:

\begin{equation}
m' 	 =   m^{\prime\prime} \left( 1- \frac{2GM}{xc^2} \right)^{\frac{1}{2}} \left( 1- \frac{2GM}{(x+h)c^2} \right)^{- \frac{1}{2}} 
\end{equation}
which, in the limit $h \longrightarrow \infty$, becomes 
\begin{equation}
m'=m^{\prime\prime}\left( 1- \frac{2GM}{xc^2} \right)^{\frac{1}{2}} .
\end{equation}
Similarly, for the case of clocks with continuous time values equation (\ref{evrelpot}) implies that, when $A$ clicks the time state $\ket{\tau_{f^{\prime\prime}}}_A$, the clock $B$ is clicking the time state $\ket{\theta_{f'}}_B$ with:

\begin{equation}
f' 	 =   f^{\prime\prime} \left( 1- \frac{2GM}{xc^2} \right)^{\frac{1}{2}} \left( 1- \frac{2GM}{(x+h)c^2} \right)^{- \frac{1}{2}} 
\end{equation}
which, in the limit $h \longrightarrow \infty$, again becomes 
\begin{equation}
f'=f^{\prime\prime}\left( 1- \frac{2GM}{xc^2} \right)^{\frac{1}{2}} .
\end{equation} 
In conclusion, the use of the relativistic gravitational potential in the global Hamiltonian 
leads to a time dilation effect which is in agreement with the exact Schwarzschild solution. 
We notice that, taking the first order expansion of the relativistic gravitational potential, we recover the results of Sections 7.2 and 7.3.

\section{Conclusions}
In this Chapter we have studied behavior of quantum clocks when interacting with a gravitational field. We have considered two clocks, $A$ and $B$, placed within the field and we have looked at their evolution with respect to the clock $C$, taken as a far-away reference via the PaW mechanism. Our investigation has been performed in the case of $A$ and $B$ with discrete and continuous time values. 

In both cases we started considering clocks not perturbed by the field and we have shown that they evolve synchronously over time. Then we have placed the clocks in the gravitational field at different distances from the center of the large mass $M$, source of the field. In this scenario we have shown that, as time $t$ (read by the far-away clock $C$) goes on, the time states of $A$ and $B$ suffer a different delay in the phases which translates into a different ticking rate for the clocks. This time dilation effect is in agreement with the first order expansion of the gravitational time dilation as derived from the Schwarzschild metric (the emergence of the gravitational redshif is also treated). Finally, we have introduced the relativistic gravitational potential within the global Hamiltonian of our system and we have shown that, with this choice, the agreement with the exact Schwarzschild solution is obtained.


\section{Appendices}

\subsection*{7.6.A \: Conditional Probabilieties and Main Values of the Observables for Clocks with Discrete Time Values}
We investigate here the behavior of specific quantities, as conditional probabilities and mean values of the observables. The study of such quantities gives us insights into how the gravitational field affects the clocks, in our case described by Pegg's time observables. We first study the case of $A$ and $B$ as free clocks, then we compare the results with those obtained by considering $B$ within the gravitational field and $A$ placed far away, not perturbed by the field.
In this Appendix we work with cloks with discrete time values, thus starting from the formalism described in Section 7.2.1 and moving then to formalism of Section 7.2.3.  

\subsubsection*{$A$ and $B$ free clocks}
We recall here that we are working in the global space $\mathcal{H} = \mathcal{H}_C \otimes \mathcal{H}_S = \mathcal{H}_C \otimes \mathcal{H}_A  \otimes \mathcal{H}_B$, so the global state satisfying constraint (\ref{7wdw}) is: 

\begin{equation}
\begin{split}
\ket{\Psi} &= \frac{1}{T_C} \int_{0}^{T_C} dt \ket{t}_C \otimes \ket{\psi(t)}_S =
\\& = \frac{1}{T_C} \int_{0}^{T_C} dt \ket{t}_C \otimes \ket{\varphi(t)}_A\otimes \ket{\phi(t)}_B .
\end{split}
\end{equation}
We assume that the clock in the $C$ subspace reads time $t=\tau_m$ and we work with the state of the system $S$: $\ket{\psi(t=\tau_m)}_S=\braket{t=\tau_m|\Psi}= \ket{\varphi(t=\tau_{m})}_A\otimes \ket{\phi(t=\tau_m)}_B$. We consider two quantities, namely the conditional probability $P(\theta_l | \tau_m)$ of obtaining $\theta_l$ on $B$ conditioned to having found $\tau_m$ on $A$, and the mean value of $\langle \theta \rangle(\tau_m)$. We have for the probability $P(\theta_l | \tau_m)$: 

\begin{equation}
\begin{split}
P(\theta_l | \tau_m) 
= \frac{\braket{\psi(t=\tau_{m})|\tau_m}\bra{\tau_m} \otimes \ket{\theta_l}\braket{\theta_l|\psi(t=\tau_{m})}}{\braket{\psi(t=\tau_{m})|\tau_m}\braket{\tau_m|\psi(t=\tau_{m})}} 
\end{split}
\end{equation}
which leads to the result
\begin{equation}\label{prob}
P(\theta_l | \tau_m) = \frac{1}{d^2} \sum_{n=0}^{d-1}\sum_{k=0}^{d-1} e^{-i(\theta_l - \tau_m )\frac{2\pi}{T}(n - k)} .
\end{equation}
This probability is well defined, indeed it easy to verify that $\sum_{l=0}^{d-1} P(\theta_l | \tau_m) = 1$ for each given $\tau_{m}$. Furthermore, through probability (\ref{prob}), we can calculate the mean value $\langle \theta \rangle(\tau_m)$ as follows:
\begin{equation}\label{valmedio}
\langle \theta \rangle(\tau_m) = \sum_{l=0}^{d-1} \theta_l P(\theta_l | \tau_m) =
\frac{T}{d^3}\sum_{n=0}^{d-1}\sum_{k=0}^{d-1} f(n-k) e^{i \tau_m \frac{2\pi}{T}(n - k)}
\end{equation}
where $f(n-k)= \sum_{l=0}^{d-1} l e^{-i \frac{2\pi l }{d} (n-k)}$. 
In this particular case in which the two clock are identical and both non-perturbed we easily find
\begin{equation}\label{1}
P(\theta_m | \tau_m) = 1 
\end{equation}
and
\begin{equation}\label{2}
\langle \theta \rangle(\tau_m) = \tau_m = \theta_m .
\end{equation}
Equations (\ref{1}) and (\ref{2}) indicate what we had already seen in (\ref{5evoluzioneD}), namely that the two clocks, as expected, run together over time.


\subsubsection{$B$ interacting with the gravitational field}

We consider now the case of clock $B$ placed at a distance $x$ from the center of a large, spherical mass $M$ while $A$ is placed far away, at infinite distance from the spherical mass. The time states of the clocks are therefore given by equations (\ref{nuovo1}) and (\ref{nuovo3}) for the clocks $A$ and $B$ respectively.

We assume now again that the clock in the $C$ subspace reads time $t=\tau_m$ and we work with the state of the system $S$: $\ket{\psi(t=\tau_m)}_S=\braket{t=\tau_m|\Psi}= \ket{\varphi(t=\tau_{m})}_A\otimes \ket{\phi(t=\tau_m)}_B$. We look at the quantities (\ref{prob}) and (\ref{valmedio}) in this new case where $B$ is placed within the gravitatinal field. We have:
\begin{equation}\label{prob2}
P(\theta_l | \tau_m) = \frac{1}{d^2} \sum_{n=0}^{d-1}\sum_{k=0}^{d-1} e^{-i(\theta_l - \tau_m)\frac{2\pi}{T'}(n - k)} 
\end{equation}
and
\begin{equation}\label{valmedio2}
\begin{split}
\langle \theta \rangle(\tau_m) = \sum_{l=0}^{d-1} \theta_l P(\theta_l | \tau_m) 
= \frac{T'}{d^3}\sum_{n=0}^{d-1}\sum_{k=0}^{d-1} f(n-k) e^{i \tau_m \frac{2\pi}{T'}(n - k)}
\end{split}
\end{equation}
where $f(n-k)$ is again $f(n-k)= \sum_{l=0}^{d-1} l e^{-i \frac{2\pi l }{d} (n-k)}$ and, we recall, $T'=T/(1- \frac{GM}{xc^2})$. 
Through (\ref{prob2}) and (\ref{valmedio2}) we can again see the effect of the gravitational field in the clock $B$ with respect to clock $A$. Let us consider two simple examples: we will look at clocks $A$ and $B$ in the case $d=2$ and $d=3$.

In the case $d=2$ we have:
\begin{equation}
P(\theta_l|\tau_m) = \frac{1}{2} \left[  1 + \cos(\pi(l-m(1- \frac{GM}{xc^2})))   \right]
\end{equation}
that leads to $P(\theta_m|\tau_m) = \frac{1}{2}[1+\cos(\pi m \frac{GM}{xc^2})]$. So, assuming $\cos(z)\simeq 1 - \frac{z^2}{2}$, we find: $P(\theta_m|\tau_m) \simeq 1 - \frac{m^2 \pi^2}{4}\left( \frac{GM}{xc^2}\right)^2 $
which depends on $m$ (a larger $m$ implies a reduced probability) and on the intensity of the gravitational field.

Going instead to calculate $\langle \theta \rangle(\tau_m)$ we have:
	\begin{equation}
	\begin{split}
	\langle \theta \rangle(\tau_m) & = \frac{T'}{4} \left[ 1 - \cos(m\pi (1- \frac{GM}{xc^2}))\right] \\ \\& \Rightarrow \langle \theta \rangle(\tau_0) = 0 = \theta_0 \\& \Rightarrow \langle \theta \rangle(\tau_1) \simeq \frac{T'}{2}\left[ 1 - \frac{\pi^2}{4}\left( \frac{GM}{xc^2}\right)^2 \right] = \theta_1\left[ 1 - \frac{\pi^2}{4}\left( \frac{GM}{xc^2}\right)^2 \right] .
	\end{split}
	\end{equation}
	We can notice that, during the period of time in which the main value of $A$ moved from $\tau_0=0$ to $\tau_1$, the mean value $\langle \theta \rangle$ of the clock $B$ moved to $\theta_1\left[ 1 - \frac{\pi^2}{4}\left( \frac{GM}{xc^2}\right)^2 \right]$, that is it doesn't reach clicking the first eigenvalue $\theta_1$. So, also considering the main values, we find the clock $B$ delayed in ticking with respect to the clock $A$ by a factor that depends again on the intensity of the gravitational potential.
	
	The same conclusion can be reached considering slightly more complex clocks, i.e. with d=3. In this case we obtain:
	
	\begin{equation}
	P(\theta_l|\tau_m) = \frac{1}{9} \left[  3 + 4\cos(\frac{2\pi}{3}  (l-m(1- \frac{GM}{xc^2}))   ) +2\cos(\frac{4\pi}{3}   (l-m(1- \frac{GM}{xc^2}))    )  \right]
	\end{equation}
	that for $l=m$ becomes 
	\begin{equation}
	\begin{split}
	P(\theta_m|\tau_m) = \frac{1}{9}\left[3+4\cos(\frac{2\pi m}{3}\frac{GM}{xc^2})+2\cos(\frac{4\pi m}{3}\frac{GM}{xc^2})\right]. 
	\end{split}
	\end{equation}
	So, considering again the second order of approximation $\cos(z)\simeq 1 - \frac{z^2}{2}$, we have:
	
	\begin{equation}
	P(\theta_m|\tau_m) \simeq 1 - \frac{ 8\pi^2 m^2}{27}\left( \frac{GM}{xc^2}\right)^2 .
	\end{equation}
	In the same way we obtain for the mean value $\langle \theta \rangle(\tau_m)$:
	
	\begin{multline}\label{ultima}
	\langle \theta \rangle(\tau_m) = \frac{T'}{27} [ 9 - 6\cos(\frac{2\pi m}{3}(1- \frac{GM}{xc^2})) - 2\sqrt{3}\sin(\frac{2\pi m}{3}(1- \frac{GM}{xc^2})) + \\ - 3\cos(\frac{4\pi m}{3}(1- \frac{GM}{xc^2})) + \sqrt{3}\sin(\frac{4\pi m}{3}(1- \frac{GM}{xc^2})) ]  .
	\end{multline}
	Equation (\ref{ultima}) implies that:
	
	\begin{equation}
	\begin{split}
	& \Rightarrow \langle \theta \rangle(\tau_0) = 0 = \theta_0  \\& 
	\Rightarrow \langle \theta \rangle(\tau_1) \simeq  \frac{T'}{3}  \left[ 1- \frac{10\sqrt{3}}{27}\left( \frac{2\pi }{3}\frac{GM}{xc^2}\right)^3 \right] = \theta_1   \left[ 1- \frac{10\sqrt{3}}{27}\left( \frac{2\pi}{3}\frac{GM}{xc^2}\right)^3 \right]        \\&
	\Rightarrow \langle \theta \rangle(\tau_2) \simeq \frac{2T'}{3}  \left[1 - \frac{5}{36}\left( \frac{4\pi}{3}\frac{GM}{xc^2}\right)^2  \right] = \theta_2  \left[1-\frac{5}{36}\left( \frac{4\pi}{3}\frac{GM}{xc^2}\right)^2  \right] .
	\end{split}
	\end{equation}
Again we can easily see how, with the passage of time and so with respect to the $\tau_m$ values, the mean value $\langle \theta \rangle(\tau_m)$ increases its delay in ticking by a factor which depends on the intensity of the gravitational potential.

\subsection*{7.B \: Conditional Probabilieties and Main Values of the Observables for Clocks with Continuous Time Values}
We replicate here (more briefly) the discussion given in Appendix 7.A but considering the case of clocks $A$ and $B$ with continuous time values. The formalism we adopt is therefore the one described in Section 7.3.

We start considering the case of $A$ and $B$ as free clocks and the global state satisfying (\ref{7wdw}) is given by: 
\begin{equation}
\begin{split}
\ket{\Psi} &= \frac{1}{T_C} \int_{0}^{T_C} dt \ket{t}_C \otimes \ket{\psi(t)}_S =
\\& = \frac{1}{T_C} \int_{0}^{T_C} dt \ket{t}_C \otimes \ket{\varphi(t)}_A\otimes \ket{\phi(t)}_B .
\end{split}
\end{equation}
We assume that the clock in the $C$ subspace reads time $t=\tau_f$ and we work with the state of $S$: $\ket{\psi(t=\tau_f)}_S=\braket{t=\tau_f|\Psi}= \ket{\varphi(t=\tau_f)}_A\otimes \ket{\phi(t=\tau_f)}_B$. We search the conditional probability density $P(\theta_g| \tau_f)$ of obtaining $\theta_g$ on $B$ conditioned to having found $\tau_f$ on $A$, and the mean value of $\langle \theta \rangle(\tau_f)$. The probability $P(\theta_g | \tau_f)$ can be derived as: 
\begin{equation}
P(\theta_g | \tau_f) 
=  \frac{1}{T} \frac{\braket{\psi(t=\tau_f)|\tau_f}\bra{\tau_f} \otimes \ket{\theta_g}\braket{\theta_g|\psi(t=\tau_f)}}{\braket{\psi(t=\tau_f)|\tau_f}\braket{\tau_f|\psi(t=\tau_f)}} 
\end{equation}
which, performing the calculations, leads to the result

\begin{equation}\label{p1}
P(\theta_g | \tau_f) = \frac{1}{Td} \sum_{n=0}^{d-1}\sum_{k=0}^{d-1} e^{-i(\theta_g - \tau_f )\frac{2\pi}{T}(n - k)} = \frac{1}{Td} \sum_{n=0}^{d-1}\sum_{k=0}^{d-1} e^{-i 2\pi (g - f )(n - k)} .
\end{equation}
This probability density is well defined (indeed we have $\int_{0}^{1}dg ~ P(\theta_g | \tau_f) = 1$ for each given $\tau_f$) and it has a maximum for $g=f$ where it take the value $P(\theta_{g=f}|\tau_f)= \frac{d}{T}$.
To beter understand the behavior of $P(\theta_g | \tau_f)$ we can consider the simple case of $d=2$, thus obtaining:
\begin{equation}\label{p2}
P(\theta_g | \tau_f ) = \frac{1}{T} \left[ 1 + \cos(2\pi (g- f))   \right]. 
\end{equation} 
indicating that the probability density $P(\theta_g|\tau_f)$ oscillates around the value $1/T$ reaching a maximum when $g=f$, namely when the clocks click equal time states.

Through the probability density (\ref{p1}), we can calculate the mean value $\langle \theta \rangle(\tau_f)$ obtaining (see Appendix 7.6.C):
\begin{equation}\label{valmedioc}
\langle\theta\rangle (\tau_f) = \int_{0}^{1}dg ~ \theta_g P(\theta_g|\tau_f) = 
\frac{T}{2} + \frac{iT}{2\pi d } \sum_{n \ne k} \frac{e^{i 2\pi f(n-k)}}{n-k} .
\end{equation}
In the simple case of $d=2$, equation (\ref{valmedioc}) becomes:
\begin{equation}\label{main2}
\langle\theta\rangle (\tau_f) 
= \frac{T}{2} \left[1- \frac{1}{\pi}\sin(2\pi f) \right]
\end{equation}
which shows 
$\langle\theta\rangle (\tau_f)$ oscillating around the value $T/2$ as a function of the value $f$.

We can now consider the case in which $B$ is placed within the gravitational field, where the time state of the clocks are given by (\ref{statitaufree3}) and (\ref{tetagrav}) for $A$ and $B$ respectively. 
The probability density (\ref{p1}) becomes:
\begin{equation}
P(\theta_g | \tau_f) = \frac{1}{T' d} \sum_{n=0}^{d-1}\sum_{k=0}^{d-1} e^{-i(\theta_g - \tau_f )\frac{2\pi}{T'}(n - k)} = \frac{1}{T' d} \sum_{n=0}^{d-1}\sum_{k=0}^{d-1} e^{-i 2\pi (g - f(1- \frac{GM}{xc^2}))(n - k)}
\end{equation}
from which we immediately notice that the maximum is obtained for $g=f(1- \frac{GM}{xc^2})$, where the probability takes the value $d/T'$. Looking for the probability that clocks $A$ and $B$ click the same time value, we easily find:
\begin{equation}\label{eq61bis}
P(\theta_{g=f}|\tau_f)= \frac{1}{T'd} \sum_{n=0}^{d-1}\sum_{k=0}^{d-1}  e^{-i2\pi f \frac{GM}{xc^2}(n-k)} .
\end{equation}
From (\ref{eq61bis}), in the case of $d=2$, follows $P(\theta_{g=f}|\tau_f)= \frac{1}{T'}\left[1+\cos(2\pi f \frac{GM}{xc^2})\right]$ 
which, approximating the cosine function as $\cos(z)\simeq 1 - \frac{z^2}{2}$, becomes:
\begin{equation}\label{ultimaspero}
P(\theta_{g=f}|\tau_f)\simeq \frac{2}{T'} \left[ 1 - \pi^2 f^2 \left( \frac{GM}{xc^2}\right)^2 \right] .
\end{equation}
Equation (\ref{ultimaspero}) shows $P(\theta_{g=f}|\tau_f)$ depending on $f$ and on the intensity of the gravitational potential.
Still working with $d=2$, we have for the main value $\langle\theta\rangle (\tau_f)$:
\begin{equation}
\langle\theta\rangle (\tau_f) 
= \frac{T'}{2} \left[1- \frac{1}{\pi}\sin(2\pi f(1- \frac{GM}{xc^2})) \right]
\end{equation}
which differs from (\ref{main2}) because of the presence of the term $(1- \frac{GM}{xc^2})$ multiplying $f$ within the sine function. The function $\langle\theta\rangle (\tau_f)$ is therefore \lq\lq stretched\rq\rq by the action of the gravitational field on clock $B$.

\subsection*{7.6.C \:  Proof of equation (\ref{valmedioc})}
We consider here $\theta,\tau \in \left[0,T\right]$ and we show that the probability density distribution $P(\theta|\tau) = \frac{1}{Td}\sum_{n=0}^{d-1} \sum_{k=0}^{d-1} e^{-i(\theta -\tau)\frac{2\pi}{T}(n-k)}$
leads to the main value:
\begin{equation}\label{dadimostrare}
\langle\theta\rangle (\tau) = \int_{0}^{T}d\theta ~ \theta P(\theta|\tau) = 
\frac{T}{2} + \frac{iT}{2\pi d } \sum_{n \ne k} \frac{e^{i\tau \frac{2\pi}{T}(n-k)}}{n-k} .
\end{equation}
We start considering:
\begin{equation}
\begin{split}
\langle\theta\rangle (\tau) & 
= \int_{0}^{T}d\theta ~ \theta  \frac{1}{Td}\sum_{n=0}^{d-1} \sum_{k=0}^{d-1} e^{-i(\theta -\tau)\frac{2\pi}{T}(n-k)} = \\&
= \frac{1}{Td}\sum_{n=0}^{d-1} \sum_{k=0}^{d-1} e^{i\tau\frac{2\pi}{T}(n-k)} \int_{0}^{T}d\theta ~ \theta e^{-i\theta \frac{2\pi}{T}(n-k)}
\end{split} 
\end{equation}
and we divide the summation into two parts, thus obtaining

\begin{equation}\label{7ggg}
	\langle\theta\rangle (\tau) = \frac{1}{Td} \sum_{n=k} \int_{0}^{T}d\theta ~ \theta +  \frac{1}{Td} \sum_{n \ne k} e^{i\tau\frac{2\pi}{T}(n-k)} \int_{0}^{T}d\theta ~ \theta e^{-i\theta \frac{2\pi}{T}(n-k)}. 
\end{equation}
The integral in the second term on the right-hand side of (\ref{7ggg}) is

\begin{equation}\label{ggg3}
\int_{0}^{T}d\theta ~ \theta e^{-i\theta \frac{2\pi}{T}(n-k)} = \left.     \frac{e^{-i\theta \frac{2\pi}{T}(n-k)} \left(  1 + i \frac{2\pi}{T}(n-k)\theta  \right)}{\left( \frac{2\pi}{T}\right)^2 (n-k)^2}          \vphantom{\big|}\right|_{0}^{T} 
\end{equation}
and, through (\ref{ggg3}), we easily obtain:

\begin{equation}\label{ggg2}
\langle\theta\rangle (\tau) = \frac{1}{d} \sum_{n=k} \frac{T}{2} + \frac{iT}{2\pi d } \sum_{n \ne k} \frac{e^{i\tau \frac{2\pi}{T}(n-k)}}{n-k} = \frac{T}{2} + \frac{iT}{2\pi d } \sum_{n \ne k} \frac{e^{i\tau \frac{2\pi}{T}(n-k)}}{n-k}
\end{equation}
which is what we needed to prove, being (\ref{ggg2}) the same of equation (\ref{dadimostrare}).

\begin{quote}
\label{bianca2}

\end{quote}
	
	\chapter{Conclusions}
\label{conclusioni}

In this work we have examined, revisited and generalized the Page and Wootters (PaW) theory which was originally introduced in order to describe the emergence of time from quantum correlations \cite{pagewootters,wootters}. Observables in quantum theory are represented by Hermitian operators with the exception of time \cite{pauli,bush} (and of a few more observables, including phases \cite{phase}). In Quantum Mechanics, as in Newtonian physics, time is an absolute ``external'' real valued parameter that flows continuously, independently from the material world. A change of perspective from this abstract Newtonian concept was introduced first in the theory of Relativity. Here time is an ``internal'' degree of freedom of the theory itself, operationally defined by  ``what it is shown on a clock'', with the clock being a wisely chosen physical system \cite{rovelli}. In the PaW theory this operational approach is somewhat recovered also for quantum theory. Indeed, motivated by ``the problem of time'' in canonical quantization of gravity \cite{dewitt,isham} and considering a quantum Universe in a stationary global state satisfying the Wheeler-DeWitt equation $\hat{H}\ket{\Psi} = 0$, PaW show that dynamics can be considered as an emergent property of entangled subsystems, with the clock provided by a quantum spin rotating under the action of an applied magnetic field.

In Chapter 2 we provided a brief overview of PaW theory, starting from the original work of Page and Wootters. We showed how the initial criticisms of the theory were overcome, we briefly illustrated the first experimental implementation of the mechanism and we focused on the work of GLM \cite{lloydmaccone} who first introduced a clock observabe conjugated to the clock Hamiltonian by considering the clock space isomorphic to the Hilber space of a particle on a line. With the intention of generalizing as much as possible the set of physical systems that can be used as clock observable, in Chapter 3 we presented the work of D. T. Pegg in which he explores the possibility of the existence of a quantity that can be regarded as the complement of the Hamiltonian also for a generic quantum system with discrete energy levels and no restrictions \cite{pegg}. We have seen that such a quantity (\lq\lq Age\rq\rq) exists and it is described by a POVM in the general case and by an Hermitian operator when considering an Hamiltonian with equally-spaced energy levels.

We incorporated Pegg's formalism in the PaW theory in Chaprter 4 \cite{nostro}. In this way we showed that it is possible to extend the PaW framework in order to consider any Hamiltonian with a discrete, bounded spectrum as clock and we have demonstrated that even if the time states are not fully distinguishable, the rest of the Universe still evolves with respect to the clock time according to the Schrödinger equation. Furthermore, we have recovered a continuous flow of time still maintaining a discrete, bounded clock Hamiltonian and we have considered also how the case of interacting clock and system \cite{interacting} applies in our framework. Furthermore we showed how, considering the particular case of equally-spaced energy values for the clock Hamiltonian, the operator Age (overcoming the Pauli objection) can be considered as a proper time operator for the rest of the Universe. 

In Chapter 5 we merged Canonical Typicality and the PaW quantum time. Considering indeed our closed quantum Universe as consisting by a small subsystem weakly interacting with a large environment, we know that for almost all the pure states in which the whole Universe can be, by tracing over the environment, the small subsystem is found in a state of equilibrium described by the canonical distribution \cite{canonical,popescu1}. At the same time, through the PaW mechanism, we have a Schrödinger-like evolution (even if corrected by a non-local term) for the relative state of the small subsystem with respect to environment which serves as a clock. These two approaches seem contradictory but can coexist considering that, in this mixed framework the action of tracing out the environment is equivalent to tracing over all times. So the novelty of considering the environment as clock is exactly this: the trace over the environment coincides with a temporal average. Still, by calculating the conditioned state of the small subsystem and thus looking at a single time, we find a pure state evolving with a time non-local Schrödinger-like equation \cite{nostro2}. We emphasize here that the crucial point, for this mixed framework to be implemented, is clearly to recognize the environment as a clock. For this reason what was shown in Chapter 4 (namely, the possibility of using every generic bounded Hamiltonian with discrete spectrum as clock Hamiltonian) is essential.
 
In Chapter 6, we intoduced the spatial degree of freedom in the discussion and we provided a description of a model of non-relativistic quantum spacetime. We started giving our own version of the PaW mechanism for space adopting and generalizing Pegg's time approach \cite{pegg}. We showed that, in a closed quantum system satisfying a global constraint on total momentum (and therefore with the absolute position totally indeterminate), a well-defined relative position emerges between entangled subsystems, where one of the two subsystems is taken as quantum spatial reference frame. In doing this we considered non-degenerate, discrete spectra for the momentum operators and we introduce POVMs in describing the spatial degrees of freedom. In this way we recovered continuous space values still assuming discrete momentum spectra (the case of momentum with continuous spectrum was then also treated). Then we introduced in the Universe an additional subsystem acting as a clock and we considered the Universe satisfying a double constraint: both on total momentum and total energy. We showed how this framework can be implemented without contradiction in the simple case of one spatial degree of freedom (considering also the case of multiple time measurements) and in the \lq\lq more physical\rq\rq case of three spatial degrees of freedom thus providing a $3+1$ dimensional quantum spacetime emerging from entanglement \cite{nostro3}. The dynamics of relativistic particles with respect to a non-relativistic, quantum reference frame is also considered. 

Finally, in Chapter 7, we investigated the behavior of quantum clocks when interacting with a Newtonian gravitational potential produced by a non-rotating, spherical mass, still adopting the PaW framework. As a quantum phenomenon, we found a time dilation effect for the clocks in agreement with the first order expansion of the gravitational time dilation obtained from the Schwarzschild metric in the theory of General Relativity. We have also discussed the possibility of using the relativistic gravitational potential within the global Hamiltonian and we have shown that this choice leads to the agreement between our time dilation effect and the exact Schwarzschild solution. In doing this we considered the case of quantum clocks with discrete time values and with continuous time values.

We now want to focus on the importance of carrying out further research within this framework. A first direction to explore certainly concerns the possibility to provide a consistent description of quantum spacetime through entanglement in the case of relativistic reference frames. Clearly, in order to give a complete relativistic generalization of our model, in addition to our discussion of Chapter 6, we need to consider relativistic reference frames and a protocol that allows to change the point of view between different observers in different reference frames, so that dilation of times and contraction of lengths can be derived (a first example considering time dilation alone is given in \cite{timedilation}). Moreover, as mentioned several times, in the quantum gravity literature, it has been suggested that quantum reference frames are needed to formulate a quantum theory of gravity \cite{dewitt,afundamental,QG1,QG2} and we hope that our work may turn out to be a small step in this direction. We note here that also in other fields of physical research (i.e. String Theory and Black Holes physics) several and significant articles suggest that entanglement may be at the basis of the emergence of spacetime (see for example \cite{mvan,ste} and related works).

Thus, from our perspective, a fascinating scenario is emerging: the possibility that time, space (with a resulting quantum theory of gravity), and many of the thermodynamic laws can all have a common origin, that is related to the most intriguing mystery of quantum reality. A deep understanding is not well established yet, but we believe that the first step in this direction has been taken.

	


\chapter*{Acknowledgements}
\label{thanks}

First of all I would like to thank Prof. Augusto Smerzi for making this work possible. I thank him especially for helping and following me with generosity, ensuring that my formation years were a great opportunity for me to learn about new and beautiful things, discovering what it means to be a physicist. The concepts of space and time are the things that, for as long as I can remember, have fascinated me the most and being able to write on this topic is a great privilege for me. 
Together with him I want to thank Prof. Francesco Tafuri for supporting me during the PhD years and Guido Celentano for helping me (several times) in so many practical aspects.

I would like to thank Lorenzo Bartolini and Marco Marinucci for useful discussions in the writing of my articles during these years. I thank Lorenzo Spina for always supporting me, and I thank Matilde, Daniele, Matteo and Marco.

I would like to sincerely thank my family: my parents, my sisters and all the others. Then, the most important, I thank my wife Chiara who has dedicated much of her time always supporting me. She has practically helped me in several issues and she has always encouraged me not to lose hope when many circumstances seemed against me. This work is for a good percentage her merit. Thank you Chiara!

Finally, as I write these thanks, I realize that it would be impossible for me to thank and mention all those friends, close or distant, who with their presence are able to accompany me in life, supporting me in my daily journey and constantly showing me how to live authentically as a men. So I thank all these friends. 

	\relax 
\thispagestyle{empty}
\begin{quote}
\label{bianca}

\end{quote}
\end{document}